\documentclass[%
aps,
prd,
amssymb,
amsmath,
amsfonts,
eqsecnum,
showpacs,
nofootinbib,
floatfix,
twocolumn,
twoside,
a4paper,
superscriptaddress
]{revtex4-1}

\usepackage[T3,T1]{fontenc}
\usepackage{mathrsfs} % For \mathscr fonts
\usepackage{bm} % For \bm fonts
\makeatletter
\@ifclassloaded{beamer}
  { % if this is a beamer document, use sans-serif fonts
    \typeout{UsePackages: Detected beamer}
    \usepackage{tgheros}
    
  }
  { % otherwise:
    \typeout{UsePackages: Did not detect beamer}
    \ifx\asybeamer\undefined % use times for articles
    \typeout{UsePackages: Detected article}
    \usepackage[varg]{txfonts} % For math
    \usepackage{tgtermes} % For text
    \else % use beamer fonts for asy documents with beamer
    \typeout{Fonts: Detected asy for beamer}
    \usepackage{tgheros}
    
    \fi
  }
\usepackage{microtype} % For nicer alignment of text
\def\MT@register@subst@font{
  \MT@exp@one@n\MT@in@clist\font@name\MT@font@list
  \ifMT@inlist@\else\xdef\MT@font@list{\MT@font@list\font@name,}\fi}
\makeatother

\usepackage{graphicx} %
\usepackage[x11names,svgnames,rgb]{xcolor} %
\usepackage{xspace}
\usepackage{braket}
\usepackage{accents}
\usepackage{siunitx}
\usepackage{mathtools}
\DeclareSymbolFontAlphabet{\mathrm}{operators}

\definecolor{CiteColor}{rgb}{0.18039, 0.18824, 0.57255}
\definecolor{UrlColor} {rgb}{0.741, 0.173, 0.000}
\definecolor{DarkUrlColor} {rgb}{0.500, 0.110, 0.000}
\definecolor{LinkColor}{rgb}{0.25098, 0.47843, 0.04706}
\makeatletter %
\newcommand{\ShowFont}{%
  \typeout{The main font is \f@encoding \space \f@family \space %
    \f@series \space \f@shape \space at \f@size pt.}%
  \typeout{The math font sizes are \tf@size pt (main), \sf@size pt %
    (script), and \ssf@size pt (scriptscript).}%
  \typeout{The linewidth is \the\linewidth}} %
\makeatother %
\usepackage{xargs}

\usepackage{graphicx}% Include figure files
\usepackage{dcolumn}% Align table columns on decimal point
\usepackage{bm}% bold math
\usepackage{tabularx}
\usepackage{multirow}
\usepackage{capt-of}
\usepackage{color}
\usepackage{url}
\usepackage[utf8]{inputenc}

\usepackage[nolist,nohyperlinks]{acronym}
\usepackage{xspace}
\usepackage[colorlinks]{hyperref}
\usepackage[normalem]{ulem}

\usepackage[caption=false]{subfig}

\DeclareMathAlphabet{\mathbfsf}{\encodingdefault}{\sfdefault}{bx}{sl}

\newcommand{\mpr}{m^{\prime}}

\newcommand{\nl}{\newline}

\newcommand{\be}{\begin{equation}}
\newcommand{\ee}{\end{equation}}
\newcommand{\bea}{\begin{eqnarray}}
\newcommand{\eea}{\end{eqnarray}}

\newcommand{\x}{\mathbf{\hat{x}}}
\newcommand{\y}{\mathbf{\hat{y}}}
\newcommand{\z}{\mathbf{\hat{z}}}

\newcommand{\hN}{{\hat{\bf{N}}}}
\newcommand{\hP}{{\hat{\bf{P}}}}
\newcommand{\hQ}{{\hat{\bf{Q}}}}

\newcommand{\hJ}{{\hat{\bf{J}}}}

\newcommand{\hLO}{{\hat{\bf{L}}}_0}
\newcommand{\hL}{{\hat{\bf{L}}}}

\newcommand{\FT}{{\rm{FT}}}
\newcommand{\IFT}{{\rm{IFT}}}

\newcommand{\twopn}{2 \rm{PN}}

\newcommand{\bfJ}{\boldsymbol{J} \xspace}
\newcommand{\bfL}{\boldsymbol{L} \xspace}
\newcommand{\bfSone}{\boldsymbol{S}_1 \xspace}
\newcommand{\bfStwo}{\boldsymbol{S}_2 \xspace}

\newcommand{\bfell}{\boldsymbol{\ell} \xspace}
\newcommand{\bflambda}{\boldsymbol{\lambda} \xspace}
\newcommand{\bfn}{\boldsymbol{n} \xspace}

\newcommand{\bfv}{\boldsymbol{v}}

\newcommand{\bfr}{\mathbf{{r}}}

\newcommand{\J}{\mathbf{J}}

\newcommand{\LO}{\mathbf{L_0}}
\newcommand{\BL}{\mathbf{L}}

\newcommand{\ppvtwo}{\textsc{IMRPhenomPv2}\xspace}
\newcommand{\phD}{\textsc{IMRPhenomD}\xspace}
\newcommand{\phX}{\textsc{IMRPhenomXAS}\xspace}
\newcommand{\phXF}{\textsc{IMRPhenomX}\xspace}
\newcommand{\phP}{\textsc{IMRPhenomP}\xspace}
\newcommand{\phHM}{\textsc{IMRPhenomHM}\xspace}
\newcommand{\phPvtwo}{\textsc{IMRPhenomPv2}\xspace}
\newcommand{\phPvthree}{\textsc{IMRPhenomPv3}\xspace}
\newcommand{\phPvthreehm}{\textsc{IMRPhenomPv3HM}\xspace}
\newcommand{\phXHM}{\textsc{IMRPhenomXHM}\xspace}
\newcommand{\phXP}{\textsc{IMRPhenomXP}\xspace}
\newcommand{\phXPHM}{\textsc{IMRPhenomXPHM}\xspace}
\newcommand{\phT}{\textsc{IMRPhenomTP}\xspace}
\newcommand{\phTP}{\textsc{IMRPhenomTP}\xspace}
\newcommand{\NRHybSur}{\textsc{NRHybSur3dq8}\xspace}
\newcommand{\NRSur}{\textsc{NRSur7dq4}\xspace}

\newcommand{\seobnrvforhm}{\textsc{SEOBNRv4HM}\xspace}
\newcommand{\seobnrvforhmrom}{\textsc{SEOBNRv4HM\_ROM}\xspace}
\newcommand{\seobnrvforphm}{\textsc{SEOBNRv4PHM}\xspace}
\newcommand{\ph}{\textsc{Phenom}\xspace}
\newcommand{\seobnrvforp}{\textsc{SEOBNRv4P}\xspace}

\newcommand{\botwo}{\frac{\beta}{2}}

\definecolor{dodgerblue}{HTML}{1E90FF}
\definecolor{viennared}{HTML}{DA0A14}
\definecolor{ctorange}{HTML}{FF6C0C}
\definecolor{granadagreen}{HTML}{078931}
\definecolor{wales}{HTML}{ff0038}
\definecolor{valenciacfred}{HTML}{ee3524}
\definecolor{barcelonafcgold}{HTML}{edbb00}
\definecolor{jam}{HTML}{A50B5E}
\definecolor{austriawien}{HTML}{441678}

\AtBeginDocument{%
  \hypersetup{
    citecolor=dodgerblue,
    linkcolor=dodgerblue,   
    urlcolor=dodgerblue}}

\newcommand{\UIB}{Departament de F\'isica, Universitat de les Illes Balears, IAC3 -- IEEC, Crta. Valldemossa km 7.5, E-07122 Palma, Spain}

\newcommand{\UoB}{School of Physics and Astronomy and Institute for Gravitational Wave Astronomy, University of Birmingham, Edgbaston, Birmingham, B15 9TT, United Kingdom}

\newcommand{\Cardiff}{School of Physics and Astronomy, Cardiff University, Queens Buildings, Cardiff, CF24 3AA, United Kingdom}

\newcommand{\Zurich}{Physik-Institut, Universit{\"a}t Zurich, Winterthurerstrasse 190, Z{\"u}rich Switzerland}

\allowdisplaybreaks[1]

\begin{document}

% ~~~~~~~~~~ Title & Abstract ~~~~~~~~~~ %
\title[IMRPhenomXP and IMRPhenomXPHM: the fast and accurate models]
{Computationally efficient models 
for the dominant and sub-dominant harmonic modes of precessing binary black holes}

% \makeatletter \@booleantrue\frontmatterverbose@sw \makeatother

\author{Geraint Pratten}
\affiliation{\UoB}\affiliation{\UIB}

\author{Cecilio Garc{\'i}a-Quir{\'o}s}
\affiliation{\UIB}

\author{Marta Colleoni}
\affiliation{\UIB}

\author{Antoni Ramos-Buades}
\affiliation{\UIB}

\author{H\'{e}ctor Estell\'{e}s}
\affiliation{\UIB}

\author{Maite Mateu-Lucena}
\affiliation{\UIB}

\author{Rafel Jaume}
\affiliation{\UIB}

\author{Maria Haney}
\affiliation{\Zurich}

\author{David Keitel}
\affiliation{\UIB}

\author{Jonathan E. Thompson}
\affiliation{\Cardiff}

\author{Sascha Husa}
\affiliation{\UIB}

\date{\today}

\begin{abstract}
We present \phXPHM, a phenomenological frequency-domain model for the gravitational-wave signal emitted by quasi-circular precessing binary black holes, which incorporates multipoles beyond the dominant quadrupole in the precessing frame. The model is a precessing extension of \phXHM, based on approximate maps between aligned-spin waveform modes in the co-precessing frame and precessing waveform modes in the inertial frame, which is commonly referred to as ``twisting up'' the non-precessing waveforms. \phXPHM includes 
\phXP as a special case, the restriction to the dominant quadrupole contribution in the co-precessing frame.
We implement two alternative mappings, one based on a single-spin PN approximation, as used in \ppvtwo \cite{Hannam:2013oca}, and one based on the double-spin MSA approach of \cite{Chatziioannou:2017tdw}.
We include a detailed discussion of conventions used in the description of precessing binaries and of all choices made in constructing the model.
The computational cost of \phXPHM is further reduced by extending the interpolation technique of \cite{Garcia-Quiros:2020qlt} to the Euler angles.
The accuracy, speed, robustness and modularity of the \phXF family will make these models productive tools for gravitational wave astronomy in the current era of greatly increased number and diversity of detected events.
\end{abstract}


\maketitle

%%%%%%%%%%%%%%%%%%%%%%%%%%%%%%%%%%%%%%%%%%%%%%%%%%%%%%%%%%%%%%%%%
%               INTRODUCTION
%%%%%%%%%%%%%%%%%%%%%%%%%%%%%%%%%%%%%%%%%%%%%%%%%%%%%%%%%%%%%%%%%

\section{Introduction}
\label{sec:Introduction}

We have recently presented \phX \cite{Pratten:2020fqn},
a phenomenological model for the $\ell=\vert m\vert = 2$ dominant quadrupole spherical harmonic modes of the gravitational wave signal emitted by coalescing black holes in quasi-circular orbits, and with spin vectors orthogonal to the orbital plane. This model improves over the \phD model \cite{Husa:2015iqa,Khan:2015jqa} that is routinely used in gravitational wave data analysis. The improvements include modifications of the phenomenological ansatz, a systematic approach to modelling the dependence of phenomenological parameters on the three-dimensional parameter space of non-precessing quasi-circular binaries of black holes \cite{Jimenez-Forteza:2016oae,Keitel:2016krm}, extending the set of numerical relativity waveforms our model is calibrated to from 19 to 461, incorporating additional numerical perturbative waveforms for mass ratios up to 1000 into the calibration data set, 
and calibrating to a more accurate description of the inspiral \cite{Bohe:2016gbl}.

Building on \phX, we have also presented \phXHM \cite{Garcia-Quiros:2020qpx}, which extends the model to the leading subdominant harmonics, in particular the $(\ell, \vert m\vert) = (2,2), (2,1), (3,3), (3,2), (4,4)$ modes, and includes mode mixing effects in the $\ell=3, \; \vert m \vert = 2$ harmonics as described in \cite{Garcia-Quiros:2020qpx}.
This extension is aimed to supersede the \phHM model
\cite{London:2017bcn}, where the subdominant harmonics are not calibrated to numerical relativity waveforms, and instead an approximate map from the $(2,2)$ to the sub-dominant harmonics is employed.
 
These models are formulated in the frequency domain, which is typically employed in matched filter calculations, in order to reduce the computational cost of gravitational wave data analysis.
In order to accelerate the evaluation of the waveform model, which is particularly important for computationally expensive applications such as Bayesian inference \cite{Veitch:2014wba,Ashton:2018jfp}, we have 
further developed the multibanding interpolation method of \cite{Vinciguerra:2017ngf}
as described in \cite{Garcia-Quiros:2020qlt}.

Phenomenological waveform models for non-precessing systems have been extended to precessing systems \cite{Hannam:2013oca,Khan:2018fmp,Khan:2019kot} by a construction that is based on an
approximate map between precessing and non-precessing systems, and is commonly referred to as ``twisting up'' \cite{Schmidt:2010it,Schmidt:2012rh,Hannam:2013oca}.  
The aim of the present paper is to revisit the twisting-up procedure, first by documenting it in detail and deriving the equations that define the model in the frequency domain, and then to extend \phX and \phXHM to precession, resulting in the \phXP and \phXPHM models, which are publicly available as implemented in the LALSuite \cite{lalsuite} library for gravitational wave data analysis.

Approximate maps between the gravitational wave signals of precessing and non-precessing systems can be constructed based on the fact that the orbital timescale is much smaller than the precession timescale, and correspondingly the amount of gravitational waves emitted due to the precessing motion is relatively small and contributes little to the phasing of the gravitational wave signal when observed in a non-inertial co-precessing frame.
Rather, the dominant effect of precession is an amplitude and phase modulation that can be approximated in terms of a time-dependent rotation of a non-precessing system \cite{Schmidt:2010it,Schmidt:2012rh}.

We will describe this rotation in terms of three time-dependent Euler angles, and our non-precessing gravitational wave signal will be described by the \phXHM model (or \phX for the dominant quadrupole modes).
The waveform for precessing binaries can thus be approximated by interpreting a non-precessing waveform as an approximation to the precessing waveform observed in a non-inertial frame that tracks the precession of the orbital plane \cite{Schmidt:2010it}.
This map is greatly simplified by the approximate decoupling between the spin components parallel and perpendicular to the orbital angular momentum $\mathbf{L}$ \cite{Schmidt:2012rh}. See however \cite{Gerosa:2015hba} for an instability for approximately opposite spins that can result in breaking this assumption in a small part of the parameter space.

In addition to the time-dependent rotation, the approximate map also requires a second element, which is to modify the final spin of the merger remnant, which is in general different from the non-precessing case, essentially due to the vector addition of the individual spins and angular momentum.
The final mass of the remnant is much less affected by precession, since the scalar quantity of radiated energy is not significantly affected by the precessing motion due to its slower time scale compared to the orbital motion.

An important shortcoming of this construction as presented here is that it does not include the asymmetries in the $(\ell, \vert m\vert) = (2,2)$ modes that are responsible for large recoils, see e.g.~\cite{Brugmann:2007zj}. For brevity we will refer to the approximations that are used in the ``twisting up'' procedure as the ``twisting approximation''.
For a recent detailed discussion of the effect of these approximations, with special consideration of the effect on sub-dominant harmonics, see \cite{Ramos-Buades:2020noq,Thomas:2020uqj}.
 
Our model currently uses two alternative descriptions for the Euler angles that characterize the approximate map: the one used previously in \cite{Hannam:2013oca,Bohe:PPv2} assumes that the spin of the smaller black hole vanishes, while the one developed in \cite{Chatziioannou:2017tdw} and previously used in \cite{Khan:2018fmp,Khan:2019kot} describes double-spin systems.
The code we have developed as part of LALSuite \cite{lalsuite} is modular, and allows to independently update different components, such as the calibrations of particular regions (inspiral, merger, or ringdown) for particular spherical harmonics, or the precession Euler angles, and supports calling particular versions of these components.

In a previous study of waveform systematics \cite{Abbott:2016wiq} it was found that while models such as \phD and \ppvtwo were sufficiently accurate for the first detection of gravitational waves \cite{Abbott:2016blz}, further improvements in accuracy were called for.
The \phXF family of waveform models addresses this, and the present work completes the \phXF family of waveform models to serve as a tool for gravitational wave data analysis that models quasi-circular systems, and to serve as a basis for extensions: e.g.~to address eccentricity and model fully generic mergers of black holes in general relativity, to address remaining shortcomings in describing quasi-circular systems, and as a basis for tests of general relativity.

The paper is organized as follows:
We first discuss our notation and conventions
in Sec.~\ref{sec:conventions} and the basic concepts of the modelling of precessing binaries in Sec.~\ref{sec:basics}. We then present the
construction of the model in Sec. \ref{sec:Model} 
and our tests of quality and computational efficiency in Sec.~\ref{sec:validation}.
This also includes Bayesian inference results with the new model on real gravitational wave data.
We conclude the paper in Sec.~\ref{sec:conclusions}.

Several appendices provide further technical details:
In appendix \ref{appendix:wignerd} we list the Wigner-d matrices we use to express rotations.
In appendix \ref{sec:conventions_no_prec}
we summarize conventions regarding non-precessing waveforms.
In appendix \ref{appendix:frames} we discuss frame transformations and the effect in the gravitational wave polarizations.
In appendix \ref{appendix:polarizations}
we discuss how our choice of polarization relates to other choices in the literature.
In appendix \ref{appendix:derivation} we spell out the derivation of the frequency domain gravitational waveform.
Appendix \ref{appendix:lal_implementation} contains details of the LALSuite implementation.
In appendix \ref{appendix:nnlo_angles} we write out the explicit post-Newtonian expressions for the next-to-next-to-leading order (NNLO) Euler-angle descriptions that we use here.
Finally in appendix \ref{app:orb_ang_mom} we write out the coefficients of the post-Newtonian approximation we use for the orbital angular momentum.

% Define conventions used
We define the mass ratio $q = m_1/m_2 \geq 1$, total mass $M=m_1+m_2$, and symmetric mass ratio $\eta = m_1 m_2 /M^2$. We use geometric units $G=c=1$ unless explicitly stated (in particular when using seconds, Hz or solar masses as units).

%%%%%%%%%%%%%%%%%%%%%%%%%%%%%%%%%%%%%%%%%%%%%%%%%%%%%%%%%%%%%%%%%
%               NOTATION AND CONVENTIONS
%%%%%%%%%%%%%%%%%%%%%%%%%%%%%%%%%%%%%%%%%%%%%%%%%%%%%%%%%%%%%%%%%

\section{Notation and Conventions}
\label{sec:conventions}

For non-precessing systems we have recently provided a detailed discussion of our conventions in \cite{Pratten:2020fqn,Garcia-Quiros:2020qpx}. Our work here is consistent with these conventions, but we drop the restriction to spins orthogonal to the orbital plane.
As the twisting construction is based on mapping non-precessing waveforms to precessing ones, the properties of non-precessing waveforms, in particular the consequences of equatorial symmetry with respect to the orbital plane, are still relevant for the map, and we summarize these conventions in appendix \ref{sec:conventions_no_prec}.

As our primary coordinate system we use a standard inertial spherical coordinate system $(t, r,\theta,\varphi)$, where $t$ is the inertial time coordinate of distant observers, $r$ is the luminosity distance to the source, and $\theta$ and $\varphi$ are polar angles in the sky of the source.
Associated with this spherical coordinate system will be a Cartesian coordinate system with axes $(\x_J , \y_J , \z_J)$. We will take the $\z_J$ axis to be the direction of the total angular momentum $\J$, and we will refer to this inertial coordinate system as the $J$-frame. In most binaries, the orbital and spin angular momenta will precess around the $\J$ \cite{Apostolatos:1994mx,Kidder:1995zr}. Here we will take the direction of $\J$ to be fixed, i.e. $\hat{\J} (t) \simeq \hat{\J}_{t \rightarrow - \infty}$. This is a limitation of the model and excludes special cases, such as transitional precession, where there is no fixed precession axis and the direction of $\J$ will evolve. 

Our final result will be the calculation of the observed gravitational wave polarizations in a frame where the $z$-axis corresponds to the direction $\hN$ of the line of sight toward the observer, which we will refer to as the $N$-frame. The observer of the gravitational wave signal will be located at the sky position $\theta = \theta_{JN}$ and $\varphi = \phi_{JN}$ in the $J$-frame.

We will use a third coordinate system to describe precession in terms of a rotating orbital plane, which is orthogonal to the Newtownian orbital angular momentum
$ \mathbf{L_N} =  \mu \, \mathbf{n} \times  \mathbf{v}$,
where $\mu$ is the Newtonian reduced mass, $\mathbf{n}$ the vector from the position of the secondary black hole to the primary, and $\mathbf{v}$ the relative velocity. In the presence of spin precession, the direction of the actual orbital angular momentum $\mathbf{L}$ will in general differ from the direction of 
$\mathbf{L_N}$ due to the presence of spin components in the orbital plane, orthogonal to $\mathbf{L_N}$, see e.g.~the discussions in \cite{Schmidt:2010it} related to Eq.~(4.6) of that paper. These corrections enter $\mathbf{L}$ at the first post-Newtonian order and modulate the rotation of the orbital axis. In our present implementation of the twisting-up approximation, we will neglect the influence of this effect on the final waveform, as has been done in previous implementations \cite{Hannam:2013oca,Bohe:PPv2,Khan:2018fmp,Khan:2019kot}.
We will refer to a coordinate system where the $z$-axis is chosen as 
$\mathbf{L}$ or $\mathbf{L_N}$ as the $L$-frame, and will discuss different choices for approximating $\mathbf{L}$ in Sec. \ref{sec:PN_L}.

When setting up initial data for numerical relativity simulations, it is common to 
choose spin components for the initial data set in the $L$-frame, where approximations for $\mathbf{L}$ may or may not be applied.
We will refer to the inertial coordinate frame, which corresponds to the $L$-frame at some initial reference time as the $L_0$-frame.

Our setup in this paper is constructed to be consistent with \cite{Schmidt:2017btt}, which discusses conventions for relating the 
$N$-frame (referred to as the wave frame) and the $L_0$-frame (referred to as the source frame), which have been adopted by the LALSuite \cite{lalsuite} framework for gravitational wave data analysis, where we have implemented our model as open source code.
Appendix \ref{appendix:frames} discusses how we use the remaining freedom
to fix the $J$, $N$, and $L$ (or equivalently $L_0$) frames, which corresponds to fixing the freedom of rotating around the $z$-axes of each frame, and to the three Euler angles that rotate a given coordinate frame into another.
%
%The same appendix also discusses the transformation between two alternative parameterizations of the spins of the two black holes, which are both used within LALSuite: Bayesian parameter estimation in LALInference~\cite{Veitch:2014wba} uses a geometric description of the spins in terms of scalar products and angles.
%But waveform models are implemented as part of the LALSimulation module, with an interface using Cartesian components of the spin vectors in the $L$-frame.

We will perform the ``twisting up'' construction of the gravitational-wave signal in terms of its decomposition into spin-weighted spherical harmonics in the $J$-frame, \cite{Thorne:1980ru}
\begin{align}\label{eq:def_polarizations_TD}
    h^J = h_{+}^J - i h_{\times}^J &= \displaystyle\sum_{\ell \geq 2} \displaystyle\sum_{m = -\ell}^{\ell} \; h_{\ell m}^J \; _{-2}Y_{\ell m}(\theta,\varphi) ,
\end{align}
where
\begin{align}
    _{-2}Y_{\ell m} (\theta, \varphi) &= \mathcal{Y}_{\ell m} \, (\theta) e^{i m \varphi}
\end{align}
 are the spin-weighted spherical harmonics of spin-weight $-2$ \cite{Goldberg:1966uu}, defined as in \cite{Wiaux:2005fm}.

%Following \cite{Goldberg:1966uu}, the Wigner $d$-matrices are defined to be
%\cg{The second factor in the sqrt had $m^{\prime}$. But this was not in agreement with XLALWignerdMatrix. In LAL they say they follow \url{http://en.wikipedia.org/wiki/Wigner_D-matrix#Wigner_.28small.29_d-matrix}. If that is true, the IMRPhenomPv2 technical notes have a typo in the second factor of the denominator.} \gp{There are a lot of different conventions. I am not convinced that PPv3 is self consistently correct. The Technical document adopts the Arun et al conventions for SWSH's which should be correct. I strongly advise that we ignore what others have done and make sure we are self consistent with our own implementation - the state of the implementations is a real mess to be honest.}
%\begin{widetext}
%\begin{align}
%    d_{m m^{\prime}}^{\ell}(\beta) &= \sum_{k=\rm{max}(0,m-m^{\prime})}^{\rm{min}(\ell+m,\ell-m^{\prime})} \frac{(-1)^{k}}{k !} \frac{\sqrt{(l+m) !\left(l-m\right) !\left(l+m^{\prime}\right) !\left(l-m^{\prime}\right) !}}{(l+m-k) !\left(l-m^{\prime}-k\right) !\left(k-m+m^{\prime}\right) !}\left(\cos \frac{\beta}{2}\right)^{2 \ell+m-m^{\prime}-2 k}\left(\sin \frac{\beta}{2}\right)^{2 k-m+m^{\prime}} .
%\end{align}
%\end{widetext}
%\newline

We adopt the LALSuite conventions for the Fourier transform of a signal $h(t)$ and its inverse
\begin{align}\label{eq:def_FFT}
    \tilde{h} (f) &= \FT \left[ h \right] (f) = \int \; h(t) \; e^{- 2 \pi i f t} dt, \\
    h (t) &= \IFT \left[ \tilde{h} \right] (t) = \int \; \tilde{h}(f) \; e^{2 \pi i f t} df .
\end{align}
With this definition of the Fourier transform we can convert Eq. ~(\ref{eq:def_polarizations_TD}) that defines the two gravitational wave polarizations in terms of the real and imaginary part of the time domain gravitational wave strain to expressions in the frequency domain,
\begin{align}
    \tilde{h}_{+} (f) &= \FT [ \, {\rm{Re}} ( h(t) ) \, ] = \frac{1}{2} \left[ \tilde{h}(f) + \tilde{h}^{\ast} (-f) \right] , \\ 
    \tilde{h}_{\times} (f) &= -\FT[ \, {\rm{Im}} ( h(t) ) \, ] = \frac{i}{2} \left[ \tilde{h}(f) - \tilde{h}^{\ast} (-f) \right] .
\end{align}

%\nl
%Adopting the conventions detailed in \phX, the time domain modes can be written in terms of a positive amplitude $a_{\rm{TD}} (t)$ and a phase $\phi_{\rm{TD}} (t)$ 
%\begin{align}
%    h_{22} (t) &= a_{\rm{TD}} (t) e^{- i \phi_{\rm{TD}} (t)} , \\ h_{2-2} (t) &= a_{\rm{TD}} (t) e^{+ i \phi_{\rm{TD}} (t)} ,
%\end{align}
%\nl 
%where we assume that the phase of the aligned-spin modes is a monotonically increasing function of $t$ 
%\begin{align}
%    \dot{\phi}_{\rm{TD}} (t) > 0.
%\end{align}
%\nl 
%In the frequency domain, \phX models the Fourier amplitude $A(f > 0)$, which is a postivie function of frequency, and a Fourier domain phase $\phi (f > 0)$, defined by
%\begin{align}
%    \tilde{h}_{2-2} (f, \vth ) &= A(f , \vth ) \; e^{- i \phi (f , \vth ) } \;\; {\rm{for}} \;\; f > 0 .
%\end{align}
%\nl 
%The symmetry of the aligned-spin modes implies that  
%\begin{align}
%    \tilde{h}_{22} (f,\vth) &= \left( \tilde{h}_{2-2} (-f, \vth) \right)^{\ast} = A(-f,\vth) e^{+i \phi (-f, \vth)} \;\; {\rm{for}} \;\; f < 0.
%\end{align}

%%%%%%%%%%%%%%%%%%%%%%%%%%%%%%%%%%%%%%%%%%%%%%%%%%%%%%%%%%%%%%%%%
%               MODELLING PRECESSING BINARIES
%%%%%%%%%%%%%%%%%%%%%%%%%%%%%%%%%%%%%%%%%%%%%%%%%%%%%%%%%%%%%%%%%

\section{Modelling Precessing Binaries}
\label{sec:basics}

\subsection{The twisting construction in terms of Euler angles}
\label{sec:frames}

One of the key breakthroughs in the modelling of precessing binaries was the insight that such models can be simplified by formulating them in a non-inertial frame that tracks the approximate motion of the orbital plane, and that the resulting waveform approximately resembles some corresponding aligned-spin waveform \cite{Schmidt:2010it}. In particular, one finds that a mode hierarchy consistent with non-precessing binaries is restored, allowing to define an approximate mapping between the seven-dimensional space of generic precessing binaries and the three-dimensional space of non-precessing binaries \cite{Schmidt:2012rh}. This identification immediately implies that the inverse procedure can be used to approximate the waveform modes of a precessing binary in the inertial frame \cite{Schmidt:2012rh,Hannam:2013oca}, namely to apply a time-dependent rotation to the aligned-spin waveform modes.  

In the conventions adopted in this paper, we define $(\alpha, \beta, \gamma)$ as the Euler angles that describe an active rotation from the inertial $J$-frame to the precessing $L$-frame in the $(z,y,z)$ convention. 
The angles $\alpha$ and $\beta$ are spherical angles that approximately track the direction of the Newtonian angular momentum. The third angle can be gauge-fixed by enforcing the minimal rotation condition \cite{Boyle:2011gg}, demanding the absence of rotation in the precessing frame about the orbital angular momentum\footnote{Note that $\epsilon = - \gamma$ is sometimes used in the literature, e.g. \cite{Bohe:PPv2}.}
\begin{align}\label{eq:minimal_rotation}
    \dot{\gamma} &= - \dot{\alpha} \, \cos \beta .
\end{align}
In the conventions adopted here, $\alpha$ will typically increase during the inspiral, while $\gamma$ will typically decrease. The gravitational-wave modes between these two frames can be related via the transformation of a Weyl scalar under a rotation ${\bf{R}} \in \rm{SO}(3)$ \cite{Goldberg:1966uu,Schmidt:2010it}
\begin{align}
\label{eq:prec_to_inertial}
    h^J_{\ell m} &= \displaystyle\sum_{m^{\prime} = - \ell}^{\ell} \mathcal{D}^{\ell \ast}_{m m^{\prime}} \, (\alpha, \beta, \gamma) \, h^L_{\ell m^{\prime}} , \\
     h^L_{\ell m^{\prime}} &= \displaystyle\sum_{m = - \ell}^{\ell} \mathcal{D}^{\ell}_{m m^{\prime}} \, (\alpha, \beta, \gamma) \, h^J_{\ell m}, 
    \label{eq:inertial_to_prec}
\end{align}
where $\mathcal{D}^{\ell}_{m \mpr}$ are the Wigner D-matrices\footnote{Note that the convention for the Wigner $d$-matrices adopted here implies \cite{Marsat:2018oam,Arun:2008kb}
\begin{align*}
   \mathcal{D}^{\ell}_{m m^{\prime}} (\alpha, \beta, \gamma) &=  \mathcal{D}^{\ell , \rm{ABFO}}_{m^{\prime} m} (\alpha, \beta, \gamma)  .
\end{align*}
}
\begin{align}
\label{eq:wignerD_matrices}
    \mathcal{D}_{m \mpr}^{\ell}(\alpha, \beta, \gamma) &= e^{i m \alpha} e^{i \mpr \gamma} d_{m \mpr}^{\ell}(\beta),
\end{align}
%\nl
and $d^{\ell}_{m \mpr}$ are the real-valued Wigner-$d$ matrices and are polynomial functions in $\cos (\beta / 2)$ and $\sin (\beta / 2)$, as detailed in Appendix~\ref{appendix:wignerd}.  Note that Eq.~(\ref{eq:inertial_to_prec})
follows from inverting Eq.~(\ref{eq:prec_to_inertial}).
We provide a Mathematica \cite{Mathematica} notebook as supplementary material, which allows to conveniently check key conventions, such as those related to the
Wigner-$d$ matrices.

Schematically, we construct precessing waveform models using the following ``twisting'' algorithm:
\begin{itemize}
    \item Model waveform modes in the precessing non-inertial $L$-frame, in our case these models are \phX and \phXHM.
    \item Perform an active rotation from the precessing $L$-frame to the inertial $J$-frame using a given model for the precession dynamics, as encoded in $(\alpha, \beta, \gamma)$. The inertial frame is defined such that $z_J = \J$, where $\J$ is approximately constant, and a full discussion of the relation between different frames and the conventions chosen to represent precessing motion is given in appendix \ref{appendix:frames}. In order to achieve closed form expressions in the Fourier domain, the stationary phase approximation (SPA) is used, with the result stated in the next section  (\ref{sec:FD_polarizations}), and a full derivation deferred to appendix \ref{appendix:derivation}.
 \item Project gravitational-wave polarizations into the $N$-frame
 as discussed in appendix \ref{appendix:frames}. 
 %using the polarization basis $(\hP,\hQ)$ as defined in \cite{Arun:2008kb}.
\end{itemize}

\subsection{Gravitational-Wave Polarizations in the Frequency Domain}\label{sec:FD_polarizations}

The frequency-domain expressions for the gravitational-wave polarizations in the inertial $J$-frame $\tilde{h}_{+,\times}^J (f)$ in terms of spherical harmonic modes $\tilde{h}_{\ell m}^L (f)$ in the co-precessing $L$-frame are derived in Appendix~\ref{appendix:derivation}, starting from Eq.~(\ref{eq:prec_to_inertial}), and performing Fourier transformations with the stationary phase transformation (SPA)
\cite{Finn:1992xs,Cutler:1994ys,Droz:1999qx}.
The result for the gravitational-wave polarizations in terms of modes in the precessing $L$-frame reads
\begin{widetext}
%\begin{subequations}
\begin{align}
\label{eq:polarizations_1}
    \tilde{h}_{+}^{J} (f > 0) &= \frac{1}{2} \displaystyle\sum_{\ell \geq 2} \displaystyle\sum_{m^{\prime} > 0}^{l} \tilde{h}^{L}_{\ell  -m^{\prime}} (f) e^{ i \mpr \gamma} \displaystyle\sum_{m=-\ell}^{\ell}\left[ A^{\ell}_{m \, -\mpr} + (-1)^{\ell} A^{\ell \, \ast}_{m \, \mpr} \right], 
    \\
    \label{eq:polarizations_2}
    \tilde{h}_{\times}^{J} (f > 0) &= \frac{i}{2} \displaystyle\sum_{\ell \geq 2} \displaystyle\sum_{\mpr > 0}^{\ell} \tilde{h}^{L}_{\ell -\mpr } (f) e^{i \: \mpr \gamma} \displaystyle\sum_{m=-\ell}^{\ell} \left[ A^{\ell}_{m - \mpr} - (-1)^{\ell} A^{\ell\: \ast}_{m \mpr} \right],
\end{align}
%\end{subequations}
\end{widetext}
%\nl 
where we have introduced mode-by-mode transfer functions 
\begin{equation}
   A^{\ell}_{m \, \mpr} =  e^{-i \: m \:\alpha} d^{\ell}_{m \mpr}(\beta) \; _{-2}Y_{\ell m}. 
\end{equation}
%\nl 
The modes in the precessing $L$-frame can be approximated with non-precessing waveform modes \cite{Schmidt:2010it,Schmidt:2012rh,Hannam:2013oca}. Here we use \phXHM \cite{Pratten:2020fqn,Garcia-Quiros:2020qpx}, which contains the $(\ell , |m|) = (2,2), (2,1), (3,3), (3,2), (4,4)$ modes.
Note that, as discussed in \cite{Ramos-Buades:2020noq}, our treatment of mode-mixing in the non-precessing case does not strictly carry over to precession, as one would need to consider mode mixing in the inertial frame, and not in the co-precessing frame corresponding to the aligned-spin waveform. An analysis of the shortcomings 
of our treatment of mode-mixing and further improvements of the model will be the subject of future work.

%%%%%%%%%%%%%%%%%%%%%%%%%%%%%%%%%%%%%%%%%%%%%%%%%%%%%%%%%%%%%%%%%
%               CONSTRUCTING THE MODEL
%%%%%%%%%%%%%%%%%%%%%%%%%%%%%%%%%%%%%%%%%%%%%%%%%%%%%%%%%%%%%%%%%

\section{Constructing the Model}
\label{sec:Model}

A core ingredient in modelling precessing binaries is a description for the precession dynamics in terms of the three Euler angles $\lbrace \alpha, \beta, \gamma \rbrace$ describing the active transformation from the precessing to the inertial frame. For \phXPHM, we have implemented two different prescriptions for the precession angles. The first model, described below in Sec.~\ref{sec:single_spin_angles}, is based on the NNLO single-spin PN expressions used in \ppvtwo \cite{Hannam:2013oca,Bohe:PPv2}. The second model,
described in Sec.~\ref{sec:double_spin_angles},
is based on the 2PN expressions from \cite{Chatziioannou:2017tdw}, derived using a multiple scale analysis (MSA). Such modularity will help us to gauge systematics in modelling precession and its implications for gravitational-wave data analysis. 

%%%%%%%%%%%%%%%%%%%%%%%%%%%%%%%%%%%%%%%%%%%%%%%%%%%%%%%%%%%%%%%%%
%               NNLO ANGLES
%%%%%%%%%%%%%%%%%%%%%%%%%%%%%%%%%%%%%%%%%%%%%%%%%%%%%%%%%%%%%%%%%

\subsection{Post-Newtonian NNLO Euler Angles}
\label{sec:single_spin_angles}
The single-spin description of the Euler angles is based on a post-Newtonian re-expansion setting $\bfStwo = 0$, and restricting to spin-orbit interactions \cite{Bohe:PPv2}. This framework was implemented in \ppvtwo \cite{Hannam:2013oca,Bohe:PPv2} and has been actively used in the analysis of gravitational-wave data \cite{LIGOScientific:2018mvr}. 

In the PN framework, we first solve for the evolution equations of the moving triad $\lbrace \bfn, \bflambda, \bfell \rbrace$ at a given PN order in the conservative dynamics before re-introducing radiation-reaction effects. The angular momentum $\bfJ$, neglecting radiation reaction effects, is approximately conserved and can be used to define a fixed direction $\bf{z}$. Completed with two constant unit vectors $\bf{x}$ and $\bf{y}$, this forms an orthonormal triad. We can define a separation vector $\bfn$ between the two black holes such that $\bfr = r \bfn$ with $\bfr = \bfr_1 - \bfr_2$.
The unit normal to the instantaneous orbital plane, $\bfell$, is defined by $\bfell = \bfn \times \bfv / | \bfn \times \bfv |$, where $\bfv = \bfv_1 - \bfv_2$ is the relative velocity. Finally, the orthonormal triad is completed by $\bflambda = \bfell \times \bfn$. The evolution of the triad is given by \cite{Blanchet:2011zv,Marsat:2013wwa}
\begin{subequations}
\label{eq:dtriaddt}
\begin{align}
    \label{eq:dfndt}
    \frac{d \bfn}{dt} &= \omega \bflambda , \\
    \label{eq:dflambdadt}
    \frac{d \bflambda}{d t} &= - \omega \bfn + \bar{\omega} \bfell , \\
    \label{eq:dfelldt}
    \frac{d \bfell}{d t} &= - \bar{\omega} \bflambda ,
\end{align}
\end{subequations}
\newline 
where $\bar{\omega}$ is the precession angular frequency. Introducing an orthonormal basis such that the $z$-axis points along $\bfJ$, as we do in Appendix \ref{appendix:frames}, we can introduce the Euler angles $\lbrace \alpha, \beta, \gamma \rbrace$ to track the position of $\bfL$ with respect to this fixed basis. The evolution of the Euler angles follows from Eqs.~(\ref{eq:dtriaddt}):
\begin{subequations}
\begin{align}
    \frac{d \alpha}{d t} &= -\frac{\bar{\omega}}{\sin \beta} \frac{J_n}{\sqrt{J^2_n + J^2_{\lambda}}} , \\
    \frac{d \beta}{d t} &= \bar{\omega} \frac{J_{\lambda}}{\sqrt{J^2_n + J^2_{\lambda}}}, \\
    \frac{d \gamma}{d t} &= - \dot{\alpha} \cos \beta .
\end{align}
\end{subequations}
The only assumption made in deriving these equations is that the direction of the total angular momentum is approximately constant and that we may neglect radiation reaction effects \cite{Blanchet:2011zv}. 

In the regime of simple precession, in which the total angular momentum is not small compared to the orbital and spin angular momenta, the opening angle $\beta$ of the precession cone is approximately constant and is constrained by the minimal rotation condition \cite{Boyle:2011gg}, as in Eq.(\ref{eq:minimal_rotation}).
Under the approximation that the direction of the total angular momentum, $\hat{\bfJ}$, is constant throughout the evolution, the angle $\beta$ can be determined using the closure relation $\bfJ = \bfL + \bfSone$,
yielding
\begin{align}
    \cos \beta &= \bfell \cdot \frac{\bfJ}{| \bfJ |} = \frac{J_{\ell}}{\sqrt{ J^2_{\ell} + J^2_n + J^2_{\lambda} }} .
\end{align}
With a PN re-expansion of the right hand side and decomposing the spin into contributions parallel and perpendicular to the orbital angular momentum, $\boldsymbol{S} = \boldsymbol{S}_{\parallel} + \boldsymbol{S}_{\perp}$, the expression for $\beta$ reduces to \cite{Bohe:PPv2}
\begin{equation}\label{eq:cos_beta}
    \cos \beta = \frac{L + S_{\parallel}}{\sqrt{ (L + S_{\parallel})^2 + S^2_{\perp}}} = \pm (1 + s^2)^{-1/2},
\end{equation}
%\n 
where $s = S_{\perp} / (L + S_{\parallel})$ and the overall sign is dependent on the sign of $L + S_{\parallel}$. This approximation was used in \ppvtwo, coupled with a 2PN non-spinning approximation of the orbital angular momentum $L$ \cite{Kidder:1995zr}. 
In \cite{phenomt} we discuss the use of a numerical fit to the orbital angular momentum in a non-precessing merger,
and here we use different alternative post-Newtonian approximations as discussed in Sec.~\ref{sec:PN_L}.

The dynamics for $\dot{\alpha}$ are determined using the results obtained in \cite{Blanchet:2011zv} together with the NNLO spin-orbit contributions derived in \cite{Marsat:2013wwa}:
\begin{align}
    \label{eq:dalphadt}
    \frac{d \alpha}{d t} &= - \frac{\bar{\omega}}{\sin \beta} \, \frac{J_n}{\sqrt{J^2_n + J^2_{\lambda}}} , 
\end{align}
%\n 
where $\bar{\omega}$ is defined by Eq.~(\ref{eq:dtriaddt}). The right-hand side of Eq.~(\ref{eq:dalphadt}) is PN re-expanded and orbit-averaged in order to re-express the spin components $S_{1,\lambda}$ and $S_{1,n}$ in terms of an effective spin parameter $\chi_p$ \cite{Schmidt:2014iyl}. Radiation reaction effects can be incorporated by using the evolution of the frequency, $\dot{\omega} / \omega^2$, to re-express Eq.~(\ref{eq:dalphadt}) as
\begin{align}
\label{eq:dalphadomega}
    \frac{d \alpha}{d \omega} &= \frac{1}{\dot{\omega}} \, \frac{d \alpha}{d t} .
\end{align}
%\n 
In \ppvtwo, corrections up to 3.5PN in both the orbital and spin-orbit sectors were used for the evolution of $\omega$. Equation (\ref{eq:dalphadomega}) is re-expanded as a PN series in $\omega$ and is integrated term-by-term to yield a closed-form expression for $\alpha(\omega)$, up to an overall constant of integration. A similar strategy is employed for deriving $\gamma (\omega)$ using the relation in Eq.~(\ref{eq:minimal_rotation}).
%\begin{align}
%    \dot{\gamma} &= - \dot{\alpha} \cos \beta .
%\end{align}
The constants of integration arising when solving Eqs.~(\ref{eq:dtriaddt}) are fixed by the orientation of the spins at a given reference frequency $f_{\rm{ref}}$
as discussed in appendix \ref{appendix:frames}. 

The effective spin precession parameter $\chi_p$ \cite{Schmidt:2014iyl} provides a mapping from the four in-plane spin degrees of freedom to a single parameter that captures the dominant effect of precession. As discussed in \cite{Schmidt:2014iyl}, the magnitudes of the in-plane spins $S_{i,\perp}$ will oscillate about a mean value, with the relative angle between the spin vectors changing continuously. Averaging over many precession cycles, the average magnitude of the spins can be written as \cite{Schmidt:2014iyl}
\begin{align}\label{eq:avNNLOSperp}
    S_p &= \frac{1}{2} \left( A_1 S_{1,\perp} + A_2 S_{2,\perp} + | A_{1} S_{1,\perp} - A_2 S_{2,\perp} |\right) , \\
    &= {\rm{max}} \left( A_1 S_{1,\perp} , A_2 S_{2,\perp} \right),
\end{align}
where $A_1 = 2 + 3 / (2 q)$.
For most binaries, the spin of the larger black hole will dominate and $S_p$ reduces to $A_1 S_{1,\perp}$. As the in-plane spin of the smaller black hole will become  increasingly negligible at higher mass ratios, the dimensionless effective precession parameter is defined by assigning the precession spin to the larger black hole \cite{Schmidt:2014iyl}
\begin{align}\label{def:chip}
    \chi_p &= \frac{S_p}{A_1 m^2_1} .
\end{align}
\newline 
In \ppvtwo, a choice was made to calculate the PN Euler angles by placing all of the spin on the larger black hole, i.e. $\chi_2 = 0$.
\phXPHM inherits this choice in this first prescription for the Euler angles.

%%%%%%%%%%%%%%%%%%%%%%%%%%%%%%%%%%%%%%%%%%%%%%%%%%%%%%%%%%%%%%%%%
%               MSA ANGLES
%%%%%%%%%%%%%%%%%%%%%%%%%%%%%%%%%%%%%%%%%%%%%%%%%%%%%%%%%%%%%%%%%

\subsection{MSA Euler Angles}
\label{sec:double_spin_angles}

The second formulation of the precession angles that we implement is based on the application of \textit{multiple scale analysis} (MSA) \cite{Bender:1999aa} to the post-Newtonian equations of motion \cite{Racine:2008qv,Klein:2013qda,Kesden:2014sla,Gerosa:2015hba,Chatziioannou:2013dza,Chatziioannou:2017tdw}. This approach employs a perturbative expansion in terms of the ratios of distinct characteristic scales in the system. For precessing binaries, a natural hierarchy of timescales can be identified with the orbital time scale being shorter than the precessing time scale, which is again shorter than the radiation reaction timescale. In \cite{Chatziioannou:2017tdw}, the time-domain PN precession equations are solved by incorporating radiation reaction effects through the multiple scale analysis. The resulting time domain waveforms are then transformed to the frequency domain using \textit{shifted uniform asymptotics} \cite{Klein:2014bua}, helping to overcome a number of limitations and failures in the more conventional SPA approach. By decomposing the waveform into Bessel functions, the resulting Fourier integral can be evaluated term-by-term using SPA and resummed using the exponential shift theorem. The concomitant frequency domain inspiral waveforms contain spin-orbit and spin-spin effects at leading order in the conservative dynamics and up to $3.5$PN order in the dissipative dynamics, neglecting spin-spin terms. The MSA formulation of the post-Newtonian Euler angles enables double-spin effects, as recently first incorporated into the phenomenological framework in \cite{Khan:2018fmp,Khan:2019kot}. The SPA treatment used here corresponds to the leading order, local-in-frequency correction and is equivalent to the zeroth order of the SUA \cite{Klein:2014bua,Marsat:2018oam}. 

The precession angles are given in terms of a PN series plus an additional MSA correction. The full solution for $\alpha$ in the MSA approach is \cite{Chatziioannou:2017tdw}\footnote{Note that \cite{Chatziioannou:2017tdw} adopts the notation $\phi_z \rightarrow \alpha, \zeta \rightarrow -\gamma$ and $\cos \theta_L \rightarrow \cos \beta$.}
\begin{align}
    \alpha &= \alpha_{-1} + \alpha_0 + \mathcal{O}(\epsilon) ,
\end{align}
where
\begin{align}
    \alpha_{-1} &= \displaystyle\sum^5_{n = 0} \langle \Omega_{\alpha} \rangle^{(n)} \, \alpha^{(n)} + \alpha^{0}_{-1}
\end{align}
is the leading order MSA correction given in Eq. 66 of \cite{Chatziioannou:2017tdw}
and $\alpha_{0}$ is the first-order correction to the MSA given by Eq. 67 of \cite{Chatziioannou:2017tdw}. A similar approach is taken for $\gamma$, with the leading order MSA correction given by Eq. F5 of \cite{Chatziioannou:2017tdw} and the first-order MSA correction given in Eq. F19 of Appendix F of \cite{Chatziioannou:2017tdw}.

Again, the Euler angles are fixed by the orientation of the spins at a given reference frequency $f_\mathrm{ref}$
as discussed in appendix \ref{appendix:frames}.
Similarly to the PN re-expanded results used in \ppvtwo, the identification of $\hat{\bfJ}$ as an approximately conserved quantity leads to the preferred coordinate frame in which $\hat{z} = \hat{\bfJ}$ with the angle $\beta$ defining the precession cone, as in Eq. 8 of \cite{Chatziioannou:2017tdw}:
\begin{align}
    \cos \beta &= \hat{\bfJ} \cdot \hat{\bfL} = \frac{J^2 + L^2 - S^2}{2 J L} .
\end{align}
Together with the expressions for $\alpha$ and $\gamma$, this defines the complete set of equations describing the Euler angles in the MSA framework. 

\begin{figure}[htbp!]
\begin{center}
\includegraphics[width=0.9\columnwidth]{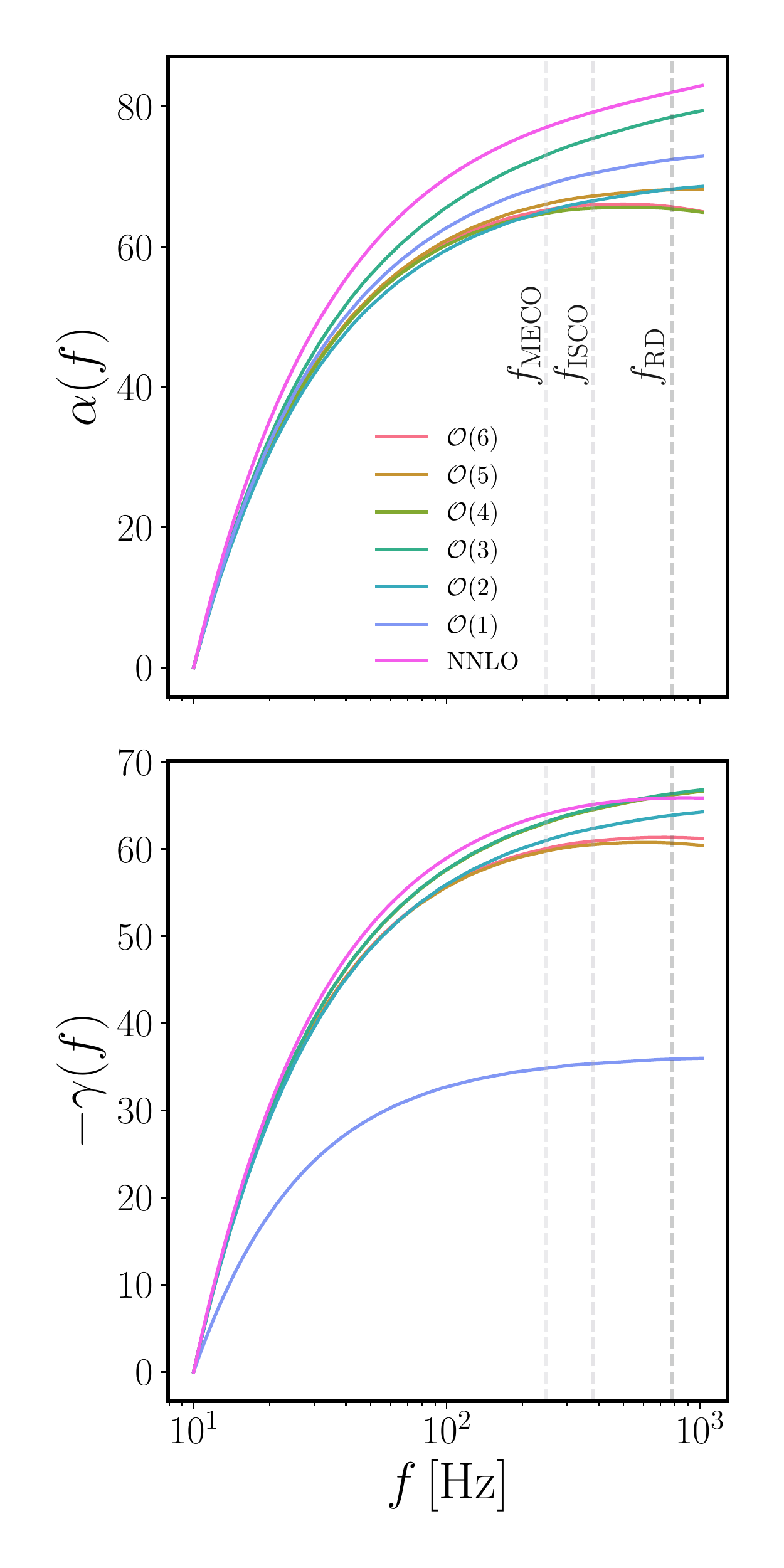}
\caption{The Euler angles $\alpha$ and $\gamma$ obtained by solving the MSA system of \cite{Chatziioannou:2017tdw} and the NNLO PN equations for a binary of mass $M = 20 M_{\odot}$, $q = 10$, $\chi_1 = \left( 0.4,0,0.4 \right)$ and $\chi_2 = \left( 0.3,0,-0.3 \right)$, with a starting frequency of $10$Hz. The vertical dashed lines correspond to the MECO \cite{Cabero:2016ayq}, ISCO and ringdown frequency in increasing order. The multiple shaded lines denote the Euler angles evaluated using the NNLO angles and different orders for the MSA correction used in the MSA system where the default order is $\mathcal{O}(5)$. 
}
\label{fig:msaeuler}
\end{center}
\end{figure}

The MSA expressions are series expansions in terms of the orbital velocity or, equivalently, the GW frequency $f^{-4/3 + n/3}$, where $n \in \left[ 1 , 6 \right]$ denotes the order of the series. As discussed in \cite{Chatziioannou:2017tdw,Khan:2018fmp}, $\gamma$ is fully PN re-expanded whereas $\alpha$ involves both PN re-expanded and un-expanded terms. This choice was motivated by solutions to the exact precession-averaged equations $\langle \dot{\alpha} \rangle_{\rm{pr}}$ leading to $\alpha_{-1}$ being ill-behaved in the equal mass ratio limit and divergences when the total spin angular momentum is (anti)aligned with the orbital angular momentum \cite{Chatziioannou:2017tdw}. A more detailed discussion can be found in Sec IV D 1 and Appendix E of \cite{Chatziioannou:2017tdw}. The order to which we retain terms in the MSA series can be controlled and, as in \cite{Khan:2018fmp}, by default we drop the highest-order terms in the expansion, working to $\mathcal{O}(5)$. The impact of the expansion order on the Euler angles is highlighted in Fig.~\ref{fig:msaeuler} where the NNLO angles from Sec.~\ref{sec:single_spin_angles} are shown for comparison. 

The subset of equal-mass binaries present a number of qualitative features that distinguish them from the generic unequal mass ratio cases. In particular, by setting $q = 1$ in the MSA framework we find that the expressions lead to singular behaviour in various aspects of the model \cite{Kesden:2014sla,Gerosa:2015hba,Chatziioannou:2017tdw}. Of particular note is the singular behaviour of the precession-averaged spin couplings \cite{Chatziioannou:2017tdw}, which are used in the construction of the final spin estimate detailed in Sec.~\ref{sec:final_state}. These terms must be regularized in the equal-mass limit to avoid singular behaviour.    

As discussed above, the MSA system of equations is known to result in numerical instabilities when $\boldsymbol{S}$ and $\boldsymbol{L}$ are nearly mis-aligned. Such instabilities result in a failure at the waveform generation level. In order to help alleviate these situations, we have opted to use the NNLO angles described in Sec.~\ref{sec:single_spin_angles} as a fallback in the default LALSuite implementation, though the end-user can still demand a terminal error for these cases. 

\begin{figure}[htbp!]
%\begin{center}
\includegraphics[width=0.9\columnwidth]{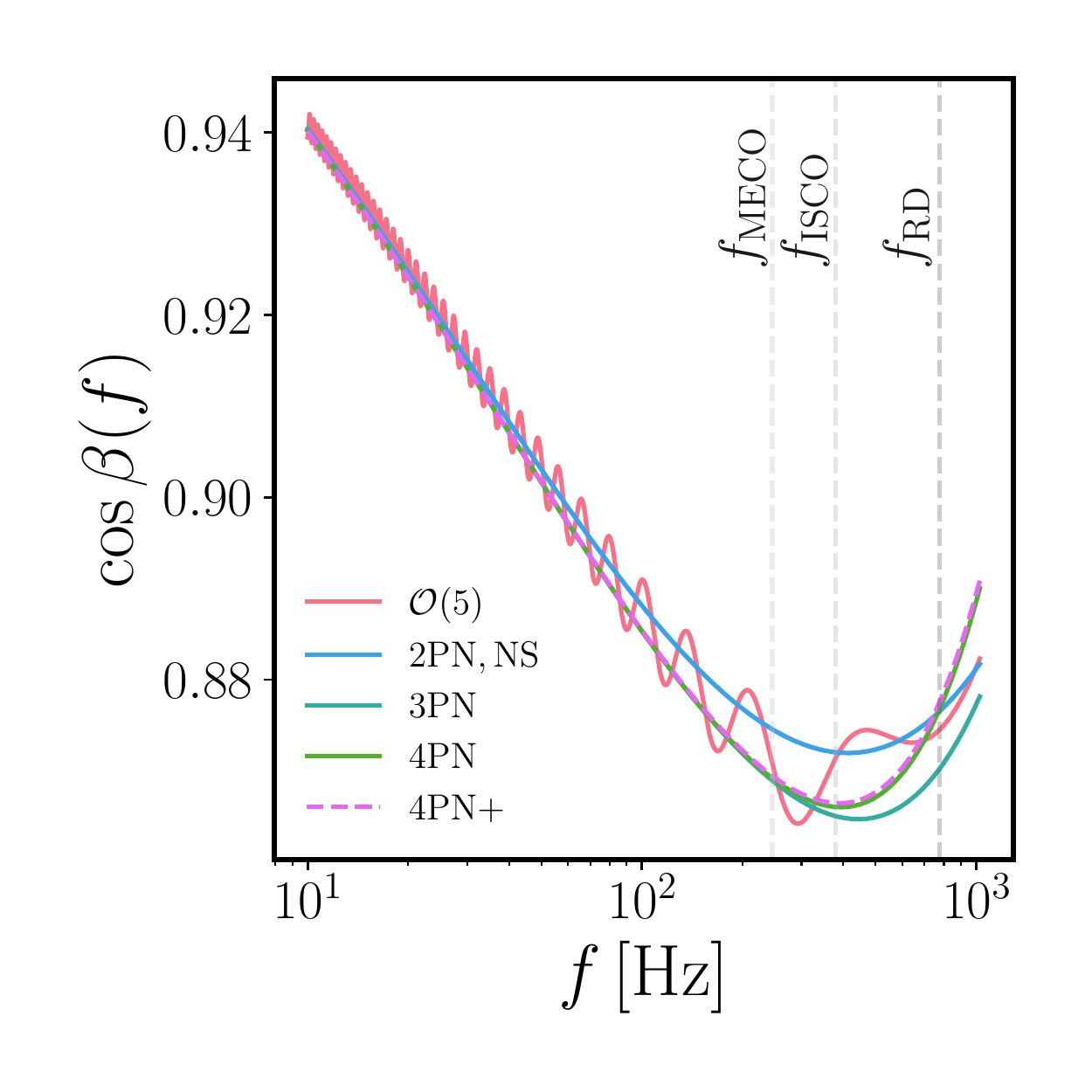}
\caption{Comparison of the NNLO Euler angle $\cos \beta$ against the MSA version ($\mathcal{O}(5)$), for the same binary as shown in Fig.~\ref{fig:msaeuler}. We show the impact of different PN orders for the orbital angular momentum $\bfL$ on the determination of the opening angle for the NNLO prescription.
}
\label{fig:pnmsabeta}
%\end{center}
\end{figure}

\subsection{Post-Newtonian Orbital Angular Momentum}
\label{sec:PN_L}

In order to calculate the orbital angular momentum, we use an aligned-spin 4PN approximation  \cite{Blanchet:2006zz,LeTiec:2011ab,Bohe:2012mr,Damour:2014jta,Bernard:2017ktp,Blanchet:2017rcn} 
\begin{align}\label{eq:L4PN}
    L &= \frac{\eta}{\sqrt{x}} \Bigg[ L_0 + L_1 x + L_2 x^2 + L_3 x^3 + L_4 x^4 + \\ \nonumber &+ L^{\rm{SO}}_{1.5} x^{3/2} + L^{\rm{SO}}_{2.5} x^{5/2} + L^{\rm{SO}}_{3.5} x^{7/2} \Bigg],
\end{align}
where $L_{a}$ are the orbital coefficients at $a$-PN order, $L^{\rm{SO}}$ are the spin-orbit contributions and we neglect spin-spin terms. The coefficients are given in Appendix \ref{app:orb_ang_mom}. 
This is in contrast to \ppvtwo, which used a non-spinning 2PN approximation to the orbital angular momentum
\begin{align}\label{eq:L2PN}
    L_{\twopn} &= \frac{\eta}{\sqrt{x}} \left[ 1 + \left( \frac{3}{2} \; + \; \frac{\eta}{6} \right) x + \left( \frac{27}{8} \; - \; \frac{19 \eta}{8} \; + \; \frac{\eta^2}{24} \right) x^2 \right] .
\end{align}
Our implementation in the LALSuite code also supports dropping various terms in the 4PN expression of Eq.~(\ref{eq:L4PN}), including reducing the approximation to Eq.~(\ref{eq:L2PN}). In addition, we have also implemented the option to incorporate spin effects at leading post-Newtonian order at all orders in spin \cite{Marsat:2014xea,Vines:2016qwa,Siemonsen:2017yux}, as given in Appendix \ref{app:orb_ang_mom}.
Note that, consistent with our approximation of the co-precessing dynamics and waveform with the corresponding aligned-spin quantities we neglect contributions to the angular momentum of the spin components in the orbital plane. Modelling of the orbital angular momentum is most relevant for calculating the opening angle $\beta$. The impact of different orders for $L_{\rm PN}$ is shown in Fig.~\ref{fig:pnmsabeta}, where we observe significant differences between the non-spinning 2PN approximation used in \cite{Bohe:PPv2} and higher-order PN approximants that incorporate aligned-spin contributions.

%%%%%%%%%%%%%%%%%%%%%%%%%%%%%%%%%%%%%%%%%%%%%%%%%%%%%%%%%%%%%%%%%
%              MODELLING FINAL STATE
%%%%%%%%%%%%%%%%%%%%%%%%%%%%%%%%%%%%%%%%%%%%%%%%%%%%%%%%%%%%%%%%%

\subsection{Modelling the Final State}\label{sec:final_state}

In our twisting construction, approximating a precessing waveform with a non-precessing one implies that the radiated energy and the radiated angular momentum orthogonal to the orbital plane are identical to the 
non-precessing values.
Indeed,  comparisons of the final mass from precessing NR simulations with fits for the final mass resulting from non-precessing mergers, see e.g.~\cite{T1600168,Varma:2018aht}, show only a weak dependence of the final mass on precession.

We do however need to take into account the dependence of the final spin on precession, which is essentially due to the vector addition of the individual spins and angular momentum, as will be discussed below. 
A surrogate model for the final spin of precessing mergers for a limited range in mass ratio and spins has been produced recently \cite{Varma:2018aht}. Here, however, we will proceed differently in order not to compromise the simplicity and domain of validity of our model and employ a simple estimate for the final spin magnitude based on accurate fits for the final spin of non-precessing mergers, simple geometric arguments, and our assumptions related to those underlying the twisting approximation.  

In order to incorporate precession into the final spin prediction, we can argue as follows (compare also to  \cite{Rezzolla:2007rz}):
we first write the total angular momentum ${\mathbf J}$ as the sum of individual spins ${\mathbf S_i}$ and orbital angular momentum ${\mathbf L}$:
$$ 
\mathbf{J} = \mathbf{S_1}  + \mathbf{S_2}  + \mathbf{L}.
$$
We can now apply this equation to compute the final angular momentum $\mathbf{S_{fin}}$ of the remnant black hole, interpreting the quantities $(\mathbf{S_1}, \mathbf{S_2}, \mathbf{L})$ as the values at merger, where the further emission of angular momentum effectively shuts off.
We split the spin vectors of the individual black holes into their orthogonal components parallel (or anti-parallel) and orthogonal to the unit vector in the direction of the orbital angular momentum  $\mathbf{{\hat L}}$, and introduce the quantities (for $i=1,2$):
\begin{eqnarray}
S_{i,\vert\vert} &=&  \mathbf{S_{i}}\cdot \mathbf{\hat L}, \quad S_{\vert\vert} = S_{1,\vert\vert} + S_{2,\vert\vert}, \\
\mathbf{S_{i,\perp}} &=& \mathbf{S_{i}} - S_{i,\vert\vert} \mathbf{\hat L}, \quad
\mathbf{S_{\perp}} = \mathbf{S_{1,\perp}}+\mathbf{S_{2,\perp}}.
\end{eqnarray}

We can now compute the magnitude of the final angular momentum in terms of the vector-sum of 2 orthogonal components, and then the magnitudes of the final spin $S_\mathrm{fin}$ and final Kerr parameter $a_\mathrm{fin}$ are given as
\begin{equation}
    \label{eq:afin_nolambda}
\vert S_{\mathrm{fin}}  \vert \equiv  M_\mathrm{fin}^2 \vert a_\mathrm{fin} \vert = \sqrt{\mathbf{S_\perp}^2 + (S_{\vert\vert} +L_\mathrm{fin})^2}.
\end{equation}
Here, $M_\mathrm{fin}$ is the final mass, and $L_\mathrm{fin}$ is defined in terms of the final mass and spin as
\begin{equation}
    \label{eq:afin_L_fin}
S_{\vert\vert} + L_\mathrm{fin} = M_\mathrm{fin}^2 a_\mathrm{fin}^{\vert\vert}, 
\end{equation}
where $a_\mathrm{fin}^{\vert\vert}$ is the final Kerr parameter in the corresponding non-precessing configuration.
We compute $a^\parallel_\mathrm{fin}$ and $M_\mathrm{fin}$ as functions of the symmetric mass ratio and the spin projections in the direction of $\mathbf{L}$, using the same fit to numerical relativity data \cite{Jimenez-Forteza:2016oae} that was used in the non-precessing \phX and \phXHM models.
Note that the fit for $a^\parallel_\mathrm{fin}$ is in fact constructed as a fit for $L_\mathrm{fin}$, and $a_\mathrm{fin}$ is then computed using Eq.~(\ref{eq:afin_L_fin}).

In the twisting approximation we assume that
in a co-moving frame the waveform is well approximated by a twisted-up non-precessing waveform. In addition, one usually assumes for simplicity that the total spin magnitudes, as well as the magnitudes of the projections of $\mathbf{S_1}  + \mathbf{S_2}$ onto ${\mathbf L}$ ($S_{\vert\vert}$) and orthogonal to it ($S_{\perp}$) are preserved as well. 

The spin components $S_\perp$ and $S_{\vert\vert}$ that enter the final spin estimate in Eq.~(\ref{eq:afin_nolambda}) can now be computed in different ways.
The simplest choice is to use the non-precessing value for $S_{\vert\vert}$, and the appropriately averaged value of $S_\perp$, which enters our inspiral descriptions. For the NNLO angle description summarised in Sec.~\ref{sec:single_spin_angles}, which is an effective single spin description, the quantity $\chi_p$ (\ref{eq:avNNLOSperp}-\ref{def:chip}) acts as an average in-plane spin, and can be used to 
estimate $S_{\perp}$ at merger.
This is the choice that has been made for the \phP \cite{Hannam:2013oca}
and \ppvtwo \cite{Bohe:PPv2} models, and it will be the default choice we have implemented when using one of the NNLO angle descriptions.
For the double spin MSA description outlined in Sec.~\ref{sec:double_spin_angles}, one can rely on the precession-averaged spin couplings of Eqs.\,(A9-A11) in \cite{Chatziioannou:2017tdw}. This can be best seen by rewriting Eq.~(\ref{eq:afin_nolambda}) more explicitly as
\begin{eqnarray}
    \label{eq:afin_nolambda_doubleSpin}
S_\mathrm{fin}^2 &=& S_{1}^2+S_{2}^2
 + 2 S_{1}\cdot S_{2} +  L_\mathrm{fin}^2 +\nonumber\\
 &+&2 L_\mathrm{fin} \left(\mathbf{S_{1}}\cdot \mathbf{\hat L}+\mathbf{S_{2}}\cdot \mathbf{\hat L} \right).
\end{eqnarray}
Averaging over one precession cycle, the above equation can be rewritten as: 
\begin{equation}\label{eq:MSA_finalSpin}
\langle S_\mathrm{fin}^2 \rangle_\mathrm{pr}=S^{2}_\mathrm{av}+ L_\mathrm{fin}^2+2 L_\mathrm{fin}(\langle\mathbf{S_{1}}\cdot \mathbf{\hat L}\rangle_\mathrm{pr}+\langle\mathbf{S_{2}}\cdot \mathbf{\hat L}\rangle_\mathrm{pr} ) \,.
\end{equation}
This will serve as our default choice when using the MSA formulation to compute the Euler angles.

Assuming that the spin components at merger are equal to the average quantities during the inspiral has the advantage of providing unambiguous values.
However, this neglects the two facts that the averaged quantities do not predict the value of the spin components at any particular time, and that they do not accurately describe the spin dynamics shortly before and at merger.
We therefore also discuss alternative descriptions, which contain additional freedom to approximately account for the unmodeled information about the spin components at merger.

We will first discuss the simpler single-spin case,
assuming that only the larger black hole, labelled with index $i=1$, carries spin, and the spin of the smaller black hole vanishes. We first rewrite  Eq.~(\ref{eq:afin_nolambda}) in the form 
\begin{equation}
    \label{eq:afin_single_spin_flexible}
\vert S_\mathrm{fin} \vert = \sqrt{S_{1}^2 +  L_\mathrm{fin}^2 + 2 S_{1,\vert\vert} L_\mathrm{fin}} \,.
\end{equation}
We assume that the total spin $\mathbf{S_{1}}$ does not change during the coalescence process, and that $L_\mathrm{fin}$ is given by the non-precessing value as in Eq.~(\ref{eq:afin_L_fin}), where for $S_{\vert\vert}$ we take its initial value, i.e.~the value we have used during the inspiral.
In previous work \cite{Hannam:2013oca,Bohe:PPv2} this initial value of $S_{\vert\vert}$ has also been used
in the final spin estimate of Eq.~(\ref{eq:afin_single_spin_flexible}),
consistent with the approximation that  $S_\perp$ and $S_{\vert\vert}$ are approximately preserved during the inspiral.
Due to the strong spin interaction close to merger, this approximation may however not be accurate, and alternatively we may only assume that the spin magnitude is preserved and treat the value of $S_{1,\vert\vert}$ as unknown. We can then determine the value of 
$S_{1,\vert\vert}$ that best fits a given precessing waveform subject to the condition $0 \leq \vert S_{1,\vert\vert}\vert \leq S$. We currently do not provide this option in our LALSuite code, in order to avoid book-keeping of extra parameters that are not typically used in parameter estimation.

Instead we provide a toy model solution for the single-spin case, where $\chi_p$ is replaced by $\chi_{1x}$, i.e.~the $x$-component of the spin of the larger black hole.
This particular choice of toy model has been implemented to facilitate comparisons with an earlier version of the \phPvthreehm model~\cite{Khan:2019kot}.\footnote{This earlier version of \phPvthreehm with $\chi_{1x}$ passed to the final spin function was introduced in version 2f1596262c3af9832dfe2a52944472cb3be81e0a of the \url{https://git.ligo.org/lscsoft/lalsuite/} repository and changed to $\chi_p$  in  b60bec3aef3be3c346fd349ddd738e55a2af4b6d.}
The rotational freedom in the in-plane spin then allows to vary the in-plane spin component that enters the final spin estimate between zero and the magnitude of the in-plane spin.

Note that in \cite{Hannam:2013oca,Bohe:PPv2} a free parameter $\lambda$ was introduced as
\begin{equation}\label{eq:afin}
\vert a_\mathrm{fin} \vert = \sqrt{\left(S_\perp^2 \frac{\lambda}{M_\mathrm{fin}^2}\right)^2+ {a_{\rm fin}^{\vert\vert}}^2},
\end{equation}
and was set to the ad-hoc value $\lambda = M_\mathrm{fin}^2$, consistent with \cite{Rezzolla:2007rz}, in order to reduce the residuals of the final spin estimate when comparing with NR data sets.

We now consider the double-spin case, where we also have to take into account the time-dependent angle between the in-plane components of the spins. We can write Eq.~(\ref{eq:afin_nolambda}) in a form similar to Eq.~(\ref{eq:afin_single_spin_flexible}) as
\begin{subequations}
 \label{eq:afin_nolambda_doubleSpin_expanded}
\begin{align}
S_\mathrm{fin}^2 = &\: (\mathbf{S_{1,\perp}}+\mathbf{S_{2,\perp}})^2 + (S_{1,\vert\vert} +S_{2,\vert\vert} +L_\mathrm{fin})^2 \\
 = &\: S_{1}^2+S_{2}^2 + L_\mathrm{fin}^2
 + 2 S_{1,\perp} S_{2,\perp} \cos{\phi_{12}}\\
 & + 2 S_{1,\vert\vert} S_{2,\vert\vert}
 + 2 L_\mathrm{fin}(S_{1,\vert\vert} + S_{2,\vert\vert} ).
\end{align}
\end{subequations}
One could now choose the unmodeled parameters in this equation and fit them to the best values in a given data set: e.g. one could leave the parallel components free analogous to Eq.~(\ref{eq:afin_single_spin_flexible}), or simply neglect the tilt of the spins at merger and use 
$\cos{\phi_{12}}$ as a free parameter. We reserve these options for future work, as they would require to perform Bayesian parameter estimation with a different parameterization than usual within LALSuite. Instead, we provide the option to model $S_{\perp}$ as 
\begin{equation}\label{eq:Sperp_total}
   S_{\perp}^2 = (\mathbf{S_{1,\perp}}+\mathbf{S_{2,\perp}})^2 ,
\end{equation}
which provides the freedom for cancellations between the two spin components.

Following the discussion above, in our LALSuite code we currently provide four options to set the magnitude of the final spin, see also Appendix \ref{appendix:lal_implementation}. We either proceed
 in analogy with Eq.~(\ref{eq:afin}) and set
\begin{equation}\label{eq:afin_LAL}
\vert a_\mathrm{fin} \vert = \sqrt{\left(\bar\chi_p^2 \frac{m_1^2}{(m_1+m_2)^2}\right)^2+ {a_\mathrm{fin}^{\vert\vert}}^2},
\end{equation}
where $\bar\chi_p$ can be chosen as one of three alternatives,
\begin{align}
   \bar\chi_p  &=  \chi_p, \label{eq:finalspin_0}\\
   \bar\chi_p  &=  \chi_{1x}, \label{eq:finalspin_1}\\
   \bar\chi_p  &=  \frac{S_{\perp} (m_1 + m_2)^2}{m_1^2}, \label{eq:finalspin_2}
\end{align}
or by setting
\begin{equation}\label{eq:finalspin_3}
 \vert a_\mathrm{fin}\vert = \frac{\sqrt{\left\langle S_\mathrm{fin}^2 \right\rangle_{\rm pr}}}{M^2},
\end{equation}
where $S_{\perp}$ in Eq.~(\ref{eq:finalspin_2}) is defined as in Eq.~(\ref{eq:Sperp_total}) and for Eq.~(\ref{eq:finalspin_3}) we have used
Eq.~(\ref{eq:MSA_finalSpin}).
Here Eq.~(\ref{eq:finalspin_0}) corresponds to the choice of \phP \cite{Hannam:2013oca} and  \phPvtwo \cite{Bohe:PPv2}
and is the default choice when using the NNLO description of the Euler angles,
Eq.~(\ref{eq:finalspin_1}) has been implemented to compare with a previous version of \phPvthreehm, and
Eq.~(\ref{eq:finalspin_3}) is the default choice when using the MSA description of the Euler angles.

In Sec.~\ref{sec:validation} we will provide results for different final spin choices. A detailed 
investigation of the differences between final spin estimates is beyond the scope of the present paper and will be investigated in future work, along with further improvements.

Note that so far we have only discussed estimates for the final spin magnitude and not its direction. It is well known that when the precession cone (i.e.~the Euler angle $\beta$) is sufficiently small, then the final spin will point approximately in the direction of the initial total angular momentum ${\mathbf J}$. For larger mass ratios however the situation can become more complicated when the orbital angular momentum becomes smaller than the (sum of the) component spins. In this situation ${\mathbf J}$ and ${\mathbf L}$ may end up pointing in opposite directions  (i.e.~their scalar product becomes negative), and ${\mathbf J}$ may end up pointing in the opposite direction compared with its initial value. The latter situation is also known as transitional precession \cite{Apostolatos:1994mx, Kidder:1995zr, Schmidt:2012rh}, as opposed to simple precession, when ${\mathbf J}$ at least approximately maintains its direction.

The final spin of the remnant is then the final value of the total angular momentum ${\mathbf J}$, and it is thus possible that it ``flips over'' with respect to the direction of the initial total angular momentum.  In the context of our ``twisting-up'' procedure we need to track the sign of the final spin with respect to the $L$-frame, where it corresponds to the sign of the final spin of the corresponding non-precessing waveform. We thus proceed as follows: We estimate the magnitude of the final spin as described above, and we determine the sign of the final spin with respect to the $L$-frame to coincide with the sign of the non-precessing final spin. This signed value of the final spin is used to determine the complex ringdown frequency
of our waveform in the $L$-frame, which is then rotated into the inertial frame by our ``twisting up'' procedure.
Situations when ${\mathbf J}$ ``flips over'' or the precession cone angle $\beta$ becomes large are however 
challenging to model: First, approximate constancy of ${\mathbf J}$ or a small precession cone are standard assumptions for post-Newtonian expansions. Second, such configurations have not yet been well explored by numerical relativity. When our model is used in parameter estimation and significant support for the posterior distribution is obtained for a ``flipped ${\mathbf J}$'' configuration, or a value of $\beta$ that comes close to or larger than $\pi/2$, we thus suggest to proceed with caution and test the robustness of results by comparing the NNLO and MSA angles for the precession implementation, and the different approximations for the magnitude of the final spin, and if possible with the results for other waveform models. Future work will aim to improve the robustness of our model for such situations.

%%%%%%%%%%%%%%%%%%%%%%%%%%%%%%%%%%%%%%%%%%%%%%%%%%%%%%%%%%%%%%%%%
%               MODEL PERFORMANCE & VALIDATION
%%%%%%%%%%%%%%%%%%%%%%%%%%%%%%%%%%%%%%%%%%%%%%%%%%%%%%%%%%%%%%%%%

\section{Model Performance and Validation}
\label{sec:validation}

In this section, we perform various tests of our model,
ranging from comparisons against numerical relativity to real-world parameter estimation applications.

\subsection{Comparison of Euler angles with Numerical Relativity}
\label{sec:compare_eulerangles_with_NR}

\begin{figure*}[thbp]
\begin{center}
\includegraphics[width=\columnwidth]{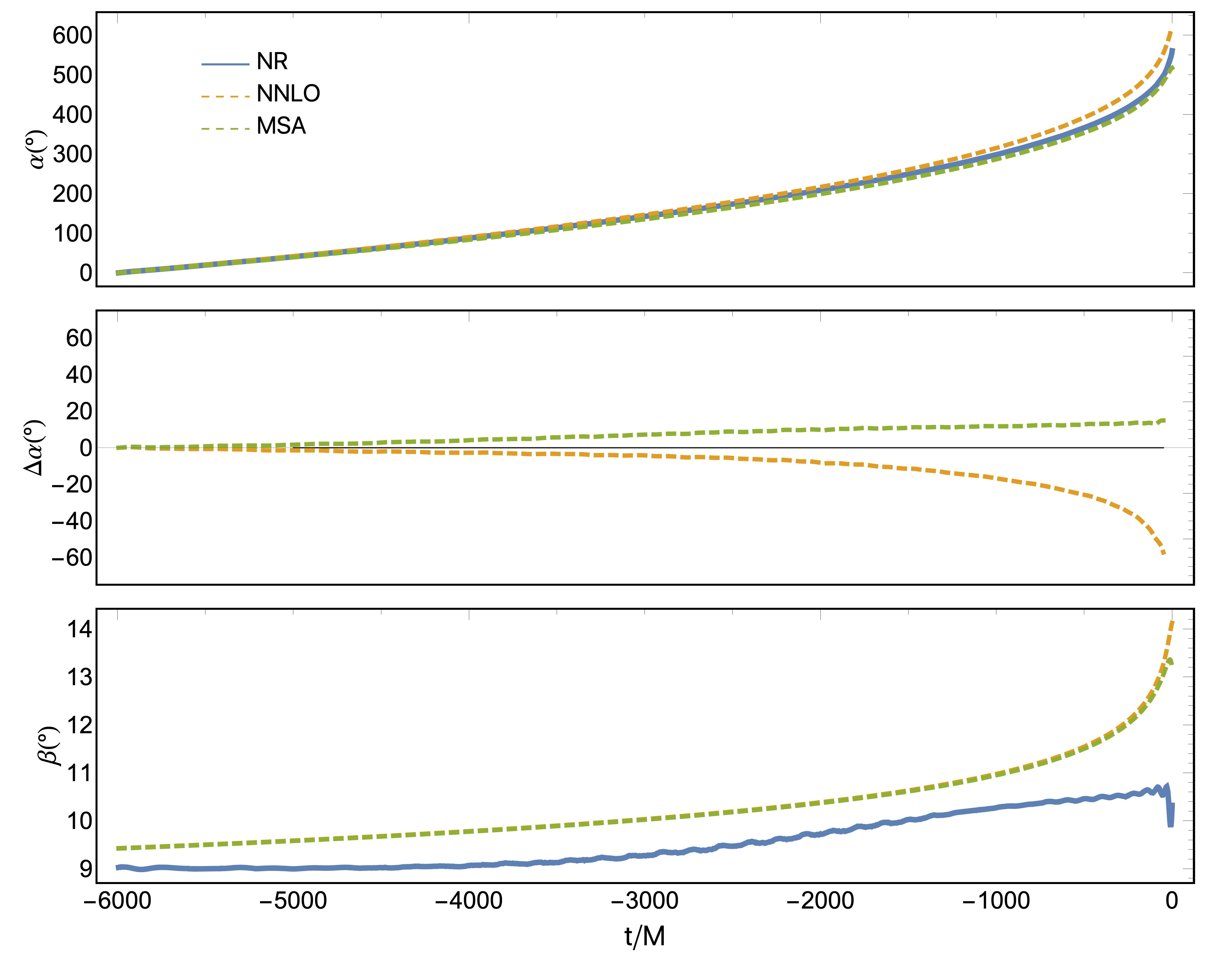}
\includegraphics[width=\columnwidth]{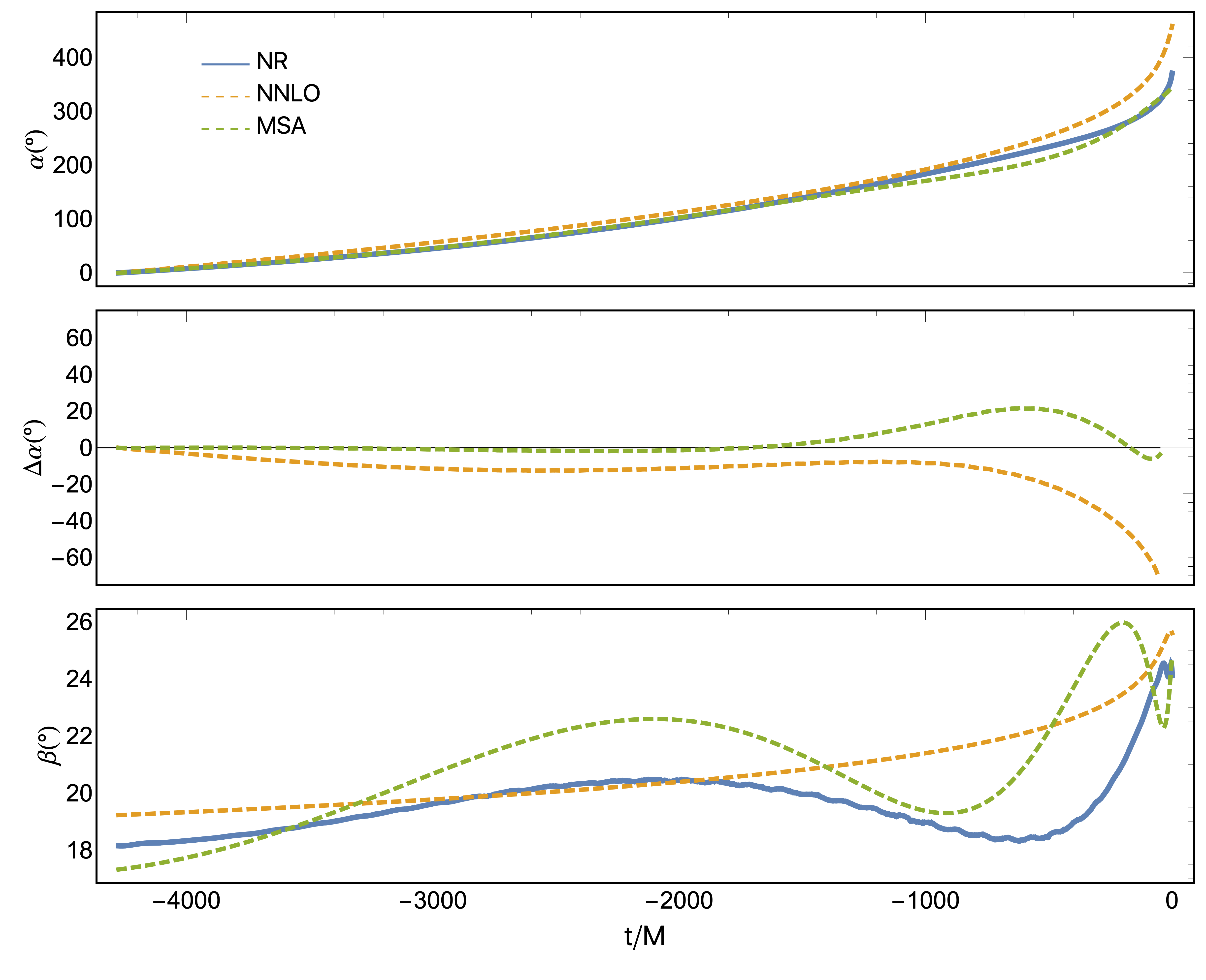}
\caption{Comparison of Euler angles with Numerical Relativity. Left: SXS:BBH:0094 ($q=1.5$, $\boldsymbol{\chi}_1=(0.5,0,0)$, $\boldsymbol{\chi}_2=(0,0,0)$). Right: SXS:BBH:0053 ($q=3$, $\boldsymbol{\chi}_1=(0.5,0,0)$, $\boldsymbol{\chi}_2=(-0.5,0,0)$). Solid blue: Quadrupole-aligned Euler angles extracted from NR simulation. Dashed orange: NNLO implementation (version 102). Dashed green: MSA implementation (version 223).}
\label{fig:EulerAnglesNR}
\end{center}
\end{figure*}

Both descriptions for the precession angles implemented in our model, and described in Sec. \ref{sec:single_spin_angles} and Sec. \ref{sec:double_spin_angles}, are based on PN analytical approximations to the solution of the angular momenta evolution equations and therefore are expected to lose accuracy when the assumptions of the PN formalism start to fail as the frequency becomes too high.
A full systematic understanding of the limitations of both descriptions is out of the scope of this work, but to illustrate the differences between both descriptions, in Fig. \ref{fig:EulerAnglesNR} we compare them with two precessing simulations from the SXS catalog \cite{Boyle:2019kee,SXS:catalog}: a single-spin simulation [SXS:BBH:0094 with mass ratio $q=1.5$ and initial dimensionless spin vectors $\boldsymbol{\chi}_1=(0.5,0,0)$ and $\boldsymbol{\chi}_2=(0,0,0)$] and a double-spin simulation [SXS:BBH:0053 with $q=3$, $\boldsymbol{\chi}_1=(0.5,0,0)$ and $\boldsymbol{\chi}_2=(-0.5,0,0)$].

Here the Euler angles of the NR simulations are computed with the ``quadrupole alignment'' procedure, see \cite{Schmidt:2010it,Boyle:2011gg} and \cite{Ramos-Buades:2020noq} for a recent discussion in the context of waveform modelling.
For the NNLO description outlined in Sec.~\ref{sec:single_spin_angles}, the in-plane spin is described by the single constant quantity $\chi_p$ defined in Eq.~(\ref{def:chip}).
In contrast, the MSA description (summarized in Sec.~\ref{sec:double_spin_angles}) contains information about both individual spins and is able to predict the evolution of the norm of the total spin $\mathbf{S}=\mathbf{S}_1+\mathbf{S}_2$, which allows it to capture the time/frequency dependent oscillations of the Euler angles on the precession timescale caused by the evolution of the norm of the total spin.

For the single-spin simulation shown in the left panel of Fig. \ref{fig:EulerAnglesNR}, both descriptions present a very similar behaviour for the opening angle $\beta$ and for the inspiral cycles in the precessing angle $\alpha$, though the MSA description seems to remain closer to NR as the end of the inspiral is approached. For the double-spin case in the right panel of Fig.~\ref{fig:EulerAnglesNR}, one can see that the behaviour of the precessing angle $\alpha$ during the inspiral is better reproduced by the MSA description and the MSA opening angle $\beta$ can also reproduce the oscillatory structure observed in the NR simulation. The oscillations due to double-spin effects dephase however relative to NR as the end of the inspiral is approached, which can even lead to a worse description of the late inspiral than the one provided by the NNLO single-spin description, as seen in the example for the precessing angle $\alpha$. Strategies to improve the behaviour of the PN precessing angles descriptions in the high-frequency regime will be addressed in future work.

%%%%%%%%%%%%%%%%%%%%%%%%%%%%%%%%%%%%%%%%%%%%%%%%%%%%%%%%%%%%%%%%%
%               TD WAVEFORMS
%%%%%%%%%%%%%%%%%%%%%%%%%%%%%%%%%%%%%%%%%%%%%%%%%%%%%%%%%%%%%%%%%

\subsection{Time Domain Waveforms}\label{sec:TD_waveforms}

\begin{figure*}[htbp]
\centering
\includegraphics[width=2\columnwidth]{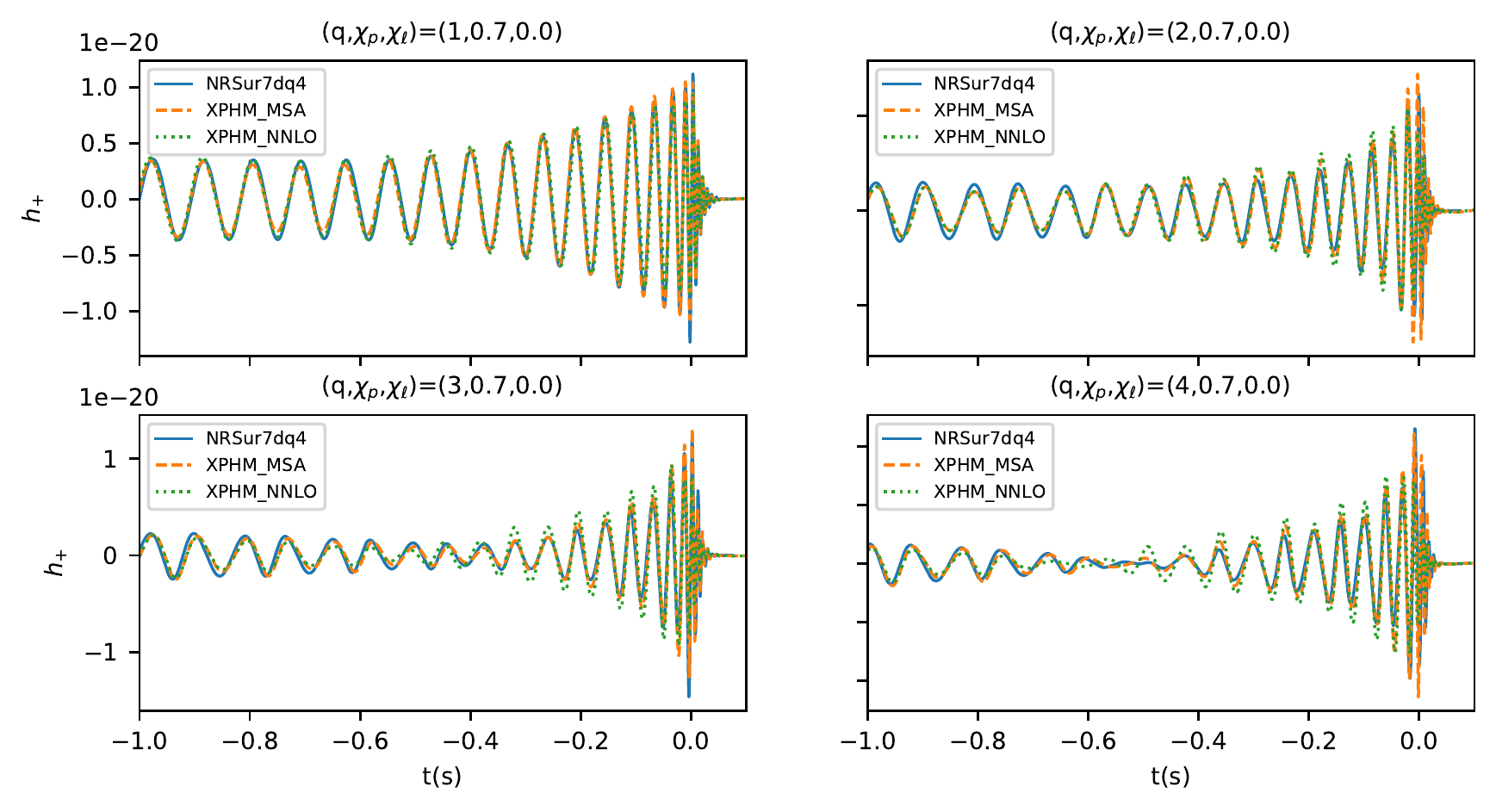}
\caption{We compare the relative performance of NNLO and MSA angles against \NRSur, by plotting the plus polarization returned by the different models for a double-spin configuration with high in-plane spins. Each panel refers to the same spin configuration, but we allow the mass ratio to vary from 1 to 4, which is the upper limit of the calibration region of \NRSur. Overall NNLO angles perform well, although we do observe some disagreement with respect to \NRSur as we increase the mass ratio (lower panels).}
\label{fig:table-TD_wf}
\end{figure*}

\begin{figure*}[htbp]
\centering
\includegraphics[width=2\columnwidth]{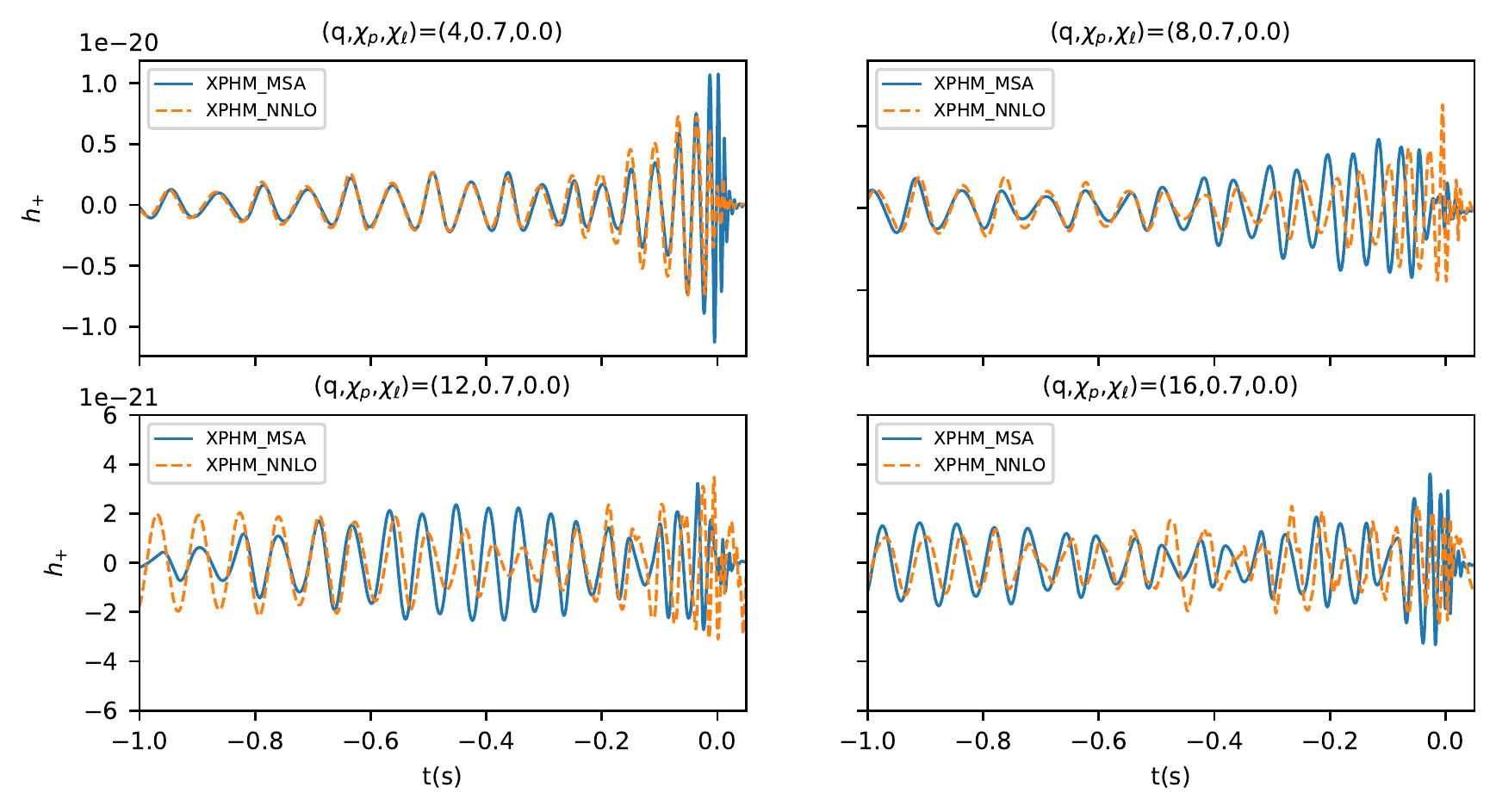}
\caption{We compare the behaviour of the two twisting-up methods implemented in \phXPHM for the same spin configuration chosen in Fig.~\ref{fig:table-TD_wf}, varying the mass ratio between 4 and 16. For $q=4$ we observe good agreement between the two angle descriptions, especially during the inspiral. However, as the mass ratio increases the agreement degrades and NNLO angles tend to produce non-smooth features in the waveform. }
\label{fig:table-TD_high_q_wf}
\end{figure*}

To best appreciate the differences between precessing waveforms constructed using NNLO and MSA Euler angle prescriptions, we have generated time-domain waveforms with \phXPHM for both versions of the "twisting-up" angles, and compared with the precessing surrogate model \NRSur~\cite{Varma:2019csw}. In Fig.~\ref{fig:table-TD_wf} we plot the cross polarization for a double-spin configuration with high in-plane spins, varying the mass ratio and aligning them in time and phase.
For increasing $q$, the MSA description tends to stay closer to \NRSur. The differences between the two descriptions become particularly strong for high mass-ratio systems, as shown in Fig. \ref{fig:table-TD_high_q_wf}, with the MSA description appearing to be more stable in this regime. This is particularly evident in the lower panels, where we show a $q=12$ and a $q=16$ configuration. Notice that the non-smoothness of the NNLO angles in some regions of the parameter space led us to impose a more stringent threshold on the multibanding of the Euler angles for this precession version (see Sec.\ref{subsec:multibanding} for a detailed discussion).

%%%%%%%%%%%%%%%%%%%%%%%%%%%%%%%%%%%%%%%%%%%%%%%%%%%%%%%%%%%%%%%%%
%                    MATCHES
%%%%%%%%%%%%%%%%%%%%%%%%%%%%%%%%%%%%%%%%%%%%%%%%%%%%%%%%%%%%%%%%%

\subsection{Match Calculations for Precessing Waveforms}\label{sec:match_algorithm}

In order to check the agreement between our waveform model and other descriptions we follow standard practice and compute matches between waveforms across a portion of the parameter space.
In Sec.~\ref{sec:NR_matches} we present matches between our model and numerical relativity waveforms, and
in Sec.~\ref{sec:model_matches} we compare with other waveform models. 
As in our previous work \cite{Pratten:2020fqn,Garcia-Quiros:2020qpx} we use the standard definition of the inner product (see e.g.~\cite{Cutler:1994ys}),
\begin{equation}
\Braket{h_1, h_2} = 4 \Re \int_{f_{\min}}^{f_{\max}} \frac{\tilde{h_1}(f) \:\tilde{h^*_2}(f)}{S_\mathrm{n}(f)},
\end{equation}
where $S_\mathrm{n}(f)$ is the one-sided power spectral-density of the detector noise. 
The \textit{match} is defined as this inner product divided by the norm of the two waveforms and maximized over relative time and phase shifts between both of them,
\begin{equation}
\mathcal{M}(h_1,h_2) = \max_{t_0, \phi_0} \frac{\Braket{h_1, h_2}}{\sqrt{\Braket{h_1, h_1}}\sqrt{\Braket{h_2, h_2}}}.
\end{equation}
It is advantageous to visualize deviations between waveforms in terms of the mismatch rather than the match, where the mismatch is defined as
\begin{equation}
\mathcal{MM}(h_1,h_2)=1-\mathcal{M}(h_1,h_2).
\end{equation}
We use the Advanced-LIGO~\cite{TheLIGOScientific:2014jea} design sensitivity Zero-Detuned-High-Power Power Spectral Density (PSD) \cite{adligopsd} with a lower cutoff frequency for the integrations of 20 Hz and an upper cutoff at 2048 Hz. 
We analytically optimize over the template polarization angle, following \cite{PhysRevD.94.024012}, and numerically optimize over reference phase and rigid rotations of the in-plane spins at the reference frequency. We do this rather than optimizing over the reference frequency as suggested in \cite{Khan:2019kot}, as this allows to set unambiguous bounds for the parameters involved in the optimization. In order to perform the numerical optimization we use the dual annealing algorithm as implemented in the {\tt SciPy} {\tt Python} package \cite{2020SciPy-NMeth}.

\subsubsection{Matches Against SXS Numerical Relativity Simulations}\label{sec:NR_matches}

\begin{figure}
 {   \centering
    \includegraphics[width=\columnwidth]{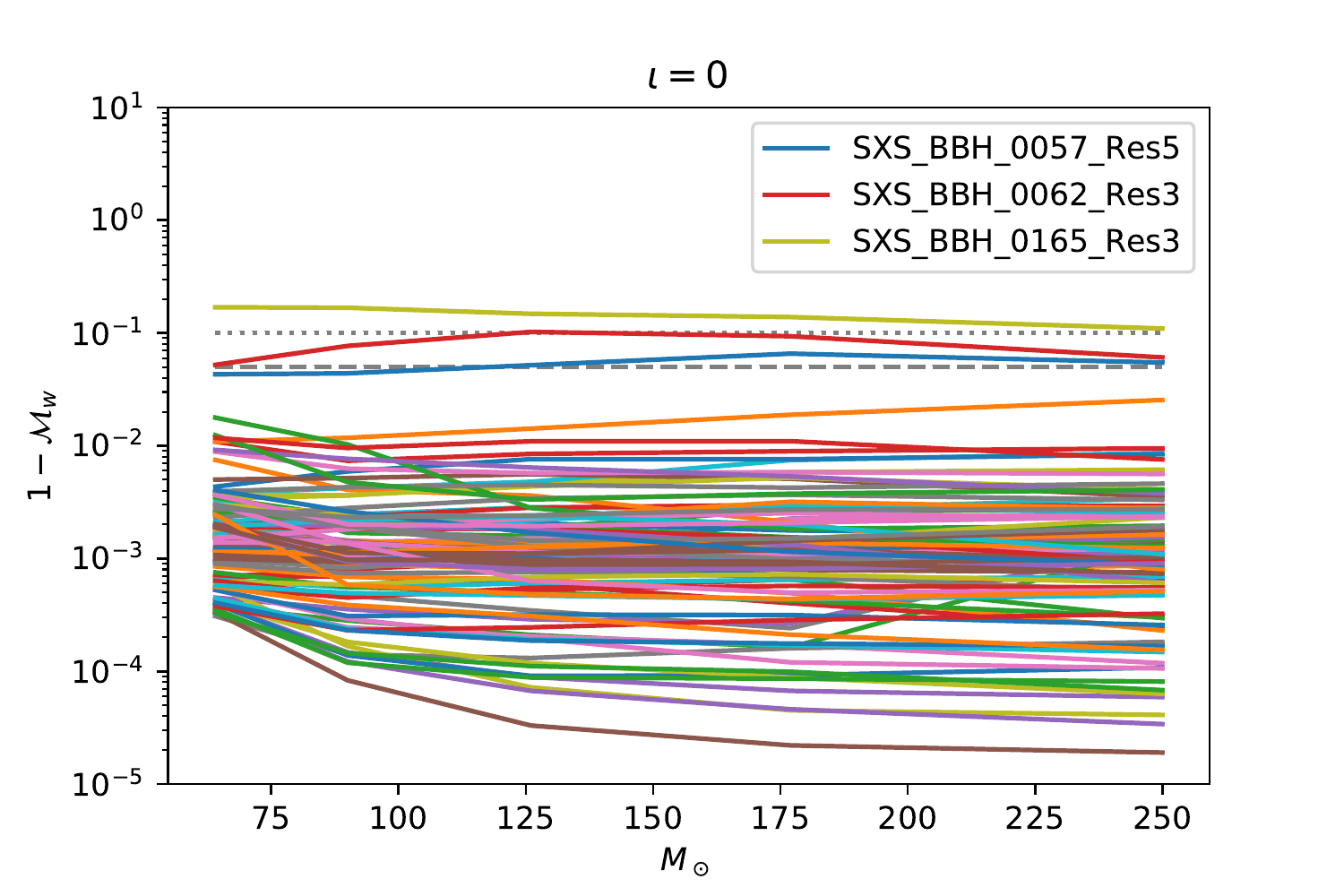}\\
    \includegraphics[width=\columnwidth]{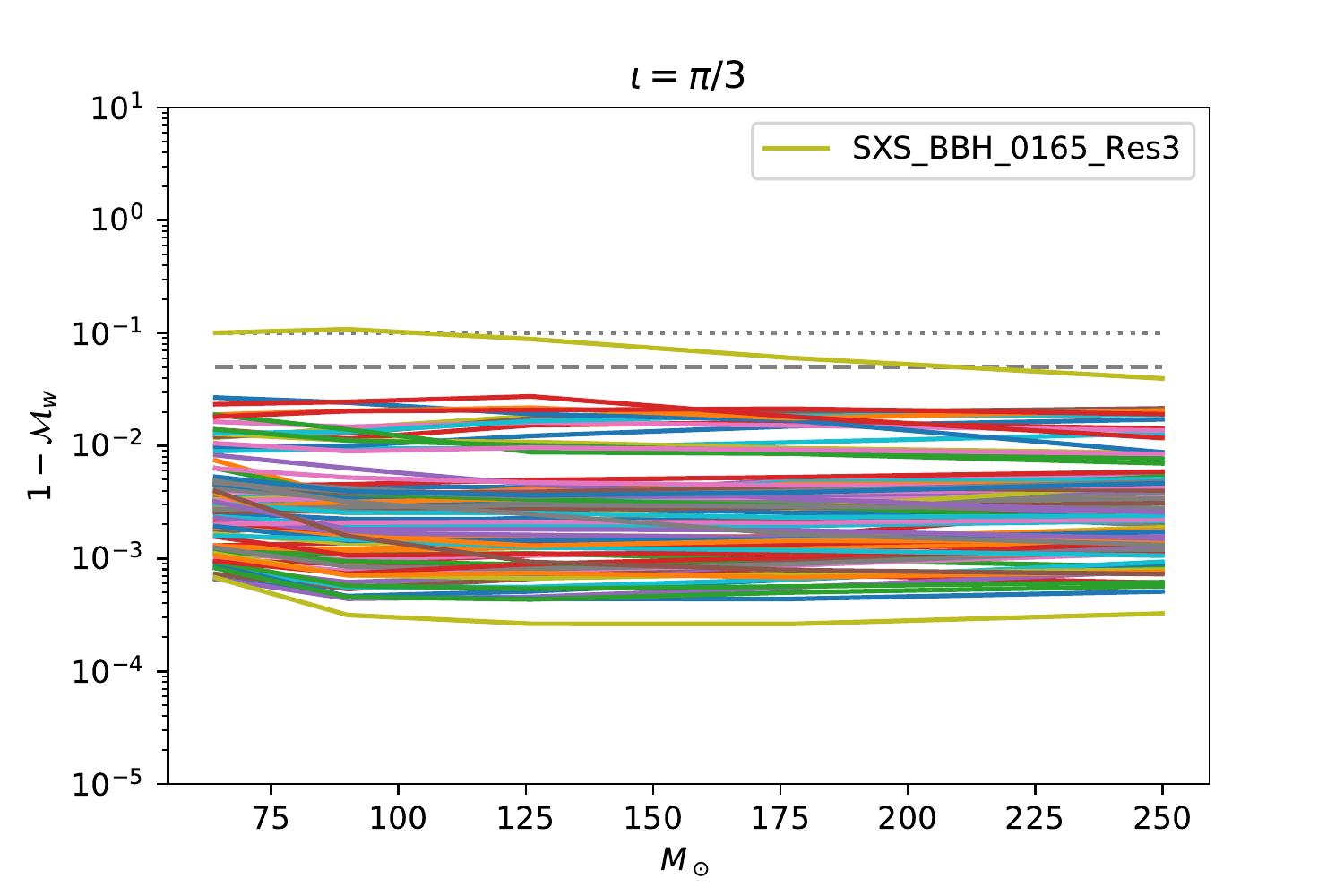}\\
    \includegraphics[width=\columnwidth]{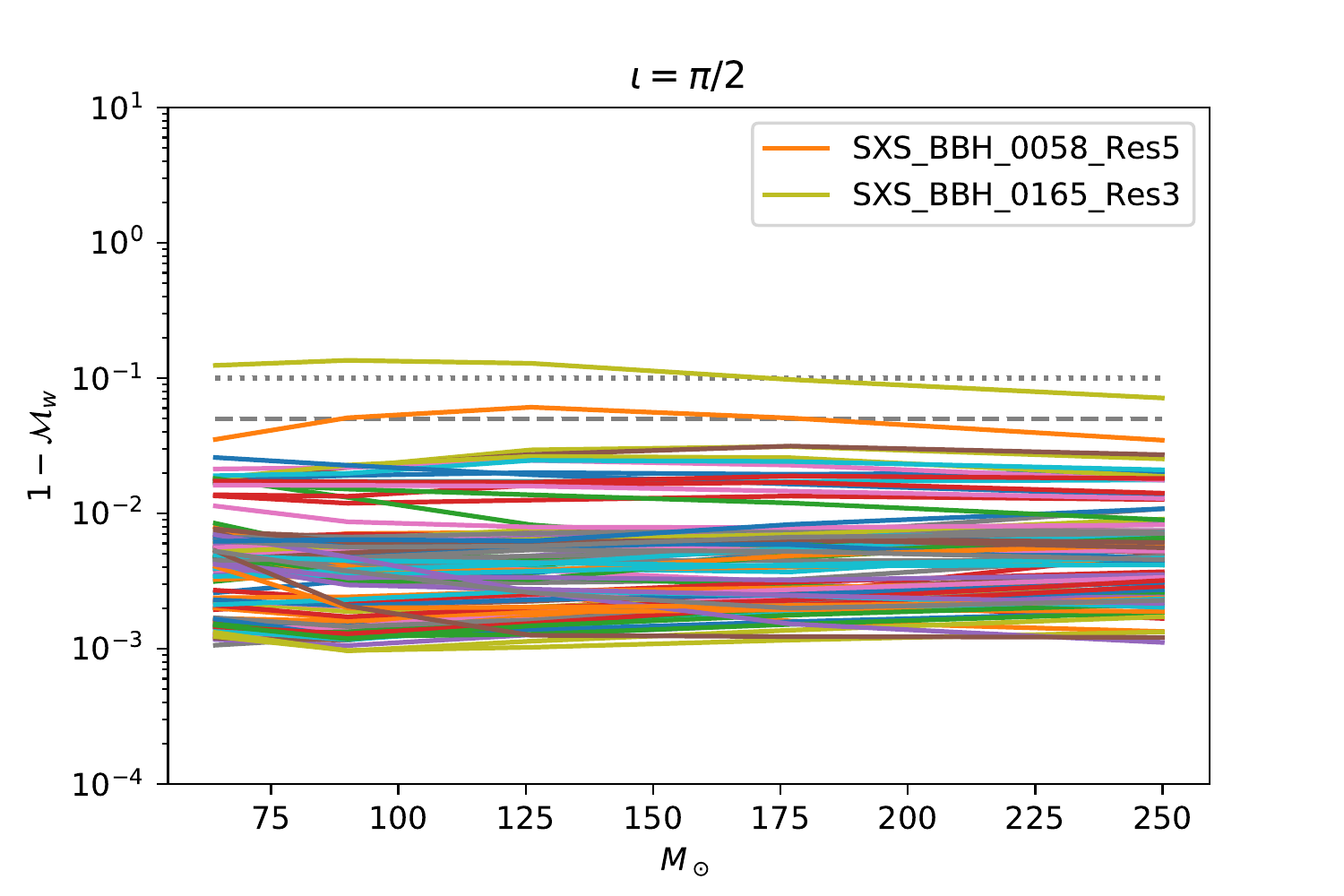}
  }
    \caption{In the plots above, we show the SNR-weighted mismatch $1-\mathcal{M}_w$ of \phXPHM against 99 SXS precessing waveforms of the $\texttt{lvcnr}$ catalog, in order of ascending inclination. A dashed and a dotted line mark the $5\%$ and $10\%$ mismatch thresholds respectively.}
    \label{fig:SXS_matches}
\end{figure}

We have computed mismatches for \phXPHM against 99 precessing SXS waveforms~\cite{Boyle:2019kee,SXS:catalog}, picking for each binary configuration the highest resolution available in the \texttt{lvcnr} catalog \cite{Schmidt:2017btt}. As a lower cutoff for the match integration, we took the minimum between 20 Hz and the starting frequency of each NR waveform. We repeated the calculation for three representative inclinations between the orbital angular momentum and the line of sight $(0, \pi/3, \pi/2)$ and total masses ranging from 64 $M_\odot$ to 250 $M_\odot$. 
As low matches tend to be correlated with low signal-to-noise ratio (SNR) and, therefore, with a lower probability for the signal to be observed, we compute here the SNR-weighted match $\mathcal{M}_{\mathrm{w}}$ ~\cite{Harry:2016aa}
\begin{equation}
\mathcal{M}_{\mathrm{w}}=\left(\frac{\sum_i\mathcal{M}_i^3 \Braket{h_{i,\mathrm{NR}}, h_{i,\mathrm{NR}}}^{3/2}}{\sum_i{\Braket{h_{i,\mathrm{NR}},h_{i,\mathrm{NR}}}^{3/2}}}\right)^{1/3},
\end{equation}
where the subscript $i$ refers to different choices of polarization and reference phase of the source i.e. in our case of the NR waveform.  
The results are shown in Fig.~\ref{fig:SXS_matches}.
The large majority of the cases considered here resulted in mismatches between $10^{-3}$ and $10^{-2}$, with a consistent number of cases below $10^{-3}$ for face-on sources.

We observed, however, three cases for which matches are below $95\%$ for at least one value of the inclination (\texttt{SXS:0057, SXS:0058, SXS:0062}) and one case where this happens for all the inclinations (\texttt{SXS:0165}). These all correspond to high mass-ratio, strongly precessing binaries: \texttt{SXS:0057, SXS:0058, SXS:0062} are $q=5$ simulations with $\chi_p\geq 0.4$ and \texttt{SXS:0165} is a $q=6$ simulation with $\chi_p=0.78$. 
For this type of systems, the complex interaction between different waveform multipoles can result in a non-trivial dependence of the SNR on the orientation of the source, with face-on configurations not being necessarily favoured (see, for instance, \cite{PhysRevD.95.104038} for a related discussion). We observe this in \texttt{SXS:0057, SXS:0062, SXS:0165}, for which the highest values of SNR do not necessarily concentrate around zero inclination. This explains why for these simulations the match increases, rather than decreases, with the inclination of the source. 

\subsubsection{Matches Against other Models}\label{sec:model_matches}

We now turn to computing the mismatch with other waveform models. In contrast to the comparison with numerical relativity waveforms shown in Sec.~\ref{sec:NR_matches}, where SNR-weighted mismatches are presented, we show ``raw'' mismatches between models, without weighting them. We compute matches in the calibration regime of the \NRSur model, $ 1\leq m_1/m_2 \leq 4$ and dimensionless spin magnitudes up to $0.8$.

We compare against a number of other waveform models, which are routinely used for gravitational wave data analysis:
\begin{itemize}
    \item Previous models from the phenomenological waveform family including \phD~\cite{Husa:2015iqa,Khan:2015jqa}, \phHM~\cite{London:2017bcn}, \phPvthree~\cite{Khan:2018fmp} and \phPvthreehm~\cite{Khan:2019kot}, and the spin-aligned basis waveforms of the new \phX family: \phX~\cite{Pratten:2020fqn} and \phXHM~\cite{Garcia-Quiros:2020qpx}.
    \item A NR surrogate model \NRSur~\cite{Varma:2019csw} that interpolates between NR waveforms, calibrated for precessing simulations up to mass ratios of $q=4$ and spin magnitudes up to $0.8$.
    \item A similar non-precessing surrogate model \NRHybSur \cite{Varma:2018mmi}, calibrated to aligned-spin hybrid waveforms up to mass ratios of $q=8$ and spin magnitudes up to $0.8$.
    \item A non-precessing, higher-modes model \seobnrvforhmrom \cite{Cotesta:2020qhw} which is a reduced order model of \seobnrvforhm \cite{Bohe:2016gbl,Cotesta:2018fcv}, an effective-one-body model calibrated to numerical relativity waveforms.
\end{itemize}

\begin{figure}[htbp]
%\begin{center}
\includegraphics[width=\columnwidth]{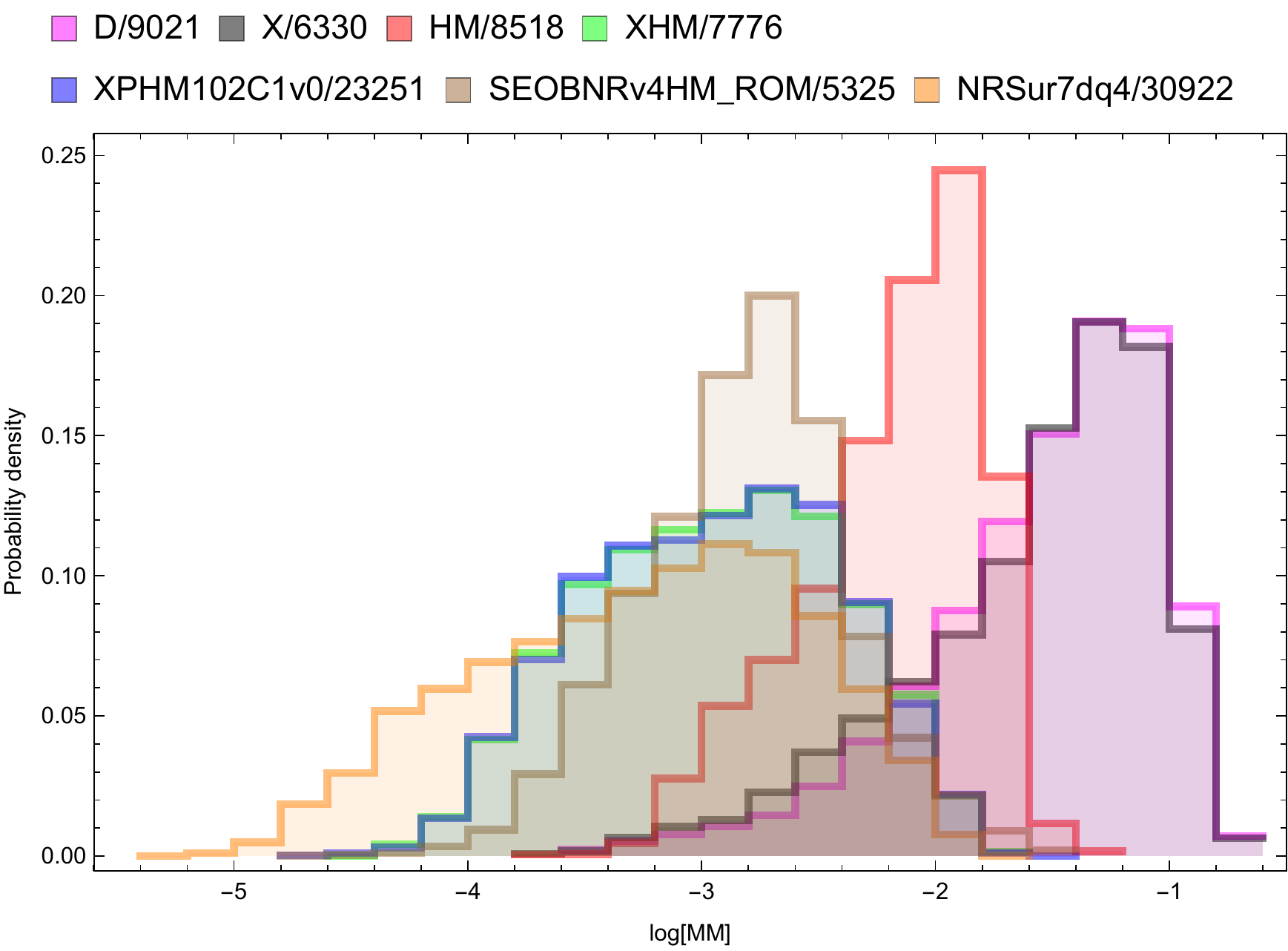}
\includegraphics[width=\columnwidth]{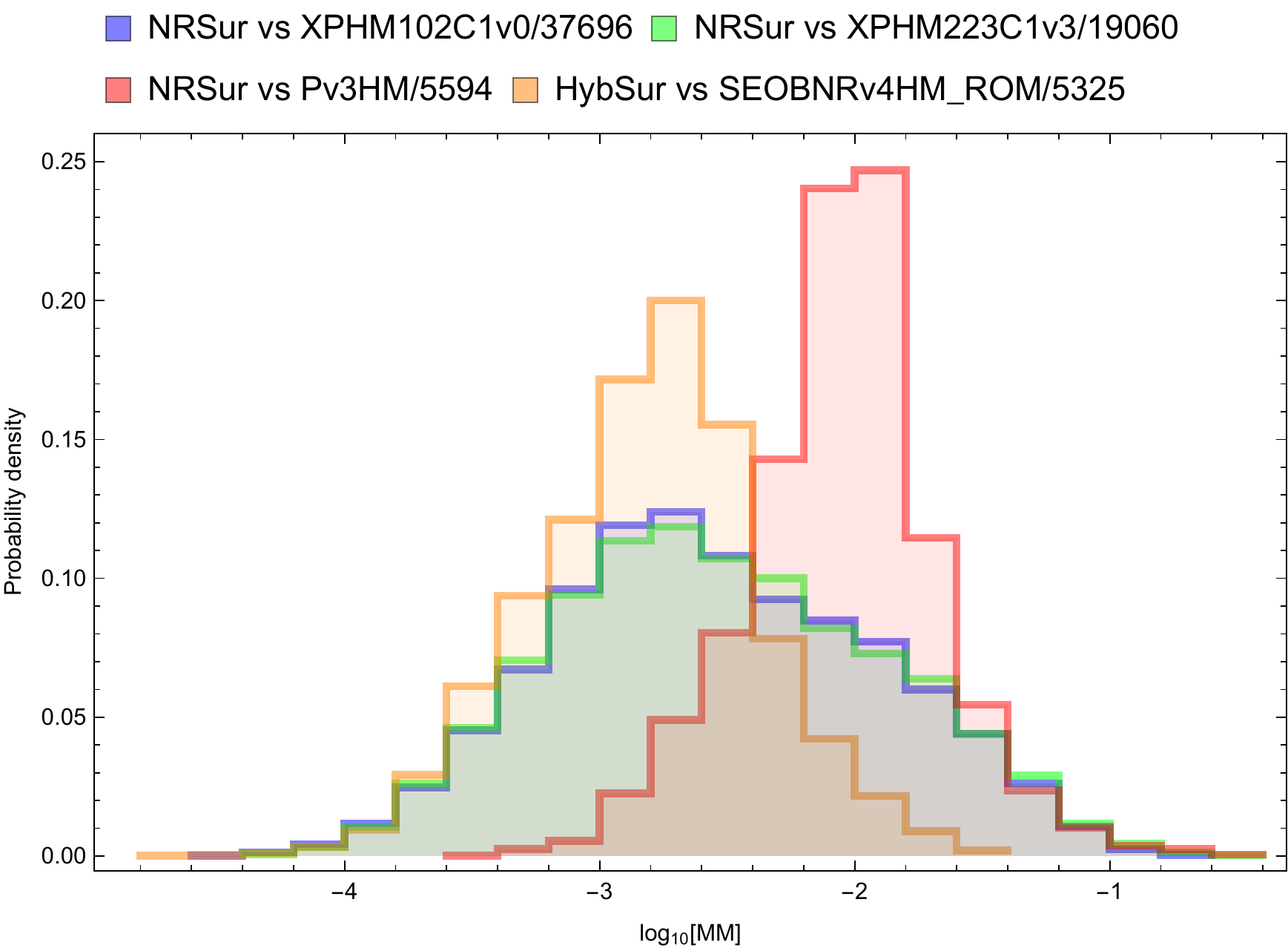}
\caption{In the upper panel we compare the non-precessing models \phD, \phX, \phHM, \phXHM, \seobnrvforhmrom
and the precessing \phXPHM  and \NRSur models with the non-precessing \NRHybSur model as discussed in the main text. In the lower panel we compare two versions of our \phXPHM model (NNLO-based version 102 with final spin version 0 and MSA-based version 223 with final spin version 3, see Table~\ref{tab:prec_version}), and  \phPvthreehm  with \NRSur, as discussed in the main text. For comparison we show also the mismatch between \NRHybSur and \seobnrvforhmrom as displayed in the upper panel. The number of samples for each comparison is indicated in the legend.}
\label{fig:ModelMatches}
%\end{center}
\end{figure}

We choose \NRSur as the reference model for high mass precessing waveforms, where higher mode contributions are significant, since this is still the only precessing model calibrated to precessing NR waveforms. Due to the limited length of the NR waveforms used to calibrate the model (4000 total mass units), we restrict to large masses above 90 solar masses and compute the mismatch for random values of the total mass taken from the list $(90, 126, 177, 249, 350) M_{\odot}$. Note that for large masses, the impact of higher mode effects and precession effects in the strong field regime on the waveform is more pronounced. 
In Fig.~\ref{fig:ModelMatches} we show mismatches both with the precessing \NRSur \cite{Varma:2019csw} and the non-precessing \NRHybSur \cite{Varma:2018mmi}. The comparisons with the latter allow to put the mismatches we see for precessing higher-modes models into the context of mismatches in the non-precessing case, where waveform models are significantly more mature.

In the upper panel of Fig.~\ref{fig:ModelMatches}
we also show the comparison of models that only contain the $\ell=|m|=2$ modes with  \NRHybSur. 
One can see that while \phX is significantly more accurate than \phD as discussed in \cite{Pratten:2020fqn},
this only yields a small advantage when comparing raw mismatches with a higher-modes model.
A model that does include higher modes, even when those are not calibrated to NR, such as \phHM, gains significant accuracy. The relative gain from calibrating higher modes is however comparable. The difference between 
the \phXHM and \seobnrvforhmrom \cite{Cotesta:2020qhw} models is small, in particular considering that they do not describe the same set of sub-dominant harmonics, with \phXHM having a larger fraction of very accurate waveforms. We have also included a variant of our precessing \phXPHM model (variant 102 based on NNLO angles and final spin version 0, see Table~\ref{tab:prec_version}). One can see that results are consistent with the manifestly non-precessing model \phXHM (up to sampling errors), which provides an end-to-end test of consistent behaviour of our new model in the aligned-spin limit. A number of more stringent tests of the appropriate aligned-spin limit have been carried out as part of the LALSuite code review.
Finally, \NRSur is most consistent with \NRHybSur, but the advantage is not very pronounced, and is likely to be significantly reduced by adding further harmonics to \phXPHM, in particular the $\ell=|m|=5$ modes already present in \seobnrvforhmrom and the $\ell=4, |m|=3$ modes present in \phHM.

In the lower panel of Fig.~\ref{fig:ModelMatches} we finally show mismatches against the precessing \NRSur model. One can see that the distributions of mismatches are roughly similar to the non-precessing case, but with a tail of high mismatches, which is similar to \phPvthreehm. The tail of small mismatches is similar to that when comparing the two non-precessing models \seobnrvforhmrom and \NRHybSur, while in the bulk \phXPHM clearly outperforms \phPvthreehm,
which is not calibrated to numerical data for subdominant harmonics.

%%%%%%%%%%%%%%%%%%%%%%%%%%%%%%%%%%%%%%%%%%%%%%%%%%%%%%%%%%%%%%%%%
%                    MULTIBANDING
%%%%%%%%%%%%%%%%%%%%%%%%%%%%%%%%%%%%%%%%%%%%%%%%%%%%%%%%%%%%%%%%%

\subsection{Multibanding and Euler Angles}
\label{subsec:multibanding}
In \cite{Garcia-Quiros:2020qlt} we have discussed our implementation of an algorithm to accelerate waveform evaluation by first evaluating the waveform on a coarse unequispaced grid,
before linear interpolation to an equispaced fine grid, following \cite{Vinciguerra:2017ngf}. The grid spacing on the coarse grid is chosen to satisfy a given error threshold for linear interpolation (a different criterion to set the grid spacing has previously been used in \cite{Vinciguerra:2017ngf}). An iterative expression can then be used to accelerate the evaluation of computationally expensive trigonometric expressions, such as those required to compute the strain from the phase (and amplitude).

In \cite{Garcia-Quiros:2020qlt} we derived simple estimates to set the grid spacing in terms of the phase errors and relative amplitude errors as a function of the grid spacing, and we have implemented a conservative default threshold of $10^{-3}$ radians of local phase error and of relative amplitude error  $10^{-3}$.

Here we apply the same idea to the Euler angles. For the inspiral, in \cite{Garcia-Quiros:2020qlt} we have derived the required grid spacing for accurate linear interpolation from  
the leading singular term of the TaylorF2 phase expression for the gravitational wave phase of spherical harmonic mode $h_{\ell m}$, which reads
\cite{Buonanno:2009zt},
\begin{equation}
\Phi_{\ell m} = \frac{m}{2} \frac{3}{128 \eta} \left(\frac{2\pi f}{m}\right)^{-5/3},
\end{equation}
where $\eta$  is the symmetric mass ratio and constants of integration that do not affect the second derivative, and thus the error estimate, have been dropped. 
The leading term for the NNLO angles $\alpha$ and $\epsilon$ is the same
\begin{equation}
    \alpha  = \left(-\frac{5 \delta }{64 m_1}-\frac{35}{192}\right) (\pi f)^{-1} + \mbox{higher order terms},
\end{equation}
see appendix \ref{appendix:nnlo_angles}.
Similar to the evaluation of the inspiral gravitational wave phase, we need to 
evaluate expressions of the type
\begin{equation}\label{eq:alpha_exponential}
e^{i m \alpha(2\pi f/m)},
\end{equation}
for spherical harmonic modes $h_{\ell m}$,
see Appendix~\ref{appendix:derivation}, and we thus have to apply multibanding interpolation to the arguments of the complex exponentials of the type in Eq.~(\ref{eq:alpha_exponential}).
The ratio of the maximal allowed step sizes for achieving the same interpolation error for the gravitational wave phase and Euler angles is thus given by the (inverse) ratio of second derivatives with respect to the frequency $f$, which evaluates to
\begin{equation}
 \frac{df_{\Phi}}{df_{\alpha}} = \left\vert \frac{m \alpha''(2\pi f/m)}{\Phi''_{\ell m} } \right\vert = \frac{(\pi f)^{2/3} \left(13 \sqrt{1-4 \eta }+7\right) \eta
   }{\sqrt{1-4 \eta }+1},
\end{equation}
which is smaller than unity during the inspiral ($f \leq 0.1$) and vanishes both in the low-frequency and extreme-mass-ratio limits. The third Euler angle $\beta$ is a regular function during the inspiral, and thus does not require high resolution for accurate interpolation.

\begin{figure}[thpb]
    \centering
    \includegraphics[width=\columnwidth]{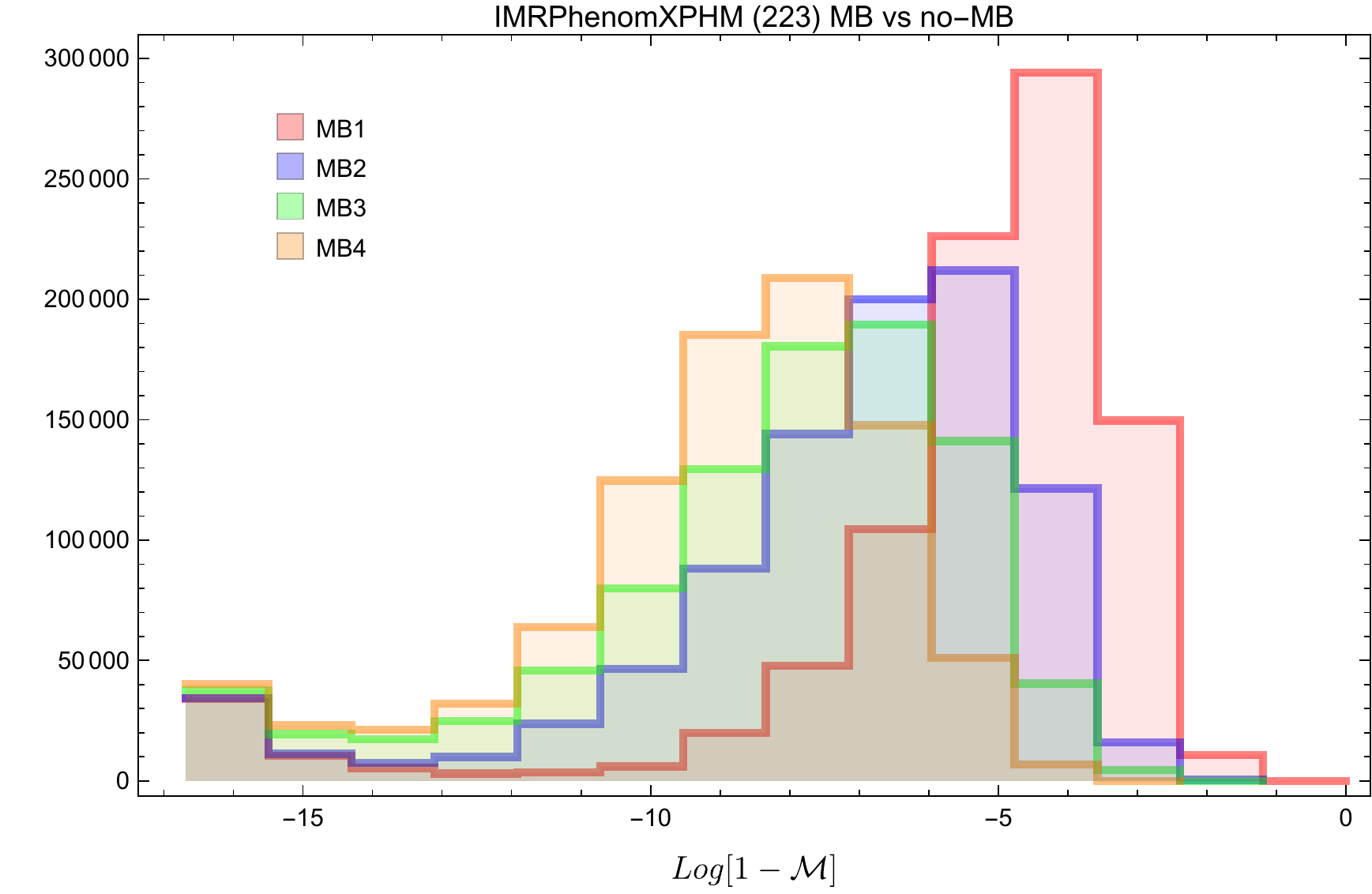}
    \includegraphics[width=\columnwidth]{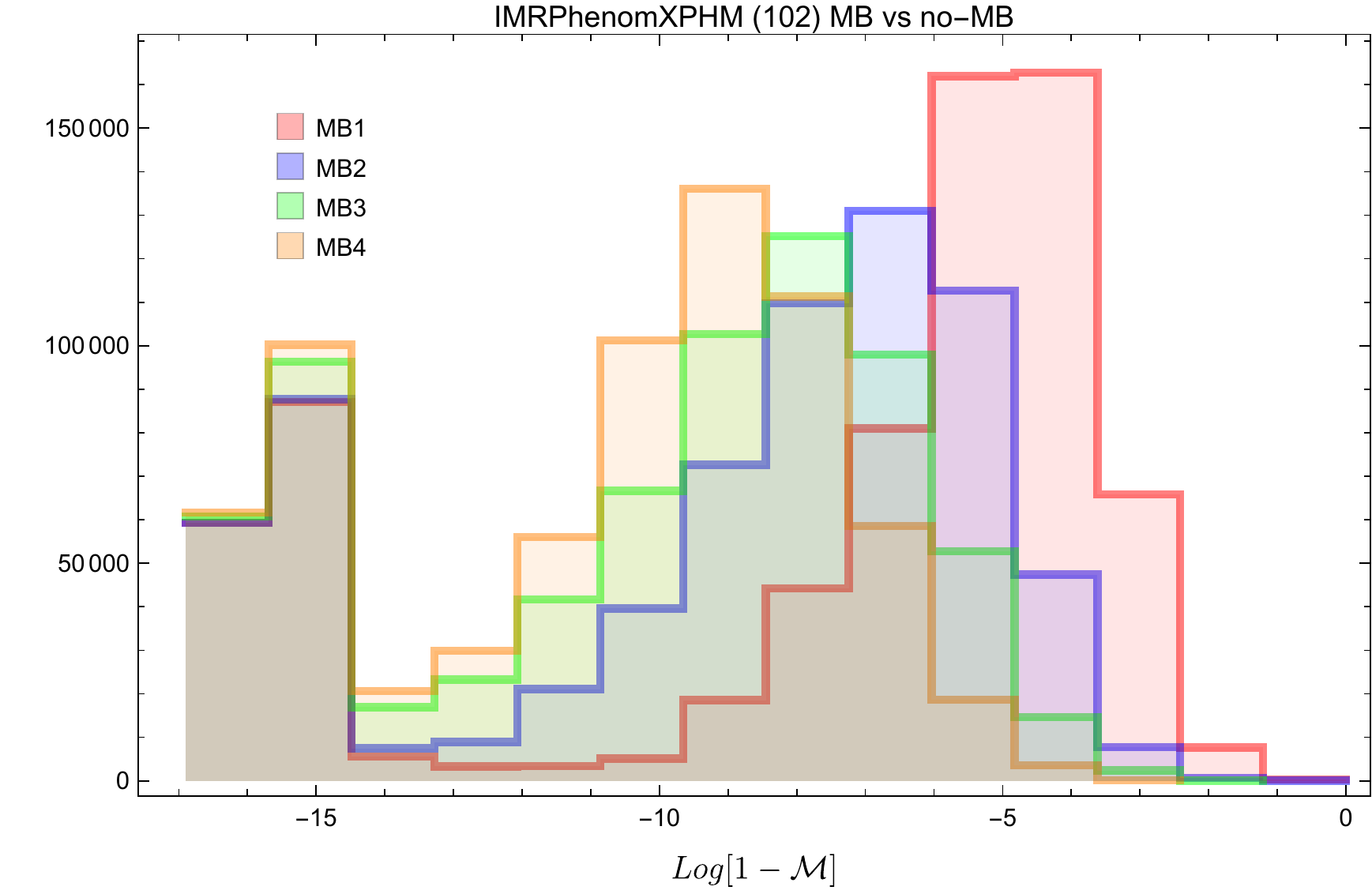}
\caption{Histograms of a mismatch calculation for $h_+$ between versions of \phXPHM without using multibanding for the Euler angles, and with four different levels of multibanding threshold: $10^{-1}$, $10^{-2}$, $10^{-3}$, $10^{-4}$.
The top panel show the results for the MSA prescription (version 223) and the bottom panel shows the result for NNLO angles (version 102).
$10^6$ waveforms were generated across a parameter space as described in the main text. The threshold $10^{-3}$ (MB3 in the plot) has been chosen as the default value in the LALSuite implementation. 
}
    \label{fig:MBerrors}
\end{figure}

During the merger and ringdown the angles have a simpler functional form than the gravitational wave phase, which is characterized by a Lorentzian. The exponential falloff of the mode amplitudes in the ringdown phase also requires significant resolution. The Euler angles in turn carry significant systematic errors, e.g. due to applying the SPA approximation for the whole waveform. 
Note that the MSA prescription for the angles causes oscillations in the angle $\beta$, however the angle prescriptions still broadly agree with the NNLO description.

Future work will attempt to calibrate the angle descriptions to numerical relativity and better understand the phenomenology during the merger and ringdown, which in turn will require improved estimates for the required grid spacing in order to not loose accuracy due to multibanding.

At the current level of accuracy produced from the precession angle models, it does not seem necessary to attempt more precise prescriptions to apply multibanding to the Euler angles. For simplicity we thus use the same coarse grid for each spherical harmonic mode that we have utilized in \cite{Garcia-Quiros:2020qlt}. To quantitatively assess the impact that multibanding of the Euler angles has on the precessing waveforms, we compute matches between the original waveform, generated without angle multibanding, and waveforms produced with the identical parameters except with multibanding, varying the multibanding threshold between 0.1 and 0.0001.

The results of this comparison are shown in Fig.~\ref{fig:MBerrors} for waveforms twisted up using the MSA angles (top panel) and the NNLO angles (bottom panel). 
%and generated over the broad parameter space range of \phXPHM (symmetric mass ratio $\eta \geq 0.01$ total mass $0 \leq M \leq 500$,
They are generated over a broad parameter space  range with $1 \leq m_{1,2} \leq 500$ and dimensionless spin magnitudes up to unity, corresponding to the extreme Kerr limit. The frequencies span from 10 to 1024 Hz and the grid spacing $df$ ranges from 0.01 to 0.3 Hz. In typical Bayesian inference applications, the value of $df$ is not chosen randomly but adjusted to the segment length of the data to be analyzed, which is itself adjusted to the time a signal is observable in the sensitive band of the detector. Here we have chosen to use a random $df$ which could lead to downsampled waveforms and hence worse matches, however the random $df$ allows us to stress-test the robustness of the multibanding algorithm and check that any kind of uniform frequency grid is supported. 

The results in Fig.~\ref{fig:MBerrors} show that indeed the lower the threshold the better is the match (at the expense of loosing speed). There is a tail of very low mismatches which is much more pronounced for version 102 than for 223; this tail corresponds to cases where the multibanding was switched off automatically by the code and hence the match is close to machine precision. The multibanding is automatically switched off in the following cases:
\begin{itemize}
    \item For total mass $M_{\mathrm{tot}}$ higher than 500 $M_{\odot}$. This cutoff is already present in the non-precessing model \phXHM and is motivated by the short length of the waveform in the frequency band of the detector for these massive systems, which renders multibanding less efficient but also unnecessary.
    \item When using MSA angles: for $q > 50$ and $M_{\mathrm{tot}} > 100 \:M_{\odot}$. This corner of the parameter space corresponds to cases where the MSA angles do not have a mild behaviour and lead to `noisy' waveforms. Applying multibanding to these cases would amplify errors, and is thus switched off.
    \item When using NNLO angles: for $q > 8$. It is well known that the NNLO angles can behave badly for high mass ratios and can even be pathological, see e.g.~our discussion in Sec.~\ref{sec:TD_waveforms}. Once again the multibanding would not properly work for theses case and is switched off. 
\end{itemize}
The veto for the multibanding in the NNLO angles is much broader than for MSA, leading to the more pronounced tail of lower mismatches.

\begin{figure*}[htpb]
    \centering
    \includegraphics[scale=.39]{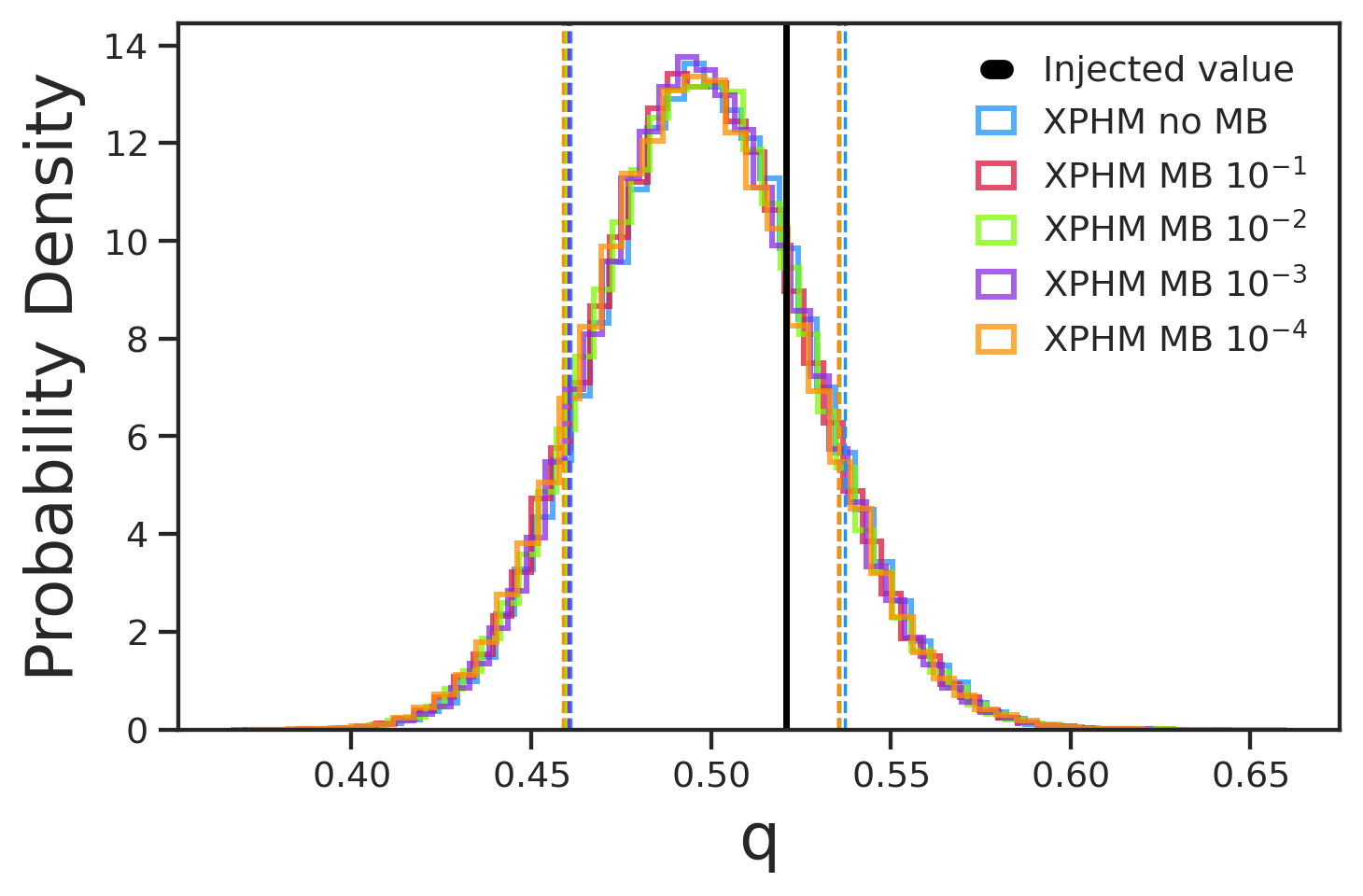}
    \includegraphics[scale=.39]{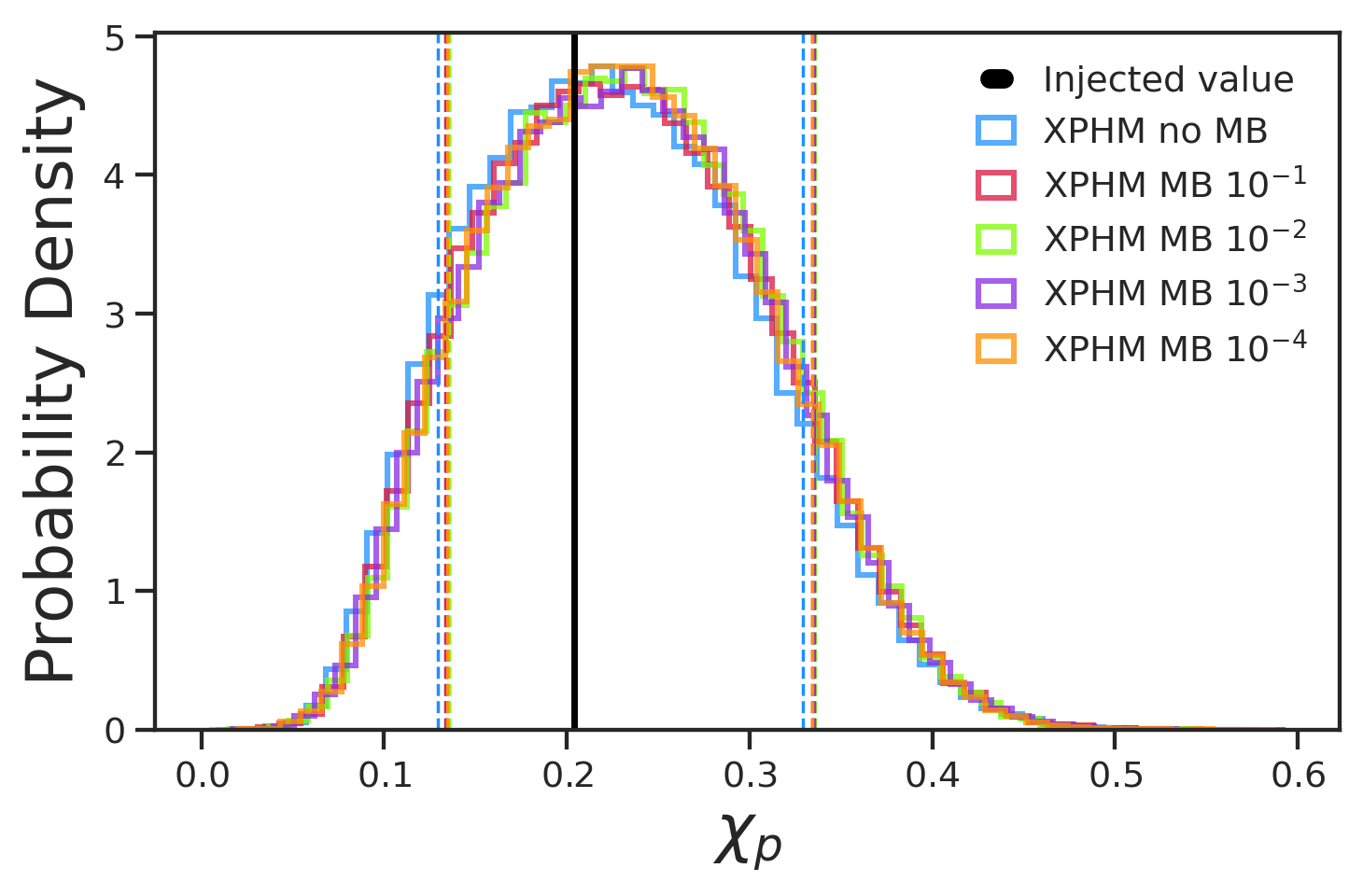}
    \includegraphics[scale=.39]{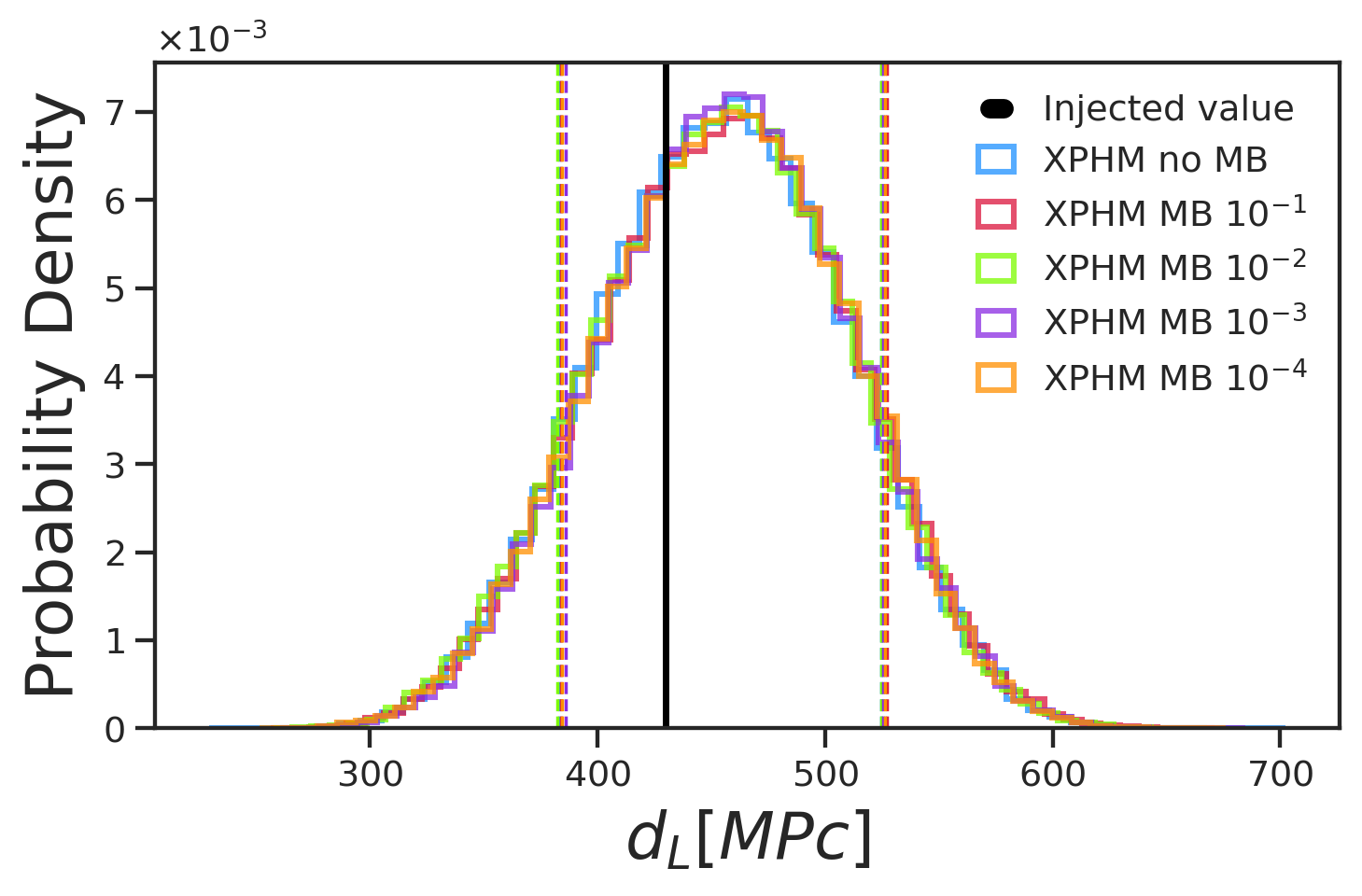}
    \caption{Injection recovery results for \texttt{SXS:BBH:0143} for \phXPHM without multibanding in the Euler angles and with four different thresholds  ($10^{-1}, 10^{-2}, 10^{-3}, 10^{-4}$) for multibanding in the angles. No appreciable differences arise in the posteriors, meaning that a more relaxed threshold than the default of $10^{-3}$ could be used in parameter estimation studies, reducing even more the computational cost of the runs.}
    \label{fig:pe_mband}
\end{figure*}

\begin{figure*}[thpb]
    \centering
    \includegraphics[width=\columnwidth]{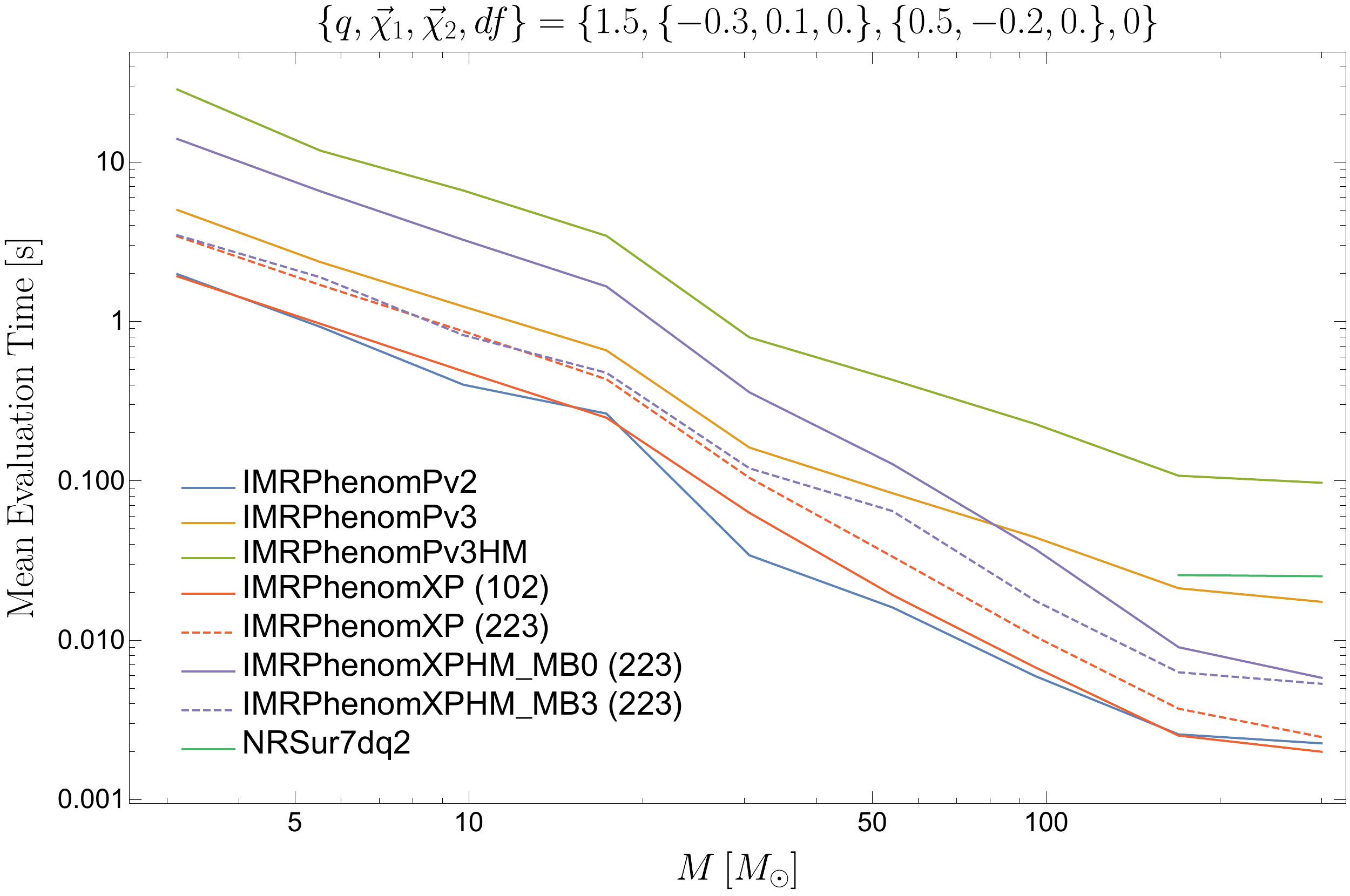}
    \includegraphics[width=\columnwidth]{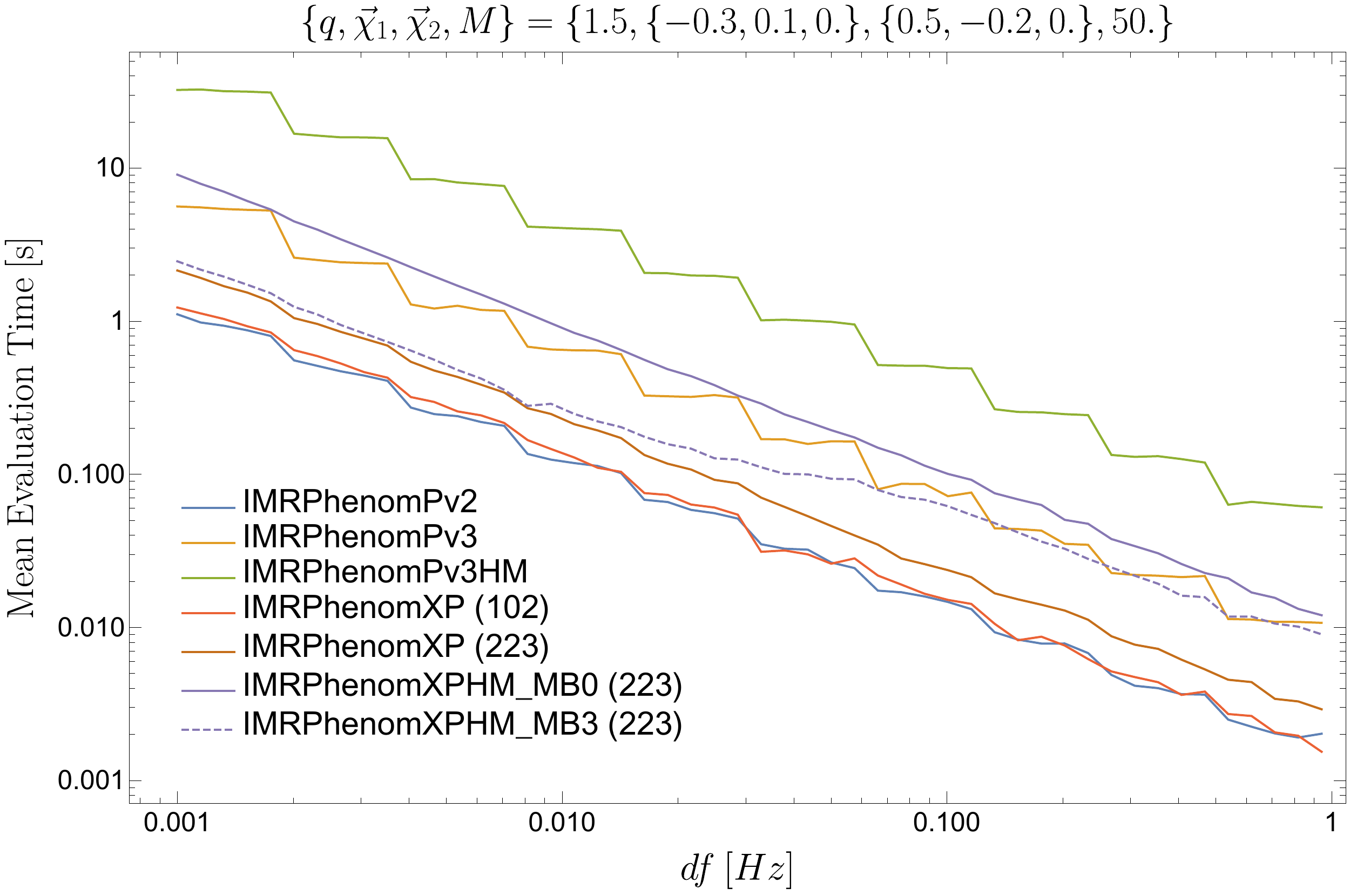}
    \caption{Mean evaluation time for different precessing models as a function of the total mass (left panel) and as a function of the spacing of the frequency grid (right panel). The \NRSur model can not be evaluated for such low frequencies as the \ph models; hence it only appears in the left panel for high masses, but not in the right panel, where we are using a total mass of $50M_\odot$.}
    \label{fig:benchmarks}
% \end{figure*}

% \begin{table*}[htpb]
\begin{center}
%\resizebox{10.cm}{!}{
 \def\arraystretch{1.3 }
\begin{tabular}{  c  c  c  c  c  c c   c   c   c }
\hline
\hline
 $M_{\min}[M_\odot]$& $\Delta T$ & \fontsize{7}{104}\selectfont IMRPhenomXP & \fontsize{7}{104}\selectfont IMRPhenomPv2 & \fontsize{7}{104}\selectfont  IMRPhenomPv3 & \fontsize{7}{104}\selectfont SEOBNRv4P & \fontsize{7}{104}\selectfont IMRPhenomXPHM & \fontsize{7}{104}\selectfont IMRPhenomPv3HM   &  \fontsize{7}{104}\selectfont  SEOBNRv4PHM  &  \fontsize{7}{104}\selectfont  NRSurd7q4\\
\hline
\multirow{2}{*}{20} & 4\,s         & $8.6$       & $ 5.7 $      & $29.1 $      &  $2691.4$  &  $31.8$       &   $160.3$         &   $4259.9$ & - \\
                    & 8\,s         & $16.8 $     & $ 11.2$       & $56.4$       &  $2976.6$  &  $52.8$       &   $311.7$        &   $4540.9$  & - \\
 \hline
\multirow{2}{*}{60} & 4\,s         & $5.8$       & $ 4.1 $      & $29.4 $      &  $1492.1$  &  $21.4$       &   $161.7$          &   $3016.2$  & $60.5$ \\
                    &  8\,s        & $11.4 $     & $ 8.0$       & $56.6$       &  $1483.9$  &  $36.3$       &   $312.8$          &   $2951.5$ &  $59.9$\\
\hline
\hline
 \end{tabular}
% }
\end{center}
\captionof{table}{Mean likelihood evaluation time in milliseconds for several precessing models for equal masses. The numbers represent averages over a mass range of $[M_{min},100]\,M_\odot$ with $M_{min}=20,60 M_\odot$ and random spin orientations and magnitudes. The first column indicates the data analysis segment length in seconds.}
\label{tab:tabBench}
% \end{table*}
\end{figure*}

We also perform a parameter estimation study with different multibanding thresholds, to test the effect on recovered posterior distributions.
We perform the same NR injections as described in Sec.~\ref{sec:pe_nr_injections} with version 223 for the MSA angles and compare the results between thresholds of $0, 10^{-1}, 10^{-2}, 10^{-3}, 10^{-4}$.
As seen in Fig.~\ref{fig:pe_mband} the results are highly consistent.
Considering these results together with the benchmarking results shown in Fig.~\ref{fig:benchmarks} and discussed in the next section, we however make a conservative choice for the default multibanding threshold for the Euler angles and set the value to $10^{-3}$.
This can be changed  as described in Appendix \ref{appendix:lal_implementation}. %on the multibanding of the MSA angle description and a more stringent threshold of \(10^{-4}\) for the NNLO angles.

%%%%%%%%%%%%%%%%%%%%%%%%%%%%%%%%%%%%%%%%%%%%%%%%%%%%%%%%%%%%%%%%%
%                   BENCHMARKS
%%%%%%%%%%%%%%%%%%%%%%%%%%%%%%%%%%%%%%%%%%%%%%%%%%%%%%%%%%%%%%%%%

\subsection{Benchmarking}\label{sec:benchmarks}

In Fig.~\ref{fig:benchmarks} we show benchmarking results for one precessing case in a frequency range from 10 to 2048 Hz comparing the previous precessing \ph models with different settings of \phXP and \phXPHM. The timing is carried out with the executable \texttt{GenerateSimulation} (included in LALSuite/LALSimulation), averaging over 100 repetitions. In the top panel we show the dependency on total mass. The frequency grid spacing $df$ is computed automatically by the \texttt{SimInspiralFD} interface to take into account the length of the waveform in the time domain for the given parameters. In the bottom panel instead we show the dependency on the frequency grid spacings a function of the total mass, where the frequency spacing is computed as $df=1/T$, where $T$
is a simple estimate of the duration of the signal, see the discussion of Fig.~5 in \cite{Garcia-Quiros:2020qlt}. In the labels for \phXP and \phXPHM, the numbers between brackets refer to the version of Euler angles: 102 uses the NNLO description, while 223 uses the MSA description (for a complete list of options see Table~\ref{tab:prec_version}). For \phXPHM we also show the result without applying multibanding in the Euler angles (MB0) and when using multibanding with threshold $10^{-3}$ (MB3).

For both plots the conclusion is the same, \phXPHM without multibanding in the Euler angles is already faster than its counterpart \phPvthreehm, and when multibanding is included it is even faster than the ``22-mode only'' version \phPvthree. The threshold of the multibanding used here is the default $10^{-3}$, however the user can modify this parameter at will. Higher values of the threshold will accelerate the evaluation further at the price of decreased accuracy, which may be found acceptable based on the signal-to-noise ratio of a given event that is analyzed.

We have also estimated the efficiency of our models  \phXP and \phXPHM compared to other precessing models by computing their mean likelihood evaluation time in the LALInference Bayesian parameter estimation code \cite{Veitch:2014wba}.
To perform this test, we have chosen an equal-mass configuration, 100 different total masses in the range $[M_{min},100] \, M_\odot$ with  $M_{\min}=20$, except for the precessing surrogate model \NRSur, where we have set a higher minimum total mass of $M_{\min}= 60 M_\odot$, due to its limitations in start frequency.
Dimensionless spin magnitudes are distributed randomly between $[0,0.99]$ with a random isotropic distribution of spin vectors, and a reference frequency of $20$ Hz. 
Two different segment lengths of $\Delta T= 4$\,s, 8\,s are studied as they are typical for the currently detected BBH GW signals \cite{LIGOScientific:2018mvr}. As for our match calculations in Secs.~\ref{sec:NR_matches} and \ref{sec:model_matches} the Advanced LIGO zero detuned power spectral density \cite{adligopsd} is used here for likelihood evaluations.
For each total mass we perform 100 likelihood evaluations with randomly chosen spin configurations. The average of these $10^4$ likelihood evaluations for each model is shown in Table~\ref{tab:tabBench}.

The results confirm that the \phXPHM model is the most efficient precessing waveform model with higher harmonics: $\sim 5$ times faster than \phPvthreehm \cite{Khan:2019kot} for $\Delta T= 4$\,s and $\sim 133 $ times faster than \seobnrvforphm  for $\Delta T= 4$\,s. \seobnrvforphm~\cite{SEOBNRv4PHM:inprep} is a precessing extension to the \seobnrvforhm model~\cite{Cotesta:2018fcv} and is predicated on the numerical integration of computationally expensive ODEs, making the waveform slow to evaluate. Though we note that there has been significant work on improving waveform generation costs of EOB models, such as the post-adiabatic scheme introduced in \cite{Nagar:2018gnk} or reduced order models \cite{Purrer:2015tud}. Regarding precessing models including only the $(2,\pm 2)$ mode in the co-precessing frame, \phXP is slightly slower than \ppvtwo %\cite{Hannam:2013oc}
as a trade-off of the inclusion of the double-spin effects in the Euler angles, although it is much faster than the other phenomenological and SEOB models: $\sim 3.4 $ times faster than \phPvthree %\cite{Khan:2018fmp}
for $\Delta T= 4$\,s and $\sim 312 $ times faster than \seobnrvforp for $\Delta T= 4$\,s.
While an increase of the segment length increases the mean evaluation time for all models, the relative differences in evaluation costs at  $\Delta T= 8$\,s are still similar to those at $4$\,s.
The numbers reported  in Table~\ref{tab:tabBench} illustrate the huge impact in efficiency that our new precessing models may have on data analysis studies like parameter estimation, where millions of likelihood evaluations are performed per run.

Finally, we note that computational cost of Bayesian inference can be significantly reduced through the use of reduced order quadratures (ROQ) \cite{Antil:2012wf,Canizares:2013ywa,Canizares:2014fya}. This framework has been applied to a number of waveform models, including \ppvtwo \cite{Smith:2016qas}. We note that our model is amenable to such an approach following the methodology detailed in \cite{Smith:2016qas}.

%%%%%%%%%%%%%%%%%%%%%%%%%%%%%%%%%%%%%%%%%%%%%%%%%%%%%%%%%%%%%%%%%
%               PARAMETER ESTIMATION
%%%%%%%%%%%%%%%%%%%%%%%%%%%%%%%%%%%%%%%%%%%%%%%%%%%%%%%%%%%%%%%%%

\subsection{Parameter estimation}\label{sec:pe}
We use coherent Bayesian inference methods to determine the posterior distribution $p({\theta} | {d})$ for the parameters $\theta$ that characterize a binary, given some data $d$. From Bayes' theorem, we have
\begin{align}
    p(\theta | d) &= \frac{\mathcal{L}(d | \theta) \, \pi (\theta)}{\mathcal{Z}} ,
\end{align}
\newline 
where $\mathcal{L}(d | \theta)$ is the Gaussian noise likelihood \cite{Veitch:2008wd,Veitch:2009hd,Veitch:2014wba}, $\pi (\theta)$ the prior distribution for $\theta$ and $\mathcal{Z}$ the evidence 
\begin{align}
    \mathcal{Z} &= \int d \theta \, \mathcal{L} (d | \theta) \, \pi (\theta) . 
\end{align}
\newline 
For the analysis here, we use both the nested sampling~\cite{Skilling:2004ns} algorithm implemented in \texttt{LALInference} \cite{Veitch:2014wba} and the nested sampling algorithm \texttt{Dynesty} \cite{Speagle:2020spe} implemented in \texttt{Bilby} \cite{Ashton:2018jfp} and \texttt{Parallel Bilby} \cite{Smith:2019ucc}. We use the public strain data from the Gravitational Wave Open Science Center (GWOSC) \cite{GWOSC,Abbott:2019ebz,gwtc1calib,gwtc1psd}. Following \cite{LIGOScientific:2018mvr}, we marginalize over the frequency-dependent spline calibration envelopes that characterize the uncertainty in the detector amplitude and strain~\cite{Cahillane:2017vkb,Viets:2017yvy,Kissel:2018cal}.

\subsubsection{GW150914}\label{sec:GW150914}

\begin{figure*}[thpb]
    \centering
    \includegraphics[width=\textwidth]{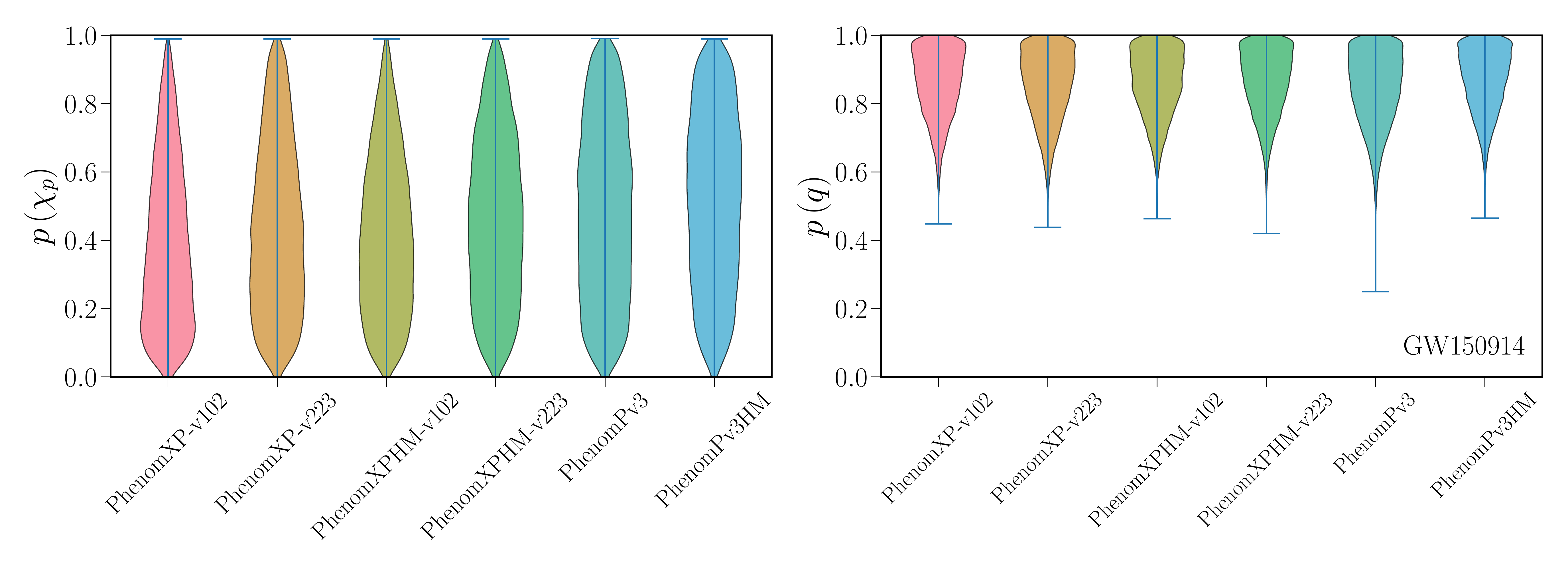}
    \caption{Bayesian inference results for GW150914: one-dimensional posterior probability distributions for the effective precessing spin parameter $\chi_p$ and the mass ratio $q$. (Here $q$ is the inverse of our definition in Sec.~\ref{sec:Introduction}, following the LALInference convention.)
    We show results for \phXPHM with and without higher modes, using NNLO and MSA angles respectively, as discussed in Sec.~\ref{sec:Model}. For comparison, we also show results for \phPvthree and \phPvthreehm. The labels of the different versions of IMRPhenomXPHM are explained in Appendix~\ref{appendix:lal_implementation}.}
    \label{fig:GW150914_2}
\end{figure*}

\begin{figure}[thpb]
    \centering
    \includegraphics[width=\columnwidth]{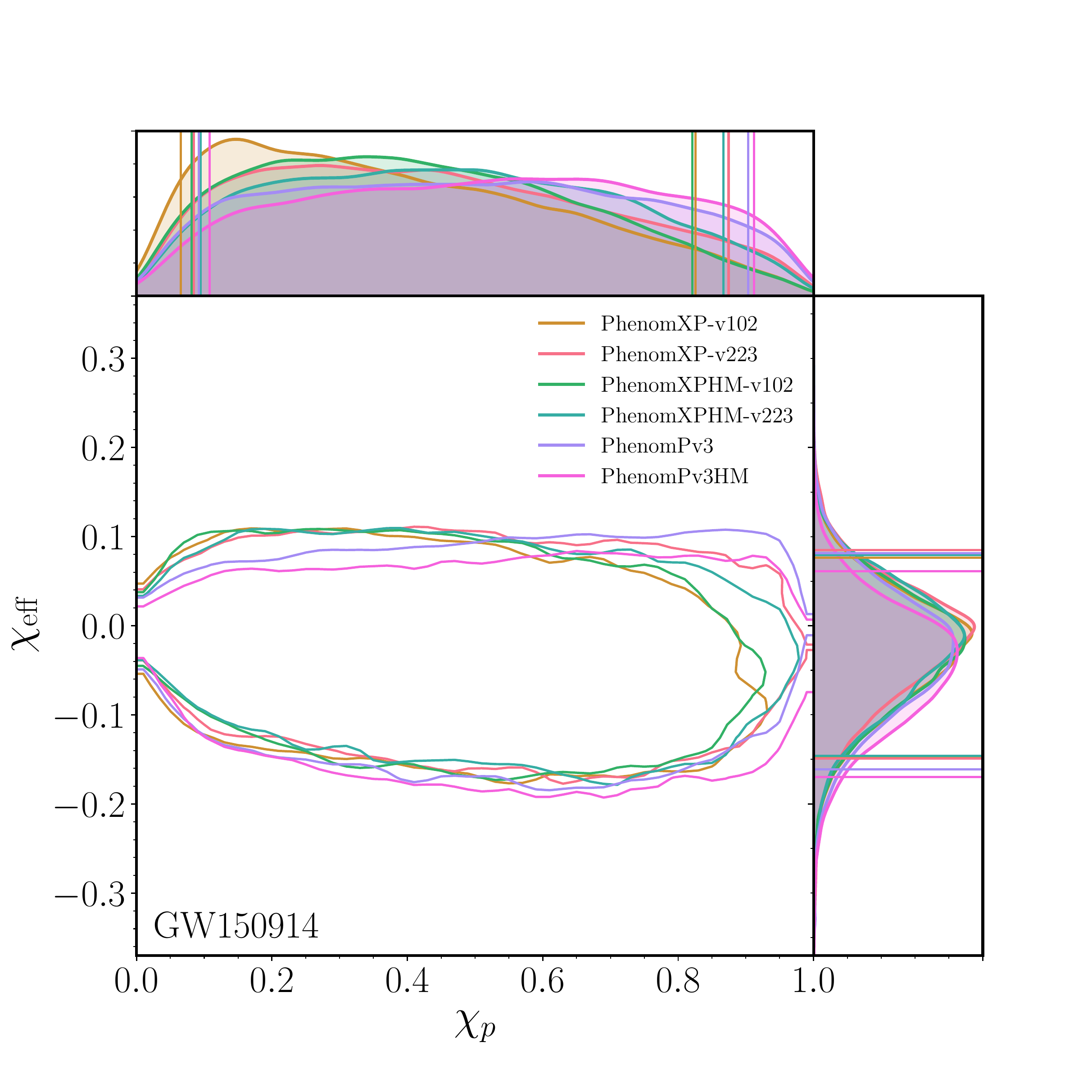}
    \caption{Bayesian inference results for GW150914: posterior probability distributions for the effective spin parameters $\chi_{\rm{eff}}$ and $\chi_p$ with two-dimensional 90\% credible intervals.
    Here we show results for \phXPHM with and without higher modes, using NNLO and MSA angles respectively, as discussed in Sec.~\ref{sec:Model}. For comparison, we also show results for \phPvthree and \phPvthreehm.
    }
    \label{fig:GW150914_1}
\end{figure}

As a prototypical example of the application of \phXPHM to GW data analysis we re-analyze GW150914, the first direct observation of GWs from the merger of two black holes~\cite{Abbott:2016blz}. For GW150914 we use the nested sampling algorithm implemented in LALInference \cite{Veitch:2014wba}. Our parameter estimation uses 2048 live points and coherently analyzes 8s of data. We use priors as detailed in Appendix C of \cite{LIGOScientific:2018mvr} and use the PSDs \cite{gwtc1psd} and detector calibration envelopes \cite{gwtc1calib} as available on GWOSC \cite{GWOSC}.

Using the inherent modularity of \phXPHM, we can try to gauge the impact of systematics arising from the modelling of spin-precession effects by performing coherent Bayesian parameter estimation using the different prescriptions for the Euler angles discussed in Sec.~\ref{sec:Model}. The final spin descriptions used here are the ones based on averaged in-plane spin for the NNLO and MSA Euler angle formulations, i.e. final spin version 0 for model version 102 (NNLO) and final spin version 3 for model version 223 (MSA), see appendix \ref{appendix:lal_implementation} for details. 

As can be seen in Figures~\ref{fig:GW150914_2} and~\ref{fig:GW150914_1}, constraints on parameters such as the effective aligned-spin parameter $\chi_{\rm eff}$ and mass ratio $q$ are consistent between the different waveform models whereas the effective precessing spin $\chi_p$ is not meaningfully constrained. This is in agreement with studies detailing the impact of waveform systematics on the analysis of GW150914 \cite{Abbott:2016wiq}, which conclude that systematic errors and biases are small compared to statistical errors.

%%%%%%%%%%%%%%%%%%%%%%%%%%%%%%%%%%%%%%%%%%%%%%%%%%%%%%%%%%%%%%%%%
%                      GW170729
%%%%%%%%%%%%%%%%%%%%%%%%%%%%%%%%%%%%%%%%%%%%%%%%%%%%%%%%%%%%%%%%%
\subsubsection{GW170729}

We now turn our attention to the analysis of GW170729, the BBH GW signal with the highest mass detected during the O1 and O2 LIGO-Virgo observing runs  \cite{LIGOScientific:2018mvr}. Both the high mass and the significant posterior support for a mass ratio different from unity makes it a good candidate to test the impact of higher-order modes on the estimation of its parameters.
 
\begin{figure*}[htpb]
    \begin{minipage}{.5\linewidth}
        \centering
       \includegraphics[scale=.35]{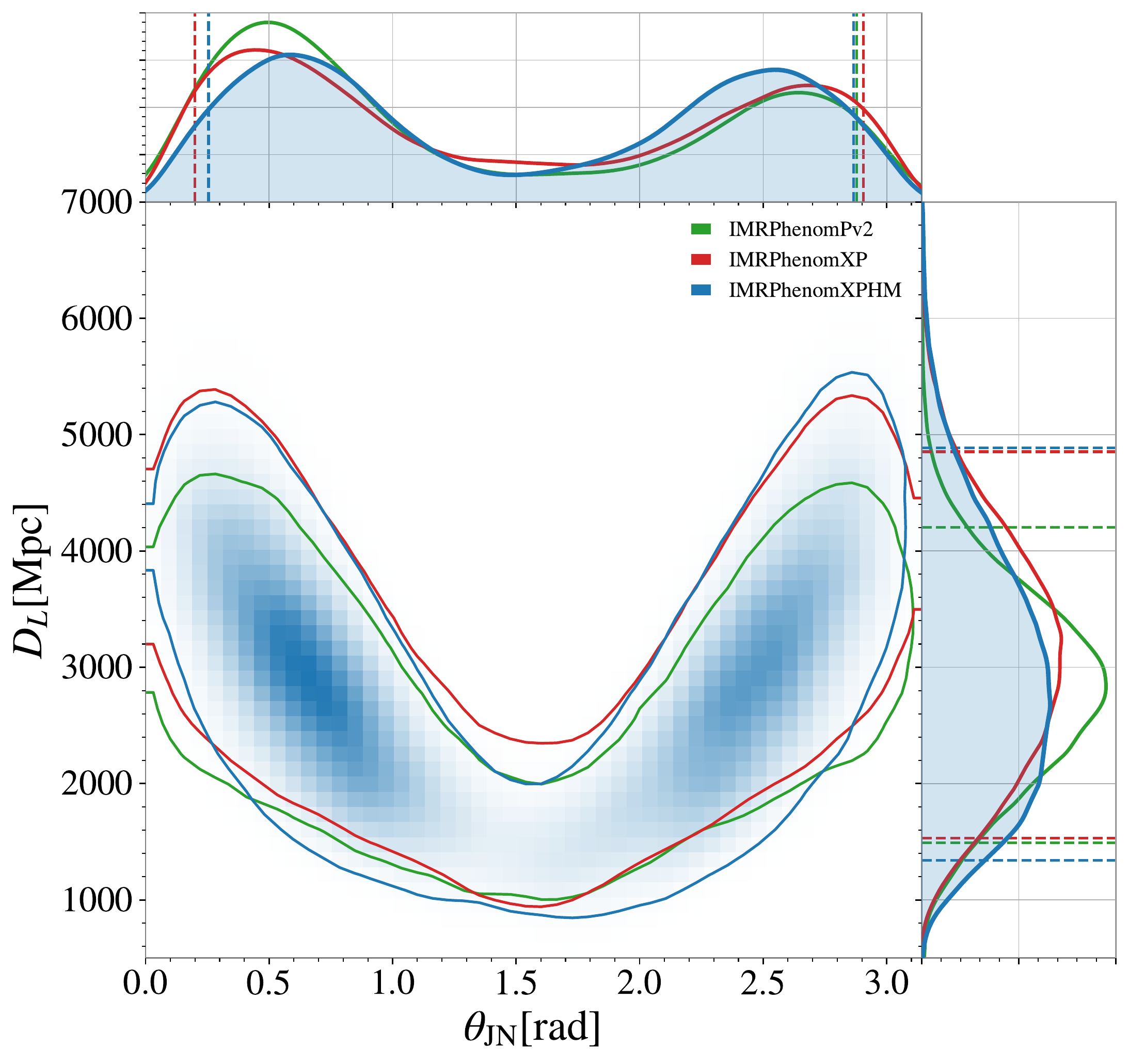}
        \end{minipage}%
        \begin{minipage}{.5\linewidth}
        \centering
        \includegraphics[scale=.35]{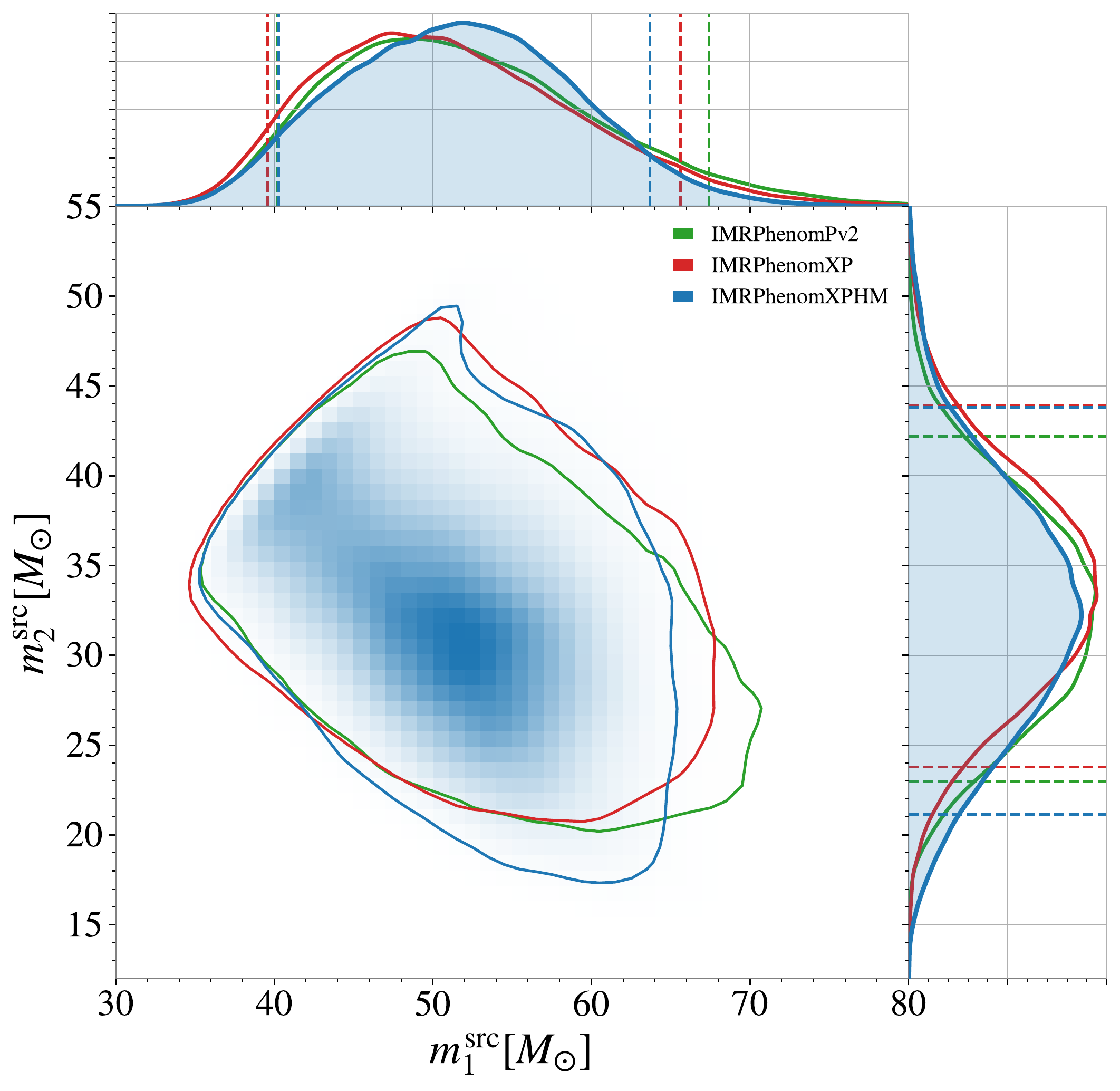}
        \end{minipage}\par\medskip
        \centering
       \includegraphics[scale=.35]{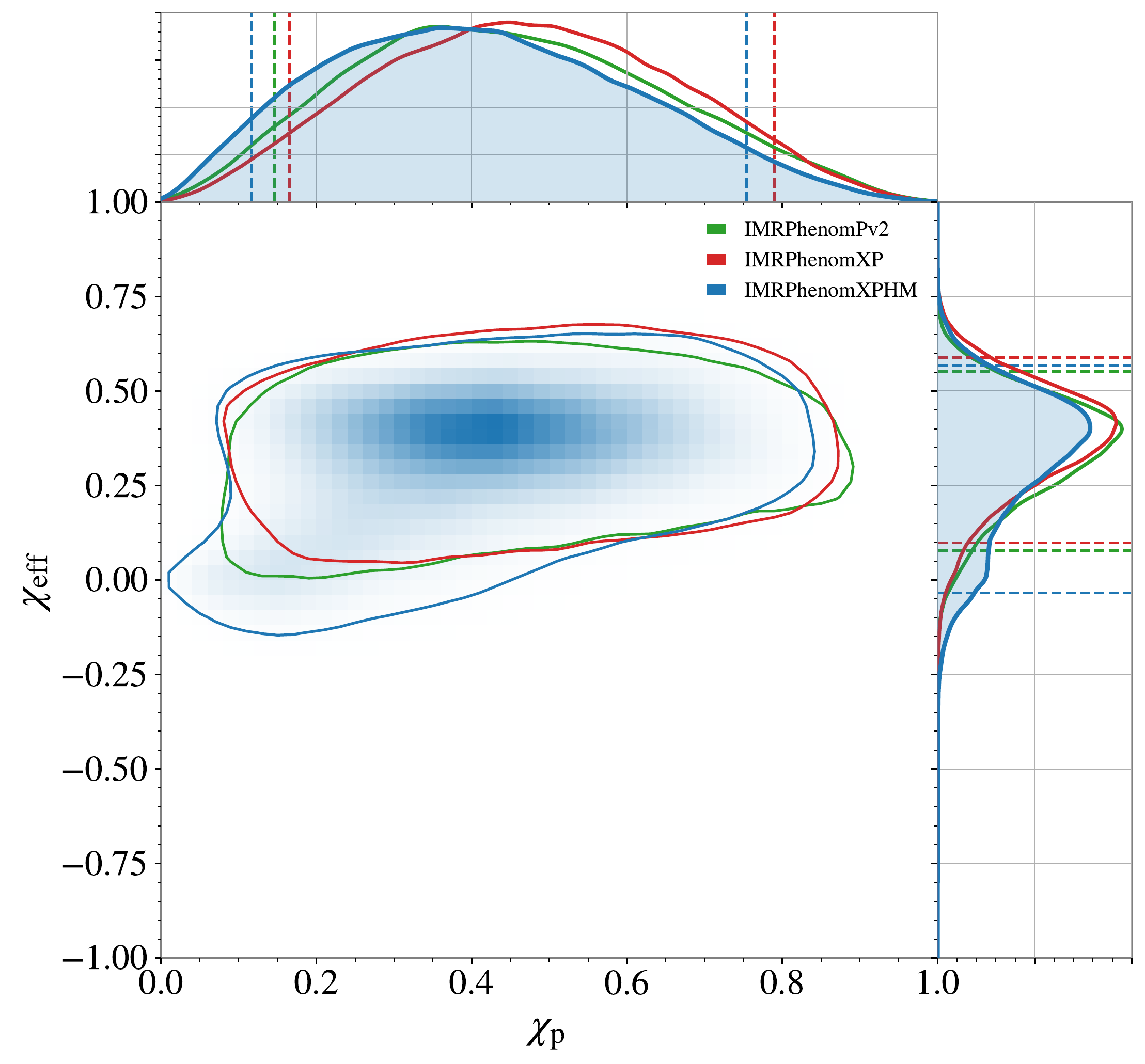}
        \caption{Bayesian inference results for GW170729: Posterior probability distributions for the effective spin parameters $\chi_\mathrm{eff}$ and $\chi_p$, the component masses $m_1$ and $m_2$, distance $D_L$ and the angle $\theta_{JN}$ between angular momentum and line of sight. The 90 \% credible intervals are represented by vertical (contour) lines in the 1D (2D) plots.}
     \label{fig:GW170729}
\end{figure*}

This fact has motivated several studies of this event in the literature with non-precessing higher-order modes models \cite{Chatziioannou:2019dsz,Payne:2019wmy} like the  phenomenological \phHM \cite{London:2017bcn}, the effective-one-body  \seobnrvforhm \cite{Cotesta:2018fcv}, and the numerical relativity surrogate  \cite{Varma:2018mmi}. We also reanalyzed this event with the upgraded version of the phenomenological non-precessing models \phXHM in \cite{Garcia-Quiros:2020qlt} and found consistency with the results in \cite{Chatziioannou:2019dsz}. Furthermore, there have been investigations of this event with precessing waveform models, in  \cite{LIGOScientific:2018mvr} with \ppvtwo %\cite{Hannam:2013oca,Bohe:PPv2},
and in \cite{Khan:2018fmp,Khan:2019kot} with \phPvthree and \phPvthreehm.

Here we report on the analysis of GW170729 with our new precessing  \phXPHM model, which upgrades \phPvthreehm. For our analysis we use $4$\,s of the publicly available strain data from the Gravitational Wave Open Science Center (GWOSC) \cite{GWOSC,Abbott:2019ebz} with a lower cutoff frequency of 20 Hz. This data is calibrated by a cubic spline and we use the same PSDs utilized in  \cite{LIGOScientific:2018mvr}. We analyze the strain with the Python-based Bayesian inference framework \texttt{Parallel Bilby} \cite{Smith:2019ucc}, which uses a parallel version of the nested sampling code \texttt{Dynesty} \cite{Speagle:2020spe}. We carry out the parameter estimation runs using 4096 live points, choose the maximum number of Markov chain Monte Carlo (MCMC) steps to take 
as $10^4$, and require 10 auto-correlation times (ACT) before accepting a point. We merge results from four different seeds in order to get a single posterior distribution. The simulations are performed for the default options of the LALSuite implementation of \phXPHM (the precessing version 223, final spin version 3 and convention 1, see Appendix \ref{appendix:lal_implementation}). The priors are the same as used in \cite{Chatziioannou:2019dsz} but adapted to precessing models.

\begin{table*}[bhpt]
\begin{center}
%\resizebox{13.cm}{!}{
 \def\arraystretch{1.4 }
\begin{tabular}{  c   c   c   c   c  c  c  c   c  c c   }
\hline
\hline
 NR Simulation & Version &  $m_1/M_\odot$&  $m_2/M_\odot$& $\mathcal{M}_c/M_\odot$&  $q$ &     $D_L/$Mpc&  $\chi_{\text{eff}}$ & $\chi_{\text{p }}$ &$\theta_{\text{JN}}$ (rad)  \\ %& $1-\mathcal{M}$ &$\rho_{\text{Net}}$\\
\hline
\multirow{5}{*}{{\tt SXS:BBH:0143}} 
& v102 FS0 & $65.26^{+1.98}_{-2.05}$  & $32.78^{+1.73}_{-1.65}$  & $39.79^{+0.96}_{-0.89}$  & $0.50^{+0.03}_{-0.04}$  & $465.32^{+72.92}_{-71.69}$  & $0.24^{+0.05}_{-0.04}$  & $0.20^{+0.05}_{-0.07}$  & $1.02^{+0.10}_{-0.10}$  \\  & v102 FS2 & $65.17^{+1.96}_{-2.06}$  & $32.76^{+1.70}_{-1.71}$  & $39.76^{+0.95}_{-0.90}$  & $0.50^{+0.03}_{-0.04}$  & $463.99^{+71.83}_{-70.47}$  & $0.24^{+0.05}_{-0.04}$  & $0.20^{+0.06}_{-0.07}$  & $1.03^{+0.11}_{-0.09}$  \\  & v223 FS2 & $65.12^{+1.96}_{-2.05}$  & $32.36^{+1.70}_{-1.71}$  & $39.48^{+0.91}_{-0.89}$  & $0.50^{+0.04}_{-0.04}$  & $456.00^{+72.82}_{-70.12}$  & $0.23^{+0.04}_{-0.04}$  & $0.23^{+0.10}_{-0.11}$  & $1.07^{+0.11}_{-0.09}$  \\  & v223 FS3 & $65.11^{+2.01}_{-2.10}$  & $32.36^{+1.70}_{-1.73}$  & $39.47^{+0.88}_{-0.91}$  & $0.50^{+0.04}_{-0.04}$  & $455.17^{+72.76}_{-69.63}$  & $0.23^{+0.04}_{-0.04}$  & $0.23^{+0.09}_{-0.11}$  & $1.07^{+0.11}_{-0.10}$  \\ & Injected & $65.77$ & $34.26$ & $40.88$ & $0.52$ & $430.00$ & $0.26$ & $0.20$ & $1.08$ \\ 
\hline
\hline
 \end{tabular}
 %}
\end{center}
\caption{Black hole binary recovered parameters for the injected NR waveform from
Fig.~\ref{fig:NRInj143}.
% Figs.~\ref{fig:NRInj143} and \ref{fig:NRInj165}. 
The values correspond to the mean recovered values with their 90\% credible interval. The first column shows the identifier of the injected NR waveform, then we specify the version of the \phXPHM model, the component masses ($m_1$, $m_2$), chirp mass $\mathcal{M}_c$, mass ratio $q(=m_2/m_1)$, luminosity distance $D_L$, effective spin parameter $\chi_{\text{eff}}$, effective precessing spin parameter  $\chi_{\text{p }}$, and the angle between the total angular momentum and the line of sight, $\theta_{\text{JN}}$. %For each NR waveform 
The injected values are also displayed.}
\label{tab:tabInj}
\end{table*}

We have analyzed this event with the non-precessing \phX and \phXHM models in \cite{Garcia-Quiros:2020qpx}, where we have also compared with results available in the literature and obtained with other non-precessing models. Our results based
on the MSA versions of \phXP and \phXPHM are shown in Fig.~\ref{fig:GW170729}, and compared with the \phPvtwo model that is routinely used for parameter estimation, but lacks higher modes.
We show posteriors for the effective spin parameters $\chi_\mathrm{eff}$ and $\chi_p$, the component masses, distance $D_L$ and the angle $\theta_{JN}$ between total angular momentum and line of sight.

The results show agreement between the new \phXP model and the old \phPvtwo model, although small differences in the shape of the posterior distributions are due to the inclusion of double-spin effects in \phXP. The inclusion of higher-order modes produces a shift in the posterior distributions of some quantities like the primary component mass. These changes in some parameters due to the inclusion of precessing higher-order modes are consistent with those observed in \cite{Khan:2019kot} for this particular event.

%%%%%%%%%%%%%%%%%%%%%%%%%%%%%%%%%%%%%%%%%%%%%%%%%%%%%%%%%%%%%%%%%
%               NR INJECTIONS
%%%%%%%%%%%%%%%%%%%%%%%%%%%%%%%%%%%%%%%%%%%%%%%%%%%%%%%%%%%%%%%%%

\subsubsection{Numerical relativity injections}
\label{sec:pe_nr_injections}

%We investigate parameter estimation biases that might affect Bayesian inference analyses with \phXPHM by performing zero-noise injections of two public binary black hole numerical relativity simulations from the first SXS waveform catalogue \cite{SXS:catalog}. We choose \texttt{SXS:BBH:0143} corresponding to a mass ratio %1/2 
%2 simulation with positive  $\chi_\mathrm{eff}$ and small $\chi_p$, and {\tt SXS:BBH:0165}, a mass ratio 6 %1/6
%simulation with negative $\chi_\mathrm{eff}$ and very high $\chi_p$. 
%The latter corresponds to the worst mismatch case in the study of Sec.~\ref{sec:NR_matches}. Based on those results, we can expect significant biases in the recovered parameters for {\tt SXS:BBH:0165}. Nonetheless, we consider illustrative to show an explicit example where the performance of \phXPHM is below average, and examine which are the most affected parameters and how large their bias is. This will help to develop an idea of the regions of parameter space where the model should not be particularly trusted.
%We set the total mass of both injected signals to be $100~M_{\odot}$ and their luminosity distance at $430$ Mpc, yielding signal-to-noise ratios of $52$ and $26$ respectively. We analyze a 4\,s segment of data with a lower cutoff frequency of $20$\,Hz.
%The parameters of the injected waveforms are listed in Table~\ref{tab:tabInj}, together with their estimated values, as discussed below.
We investigate parameter estimation biases that might affect Bayesian inference analyses with \phXPHM by performing a zero-noise injection of a public binary black hole numerical relativity simulation from the first SXS waveform catalogue \cite{SXS:catalog}. We use \texttt{SXS:BBH:0143}, a mass ratio 
2 simulation with positive $\chi_\mathrm{eff}$ and small $\chi_p$, broadly consistent with the population of BBHs observed to date \cite{Abbott:2020niy,Abbott:2020gyp}.
We set the total mass of the injected signal to be $100~M_{\odot}$ and its luminosity distance at $430$ Mpc, yielding a signal-to-noise ratio of $52$. We analyze a 4\,s segment of data with a lower cutoff frequency of $20$\,Hz.
The parameters of the injected waveform are listed in Table~\ref{tab:tabInj}, together with their estimated values, as discussed below.

\begin{figure*}[htpb]
    \centering
       \includegraphics[scale=.39]{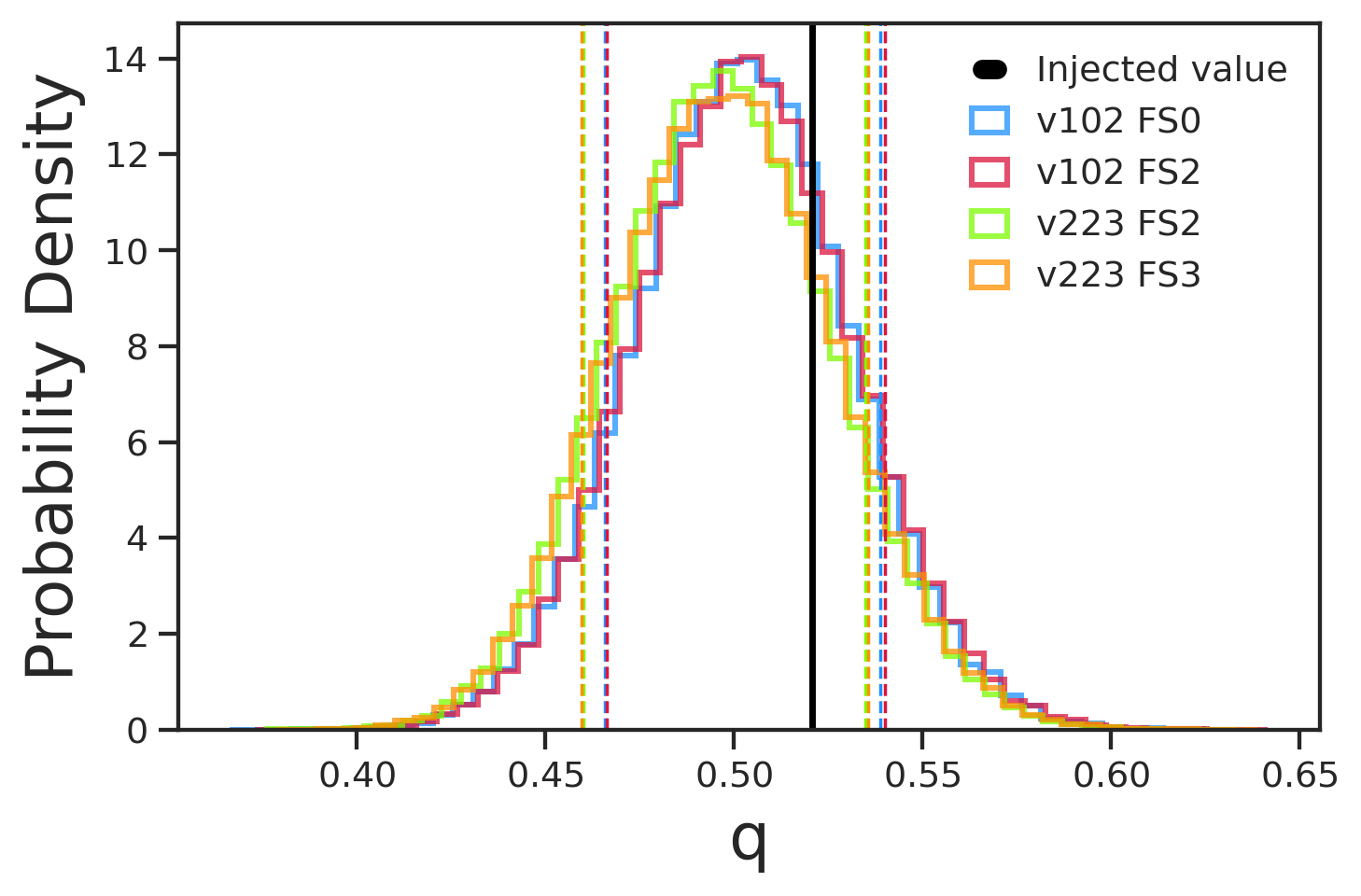}
       \includegraphics[scale=.39]{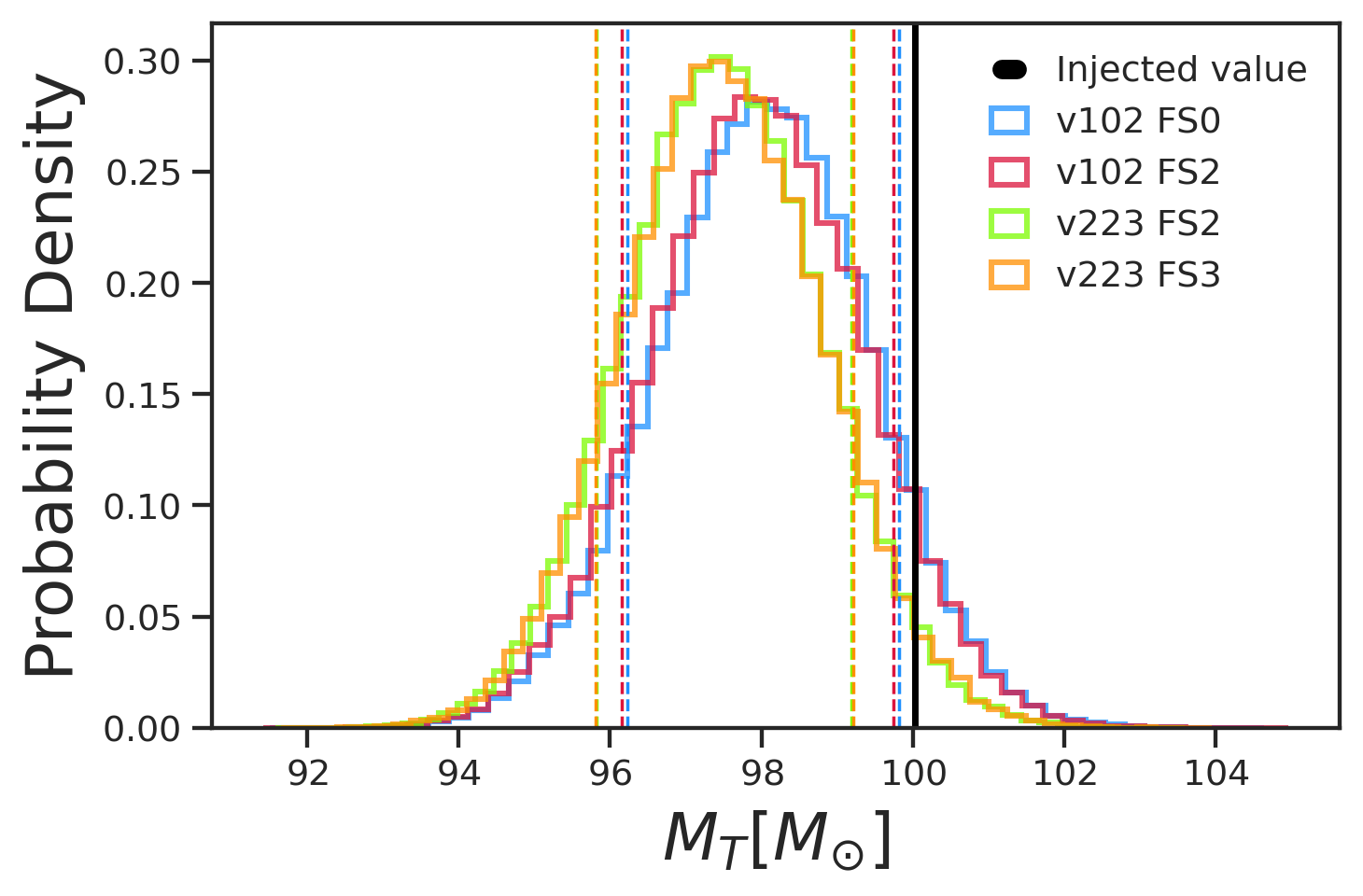}
       \includegraphics[scale=.39]{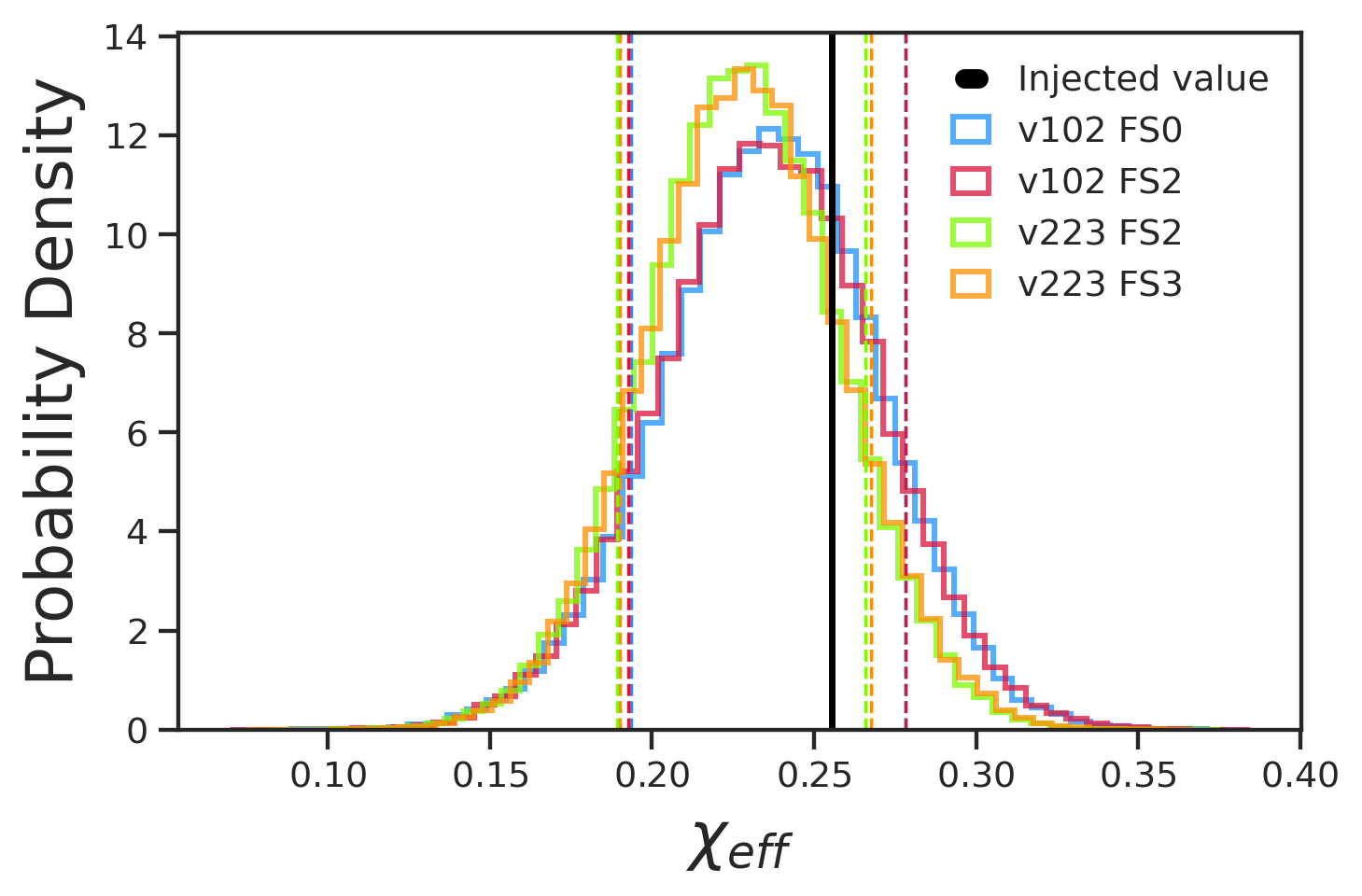}
       \includegraphics[scale=.39]{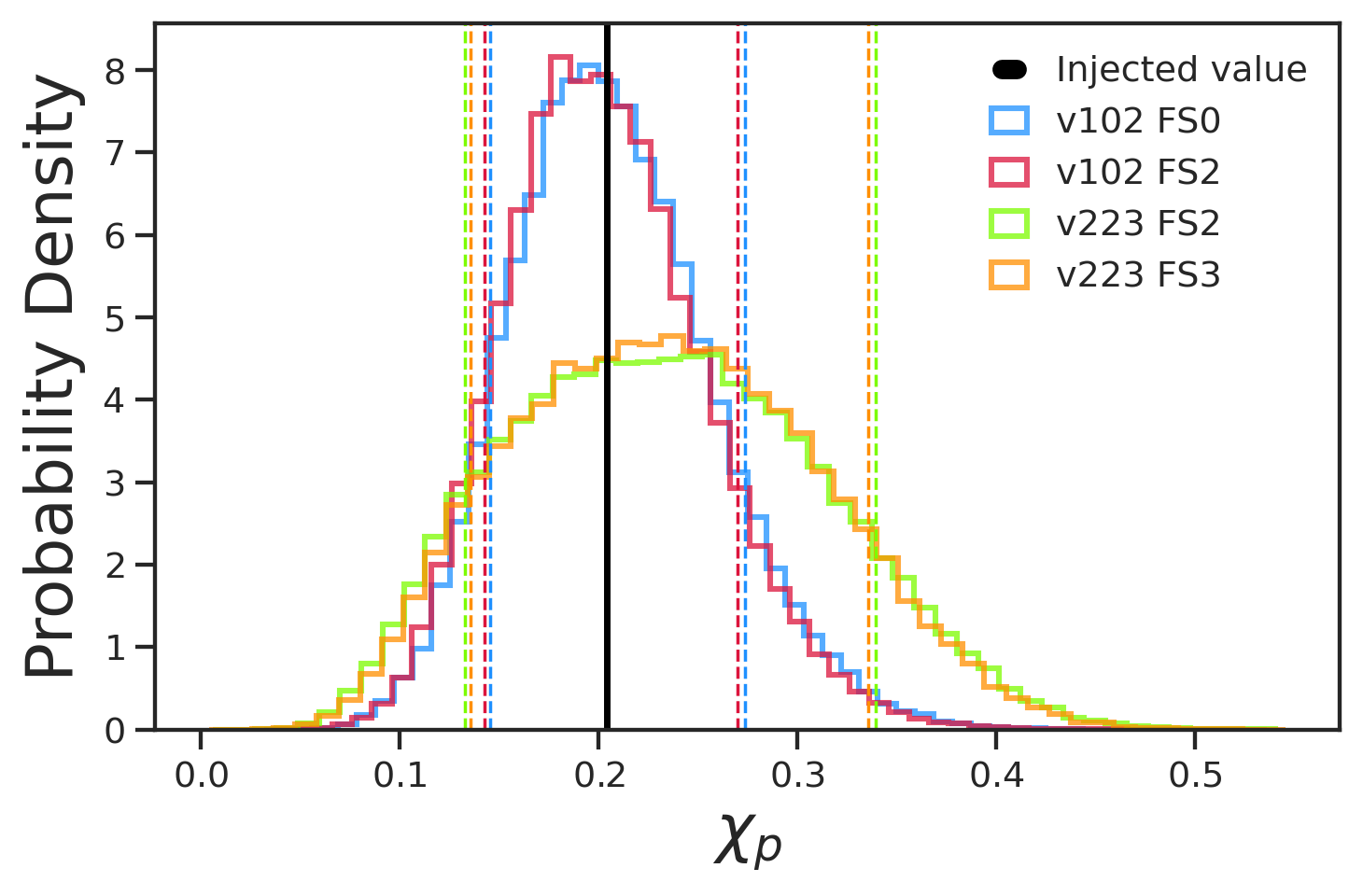}
       \includegraphics[scale=.39]{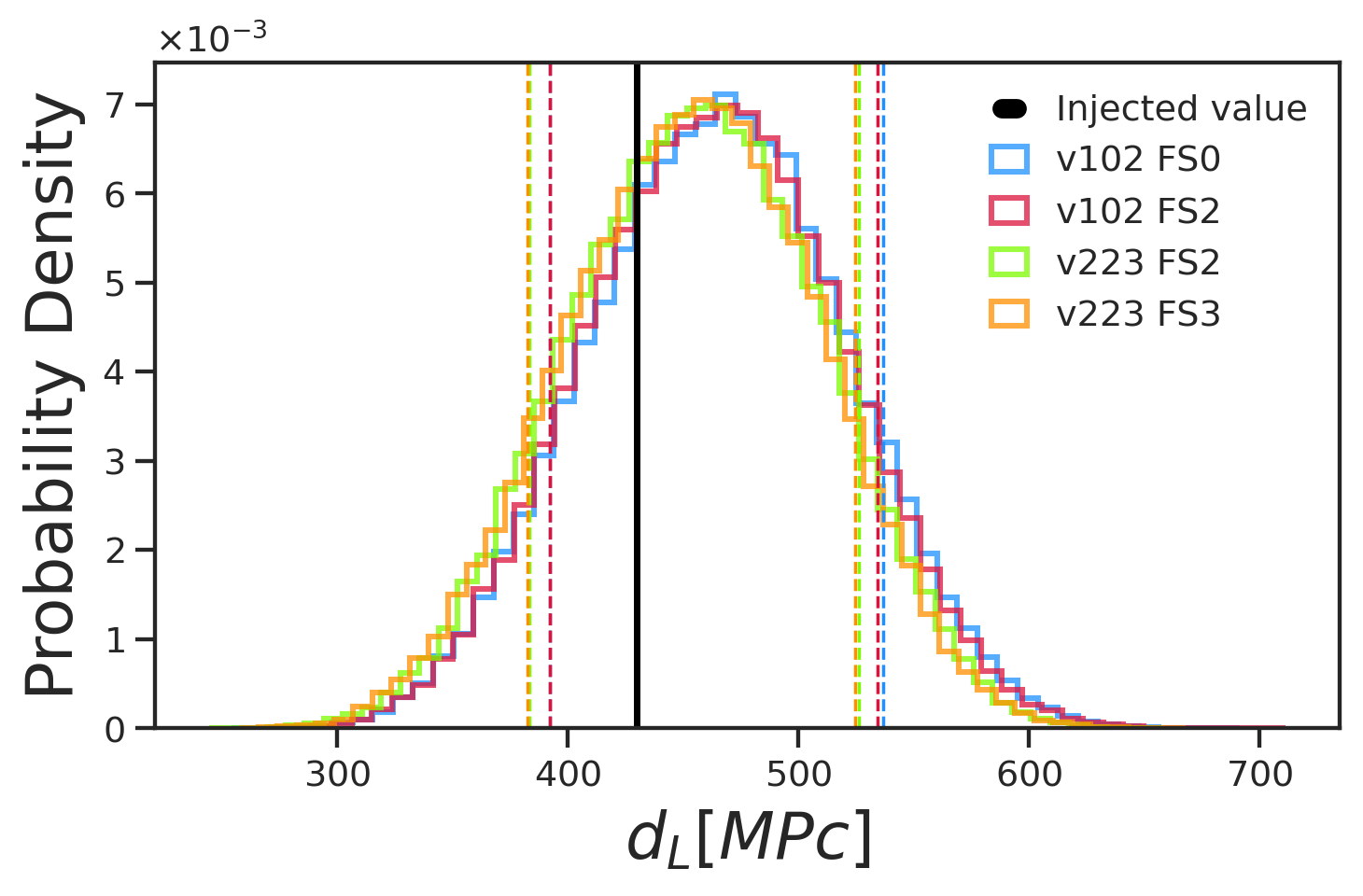}
       \includegraphics[scale=.39]{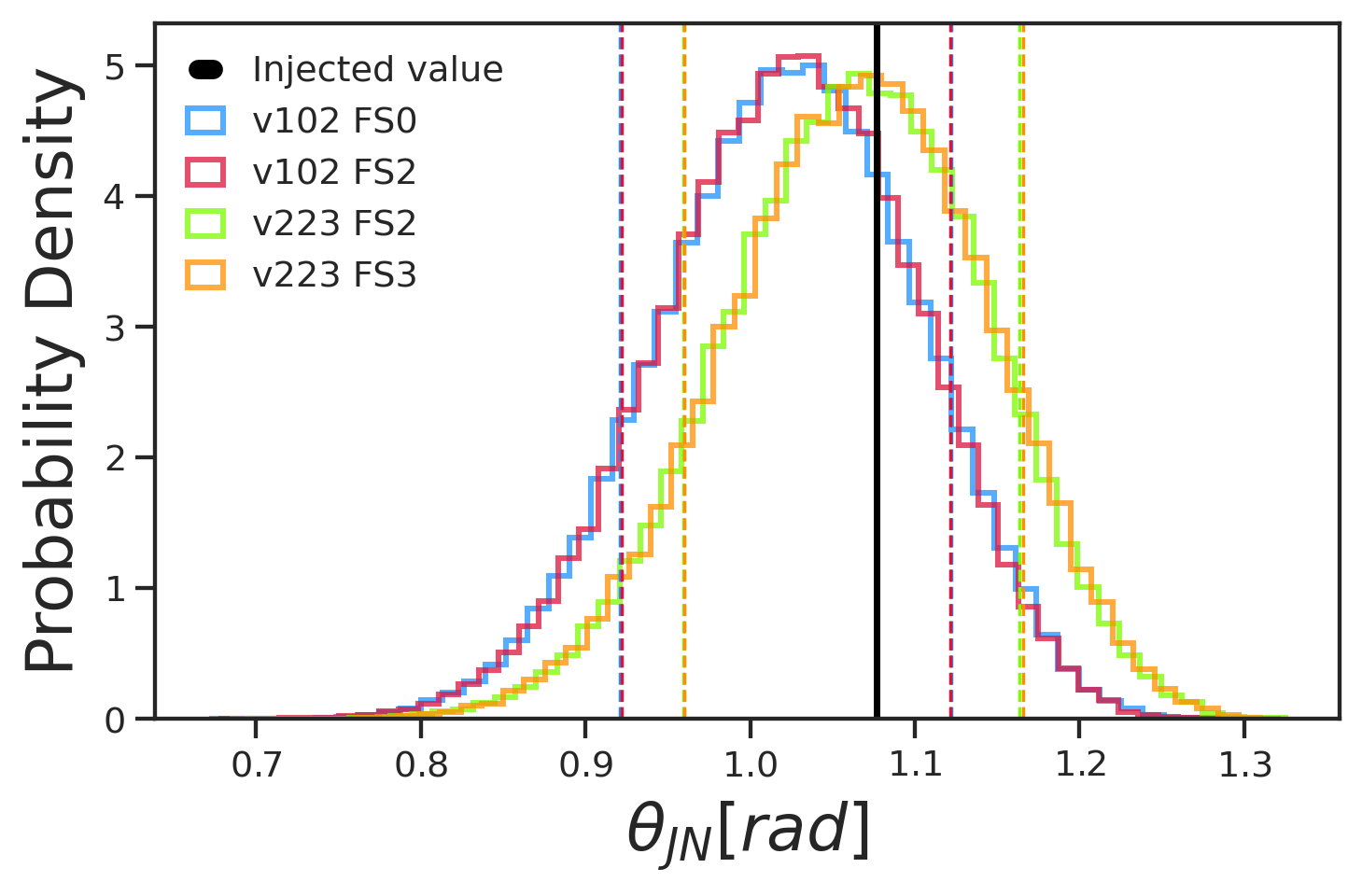}
        \caption{Injection recovery results for \texttt{SXS:BBH:0143}.
        Top row: From left to right, posterior probability distributions for the mass ratio ($q=m_2/m_1$), total mass in the detector frame and $\chi_{\text{eff}}$.
        Bottom row: From left to right, posterior probability distributions for $\chi_p$, luminosity distance and the angle $\theta_{\text{JN}}$ between the total angular momentum and line of sight.
        The dashed vertical lines represent 90 \% credible intervals, while the thick black lines represent the injected value. The notation for the different versions of \phXPHM is described in Appendix \ref{appendix:lal_implementation}. 
        }
     \label{fig:NRInj143}
% \end{figure*}

% \vspace{\baselineskip}

% % \begin{figure*}[htpb]
%     \centering
%       \includegraphics[scale=.39]{Plots/SXS0049Posterior_chirp_mass.png}
%       \includegraphics[scale=.39]{Plots/SXS0049Posterior_total_mass.png}
%       \includegraphics[scale=.39]{Plots/SXS0049Posterior_chi_eff.png}
%       \includegraphics[scale=.39]{Plots/SXS0049Posterior_chi_p.png}
%       \includegraphics[scale=.39]{Plots/SXS0049Posterior_luminosity_distance.png}
%       \includegraphics[scale=.39]{Plots/SXS0049Posterior_theta_jn.png}
%         \caption{Injection recovery results for \texttt{SXS:BBH:0049}.
%         All panels are the same as in Fig.~\ref{fig:NRInj143}.}
%      \label{fig:NRInj049}

% \vspace{\baselineskip}     
%       \centering
%       \includegraphics[scale=.39]{Plots/SXS0165Posterior_mass_ratio.png}
%       \includegraphics[scale=.39]{Plots/SXS0165Posterior_total_mass.png}
%       \includegraphics[scale=.39]{Plots/SXS0165Posterior_chi_eff.png}
%       \includegraphics[scale=.39]{Plots/SXS0165Posterior_chi_p.png}
%       \includegraphics[scale=.39]{Plots/SXS0165Posterior_luminosity_distance.png}
%       \includegraphics[scale=.39]{Plots/SXS0165Posterior_theta_jn.png}
%         \caption{Injection recovery results for \texttt{SXS:BBH:0165}.
%         All panels are the same as in Fig.~\ref{fig:NRInj143}.}
%      \label{fig:NRInj165}
\end{figure*}

As for our analysis of GW170729 above, we use the \texttt{Parallel Bilby} code \cite{Smith:2019ucc} with the \texttt{Dynesty} \cite{Speagle:2020spe} nested sampler. 
We perform these runs using $2048$ live points and 5 ACTs without any kind of marginalization and 4 independent seeds which are then merged together. The noise spectral density used for evaluating the likelihood function is the projected sensitivity for Advanced LIGO obtained from simulations of O4 data \cite{PSDs_NRinj}. The orientation of the detector at the time of injection is specified through the injection time (within 3 minutes of the event GW150914) and at right ascension and declination angles  $\alpha$=1.375 rad and  $\delta$=-1.2108 rad (consistent with GW150914 \cite{Abbott:2016blz}). 
%, as well as the GPS time of the injection 1126259642.413 s. 
We specify the line of sight relative to the binary at the reference frequency of 20 Hz through the angles $\iota$=1.077 and $\phi=0$, relative to the direction of the initial orbital angular momentum, as required by the LALSuite infrastructure for numerical relativity injections \cite{Schmidt:2017btt}. Note that these angles are actually strongly time dependent due to the precessing motion of the orbital angular momentum. For strongly precessing systems results will in general depend significantly on the line of sight and on the polarisation (in addition to the sky location). While this is partially accounted for in our SNR-weighted match calculation (see Sec.\ref{sec:model_matches}), a realistic assessment of the systematic errors for parameter estimation with a precessing waveform model would require a much more detailed study discussing such dependencies that would go beyond the scope of this work. Finally, we choose the polarization angle to be $\psi=0.33$ rad.

%For \texttt{SXS:BBH:0143} we use 5 ACTs, however for \texttt{SXS:BBH:0165} we needed to increase to 30 ACTs to ensure convergence.

% We set the prior distributions for the source parameters as follows:
% for the first injection (\texttt{SXS:BBH:0143}), we restrict to mass ratios up to $q=8$, and the chirp mass prior is assumed uniform between $30$ and $55~M_{\odot}$. The component masses are constrained to be between $10$ and $80~M_{\odot}$. The luminosity distance prior is uniform in volume with maximal distance at $2000$ Mpc.  For the \texttt{SXS:BBH:0165} injection, we allow mass ratios up to $q=20$ and set a chirp mass prior uniform between $15$ and $45~M_{\odot}$. The component masses are constrained to be between $5$ and $100~M_{\odot}$. The luminosity distance prior is the same than in the previous case. For both injection recoveries, dimensionless spin magnitudes are allowed up to the extreme Kerr limit.
We set the prior distributions for the source parameters as follows:
uniform prior on the mass ratio up to $q=8$ and a uniform prior on the chirp mass between $30$ and $55~M_{\odot}$. The component masses are constrained to be between $10$ and $80~M_{\odot}$. The luminosity distance prior is uniform in volume with maximal distance at $2000$ Mpc. The dimensionless spin magnitudes are allowed up to the extreme Kerr limit and the spin orientations are taken to be isotropically distributed.

% In Figs.~\ref{fig:NRInj143} and \ref{fig:NRInj165} 
In Fig.~\ref{fig:NRInj143}
we show 1D probability distributions for the main parameters of the injected waveform using different versions of the \phXPHM model as templates. We do not include all possible combinations of versions but only the most relevant ones. We also point out that the FS2 version for the final spin is more suitable for massive events where the merger is more prominent, while the default versions FS0 and FS3 (respectively for NNLO and MSA angles) are more suitable for lower-mass events where the inspiral region is more relevant. The mean values of the recovered parameters and their 90\% credible interval are reported in Table~\ref{tab:tabInj}. Note that the figures and table report the mass ratio as the inverse of our definition in the text for consistency with the typical choice in publications of the LIGO and Virgo collaborations to report results for the mass ratio in the interval (0,1].

In Fig~\ref{fig:NRInj143}, %corresponding to the injection of \texttt{SXS:BBH:0143}, 
we observe that the differences between several model versions are small except for $\chi_p$, where different prescriptions for the Euler angles (NNLO vs MSA) deliver sensibly different results. One can appreciate that, even for moderate mass ratio and small $\chi_p$ values, different modelling strategies can affect the final result and, most noticeably, our measurement of precession effects.

%For a very unequal mass and strongly precessing case such as \texttt{SXS:BBH:0165}, we observe significant biases in some parameters, as anticipated. The most affected parameter is again $\chi_p$. We also see that the differences between model versions are even more significant. We take this as an indication that current PN-based prescriptions for the Euler angles lose accuracy for very unequal-mass, strongly precessing systems. 

Since our model is built in the Fourier domain, we rely on the stationary phase approximation, which is strictly valid only in the inspiral regime. This might exacerbate biases for high total-mass events, for which the merger portion of the signal is more ``visible'' in the detector band. Injecting a lower-mass signal or a longer waveform could improve the results.
%: we note that \texttt{SXS:BBH:0165} is a very short simulation with only $\approx$6.5 orbits, what makes its recovery particularly challenging. We also note that its short duration could lead to problems in the windowing of the waveform during the Fourier transform. A thorough study of the impact of these choices on injection recovery is beyond the scope of this work.
There exist some proposals to improve the description of the Euler angles for Fourier-domain models, such as an extension of the SPA approximation through merger \cite{marsat2018fourierdomain} or a direct calibration of the Euler angles to NR simulations \cite{eleanor}. Time-domain models do not rely on the SPA approximation and are expected to perform better than their Fourier-domain counterparts. We foresee that a new precessing time-domain model based on the recently developed \textsc{IMRPhenomT} waveform family \cite{phenomt, phenomthm} will help to alleviate some of the problems discussed here. The new model, called \textsc{IMRPhenomTPHM}, allows to choose among different final spin and precession prescriptions, including a fully numerical evolution of the spin evolution equations which goes beyond the analytical MSA and NNLO approximations available in \phXPHM. %We therefore consider \texttt{SXS:BBH:0165} as a test case that pushes the model to its limits and is not indicative of the overall performance of the model. We have shown in previous examples that the model indeed performs very well in the vanilla region of parameter space where most of the detected events lay so far.

In this example, the choice of priors and sampler settings were sufficient to obtain a correct recovery of parameters. However, for more massive events, the short duration of the signal is expected to introduce degeneracies in the recovered parameters, and the sampler may struggle to find the correct posterior distribution. For more challenging events, one will need to perform a deeper study to ensure that the different PE settings do not affect the result. We performed a more detailed analysis of PE settings and convergence of runs in our reanalysis of GW190412 \cite{GW190412_uib}.

\section{Conclusions}\label{sec:conclusions}

With this paper we conclude the construction of a first version of the \phXF waveform family for coalescences of non-eccentric binary black holes. We have shown that including subdominant harmonics and precession in a waveform model does not have to come at the expense of evaluation speed.
This opens up the possibility to routinely perform parameter estimation for the larger numbers of events observed as gravitational wave detector sensitivity increases without neglecting subdominant harmonics and precession.

Our model and its LALSuite implementation have been designed with flexibility and modularity as key design elements, not only in order to incorporate future improvements, but also to allow a wide range of computational experiments -- as a first application we have shown parameter estimation studies that compare different prescriptions for the final spin.

A number of further improvements are foreseen: first, the increasing number and quality of numerical relativity waveforms will allow to calibrate further non-precessing subdominant harmonics to numerical relativity, e.g.~the $(5,5)$ and $(4,3)$ modes, and to further increase the quality of the model for non-precessing waveforms.

Regarding precession, several challenges need to be addressed (see also \cite{Ramos-Buades:2020noq}): First, it will be desirable to develop a computationally inexpensive numerical fit for the final spin of precessing coalescences. Another natural extension will be to develop a phenomenological ansatz for the Euler angle description, and calibrate it to numerical relativity. Other challenges are the development and calibration of a
phenomenological description for the complete precessing waveform for the merger and ringdown, where the SPA approximation is invalid, and which should also include the asymmetries responsible for large recoils \cite{Brugmann:2007zj}. Finally, we note that precession is currently measured in parameter estimation in terms of the parameter $\chi_p$, which is motivated by the NNLO inspiral description, and it is likely to be fruitful to also consider other parameters in the future.

Some of these challenges will be difficult to address in the frequency domain, and we foresee
synergies between the development of time domain and frequency domain approaches. As a first step we have constructed \phTP, a time domain version of the ideas underlying the \phXP model, which is aimed to serve as a tool to better understand and remedy current shortcomings of our construction, and is presented in \cite{phenT}.
We expect our code to be sufficiently flexible to serve as a basis for a future implementation of the \phT model.

Planned extensions regard the incorporation of eccentricity, and of matter effects along the lines of \cite{Dietrich:2017aum,Dietrich:2019kaq,Thompson:2020nei}. We note that the accelaration technique of multibanding is particularly important for lower-mass systems like binary neutron star and black hole--neutron star systems.
Finally we note that our code implementation leaves much room for increasing computational efficiency, e.g.~by utilizing GPUs.

\section*{Acknowledgements}

This work was supported by European Union FEDER funds, the Ministry of Science, 
Innovation and Universities and the Spanish Agencia Estatal de Investigación grants FPA2016-76821-P,        % FPA2017-90687-REDC, FPA2017-90566-REDC. 
RED2018-102661-T,    % RENATA
RED2018-102573-E,    % REDES ESTRATÉGICAS: Participación Española en Estructuras Euro... 
FPA2017-90687-REDC,  % CPAN
Vicepresid\`encia i Conselleria d’Innovació, Recerca i Turisme, Conselleria d’Educació i Universitats del Govern de les Illes Balears i Fons Social Europeu, 
Generalitat Valenciana (PROMETEO/2019/071),  
EU COST Actions CA18108, CA17137, CA16214, and CA16104, and
the Spanish Ministry of Education, Culture and Sport grants FPU15/03344 and FPU15/01319.
M.C. acknowledges funding from the European Union's Horizon 2020 research and innovation programme, under the Marie Skłodowska-Curie grant agreement No. 751492.
M.H. acknowledges support from Swiss National Science Foundation (SNSF) grant IZCOZ0-177057.
D.K. is supported by the Spanish Ministerio de Ciencia, Innovaci{\'o}n y
Universidades (ref. BEAGAL 18/00148)
and cofinanced by the Universitat de les Illes Balears.
J.T. was supported by the Science and Technology Facilities Council (STFC) grant ST/L000962/1 and by the European Research Council Consolidator Grant 647839.
The authors thankfully acknowledge the computer resources at MareNostrum and the technical support provided by Barcelona Supercomputing Center (BSC) through Grants No. AECT-2019-2-0010, AECT-2019-1-0022, AECT-2019-1-0014, AECT-2018-3-0017, AECT-2018-2-0022, AECT-2018-1-0009, AECT-2017-3-0013, AECT-2017-2-0017, AECT-2017-1-0017, AECT-2016-3-0014, AECT2016-2-0009,  from the Red Española de Supercomputación (RES) and PRACE (Grant No. 2015133131). BAM and ET simulations were carried out on the BSC MareNostrum computer under PRACE and RES (Red Española de Supercomputación) allocations and on the FONER computer at the University of the Balearic Islands. 
Authors also acknowledge the computational resources at the cluster CIT provided by LIGO Laboratory and supported by National Science Foundation Grants PHY-0757058 and PHY-0823459, as well as the cluster HAWK provided by Cardiff University and supported by STFC grant ST/I006285/1.
This research has made use of data obtained from the Gravitational Wave Open Science Center~\cite{GWOSC}, a service of LIGO Laboratory, the LIGO Scientific Collaboration and the Virgo Collaboration. LIGO is funded by the U.S. National Science Foundation. Virgo is funded by the French Centre National de Recherche Scientifique (CNRS), the Italian Istituto Nazionale della Fisica Nucleare (INFN) and the Dutch Nikhef, with contributions by Polish and Hungarian institutes.

% ~~~~~~~~~~ To-Do List ~~~~~~~~~~ %

% ~~~~~~~~~~ Appendix ~~~~~~~~~~ %
\appendix
%\clearpage
%%%%%%%%%%%%%%%%%%%%%%%%%%%%%%%%%%%%%%%%%%%%%%%%%%%%%%%%%%%%%%%%%
%               WIGNER COEFFICIENTS
%%%%%%%%%%%%%%%%%%%%%%%%%%%%%%%%%%%%%%%%%%%%%%%%%%%%%%%%%%%%%%%%%

\section{Wigner-\texorpdfstring{$d$}{d} Matrices}
\label{appendix:wignerd}

The real-valued Wigner-$d$ matrices are defined by \cite{Goldberg:1966uu,Brown:2007jx}
\begin{widetext}
\begin{align}
    d_{m m^{\prime}}^{\ell}(\beta)=\sum_{k=k_{\min }}^{k_{\max }} \frac{(-1)^{k}}{k !} \frac{\sqrt{(l+m) !(l-m) !\left(l+m^{\prime}\right) !\left(l-m^{\prime}\right) !}}{(l+m-k) !\left(l-m^{\prime}-k\right) !\left(k-m+m^{\prime}\right) !}\left(\cos \frac{\beta}{2}\right)^{2 \ell+m-m^{\prime}-2 k}\left(\sin \frac{\beta}{2}\right)^{2 k-m+m^{\prime}} ,
\end{align}
\end{widetext}
with $k_{\min }=\max\left(0,m-m^{\prime}\right)$ and $k_{\max }=\min(\ell+m,\ell-m^{\prime})$,
where analogous expressions for $d_{m -\mpr}^{\ell}$ can be constructed using the symmetry of the Wigner d-matrices
\begin{align}
    d^{\ell}_{-m -\mpr} &= (-1)^{(m-\mpr)} d^{\ell}_{m \mpr}, \\
    d^{\ell}_{m \mpr} (\beta) &= (-1)^{m + \mpr} \, d^{\ell}_{\mpr m} (\beta), \\ d^{\ell}_{m \mpr} (\beta) &= d^{\ell}_{\mpr m }(-\beta) .
\end{align} 
Here we provide explicit expressions for the Wigner $d_{m \mpr}^{\ell}(\beta)$ matrices for all modes involved in the underlying non-precessing model. We also include the $43$ mode since it is used by the option that twists \phHM.
\subsection{\texorpdfstring{$\ell = 2, \:m^{\prime} = 2$}{l=2,m'=2}}
\begin{align*}
    d^2_{22} (\beta) &= \cos^4 \botwo, \\
    d^2_{12} (\beta) &= 2 \;  \cos^3 \botwo \; \sin \botwo, \\
    d^2_{02} (\beta) &= \sqrt{6} \;  \cos^2 \botwo \; \sin^2 \botwo, \\
    d^2_{-12} (\beta) &= 2 \;  \cos \botwo \; \sin^3 \botwo, \\
    d^2_{-22} (\beta) &= \sin^4 \botwo.
\end{align*}

\subsection{\texorpdfstring{$\ell = 2, \:m^{\prime} = 1$}{l=2,m'=1}}
\begin{align*}
    d^2_{21} (\beta) &= -2 \; \cos^3 \botwo \; \sin \botwo, \\
    d^2_{11} (\beta) &= \cos^2 \botwo \left( \cos^2 \botwo - 3 \sin^2 \botwo \right), \\
    d^2_{01} (\beta) &= \sqrt{6} \; \left( \cos^3 \botwo \sin \botwo - \cos \botwo \sin^3 \botwo \right), \\
    d^2_{-11} (\beta) &= \sin^2 \botwo \left( 3 \cos^2 \botwo - \sin^2 \botwo \right), \\
    d^2_{-21} (\beta) &= 2 \; \cos \botwo \; \sin^3 \botwo.
\end{align*}

\subsection{\texorpdfstring{$\ell = 3,  \:m^{\prime} = 3$}{l=3,m'=3}}
\begin{align*}
    d^3_{33} (\beta) &= \cos^6 \botwo, \\
    d^3_{23} (\beta) &= \sqrt{6} \;  \cos^5 \botwo \; \sin \botwo, \\
    d^3_{13} (\beta) &= \sqrt{15} \;  \cos^4 \botwo \; \sin^2 \botwo, \\
    d^3_{03} (\beta) &= 2 \sqrt{5} \;  \cos^3 \botwo \; \sin^3 \botwo, \\
    d^3_{-13} (\beta) &= \sqrt{15} \;  \cos^2 \botwo \; \sin^4 \botwo, \\
    d^3_{-23} (\beta) &= \sqrt{6} \;  \cos \botwo \; \sin^5 \botwo, \\
    d^3_{-33} (\beta) &= \sin^6 \botwo.
\end{align*}

\subsection{\texorpdfstring{$\ell = 3, \:m^{\prime} = 2$}{l=3,m'=2}}
\begin{align*}
    d^3_{32} (\beta) &= -\sqrt{6}\;  \cos^5 \botwo \; \sin \botwo, \\
    d^3_{22} (\beta) &= \cos^4 \botwo \left( \cos^2 \botwo - 5 \sin^2 \botwo \right), \\
    d^3_{12} (\beta) &= \sqrt{10} \cos^3 \botwo \left( \cos^2 \botwo \sin \botwo - 2 \sin^3 \botwo \right) \\
    d^3_{02} (\beta) &= \sqrt{30} \cos^2 \botwo \sin^2 \botwo \left( \cos^2 \botwo - \sin^2 \botwo  \right), \\
    d^3_{-12} (\beta) &= \sqrt{10} \sin^3 \botwo \left( 2 \cos^3 \botwo - \cos \botwo \sin^2 \botwo  \right), \\
    d^3_{-22} (\beta) &= \sin^4 \botwo \left( 5 \cos^2 \botwo - \sin^2 \botwo \right), \\
    d^3_{-32} (\beta) &= \sqrt{6} \cos \botwo \; \sin^5 \botwo.
\end{align*}

\subsection{\texorpdfstring{$\ell = 4,  \:m^{\prime} = 4$}{l=4,m'=4}}
\begin{align*}
    d^4_{44} (\beta) &= \cos^8 \botwo, \\
    d^4_{34} (\beta) &= 2 \sqrt{2} \;  \cos^7 \botwo \; \sin \botwo, \\
    d^4_{24} (\beta) &= 2 \sqrt{7} \;  \cos^6 \botwo \; \sin^2 \botwo, \\
    d^4_{14} (\beta) &= 2 \sqrt{14} \; \cos^5 \botwo \; \sin^3 \botwo, \\
    d^4_{04} (\beta) &= \sqrt{70} \; \cos^4 \botwo \; \sin^4 \botwo, \\
    d^4_{-14} (\beta) &= 2 \sqrt{14} \;  \cos^3 \botwo \; \sin^5 \botwo, \\
    d^4_{-24} (\beta) &= 2 \sqrt{7} \;  \cos^2 \botwo \; \sin^6 \botwo, \\
    d^4_{-34} (\beta) &= 2 \sqrt{2} \;  \cos \botwo \; \sin^7 \botwo, \\
    d^4_{-44} (\beta) &= \sin^8 \botwo.
\end{align*}

\subsection{\texorpdfstring{$\ell = 4,  \:m^{\prime} = 3$}{l=4,m'=3}}
\begin{align*}
    d^4_{43} (\beta) &= -2 \sqrt{2} \sin \left(\frac{\beta }{2}\right) \cos \
^7\left(\frac{\beta }{2}\right), \\
    d^4_{33} (\beta) &= \cos ^8\left(\frac{\beta }{2}\right)-7 \sin ^2\left(\frac{\beta }{2}\right) \cos ^6\left(\frac{\beta }{2}\right), \\
    d^4_{23} (\beta) &= \sqrt{14} \sin \left(\frac{\beta }{2}\right) \cos ^7\left(\frac{\beta \
}{2}\right)-3 \sqrt{14} \sin ^3\left(\frac{\beta }{2}\right) \cos \
^5\left(\frac{\beta }{2}\right), \\
    d^4_{13} (\beta) &= 3 \sqrt{7} \sin ^2\left(\frac{\beta }{2}\right) \cos \
^6\left(\frac{\beta }{2}\right)-5 \sqrt{7} \sin ^4\left(\frac{\beta \
}{2}\right) \cos ^4\left(\frac{\beta }{2}\right), \\
    d^4_{03} (\beta) &= 2 \sqrt{35} \sin ^3\left(\frac{\beta }{2}\right) \cos \
^5\left(\frac{\beta }{2}\right)-2 \sqrt{35} \sin ^5\left(\frac{\beta \
}{2}\right) \cos ^3\left(\frac{\beta }{2}\right), \\
    d^4_{-13} (\beta) &= 5 \sqrt{7} \sin ^4\left(\frac{\beta }{2}\right) \cos \
^4\left(\frac{\beta }{2}\right)-3 \sqrt{7} \sin ^6\left(\frac{\beta \
}{2}\right) \cos ^2\left(\frac{\beta }{2}\right), \\
    d^4_{-23} (\beta) &= 3 \sqrt{14} \sin ^5\left(\frac{\beta }{2}\right) \cos \
^3\left(\frac{\beta }{2}\right)-\sqrt{14} \sin ^7\left(\frac{\beta \
}{2}\right) \cos \left(\frac{\beta }{2}\right), \\
    d^4_{-33} (\beta) &= 7 \sin ^6\left(\frac{\beta }{2}\right) \cos ^2\left(\frac{\beta }{2}\right)-\sin ^8\left(\frac{\beta }{2}\right), \\
    d^4_{-43} (\beta) &= 2 \sqrt{2} \sin ^7\left(\frac{\beta }{2}\right) \cos \
\left(\frac{\beta }{2}\right).
\end{align*}

%%%%%%%%%%%%%%%%%%%%%%%%%%%%%%%%%%%%%%%%%%%%%%%%%%%%%%%%%%%%%%%%%
%               CONVENTIONS FOR NON-PRECESSING
%%%%%%%%%%%%%%%%%%%%%%%%%%%%%%%%%%%%%%%%%%%%%%%%%%%%%%%%%%%%%%%%%

\section{Conventions for non-Precessing Modes}
\label{sec:conventions_no_prec}

In the precessing $L$-frame we can decompose the gravitational wave strain into spherical harmonic modes $h_{\ell m}$ as
\begin{equation}\label{eq:h_harmonics}
    h^L = \sum_{\ell=2, \, m=-\ell}^{4,\ell}   h_{\ell m}^L \, (t) \; _{-2}Y_{\ell m}^L,
\end{equation}
where the $Y_{\ell m}^L$ are sphericdal harmonics in the precessing $L$-frame defined in Sec.~\ref{appendix:frames}. In our twisting approximation we identify the modes $h_{\ell m}^L$ in the non-inertial precessing frame with the spherical harmonic modes described by the non-precessing \phX and \phXHM models, which are however modified by changing the complex ringdown frequencies in the waveform to be consistent with the estimate for the precessing final spin, which we have described in Sec.~\ref{sec:final_state}.

The time domain modes can be written in terms of positive amplitudes $a_{\ell m} (t)$ and phases $\phi_{\ell m} (t)$ such that 
\begin{equation}
h_{\ell m}^L (t) = a_{\ell m} (t) e^{- i \phi_{\ell m} (t)},
\end{equation}
where we assume that the phase of the aligned-spin modes is a monotonically increasing function of $t$ 
\begin{align}
    \dot{\phi}_{\ell m} (t) > 0.
\end{align}

As discussed in detail in \cite{Garcia-Quiros:2020qpx} there are only two inequivalent choices of tetrad convention that are consistent with equatorial symmetry, and for simplicity we adopt the convention that for low frequencies the time domain phases satisfy
\begin{equation}\label{eq:tetrad_convention}
    \Phi_{\ell m} \approx  \frac{m}{2}\Phi_{2 2} .
\end{equation}
This differs from the convention of Blanchet et al. \cite{Blanchet:2008je} by overall factors of $(-1) (-\iota)^m$ in front of the $h_{\ell m}^L$ modes.

The equatorial symmetry of non-precessing binaries implies
\begin{equation}\label{eq:neg_m}
    h_{\ell m}^L(t) = (-1)^\ell {h^{*}}_{\ell -m}^L(t). 
\end{equation}
With our conventions  for the Fourier transform (\ref{eq:def_FFT}),
the time domain relations between modes (\ref{eq:neg_m}) that express equatorial symmetry can be converted to the Fourier domain, where they read 
\begin{equation}\label{eq:htilde_mode_relations}
   \tilde h_{\ell m}^L(f) =  (-1)^\ell \tilde {h^{*}}_{\ell -m}^L(-f).
\end{equation}
The definitions above then also imply that $\tilde h_{\ell m}^L(f)$ (with $m>0$) is concentrated
in the negative frequency domain and $\tilde h_{\ell -m}^L(f)$ 
in the positive frequency domain. 

The Fourier amplitudes $A_{\ell m}(f > 0)$ are then non-negative functions for positive frequencies, and zero otherwise, and the Fourier domain phases $\Phi_{\ell m} (f > 0)$, defined by 
\begin{equation}
\tilde{h}_{\ell -m}^L (f) = A_{\ell m} (f) \, e^{-i \Phi_{\ell m} (f)}. 
\end{equation}

%%%%%%%%%%%%%%%%%%%%%%%%%%%%%%%%%%%%%%%%%%%%%%%%%%%%%%%%%%%%%%%%%
%               FRAMES TRANSFORMATIONS
%%%%%%%%%%%%%%%%%%%%%%%%%%%%%%%%%%%%%%%%%%%%%%%%%%%%%%%%%%%%%%%%%

\section{Frames Transformations and Polarization Basis}\label{appendix:frames}

We  construct our waveform model in terms of a 
transformation from spherical harmonics $h_{\ell m}^L$ in the precessing $L$-frame to spherical harmonics $h_{\ell m}^J$ in the inertial $J$-frame. 
The input to our model consists of intrinsic parameters that specify the masses and spin vectors of the binary system, and extrinsic parameters which describe the location of the source and its spatial orientation relative to the frame of the observer, which we have chosen to refer to as the $N$-frame, where $\hN$ is the direction from the source to the observer. 

We need to guarantee that our model returns an unambiguous waveform for given values of the intrinsic and extrinsic parameters, in particular we carefully need to specify in which coordinate system we specify the spin vectors, and how to specify the spatial orientation of the source as needed to define all extrinsic parameters. To this end, in this section we will discuss the relation between the different coordinate frames we are using, and identify a complete set of input parameters.

We will work with three inertial frames,
which we have introduced in Sec.~\ref{sec:conventions}:
the $L_0$-frame, the $J$-frame and $N$-frame or wave frame. These three frames will have their $z$-axis aligned with $\LO$,  $\J$ and $\hN$ respectively, and they do not evolve in time. The $L_0$-frame
will also be referred to as the source frame, since in accordance with the LALSuite software infrastructure the \phXP and \phXPHM models parameterize the two spin vectors of the black holes by Cartesian components in this frame.
%The LALInference \cite{Veitch:2014wba} model for Bayesian inference within LALSuite does in fact use a different parameterisation of the spin components, which emphasizes angles between vectors of physical interest, which facilitates the interpretation of observational results, and we will discuss how these two parameterizations relate to each other.

In addition to the three inertial frames we will consider the non-inertial $L$-frame that tracks the precession of the orbital plane, and coincides with the $L_0$-frame at a chosen reference frequency $f_{ref}$. Choosing a different value for $f_{ref}$ while fixing the initial spin components, which we have chosen to specify with respect to the $L_0$-frame, will thus in general corresponds to a different waveform.

We will denote the triads of our frames by $\{\x_{\LO}, \y_{\LO},\z_{\LO} \}$,
$\{\x_{\BL}, \y_{\BL},\z_{\BL} \}$,
$\{\x_{\J}, \y_{\J}, \z_{\J}\}$ and
$\{\x_{\hN}, \y_{\hN},\z_{\hN} \}$ and choose the $z$-axes as
\begin{equation}
\z_{\LO} = \hLO, \quad 
\z_{\BL} = \hL, \quad 
\z_{\J} = \hJ, \quad 
\z_{\hN} = \hN.
\end{equation}
Note that the $\x_{\LO}-\y_{\LO}$ plane correspond to initial orbital plane.

For clarity, in what follows we will provide explicit expressions for the vectors $\hLO, \hN, \J$ in the $L_0$ and $J$-frames. 
We will write the components of a generic vector $\mathbf{u}$ in a particular frame as 
\begin{equation}
    \mathbf{u} \dot= \begin{pmatrix}u_{x,frame}\\ u_{y,frame}\\ u_{z,frame} \end{pmatrix}_{frame},
\end{equation}
e.g.~the vector $\J$ will have the following components in the $L_0$ frame
\begin{equation}
    \J \dot= \begin{pmatrix} J_{x,L_0}\\ J_{y,L_0}\\ J_{z,L_0} \end{pmatrix}_{L_0}.
\end{equation}

So far we have characterized our three inertial coordinate frames by the choice of $z$-axis, and we have defined the precessing $L$-frame as the time evolution of the $L_0$-frame.
We now need to complete our definitions of the three inertial frames by fixing the freedom of rotations around the axes, which precisely corresponds to the freedom of specifying three Euler angles to fix a spatial rotation.
In the $L_0$ frame, which we have chosen as
the frame where we parametrize our input spin components, we choose  
the line of sight $\hN$  to have spherical angles $\left(\iota, \frac{\pi}{2} - \phi_{ref} \right)$, which  is chosen to adopt to the conventions of \cite{Schmidt:2010it},
where this choice corresponds to eq.~(31c) 
and setting the angle $\Phi$ in this equation to $\Phi = \phi_{ref}$.
We call $\iota$ the inclination of the system, and interpret $\phi_{ref}$
as fixing the freedom of rotations in the orbital plane.
We treat $\iota$ and $\phi_{ref}$ as input parameters that the user specifies when calling the waveform model.

In the $L_0$ frame the components of $\hN$ are thus
\begin{equation}
\label{eq:NinLframe}
        \hN \:\dot=\begin{pmatrix} \sin \iota \cos \left( \frac{\pi}{2} - \phi_{ref} \right) \\ \sin \iota \sin \left( \frac{\pi}{2} - \phi_{ref} \right) \\ \cos \iota \end{pmatrix}_{L_0},
\end{equation}
while the vector $\J$ reads
\begin{align}
\label{eq:JinLframe}
 \J \:&\dot=  \begin{pmatrix} m_1^2 \chi_{1x} + m_2^2 \chi_{2x} \\  m_1^2 \chi_{1y} + m_2^2 \chi_{2y} \\ L_0 + m_1^2 \chi_{1z} + m_2^2 \chi_{2z} \end{pmatrix}_{L_0}.
\end{align}
From the above equation it follows that the spherical angles $(\theta_{JL_0},\phi_{JL_0})$ of $\J$ in this frame are given by:
\begin{align}
    \theta_{JL_0} &= \arccos \frac{J_{z,L_0}}{|\J|} = \arccos \frac{ L_0 + m_1^2 \chi_{1z} + m_2^2 \chi_{2z} }{|\J|}, \\
    \phi_{JL_0} &= \arctan \frac{J_{y,L_0}}{J_{x,L_0}} = \arctan \frac{  m_1^2 \chi_{1y} + m_2^2 \chi_{2y} }{ m_1^2 \chi_{1x} + m_2^2 \chi_{2x} }.
\end{align}

We will now turn to describing the $J$-frame. We will fix the orientation of the axes in this frame by requiring that the line of sight $\hN$ lies in the $\x_{\J}-\J$ plane and has positive projection on the $\x_{\J}$-axis. If we denote by $\theta_{JN}$ the angle between $\hN$ and the $\z_{\J}$ axis, this choice implies
\begin{equation}
\label{eq:NinJ}
 \hN \:\dot= \begin{pmatrix} \sin \theta_{JN} \\ 0\\ \cos \theta_{JN} \end{pmatrix}_{J}.
 \end{equation}
With the above definition of $\theta_{JN}$, and using Eqs.\,(\ref{eq:NinLframe}, \ref{eq:JinLframe}) one has 
\begin{align}
    \theta_{\rm{JN}}&=\arccos\left({\frac{\bf{J}\cdot\hN}{| \J | }}\right)\\
    &=\frac{J_{x,L_0}\sin{\iota}\sin{\phi_{\rm{ref}}}+ J_{y,L_0}\sin{\iota}\cos{\phi_{\rm{ref}}}+J_{z,L_0}\cos\iota}{\sqrt{J_{x,L_0}^2+J_{y,L_0}^2+J_{z,L_0}^2}}.
\end{align}
Consequently, in the twisting-up formula (\ref{eq:def_polarizations_TD}), the spherical harmonics $_{-2}Y_{\ell m}(\theta,\phi)$ will have arguments $(\theta,\phi)=(\theta_{\mathrm{JN}},0)$.  
Having specified our convention regarding the $J$-frame, we can determine unambiguously the Euler angles relating it to the $L_0$-frame. We can identify two of these angles with $\theta_{JL_0}$ and $\phi_{JL_0}$, as two successive rotations $\mathbf{R}_{y}(-\theta_{JL_0}) \cdot \mathbf{R}_{z}(-\phi_{JL_0})$ will align $\hJ$ with the $\z_{\LO}$ axis, we call this intermediate frame as $J^\prime$-frame. In order to bring $\hN$ in the $\x_{\J}-\z_{\J}$ plane, we need to apply a further rotation around the $\z_{\LO}$-axis by an angle $-\kappa$, so that Eq.\,(\ref{eq:NinJ}) is satisfied. To this end, we first compute
\begin{align}
    \begin{pmatrix} \hat N_{x,J'}\\\hat N_{y,J'}\\\hat N_{z,J'}\end{pmatrix}_{J'} =\mathbf{R}_{y}(-\theta_{JL_0}) \cdot \mathbf{R}_{z}(-\phi_{JL_0}) \cdot \begin{pmatrix} \hat N_{x,L_0}\\\hat N_{y,L_0}\\\hat N_{z,L_0}\end{pmatrix}_{L_0},
\end{align}
and then take
\begin{align}
\kappa=\arctan{\frac{\hat N_{y,J'}}{\hat N_{x,J'}}}.
\end{align}
The three Euler angles relating the $L_0$ and $J$ frames are therefore $\kappa,\theta_{JL_0},\phi_{JL_0}$. The components of any vector $\mathbf{u}$ in the two frames are related via: 
\begin{align}
\label{eq:rotLtoJ}
    \begin{pmatrix} u_{x,J}\\{u}_{y,J}\\{u}_{z,J}\end{pmatrix}_{J} =\mathbf{R}_{z}(-\kappa) \cdot \mathbf{R}_{y}(-\theta_{JL_0}) \cdot \mathbf{R}_{z}(-\phi_{JL_0}) \cdot  \begin{pmatrix} {u}_{x,L_0}\\{u}_{y,L_0}\\{u}_{z,L_0}\end{pmatrix}_{L_0}.
\end{align}
Equivalently, a generic vector in the $J$-frame can be rotated to the $L_0$-frame by applying the inverse transformation 
\begin{equation}
\label{eq:rotJtoL}
    \begin{pmatrix} {u}_{x,L_0}\\{u}_{y,L_0}\\{u}_{z,L_0}\end{pmatrix}_{L_0} =\mathbf{R}_{z}(\phi_{JL_0}) \cdot \mathbf{R}_{y}(\theta_{JL_0}) \cdot \mathbf{R}_{z}(\kappa) \cdot  \begin{pmatrix} u_{x,J}\\u_{y,J}\\u_{z,J}\end{pmatrix}_{J}.
\end{equation}
%It can be checked that, with the above identifications, the vector
Notice that, in the $J$-frame, the initial angular momentum has components:
\begin{align}
\label{eq:L0inJ}
    \hLO \dot=  \begin{pmatrix} \sin \beta_0 \cos \alpha_0\\ \sin \beta_0 \sin \alpha_0\\ \cos \beta_0 \end{pmatrix}_{J}.
\end{align}
We can compute the initial value of the Euler angles $(\alpha,\beta,\gamma)$ introduced in Sec.\,\ref{sec:basics}, by identifying the product of rotations in Eq.\,(\ref{eq:rotLtoJ}) with $\mathbf{R}_z(\alpha_0) \cdot \mathbf{R}_y(\beta_0) \cdot \mathbf{R}_z(\gamma_0)$, whence it follows that
\begin{subequations}
\label{eq:def_alpha_epsilon_0}
\begin{align}
%\label{eq:def_alpha_epsilon_0}
\alpha_0&=\pi-\kappa,\\
\beta_0&=\theta_{JL_0},\\
\gamma_0&=-\epsilon_0=\pi-\phi_{JL_0}
\end{align}
\end{subequations}
It can be checked that, with the choice of offsets above, the initial angular momentum in the $J$-frame is indeed rotated to the $\z_{\LO}$-axis by the transformation of Eq.\,(\ref{eq:rotJtoL}). 

The NNLO and MSA angle prescriptions provide expressions described in Secs.~\ref{sec:single_spin_angles} and \ref{sec:double_spin_angles} for the Euler angles as functions of frequency. 
In order to initialize the angles to prescribed values $\alpha_0$, $\beta_0$, $\epsilon_0$
according to Eqs.~(\ref{eq:def_alpha_epsilon_0})
at a given reference frequency $f_{ref}$, we have to add an offset to the functional dependence of the NNLO/MSA angles in the following way 
\begin{subequations}
\label{eq:def_angle_offset}
\begin{align}
\alpha(f) &= \alpha_{\mathrm{NNLO/MSA}}(f) - \alpha_{\mathrm{offset}}
\\
\epsilon(f) &= \epsilon_{\mathrm{NNLO/MSA}}(f) - \epsilon_{\mathrm{offset}}
\end{align}
\end{subequations}
Here the offsets are constant values that
correspond to frequency-independent rotations of the system into the desired state at the reference frequency, and a typical example would be to choose the $\alpha$ offset as $-(\alpha( f_{ref}) - \alpha_0)$. 
Our code offers different options to compute these offsets, which we discuss in appendix \ref{appendix:lal_implementation} and list in Table~\ref{tab:offsets}.

Finally we fix the remaining freedom in the $N$-frame. We have previously aligned the $\z_{\hN}$ axis with $\hN$, so we just have to fix a rotation around $\hN$. Following the LALSuite convention \cite{laldoxygen,Schmidt:2017btt}
we choose the $\x_{\hN}$-axis such that the vector $\LO$ lies in the $\x_{\hN}-\z_{\hN}$ plane with positive projection over $\x_{\hN}$, and $\y_{\hN}$ so that it completes the triad. Equivalently:
\begin{equation}
\label{eq:axisNframe}
    \hLO \cdot \x_{\hN} = \sin \iota,  \:\;   \y_{\hN} = \frac{\hN \times \x_{\hN}}{|\hN \times \x_{\hN}|}.
\end{equation}
In the $N$-frame the components of $\hLO$ are then given by 
\begin{align}\label{Nframe}
    \hLO \:&\dot= \:( \sin \iota, 0, \cos \iota)_{N}.
\end{align}

The $\x_{\hN}$, $\y_{\hN}$ axes we have just introduced do not coincide with the spherical basis vectors orthogonal to $\hN$ that determine the arguments of the weighted spherical harmonics in our $J$-frame. Therefore, we have to compute the angle $\zeta$ that rotates one basis into the other. We will call the original  polarization axes $\x^{\prime}_{\hN}$ and $\y^{\prime}_{\hN}$. Note that, geometrically, these vectors can be defined as
\begin{align}
\x^{\prime}_{\hN}=\frac{\y^{\prime}_{\hN}\times \hN}{|\y^{\prime}_{\hN}\times \hN|},\ \ \ \y^{\prime}_{\hN}=\frac{\J\times \hN}{|\J\times \hN|},
\end{align}
and are therefore equivalent to the choice made in \cite{Kidder:1995zr}, as we explain in App.\,{\ref{appendix:polarizations}} below. Under a rotation by an angle $\zeta$, the polarization basis-vectors transform as:
\begin{subequations}
\label{eq:rotate_pol_basis}
\begin{align}
\x^{\prime}_{\hN}&=\cos({\zeta})\x_{\hN}-\sin({\zeta})\y_{\hN}\\
\y^{\prime}_{\hN}&=\sin({\zeta})\x_{\hN}+\cos({\zeta})\y_{\hN}.
\end{align}
\end{subequations}
Since $\zeta$ can vary from $0$ to $2\pi$, we will use the C function \texttt{atan2} to track the correct quadrant and set
\begin{equation}
    \zeta = \texttt{atan2} \left( \x_{\hN} \cdot \y^{\prime}_{\hN}, \x_{\hN} \cdot \x^{\prime}_{\hN}   \right).
\end{equation}
In the code implementation the scalar products above are computed in the $J$-frame, where the vectors $\x^{\prime}_{\hN}$ and $\y^{\prime}_{\hN}$ have components
\begin{equation}
    \x^{\prime}_{\hN} \dot= \begin{pmatrix}   \cos \theta_{JN}\\ 0\\ - \sin \theta_{JN} \end{pmatrix}_J, \:\:\:
    \y^{\prime}_{\hN} \dot= 
        \begin{pmatrix}  0\\0\\1\end{pmatrix}_J.
\end{equation}
The components of $\x_{\hN}$ in the $J$-frame can be computed by applying the transformation (\ref{eq:rotLtoJ}) to $\x_{\hN}$ expressed in the $L_0$-frame, giving 
\begin{equation}
 \x_{\hN} \dot= 
 \begin{pmatrix} - \cos \iota \:\sin \phi_{ref}\\ -\cos \iota \:\cos \phi_{ref} \\ \sin \iota\end{pmatrix}_{L_0}.
\end{equation}

%%%%%%%%%%%%%%%%%%%%%%%%%%%%%%%%%%%%%%%%%%%%%%%%%%%%%%%%%%%%%%%%%
%               CHOICES OF P AND Q
%%%%%%%%%%%%%%%%%%%%%%%%%%%%%%%%%%%%%%%%%%%%%%%%%%%%%%%%%%%%%%%%%

\section{Choices of polarization vectors P and Q}
\label{appendix:polarizations}

In the literature it is common to introduce a polarization basis 
$(\hat{P}^i, \hat{Q}^i)$ such that
the strain tensor is constructed in the usual way as \cite{Arun:2008kb,Schmidt:2017btt}
\begin{align}
    h_{+} &= \frac{1}{2} \left( \hat{P}^i \hat{P}^j - \hat{Q}^i \hat{Q}^j \right) h_{ij} , \\
    h_{\times} &= \frac{1}{2} \left( \hat{P}^i \hat{Q}^j + \hat{Q}^i \hat{P}^j \right) h_{ij} .
\end{align}

Different choices of polarization basis can be achieved through a rotation around the $\z_{\hN}$-axis, i.e. the line of sight. 
In the convention followed by Arun et al. \cite{Arun:2008kb}, the polarization basis is given by 
\begin{equation}
\label{eq:Arun}
  \hP_{\rm{ABFO}} = \frac{\hN \times \J}{|\hN \times \J|}, \:\:\: \hQ_{\rm{ABFO}} = \frac{\hN \times \hP_{\rm{ABFO}}}{|\hN \times \hP_{\rm{ABFO}}|},  
\end{equation}
whereas the basis chosen by Kidder \cite{Kidder:1995zr} is 
\begin{equation}
\label{eq:Kidder}
  \hP_{\rm{Kidder}} = \frac{\hQ_{\rm{Kidder}} \times \hN}{|\hQ_{\rm{Kidder}} \times \hN|}, \:\:\: 
  \hQ_{\rm{Kidder}} = \frac{\J \times \hN}{|\J \times \hN|}.
\end{equation}
Note that, from the above definitions, it follows that
\begin{equation}
    \hP_{\rm{ABFO}} = -\hQ_{\rm{Kidder}}, \:\:\: \hQ_{\rm{ABFO}} = \hP_{\rm{Kidder}},
\end{equation}
and, using Eqs.\,(\ref{eq:rotate_pol_basis}), one can equivalently say that the two polarization bases are related by a rotation of $\zeta = \pi/2$.
This translates into an overall sign difference in the gravitational-wave polarizations $h_{+}$ and $h_{\times}$, since these transform under a rotation by $\zeta$ around the line of sight according to
\begin{align}
    h^{\prime}_+ &= \cos (2\zeta) \: h_+ - \sin ({2\zeta}) \: h_{\times}, \\
    h^{\prime}_{\times} &= \sin (2\zeta) \: h_+ + \cos ({2\zeta}) \: h_{\times}.
\end{align}

%%%%%%%%%%%%%%%%%%%%%%%%%%%%%%%%%%%%%%%%%%%%%%%%%%%%%%%%%%%%%%%%%
%               DERIVATION FD WAVEFORM
%%%%%%%%%%%%%%%%%%%%%%%%%%%%%%%%%%%%%%%%%%%%%%%%%%%%%%%%%%%%%%%%%
\section{Derivation of the frequency domain waveform}
\label{appendix:derivation}

The waveform modes in the inertial $J$-frame and the precessing $L$-frame can be related via a time-domain transformation
\begin{align}
\label{eq:twisting_td}
    h^{J}_{\ell m}(t) &= e^{-i\:m \alpha(t)} \sum_{m^{\prime}=-\ell}^{\ell} e^{i\:m^{\prime}\epsilon(t)} d^{\ell}_{m m^{\prime}}(\beta(t))h_{\ell m^{\prime}}^{L}(t)
\end{align}
%\nl
Performing a Fourier transform, and making use of the SPA, as done in \cite{Bohe:PPv2}, we obtain
\begin{align}
\label{eq:twisting_fd}
    \tilde{h}^{J}_{\ell m}(f) &=  \sum_{m^{\prime}=-\ell}^\ell  e^{i\:m^{\prime}\epsilon \left(\frac{2\pi f}{m^{\prime}} \right)} e^{-i\:m \alpha \left(\frac{2\pi f}{m^{\prime}} \right)} d^{\ell}_{m m^{\prime}} \left(\beta\left(\frac{2\pi f}{m^{\prime}}\right)\right) \tilde{h}_{\ell m^{\prime}}^{L}(f) .
\end{align}
%\nl 
We now follow the standard paradigm and approximate the precessing frame modes with some equivalent non-precessing modes \cite{Schmidt:2010it,Schmidt:2012rh,Hannam:2013oca}. In our conventions, the positive $m$-modes are defined with support only for negative frequencies while the negative $m$-modes are defined with support for positive frequencies, i.e. $\tilde{h}_{\ell m^{\prime}>0}^{J}(f>0) = 0$ and $\tilde{h}_{\ell m^{\prime}<0}^{J}(f<0) = 0$. We can therefore re-write the above expression as 
\begin{widetext}
\begin{align}
\label{eq:twisting_fd_split}
    \tilde{h}^{J}_{\ell m}(f>0) &=  \sum_{m^{\prime}>0 }^{\ell } e^{-i\:m^{\prime}\epsilon \left(\frac{2\pi f}{m^{\prime}} \right)} e^{-i\:m \alpha \left(\frac{2\pi f}{m^{\prime}} \right)} d^{\ell}_{m -m^{\prime}} \left(\beta\left(\frac{2\pi f}{m^{\prime}}\right)\right) \tilde{h}_{\ell -m^{\prime}}^{L}(f)
    \\
    \tilde{h}^{J}_{\ell  m}(f<0) &=  \sum_{m^{\prime}>0 }^{\ell}  e^{i\:m^{\prime}\epsilon \left(\frac{-2\pi f}{m^{\prime}} \right)} e^{-i\:m \alpha \left(\frac{-2\pi f}{m^{\prime}} \right)} d^{\ell}_{m m^{\prime}} \left(\beta\left(\frac{-2\pi f}{m^{\prime}}\right)\right) (-1)^l\tilde{h}_{\ell -m^{\prime}}^{L*}(-f)
\end{align}
\end{widetext}
We now wish to construct expressions for the gravitational-wave polarizations $h_+$ and $h_{\times}$. First we start with the gravitational-wave strain 
\begin{align}
    h^{J} (t) &= h^J_+(t) - i\: h^J_{\times}(t) = \sum_{\ell \geq 2} \sum_{m=-\ell}^{m=\ell} h^J_{\ell m}(t) _{-2}Y_{\ell m}.
\end{align}
%\nl 
The individual polarizations can therefore be written as 
\begin{align}
    h^J_+(t) &= \frac{1}{2} \sum_{\ell \geq 2} \sum_{m=-\ell}^{m=\ell} \left( h^J_{\ell m}(t) \, _{-2}Y_{\ell m} + h^{J\ast}_{\ell m}(t) \, _{-2}Y^{\ast}_{\ell m} \right),
    \\
    h^J_{\times}(t) &= \frac{i}{2} \sum_{\ell \geq 2} \sum_{m=-\ell }^{m=\ell } \left( h^J_{\ell m}(t) \, _{-2}Y_{\ell m} - \, h^{J\ast}_{\ell m}(t) \, _{-2}Y^{\ast}_{\ell m} \right),
\end{align}
\nl 
which, after performing a Fourier transformation, can be written as frequency-domain functions
\begin{align}
    \tilde h^J_+(f) &= \frac{1}{2} \sum_{\ell \geq 2} \sum_{m=-\ell}^{m=\ell} \left( \tilde h^J_{\ell m}(f) \, _{-2}Y_{\ell m} + \tilde h^{J\ast}_{\ell m}(-f) \, _{-2}Y^{\ast}_{\ell m} \right),
    \\
    \tilde h^J_{\times}(f) &= \frac{i}{2} \sum_{\ell \geq 2} \sum_{m=-\ell}^{m=\ell} \left( \tilde h^J_{\ell m}(f) \, _{-2}Y_{\ell m} - \tilde h^{J\ast}_{\ell m}(-f) \, _{-2}Y^{\ast}_{\ell m} \right).
\end{align}
Now we insert Eq.~(\ref{eq:twisting_fd}) into the above expressions to expand the polarizations in terms of the non-precessing modes $h^L_{\ell m}(f)$
\begin{widetext}
\begin{align}
     \tilde h^J_+(f) &= 
     \frac{1}{2} \sum_{\ell \geq 2} \sum_{m=-\ell}^{m=\ell} \left(  \sum_{m^{\prime}=-\ell}^{\ell}  e^{i\:m^{\prime}\epsilon} e^{-i\:m \alpha } d^{\ell}_{m m^{\prime}}\left( \beta \right) \tilde{h}_{\ell m^{\prime}}^{L}(f) \, _{-2}Y_{\ell m} 
     + \sum_{m^{\prime}=-\ell}^{\ell}  e^{-i\:m^{\prime}\epsilon } e^{i\:m \alpha} d^{\ell}_{m m^{\prime}} \left(\beta\right) \tilde{h}_{\ell m^{\prime}}^{L\ast}(-f) \, _{-2}Y^{\ast}_{\ell m} \right),
     \\
     \tilde h^J_{\times}(f) &= 
     \frac{i}{2} \sum_{\ell\geq 2} \sum_{m=-\ell}^{m=\ell} \left(  \sum_{m^{\prime}=-\ell}^{\ell}  e^{i\:m^{\prime}\epsilon} e^{-i\:m \alpha } d^{\ell}_{m m^{\prime}}\left( \beta \right) \tilde{h}_{\ell m^{\prime}}^{L}(f) \, _{-2}Y_{\ell m} 
     - \sum_{m^{\prime}=-\ell}^\ell  e^{-i\:m^{\prime}\epsilon } e^{i\:m \alpha} d^{\ell}_{m m^{\prime}} \left(\beta\right) \tilde{h}_{\ell m^{\prime}}^{L\ast}(-f) \, _{-2}Y^{\ast}_{\ell m} \right).
\end{align}
\end{widetext}
%\n 
We can now use the equatorial symmetry of the non-precessing modes to relate the positive $m$ and negative $m$ modes via $\tilde h^L_{\ell m}(f) = (-1)^{\ell} \tilde h^{L\ast}_{\ell - m}(-f)$. Inserting this into the above equations, we find
\begin{widetext}
\begin{align}
\label{eq:fourier_polariz_1}
     \tilde h^J_+(f) &= 
     \frac{1}{2} \sum_{\ell \geq 2} \sum_{m=-\ell}^{m=\ell} \left(  \sum_{m^{\prime}=-\ell}^{\ell}  e^{i\:m^{\prime}\epsilon} e^{-i\:m \alpha } d^{\ell}_{m m^{\prime}}\left( \beta \right) \tilde{h}_{ \ell m^{\prime}}^{L}(f) \, _{-2}Y_{\ell m} 
     + \sum_{m^{\prime}=-\ell}^{\ell}  e^{-i\:m^{\prime}\epsilon } e^{i\:m \alpha} d^{\ell}_{m m^{\prime}} \left(\beta\right) (-1)^{\ell} \tilde{h}_{\ell - m^{\prime}}^{L}(f) \, _{-2}Y^{\ast}_{\ell m} \right),
     \\
     \label{eq:fourier_polariz_2}
     \tilde h^J_{\times}(f) &= 
     \frac{i}{2} \sum_{\ell \geq 2} \sum_{m=-\ell}^{m=\ell} \left(  \sum_{m^{\prime}=-\ell}^{\ell}  e^{i\:m^{\prime}\epsilon} e^{-i\:m \alpha } d^{\ell}_{m m^{\prime}}\left( \beta \right) \tilde{h}_{\ell m^{\prime}}^{L}(f) \, _{-2}Y_{\ell m} 
     - \sum_{m^{\prime}=-\ell}^{\ell}  e^{-i\:m^{\prime}\epsilon } e^{i\:m \alpha} d^{\ell}_{m m^{\prime}} \left(\beta\right) (-1)^{\ell} \tilde{h}_{\ell - m^{\prime}}^{L}(f) \, _{-2}Y^{\ast}_{\ell m} \right).
\end{align}
\end{widetext}
%\nl 
Since the polarizations $\tilde h^J_{+,\times}(f)$ are Fourier transforms of real functions $h^J_{+,\times}(t)$, they satisfy the property $\tilde{h}^J_{+,\times}(f) = \tilde{h}^{J,\ast}_{+,\times}(-f)$. This means that we can work with just one of the frequency regimes, i.e. positive or negative. We opt to work with the positive frequencies, following the standard convention in LALSuite. Restricting the expressions to positive frequencies, and remembering that only the negative non-precessing modes are non-zero, we find that Eqs.~(\ref{eq:fourier_polariz_1}, \ref{eq:fourier_polariz_2}) reduce to
\begin{widetext}
\begin{align}
\label{eq:fourier_polariz_positive_pre_prelim}
     \tilde h^J_+(f>0) &= 
     \frac{1}{2} \sum_{\ell \geq 2} \sum_{m=-\ell}^{m=\ell} \left(  \sum_{m^{\prime}>0}^{\ell}  e^{-i\:m^{\prime}\epsilon} e^{-i\:m \alpha } d^{\ell}_{m -m^{\prime}}\left( \beta \right) \tilde{h}_{\ell - m^{\prime}}^{L}(f) \, _{-2}Y_{\ell m} 
     + \sum_{m^{\prime}>0}^{\ell}  e^{-i\:m^{\prime}\epsilon } e^{i\:m \alpha} d^{\ell}_{m m^{\prime}} \left(\beta\right) (-1)^{\ell} \tilde{h}_{\ell - m^{\prime}}^{L}(f) \, _{-2}Y^{\ast}_{\ell m} \right),
     \\
     \tilde h^J_{\times}(f>0) &= 
     \frac{i}{2} \sum_{\ell \geq 2} \sum_{m=-\ell}^{m=\ell} \left(  \sum_{m^{\prime}>0}^{\ell}  e^{-i\:m^{\prime}\epsilon} e^{-i\:m \alpha } d^{\ell}_{m -m^{\prime}}\left( \beta \right) \tilde{h}_{\ell - m^{\prime}}^{L}(f) \, _{-2}Y_{\ell m} 
     - \sum_{m^{\prime}>0}^{\ell}  e^{-i\:m^{\prime}\epsilon } e^{i\:m \alpha} d^{\ell}_{m m^{\prime}} \left(\beta\right) (-1)^{\ell} \tilde{h}_{\ell-m^{\prime}}^{L}(f) \, _{-2}Y^{\ast}_{\ell m} \right).
\end{align}
%\end{widetext}
%\nl 
%\begin{widetext}
Rearranging terms and swapping the sums in $m$ and $m^{\prime}$ we get 
\begin{align}
\label{eq:fourier_polariz_positive_prelim}
     \tilde h^J_+(f>0) &= 
     \frac{1}{2} \sum_{\ell \geq 2} \sum_{m^{\prime}>0}^{\ell} e^{-i\:m^{\prime}\epsilon} \tilde{h}_{\ell - m^{\prime}}^{L}(f) \sum_{m=-\ell}^{m=\ell} \left(   e^{-i\:m \alpha } d^{\ell}_{m -m^{\prime}}\left( \beta \right)  \, _{-2}Y_{\ell m} 
     + e^{i\:m \alpha} d^{\ell}_{m m^{\prime}} \left(\beta\right) (-1)^{\ell} _{-2}Y^{\ast}_{\ell m} \right),
     \\
     \tilde h^J_{\times}(f>0) &= 
     \frac{i}{2} \sum_{\ell \geq 2} \sum_{m^{\prime}>0}^{\ell} e^{-i\:m^{\prime}\epsilon} \tilde{h}_{\ell - m^{\prime}}^{L}(f) \sum_{m=-\ell}^{m=\ell} \left(   e^{-i\:m \alpha } d^{\ell}_{m -m^{\prime}}\left( \beta \right) \, _{-2}Y_{\ell m} 
     -  e^{i\:m \alpha} d^{\ell}_{m m^{\prime}} \left(\beta\right) (-1)^{\ell} \, _{-2}Y^{\ast}_{\ell m} \right).
\end{align}
\end{widetext}
%\nl
We define now the transfer function $A^{\ell}_{m m^{\prime}}(f) = e^{-i m \alpha} d^{\ell}_{mm^{\prime}}(\beta)  _{-2}Y_{\ell m}$ and rewrite the above expressions in a more compact form,
%\n 
%\begin{widetext}
\begin{align}
\label{eq:fourier_polariz_positive}
     \tilde h^J_+(f>0) &= 
     \frac{1}{2} \sum_{\ell \geq 2} \sum_{m^{\prime}>0}^{\ell}  e^{-i\:m^{\prime}\epsilon} \tilde{h}_{\ell - m^{\prime}}^{L}(f) \sum_{m=-\ell}^{m=\ell} \left(  A^{\ell}_{m -m^{\prime}} + (-1)^{\ell} \: A^{\ell \ast}_{m m^{\prime}} \right),
     \\
     \tilde h^J_{\times}(f>0) &= 
     \frac{i}{2} \sum_{\ell \geq 2} \sum_{m^{\prime}>0}^{\ell}  e^{-i\:m^{\prime}\epsilon} \tilde{h}_{\ell-m^{\prime}}^{L}(f) \sum_{m=-\ell}^{m=\ell} \left(  A^{\ell}_{m-m^{\prime}} -(-1)^{\ell} \: A^{\ell \ast}_{m m^{\prime}} \right),
\end{align}
%\end{widetext}
which are equivalent to the expressions in Eqs.~(\ref{eq:polarizations_1}, \ref{eq:polarizations_2}). Note that the Euler angles are evaluated at the SPA frequencies $2\pi f / m^{\prime}$.

%%%%%%%%%%%%%%%%%%%%%%%%%%%%%%%%%%%%%%%%%%%%%%%%%%%%%%%%%%%%%%%%%
%               LAL Implementation
%%%%%%%%%%%%%%%%%%%%%%%%%%%%%%%%%%%%%%%%%%%%%%%%%%%%%%%%%%%%%%%%%
\section{LALSuite Implementation}
\label{appendix:lal_implementation}

The \phXP and \phXPHM models are implemented as part of the
{\tt LALSimIMR} package of inspiral-merger-ringdown waveform models as extensions of the non-precessing models  \phX \cite{Pratten:2020fqn} and \phXHM \cite{Garcia-Quiros:2020qpx}.
The code is implemented in the C language, and 
{\tt LALSimIMR} is 
part of the {\tt LALSimulation} collection of code for gravitational waveform and noise generation within LALSuite \cite{lalsuite}.
Online Doxygen documentation is available at {\tt https://lscsoft.docs.ligo.org/lalsuite}, with top level information for the {\tt LALSimIMR} package
provided through the {\tt LALSimIMR.h} header file.
Externally callable functions follow the \texttt{XLAL} coding standard of LALSuite.

Notes about the implementation of the \phXHM model, and on calling the code through different interfaces, in particular through \texttt{Python}, \texttt{GenerateSimulation}, \texttt{LALInference} \cite{Veitch:2014wba} and \texttt{Bilby} \cite{Ashton:2018jfp}
can be found in Appendix C of \cite{Garcia-Quiros:2020qpx}.
Here we extend this documentation to the specific properties of \phXP and \phXPHM. The LALSuite code provides the option to call the model in the time domain,  where an inverse fast Fourier transformation is applied, in addition to the native Fourier domain. The SWIG \cite{Beazley:1996swig,SWIG:code} software development tool is used to automatically create Python interfaces to all \texttt{XLAL} functions \cite{Wette:2020swig} of our code, which can be used alternatively to the C interfaces. 

In \texttt{LALSimulation} the model is called through the function \texttt{ChooseFDWaveform}, whose input parameters \texttt{f\_ref} and \texttt{phiRef} are used to define the phase of the 22 mode at some particular reference frequency. 

The user is free to specify the spherical harmonic modes in the co-precessing $L$-frame, $h^{L}_{\ell m}$, that should be used to construct the waveform. The default behaviour is to use
all the modes available: (22, 2-2, 21, 2-1, 33, 3-3, 32, 3-2, 44, 4-4), but subsets can be selected with the \textit{ModeArray} option available in \texttt{LALSimulation}. The negative modes are always included in the twisting-up procedure, even if not specified in the mode array, thus the list (22, 21, 33, 32, 44) would return the same result as the default setting. Specifying only negative modes is however not supported, e.g.~when only specifying the array (22, 2-1) only the (22, 2-2) modes would be twisted up.

Furthermore, the model implemented in LALSuite supports acceleration of waveform evaluation by interpolation of an unequispaced frequency grid broadly following the ``multibanding'' approach of \cite{Vinciguerra:2017ngf}. Our version of the algorithm is described in 
\cite{Garcia-Quiros:2020qlt} to do the evaluation faster and can also use a custom list of modes specified by the user.
The multibanding algorithm is parameterized by a threshold, which describes the permitted local interpolation error for the phase in radians. Lower values 
thus correspond to higher accuracy.
The default multibanding threshold for computing the non-precessing modes is set to a value of $10^{-3}$ and, as for \textsc{IMRPhenomXHM}, is modified through the \textit{ThresholdMband} option.
For multibanding in the Euler angles, the default threshold is $10^{-3}$ for the MSA versions and $10^{-4}$ for all NNLO versions; this can be changed through the \textit{PrecThresholdMband} option. The multibanding is only supported in the \phXPHM model not so in \phXP.

\begin{table*}[hptb]
\centering
    \begin{tabular}{ |c | c | }
\hline
\textit{PrecVersion} & Explanation \\ \hline\hline
101    & NNLO PN Euler angles and a 2PN non-spinning approximation to L \\ \hline
102     & NNLO PN Euler angles and a 3PN spinning approximation to L \\ \hline
103     & NNLO PN Euler angles and a 4PN spinning approximation to L \\ \hline
104     & NNLO PN Euler angles and a 4PN spinning approximation to L augmented with\\ & leading PN order at all order in spin terms. \\ \hline
220     &  MSA Euler angles and a 3PN spinning approximation to L. \\ & Fall back to NNLO angles with 3PN approximation to L if MSA system fails to initialize.\\ \hline
221     &  MSA Euler angles and a 3PN spinning approximation to L. \\ & Throw error message if MSA system fails to initialize.\\ \hline
222     &  MSA Euler angles close to Pv3HM implemenation. \\& Throw error message if MSA system fails to initialize.\\ \hline
223     & MSA Euler angles closer to Pv3HM implemenation. \\ (default) & Fall back to NNLO with 3PN approximation to L if MSA system fails to initialize. \\ \hline
224   & As version 220 but using the $\phi_{z,0}$ and $\zeta_{z,0}$ prescription from 223. \\ \hline
\end{tabular}
    \caption{Options in the LALSuite implementation to change between different descriptions of the Euler angles. }
    \label{tab:prec_version}
% \end{table*}

\vspace{2\baselineskip}

% \begin{table*}[hptb]
    \centering
    \begin{tabular}{|c|c|c|c|}
    \hline
         \textit{Convention} & $\alpha_{\mathrm{offset}}$ & $\epsilon_{\mathrm{offset}}$ & \texttt{phiRef} argument passed \\ & & & to non-precessing modes \\\hline\hline
         0 &  $\alpha_{\mathrm{NNLO/MSA}}(f_\mathrm{ref}) -\alpha_0^\mathrm{XP}$ &  $\epsilon_{\mathrm{NNLO/MSA}}(f_\mathrm{ref}) $  & $\phi_{JL_0}$ \\ \hline
         1 (default) &  $\alpha_{\mathrm{NNLO/MSA}}( f_\mathrm{ref})-\alpha_0^\mathrm{Pv2}$ & $\epsilon_{\mathrm{NNLO/MSA}}( f_\mathrm{ref}) -\epsilon_0^\mathrm{XP}$ & 0 \\ \hline
         5 & $-\alpha^\mathrm{Pv2}_0$ & 0 &  \texttt{phiRef} \\ \hline
         6 & $\alpha_{\mathrm{NNLO/MSA}}( f_\mathrm{ref}) - \alpha^\mathrm{XP}_0$ &  $\epsilon_{\mathrm{NNLO/MSA}}( f_\mathrm{ref}) - \epsilon^\mathrm{XP}_0$ & \texttt{phiRef} \\ \hline
         7 & $-\alpha^\mathrm{XP}_0$  & 0 & \texttt{phiRef}\\ \hline
    \end{tabular}
    \caption{The superscript $\mathrm{XP}$ means that the initial angle was computed as described in Eqs.~(\ref{eq:def_alpha_epsilon_0}) and the superscript $\mathrm{Pv2}$ following the conventions detailed in \cite{Hannam:2013oca,Bohe:PPv2}.}
    \label{tab:offsets}
% \end{table*}

\vspace{2\baselineskip}

\begin{tabular}{ |c | c | }
\hline
\textit{FinalSpinMod} & Explanation \\ \hline\hline
0     & Final spin formula based on $\chi_p$. Default value for NNLO angles. \\ \hline
1     & Final spin formula based on $\chi_{1x}$. \\ &  Not recommended, introduced to compare to \phPvthree before bug fix. \\ \hline
2     & Final spin formula based on the norm for the total in-plane spin vector\\ \hline
3     & Final spin formula based on precession-averaged couplings from MSA analysis.\\
      & Default value for MSA angles. \\ \hline
\end{tabular}
\caption{Options for changing the final spin prescription in the LALSuite implementation of \phXP and \phXPHM.}
\label{tab:finalSpin}

\vspace{2\baselineskip}

% \begin{table*}[hptb]
\centering
    \begin{tabular}{ |c | c | c| c| }
\hline
Option & Values & Default & Explanation \\ \hline\hline
\textit{TwistPhenomHM} & 0, 1 &  0 (False) &  Twist-up \phHM instead of \phXHM.\\ & & & With $Convention=5$ this  produces a faster implementation of \phPvthreehm. 
\\ \hline
\textit{PrecModes} & 0, 1 & 0 (False) & When calling the individual modes functions return \\ & & & the modified non-precessing modes before the twist-up.
\\ \hline
\textit{UseModes} & 0, 1 & 0 (False) & When computing the polarizations first call all the \\ & & & individual modes in the $J$-frame and sum them all.
\\ \hline
\textit{PrecThresholdMband} & Float & $10^{-3}$  & Threshold value for the multibanding algorithm applied to \\ & & & the Euler angles. If 0 then multibanding for angles is switched off. 
\\ \hline
\textit{MBandPrecVersion} & 0 & 0 & Control the version for the coarse grid used for the \\& & & multibanding of Euler angles. Currently there is only one implementation and the \\ & & & grid is the same than for the non-precessing model. 
\\ \hline
\end{tabular}
    \caption{Extra options in the LALSuite implementation of \phXPHM. }
    \label{tab:extra_options}
% \end{table*}

\vspace{2\baselineskip}

% \begin{table*}[hptb]
    \centering
    \begin{tabular}{|c||c|c|c|c|}
    \hline
    %   \texttt{LALSimulation} & \multicolum{4}{|c|}{ Corresponding label for the options in external code } \\
        Option & \texttt{LALInference} & \texttt{Bilby} & \texttt{PyCBC} & \texttt{GenerateSimulation} \\ \hline\hline
        \textit{ModesList} & \texttt{modeList} & \texttt{mode\_array} & \texttt{mode\_array} & \texttt{modesList} \\ \hline
        \textit{PrecVersion} &  \texttt{phenomXPrecVersion} & \texttt{phenomXPrecVersion} & \texttt{phenomXPrecVersion} & \texttt{phenomXPrecVersion} \\ \hline
        \textit{FinalSpinMod} & \texttt{phenomXPFinalSpinMod} & \texttt{phenomXPFinalSpinMod} & - & -\\ \hline
        \textit{PrecThresholdMband} & \texttt{phenomXPHMMband} & \texttt{phenomXPHMMband} & \texttt{phenomXPHMMband} & \texttt{phenomXPHMMband} \\ \hline
        \textit{UseModes} & - & - & - & \texttt{phenomXPHMUseModes}\\ \hline
    \end{tabular}
    \caption{Labels used to pass \phXPHM options to different external codes. }
    \label{tab:external_options}
\end{table*}

The \phXP and \phXPHM models add further precession-specific options to those already documented in 
Appendix C of \cite{Garcia-Quiros:2020qpx}.  The default values of these options are set up in the file {\tt LALSimInspiralWaveformParams.c}. 
Available choices for the Euler angles are listed in Table~\ref{tab:prec_version} and set by a parameter 
\textit{PrecVersion}. The principal choice is between the NNLO and MSA angle descriptions discussed in sections \ref{sec:single_spin_angles} and \ref{sec:double_spin_angles}.
In addition for the NNLO angles different post-Newtonian orders for the angular momentum can be chosen, as discussed in Sec.~\ref{sec:PN_L}.
Different implementation choices are also available for the MSA angles.

The NNLO and MSA angle prescriptions provide expressions for frequency-dependent Euler angles. 
In order to initialize the angles to prescribed values $\alpha_0$, $\beta_0$, $\epsilon_0$
at a given reference frequency $f_\mathrm{ref}$
according to Eqs.~(\ref{eq:def_alpha_epsilon_0}), appropriate offsets need to be added as in Eqs.~(\ref{eq:def_angle_offset}).
Our code offers different options to compute these offsets, which we list in Table~\ref{tab:offsets}. These conventions are changed with the option \textit{Convention}, which also controls  how the argument \texttt{phiRef} enters in the model.  The default choice is set to option 1. Note that option 7 does not set offsets for a given reference frequency. This option is implemented for its correspondence to the implementation of the \phPvthreehm \cite{Khan:2019kot} model, which sets the offset of $\alpha$ equal to $-\alpha_0$ and the offset of $\epsilon$ equal to 0; the argument \texttt{phiRef} is passed when calling the non-precessing model.

Several variants are available to compute the final spin, which are selected with the option \textit{FinalSpinMod} (see Table~\ref{tab:finalSpin}). By default, the final spin is computed by using orbit-averaged quantities
for the in-plane spin components: When choosing NNLO angles, the default spin version is set to 0, corresponding to Eq.~(\ref{eq:finalspin_0}), while for MSA angles, the default spin version is set to 3, corresponding to Eq.~(\ref{eq:finalspin_3}). In addition, two non-averaged options are provided, which allow for cancellations between spin components as discussed in Sec.~\ref{sec:final_state}:
setting the spin version to 1 corresponds to  Eq.~(\ref{eq:finalspin_1}), while version 2 selects Eq.~(\ref{eq:finalspin_2}).

Finally, in Table~\ref{tab:extra_options} we summarize further options available.
We make some of these options also callable from other codes that may use the model like \texttt{LALInference}, \texttt{Bilby}, \texttt{PyCBC} or \texttt{GenerateSimulation}. In Table~\ref{tab:external_options} we summarize the options that can be called through the different codes and the label that is used to specify their value.
Since the released versions of some of these codes do not support these features yet, we provide dedicated branches for that:
\begin{itemize}
    \item \texttt{Bilby}:
    \url{https://git.ligo.org/cecilio.garcia-quiros/bilby/-/tree/imrphenomx}
    \item \texttt{Bilby\_pipe}:
    \url{https://git.ligo.org/maite.mateu-lucena/bilby_pipe}
    \item \texttt{PyCBC}:
    \url{https://github.com/Ceciliogq/pycbc/tree/imrphenomx/pycbc}
\end{itemize}
Extensive debugging information can be enabled at compile time with the C preprocessor flag {\tt -D PHENOMXHMDEBUG}.

%%%%%%%%%%%%%%%%%%%%%%%%%%%%%%%%%%%%%%%%%%%%%%%%%%%%%%%%%%%%%%%%%
%            Post-Newtonian Results
%%%%%%%%%%%%%%%%%%%%%%%%%%%%%%%%%%%%%%%%%%%%%%%%%%%%%%%%%%%%%%%%%

\section{Post-Newtonian Results}
We consider a compact binary with masses $m_{1,2}$ and spin angular momenta $S_{1,2}$. The post-Newtonian results presented in this section can be expressed in terms of the following variables:
\begin{align}
    \delta &= \sqrt{1 - 4 \eta}, \\
m_1 &= \frac{1+ \delta}{2}, \quad
m_2 = \frac{1- \delta}{2}, \\
\chi_i &= \frac{S_i}{m^2_1}, \\
\chi_{\ell} &= m_1 \chi_{1 \ell} + m_2 \chi_{2 \ell}, \\
S_{\ell} &= m^2_1 \chi_{1\ell} + m^2_2 \chi_{2 \ell}, \\
\Sigma_{\ell} &= \chi_{2\ell} m_2 - \chi_{1\ell} m_1 .
\end{align}

\subsection{NNLO post-Newtonian Euler Angles}
\label{appendix:nnlo_angles}

For completeness we write out the explicit expressions for the Euler angles $\alpha$ and $\epsilon$, computed to NNLO accuracy for single spin systems as used in the single spin version of our model, see \ref{sec:single_spin_angles}.
Both $\alpha$ and $\epsilon$ have the same functional form as functions of the frequency $f$,
\begin{eqnarray}
\alpha_{\rm{NNLO}} \, (\omega) = \sum^1_{i=-3} \, \alpha_i \, \left( \pi f M \right)^{i/3} + \alpha_{\log} \log (\pi f M),\\
\epsilon_{\rm{NNLO}} \, (\omega) = \sum^1_{i=-3} \, \epsilon_i \, \left( \pi f M \right)^{i/3} + \alpha_{\log} \log (\pi f M).
\end{eqnarray}
The coefficients $\alpha_{0}$ and 
$\epsilon_0$ are determined by Eqs.~(\ref{eq:def_alpha_epsilon_0}). The coefficients for $\alpha$  are listed below as functions of the intrinsic parameters $(\eta, \chi_l, \chi_p)$.
%, where
%we use the abbreviations
%\begin{eqnarray}
%\delta &=& \sqrt{1 - 4 \eta}, \\
%m_1 &=& \frac{1+ \delta}{2}, \\
%m_2 &=& \frac{1- \delta}{2}, \\
%\chi_l &=& m_1 \chi_{1,l} + m_2 \chi_{2,l}.
%\end{eqnarray}
\begin{widetext}
\begin{subequations}
\begin{align}
\label{eq:phase_coeff_0}
\alpha_{-3} &= -\frac{5 \delta }{64 m_1}-\frac{35}{192}\\
\alpha_{-2} &= -\frac{5 m_1 \chi_{\ell} \left(3 \delta +7 m_1\right)}{128 \eta }\\
\alpha_{-1} &=-\frac{5 (824 \eta +1103)}{3072}-\frac{15 \delta ^2 \eta }{256 m_1^2}-\frac{15 \delta  m_1^3 \chi_p^2}{128 \eta ^2}-\frac{5 \delta  (980 \eta
   +911)}{7168 m_1}-\frac{35 m_1^4 \chi_p^2}{128 \eta ^2}\\
\alpha_{1} &= \frac{5 (36 \eta  (85568 \eta +23817)+8024297)}{9289728}+\frac{5 m_1^2 \left(3 \delta ^2 \chi_p^2+75 \delta ^2 \chi_{\ell}^2-112 \pi  \chi_{\ell}\right)}{256
   \eta } \nonumber\\
   & -\frac{15 \delta  m_1^7 \left(\chi_p^4-4 \chi_p^2 \chi_{\ell}^2\right)}{512 \eta ^4}+\frac{5 \delta  m_1^3 \left((812 \eta -97)
   \chi_p^2+20328 \eta  \chi_{\ell}^2\right)}{14336 \eta ^2}-\frac{15 \pi  \delta  m_1 \chi_{\ell}}{16 \eta }-\frac{35 m_1^8 \left(\chi_p^4-4
   \chi_p^2 \chi_{\ell}^2\right)}{512 \eta ^4} \nonumber\\
   & +\frac{m_1^4 \left(25 (92 \eta +19) \chi_p^2+52640 \eta  \chi_{\ell}^2\right)}{6144 \eta ^2}+\frac{15
   \delta ^3 \eta ^2}{1024 m_1^3}+\frac{5 \delta ^2 \eta  (784 \eta +323)}{28672 m_1^2}+\frac{5 \delta  (504 \eta  (7630 \eta -159)+5579177)}{21676032 m_1}\\
  \alpha_{\rm log} &= -\frac{5}{48} \left(7 \pi -3 \delta ^2 \chi_{\ell}\right)-\frac{5 \delta  m_1^5 \chi_p^2 \chi_{\ell}}{128 \eta ^3}+\frac{5 \delta  (7168 \eta +407) m_1 \chi
   _{\ell}}{21504 \eta }-\frac{35 m_1^6 \chi_p^2 \chi _{\ell}}{384 \eta ^3}+\frac{5 (4072 \eta +599) m_1^2 \chi _{\ell}}{9216 \eta }-\frac{5 \pi  \delta }{16 m_1}.\\
   \label{eq:alpha_coeff}
%\end{align}
%\end{widetext}
%\begin{widetext}
%\begin{align}
\epsilon_{-3} &= \alpha_{-3}\\
\epsilon_{-2} &= \alpha_{-2}\\
\epsilon_{-1} &=-\frac{5 (824 \eta +1103)}{3072}-\frac{15 \delta ^2 \eta }{256 m_1^2}-\frac{5 \delta  (980 \eta +911)}{7168 m_1}\\
\epsilon_{1} &= \frac{5 (36 \eta  (85568 \eta +23817)+8024297)}{9289728}+\frac{5 m_1^2 \chi _{\ell} \left(75 \delta ^2 \chi _l-112 \pi \right)}{256 \eta }+\frac{1815 \delta  m_1^3
   \chi _{\ell}^2}{256 \eta }\nonumber\\
   &-\frac{15 \pi  \delta  m_1 \chi _{\ell}}{16 \eta }+\frac{1645 m_1^4 \chi_{\ell}^2}{192 \eta }+\frac{15 \delta ^3 \eta ^2}{1024 m_1^3}+\frac{5
   \delta ^2 \eta  (784 \eta +323)}{28672 m_1^2}+\frac{5 \delta  (504 \eta  (7630 \eta -159)+5579177)}{21676032 m_1}\\
\epsilon_{\rm log} &= -\frac{5}{48} \left(7 \pi -3 \delta ^2 \chi _{\ell}\right)+\frac{5 \delta  (7168 \eta +407) m_1 \chi _{\ell}}{21504 \eta }+\frac{5 (4072 \eta +599) m_1^2 \chi _{\ell}}{9216
   \eta }-\frac{5 \pi  \delta }{16 m_1}.
   \label{eq:epsilon_coeff}
\end{align}
\end{subequations}
\end{widetext}

%\section{Post-Newtonian Orbital Angular Momentum}
\subsection{Orbital Angular Momentum}
The orbital angular momentum is estimated using an aligned-spin approximation with orbital terms up to 4PN and spin-orbit terms up to 3.5PN. We neglect spin-spin couplings. 
\label{app:orb_ang_mom}
\begin{widetext}
\begin{subequations}
\begin{align}
    L_0 &= 1, \\
    L_1 &= \frac{\eta }{6}+\frac{3}{2}, \\ 
    L_2 &= \frac{\eta ^2}{24}-\frac{19 \eta }{8}+\frac{27}{8}, \\ 
    L_3 &= \frac{7 \eta ^3}{1296}+\frac{31 \eta ^2}{24}+\left(\frac{41 \pi ^2}{24}-\frac{6889}{144}\right) \eta +\frac{135}{16}, \\ 
    L_4 &= -\frac{55 \eta ^4}{31104}-\frac{215 \eta ^3}{1728}+\left(\frac{356035}{3456}-\frac{2255 \pi ^2}{576}\right) \eta ^2+\eta
   \left(-\frac{64}{3} \log (16 x)-\frac{6455 \pi ^2}{1536}-\frac{128 \gamma
   }{3}+\frac{98869}{5760}\right)+\frac{2835}{128} , \\
   L_{1.5}^{\rm{SO}} &=  -\frac{35}{6} S_{\ell} - \frac{5}{2} \frac{\delta m}{m} \Sigma_{\ell}, \\
   L_{2.5}^{\rm{SO}} &= \left( -\frac{77}{8} + \frac{427}{72} \eta \right) S_{\ell} + \frac{\delta m}{m} \left( - \frac{21}{8} + \frac{35}{12} \eta  \right)  \Sigma_{\ell}, \\
   L_{3.5}^{\rm{SO}} &= \left( -\frac{405}{16} + \frac{1101}{16} \eta - \frac{29}{16} \eta^2 \right) S_{\ell} + \frac{\delta m}{m} \left( - \frac{81}{6} + \frac{117}{4} \eta - \frac{15}{16} \eta^2  \right)  \Sigma_{\ell}, \\
   L^{\rm{LO-}S^{\infty}}_{2} &= \left( \frac{1}{2} + \frac{\delta}{2} - \eta \right) \chi^2_{1\ell} + 2 \eta \chi_{1\ell} \chi_{2\ell} + \left( \frac{1}{2} - \frac{\delta}{2} - \eta \right) \chi^2_{2\ell} , \\
   L^{\rm{LO-}S^{\infty}}_{3.5} &= \chi_{1\ell}^3 \left(\frac{3 \delta  \eta }{4}-\frac{3 \eta ^2}{2}+\frac{3 \eta }{4}\right)+\chi_{1\ell}^2 \, \chi_{2\ell} \left(\frac{3 \delta  \eta
   }{4}+\frac{3 \eta ^2}{2}+\frac{3 \eta }{4}\right) + \chi_{1\ell} \, \chi_{2\ell}^2 \left(-\frac{3 \delta  \eta }{4}+\frac{3 \eta ^2}{2}+\frac{3
   \eta }{4}\right)+\chi_{2\ell}^3 \left(-\frac{3 \delta  \eta }{4}-\frac{3 \eta ^2}{2}+\frac{3 \eta }{4}\right) .
\end{align}
\end{subequations}
\end{widetext}

% Adding this for now as it the line break looked ugly...
\clearpage
\vspace{0.1in}
\vfil

% ~~~~~~~~~~ References ~~~~~~~~~~ %

% Try this if the last two columns before the bib are not breaking nicely.
\vspace{0.1in}

\vfil

\let\c\Originalcdefinition %
\let\d\Originalddefinition %
\let\i\Originalidefinition

\bibliography{phenomx,phenom_refs,eob_refs,nr_refs,postnewtonian,gravitationalwaves}

\providecommand{\noopsort}[1]{}\providecommand{\singleletter}[1]{#1}%\providecommand{\noopsort}[1]{}\providecommand{\singleletter}[1]{#1}%\providecommand{\noopsort}[1]{}\providecommand{\singleletter}[1]{#1}%\providecommand{\noopsort}[1]{}\providecommand{\singleletter}[1]{#1}%\providecommand{\noopsort}[1]{}\providecommand{\singleletter}[1]{#1}%\providecommand{\noopsort}[1]{}\providecommand{\singleletter}[1]{#1}%
\begin{thebibliography}{113}%
\makeatletter
\providecommand \@ifxundefined [1]{%
 \@ifx{#1\undefined}
}%
\providecommand \@ifnum [1]{%
 \ifnum #1\expandafter \@firstoftwo
 \else \expandafter \@secondoftwo
 \fi
}%
\providecommand \@ifx [1]{%
 \ifx #1\expandafter \@firstoftwo
 \else \expandafter \@secondoftwo
 \fi
}%
\providecommand \natexlab [1]{#1}%
\providecommand \enquote  [1]{``#1''}%
\providecommand \bibnamefont  [1]{#1}%
\providecommand \bibfnamefont [1]{#1}%
\providecommand \citenamefont [1]{#1}%
\providecommand \href@noop [0]{\@secondoftwo}%
\providecommand \href [0]{\begingroup \@sanitize@url \@href}%
\providecommand \@href[1]{\@@startlink{#1}\@@href}%
\providecommand \@@href[1]{\endgroup#1\@@endlink}%
\providecommand \@sanitize@url [0]{\catcode `\\12\catcode `\$12\catcode
  `\&12\catcode `\#12\catcode `\^12\catcode `\_12\catcode `\%12\relax}%
\providecommand \@@startlink[1]{}%
\providecommand \@@endlink[0]{}%
\providecommand \url  [0]{\begingroup\@sanitize@url \@url }%
\providecommand \@url [1]{\endgroup\@href {#1}{\urlprefix }}%
\providecommand \urlprefix  [0]{URL }%
\providecommand \Eprint [0]{\href }%
\providecommand \doibase [0]{http://dx.doi.org/}%
\providecommand \selectlanguage [0]{\@gobble}%
\providecommand \bibinfo  [0]{\@secondoftwo}%
\providecommand \bibfield  [0]{\@secondoftwo}%
\providecommand \translation [1]{[#1]}%
\providecommand \BibitemOpen [0]{}%
\providecommand \bibitemStop [0]{}%
\providecommand \bibitemNoStop [0]{.\EOS\space}%
\providecommand \EOS [0]{\spacefactor3000\relax}%
\providecommand \BibitemShut  [1]{\csname bibitem#1\endcsname}%
\let\auto@bib@innerbib\@empty
%</preamble>
\bibitem [{\citenamefont {Hannam}\ \emph {et~al.}(2014)\citenamefont {Hannam},
  \citenamefont {Schmidt}, \citenamefont {Bohé}, \citenamefont {Haegel},
  \citenamefont {Husa} \emph {et~al.}}]{Hannam:2013oca}%
  \BibitemOpen
  \bibfield  {author} {\bibinfo {author} {\bibfnamefont {M.}~\bibnamefont
  {Hannam}}, \bibinfo {author} {\bibfnamefont {P.}~\bibnamefont {Schmidt}},
  \bibinfo {author} {\bibfnamefont {A.}~\bibnamefont {Bohé}}, \bibinfo
  {author} {\bibfnamefont {L.}~\bibnamefont {Haegel}}, \bibinfo {author}
  {\bibfnamefont {S.}~\bibnamefont {Husa}},  \emph {et~al.},\ }\href {\doibase
  10.1103/PhysRevLett.113.151101} {\bibfield  {journal} {\bibinfo  {journal}
  {Phys.Rev.Lett.}\ }\textbf {\bibinfo {volume} {113}},\ \bibinfo {pages}
  {151101} (\bibinfo {year} {2014})},\ \Eprint {http://arxiv.org/abs/1308.3271}
  {arXiv:1308.3271 [gr-qc]} \BibitemShut {NoStop}%
%%CITATION = ARXIV:1308.3271;%%
\bibitem [{\citenamefont {Chatziioannou}\ \emph {et~al.}(2017)\citenamefont
  {Chatziioannou}, \citenamefont {Klein}, \citenamefont {Yunes},\ and\
  \citenamefont {Cornish}}]{Chatziioannou:2017tdw}%
  \BibitemOpen
  \bibfield  {author} {\bibinfo {author} {\bibfnamefont {K.}~\bibnamefont
  {Chatziioannou}}, \bibinfo {author} {\bibfnamefont {A.}~\bibnamefont
  {Klein}}, \bibinfo {author} {\bibfnamefont {N.}~\bibnamefont {Yunes}}, \ and\
  \bibinfo {author} {\bibfnamefont {N.}~\bibnamefont {Cornish}},\ }\href
  {\doibase 10.1103/PhysRevD.95.104004} {\bibfield  {journal} {\bibinfo
  {journal} {Phys. Rev.}\ }\textbf {\bibinfo {volume} {D95}},\ \bibinfo {pages}
  {104004} (\bibinfo {year} {2017})},\ \Eprint
  {http://arxiv.org/abs/1703.03967} {arXiv:1703.03967 [gr-qc]} \BibitemShut
  {NoStop}%
%%CITATION = ARXIV:1703.03967;%%
\bibitem [{\citenamefont {Garc{\'i}a-Quir{\'o}s}\ \emph
  {et~al.}(2020{\natexlab{a}})\citenamefont {Garc{\'i}a-Quir{\'o}s},
  \citenamefont {Husa}, \citenamefont {Mateu-Lucena},\ and\ \citenamefont
  {Borchers}}]{Garcia-Quiros:2020qlt}%
  \BibitemOpen
  \bibfield  {author} {\bibinfo {author} {\bibfnamefont {C.}~\bibnamefont
  {Garc{\'i}a-Quir{\'o}s}}, \bibinfo {author} {\bibfnamefont {S.}~\bibnamefont
  {Husa}}, \bibinfo {author} {\bibfnamefont {M.}~\bibnamefont {Mateu-Lucena}},
  \ and\ \bibinfo {author} {\bibfnamefont {A.}~\bibnamefont {Borchers}},\
  }\href@noop {} {\bibfield  {journal} {\bibinfo  {journal} {ArXiv e-prints}\ }
  (\bibinfo {year} {2020}{\natexlab{a}})},\ \Eprint
  {http://arxiv.org/abs/2001.10897} {arXiv:2001.10897 [gr-qc]} \BibitemShut
  {NoStop}%
\bibitem [{\citenamefont {Pratten}\ \emph {et~al.}(2020)\citenamefont
  {Pratten}, \citenamefont {Husa}, \citenamefont {Garcia-Quiros}, \citenamefont
  {Colleoni}, \citenamefont {Ramos-Buades}, \citenamefont {Estelles},\ and\
  \citenamefont {Jaume}}]{Pratten:2020fqn}%
  \BibitemOpen
  \bibfield  {author} {\bibinfo {author} {\bibfnamefont {G.}~\bibnamefont
  {Pratten}}, \bibinfo {author} {\bibfnamefont {S.}~\bibnamefont {Husa}},
  \bibinfo {author} {\bibfnamefont {C.}~\bibnamefont {Garcia-Quiros}}, \bibinfo
  {author} {\bibfnamefont {M.}~\bibnamefont {Colleoni}}, \bibinfo {author}
  {\bibfnamefont {A.}~\bibnamefont {Ramos-Buades}}, \bibinfo {author}
  {\bibfnamefont {H.}~\bibnamefont {Estelles}}, \ and\ \bibinfo {author}
  {\bibfnamefont {R.}~\bibnamefont {Jaume}},\ }\href@noop {} {\bibfield
  {journal} {\bibinfo  {journal} {ArXiv e-prints}\ } (\bibinfo {year}
  {2020})},\ \Eprint {http://arxiv.org/abs/2001.11412} {arXiv:2001.11412
  [gr-qc]} \BibitemShut {NoStop}%
%%CITATION = ARXIV:2001.11412;%%
\bibitem [{\citenamefont {Husa}\ \emph {et~al.}(2016)\citenamefont {Husa},
  \citenamefont {Khan}, \citenamefont {Hannam}, \citenamefont {Pürrer},
  \citenamefont {Ohme}, \citenamefont {Jiménez~Forteza},\ and\ \citenamefont
  {Bohé}}]{Husa:2015iqa}%
  \BibitemOpen
  \bibfield  {author} {\bibinfo {author} {\bibfnamefont {S.}~\bibnamefont
  {Husa}}, \bibinfo {author} {\bibfnamefont {S.}~\bibnamefont {Khan}}, \bibinfo
  {author} {\bibfnamefont {M.}~\bibnamefont {Hannam}}, \bibinfo {author}
  {\bibfnamefont {M.}~\bibnamefont {Pürrer}}, \bibinfo {author} {\bibfnamefont
  {F.}~\bibnamefont {Ohme}}, \bibinfo {author} {\bibfnamefont {X.}~\bibnamefont
  {Jiménez~Forteza}}, \ and\ \bibinfo {author} {\bibfnamefont
  {A.}~\bibnamefont {Bohé}},\ }\href {\doibase 10.1103/PhysRevD.93.044006}
  {\bibfield  {journal} {\bibinfo  {journal} {Phys. Rev.}\ }\textbf {\bibinfo
  {volume} {D93}},\ \bibinfo {pages} {044006} (\bibinfo {year} {2016})},\
  \Eprint {http://arxiv.org/abs/1508.07250} {arXiv:1508.07250 [gr-qc]}
  \BibitemShut {NoStop}%
%%CITATION = ARXIV:1508.07250;%%
\bibitem [{\citenamefont {Khan}\ \emph {et~al.}(2016)\citenamefont {Khan},
  \citenamefont {Husa}, \citenamefont {Hannam}, \citenamefont {Ohme},
  \citenamefont {Pürrer}, \citenamefont {Jiménez~Forteza},\ and\
  \citenamefont {Bohé}}]{Khan:2015jqa}%
  \BibitemOpen
  \bibfield  {author} {\bibinfo {author} {\bibfnamefont {S.}~\bibnamefont
  {Khan}}, \bibinfo {author} {\bibfnamefont {S.}~\bibnamefont {Husa}}, \bibinfo
  {author} {\bibfnamefont {M.}~\bibnamefont {Hannam}}, \bibinfo {author}
  {\bibfnamefont {F.}~\bibnamefont {Ohme}}, \bibinfo {author} {\bibfnamefont
  {M.}~\bibnamefont {Pürrer}}, \bibinfo {author} {\bibfnamefont
  {X.}~\bibnamefont {Jiménez~Forteza}}, \ and\ \bibinfo {author}
  {\bibfnamefont {A.}~\bibnamefont {Bohé}},\ }\href {\doibase
  10.1103/PhysRevD.93.044007} {\bibfield  {journal} {\bibinfo  {journal} {Phys.
  Rev.}\ }\textbf {\bibinfo {volume} {D93}},\ \bibinfo {pages} {044007}
  (\bibinfo {year} {2016})},\ \Eprint {http://arxiv.org/abs/1508.07253}
  {arXiv:1508.07253 [gr-qc]} \BibitemShut {NoStop}%
%%CITATION = ARXIV:1508.07253;%%
\bibitem [{\citenamefont {Jiménez-Forteza}\ \emph {et~al.}(2017)\citenamefont
  {Jiménez-Forteza}, \citenamefont {Keitel}, \citenamefont {Husa},
  \citenamefont {Hannam}, \citenamefont {Khan},\ and\ \citenamefont
  {Pürrer}}]{Jimenez-Forteza:2016oae}%
  \BibitemOpen
  \bibfield  {author} {\bibinfo {author} {\bibfnamefont {X.}~\bibnamefont
  {Jiménez-Forteza}}, \bibinfo {author} {\bibfnamefont {D.}~\bibnamefont
  {Keitel}}, \bibinfo {author} {\bibfnamefont {S.}~\bibnamefont {Husa}},
  \bibinfo {author} {\bibfnamefont {M.}~\bibnamefont {Hannam}}, \bibinfo
  {author} {\bibfnamefont {S.}~\bibnamefont {Khan}}, \ and\ \bibinfo {author}
  {\bibfnamefont {M.}~\bibnamefont {Pürrer}},\ }\href {\doibase
  10.1103/PhysRevD.95.064024} {\bibfield  {journal} {\bibinfo  {journal} {Phys.
  Rev.}\ }\textbf {\bibinfo {volume} {D95}},\ \bibinfo {pages} {064024}
  (\bibinfo {year} {2017})},\ \Eprint {http://arxiv.org/abs/1611.00332}
  {arXiv:1611.00332 [gr-qc]} \BibitemShut {NoStop}%
%%CITATION = ARXIV:1611.00332;%%
\bibitem [{\citenamefont {{Keitel}}\ \emph {et~al.}(2017)\citenamefont
  {{Keitel}}, \citenamefont {{Forteza}}, \citenamefont {{Husa}}, \citenamefont
  {{London}}, \citenamefont {{Bernuzzi}}, \citenamefont {{Harms}},
  \citenamefont {{Nagar}}, \citenamefont {{Hannam}}, \citenamefont {{Khan}},
  \citenamefont {{P{\"u}rrer}}, \citenamefont {{Pratten}},\ and\ \citenamefont
  {{Chaurasia}}}]{Keitel:2016krm}%
  \BibitemOpen
  \bibfield  {author} {\bibinfo {author} {\bibfnamefont {D.}~\bibnamefont
  {{Keitel}}}, \bibinfo {author} {\bibfnamefont {X.~J.}\ \bibnamefont
  {{Forteza}}}, \bibinfo {author} {\bibfnamefont {S.}~\bibnamefont {{Husa}}},
  \bibinfo {author} {\bibfnamefont {L.}~\bibnamefont {{London}}}, \bibinfo
  {author} {\bibfnamefont {S.}~\bibnamefont {{Bernuzzi}}}, \bibinfo {author}
  {\bibfnamefont {E.}~\bibnamefont {{Harms}}}, \bibinfo {author} {\bibfnamefont
  {A.}~\bibnamefont {{Nagar}}}, \bibinfo {author} {\bibfnamefont
  {M.}~\bibnamefont {{Hannam}}}, \bibinfo {author} {\bibfnamefont
  {S.}~\bibnamefont {{Khan}}}, \bibinfo {author} {\bibfnamefont
  {M.}~\bibnamefont {{P{\"u}rrer}}}, \bibinfo {author} {\bibfnamefont
  {G.}~\bibnamefont {{Pratten}}}, \ and\ \bibinfo {author} {\bibfnamefont
  {V.}~\bibnamefont {{Chaurasia}}},\ }\href {\doibase
  10.1103/PhysRevD.96.024006} {\bibfield  {journal} {\bibinfo  {journal} {Phys.
  Rev.}\ }\textbf {\bibinfo {volume} {D96}},\ \bibinfo {pages} {024006}
  (\bibinfo {year} {2017})},\ \Eprint {http://arxiv.org/abs/1612.09566}
  {arXiv:1612.09566 [gr-qc]} \BibitemShut {NoStop}%
%%CITATION = ARXIV:1612.09566;%%
\bibitem [{\citenamefont {Boh\'{e}}\ \emph {et~al.}(2017)\citenamefont
  {Boh\'{e}} \emph {et~al.}}]{Bohe:2016gbl}%
  \BibitemOpen
  \bibfield  {author} {\bibinfo {author} {\bibfnamefont {A.}~\bibnamefont
  {Boh\'{e}}} \emph {et~al.},\ }\href {\doibase 10.1103/PhysRevD.95.044028}
  {\bibfield  {journal} {\bibinfo  {journal} {Phys. Rev.}\ }\textbf {\bibinfo
  {volume} {D95}},\ \bibinfo {pages} {044028} (\bibinfo {year} {2017})},\
  \Eprint {http://arxiv.org/abs/1611.03703} {arXiv:1611.03703 [gr-qc]}
  \BibitemShut {NoStop}%
%%CITATION = ARXIV:1611.03703;%%
\bibitem [{\citenamefont {Garc{\'i}a-Quir{\'o}s}\ \emph
  {et~al.}(2020{\natexlab{b}})\citenamefont {Garc{\'i}a-Quir{\'o}s},
  \citenamefont {Colleoni}, \citenamefont {Husa}, \citenamefont {Estellés},
  \citenamefont {Pratten}, \citenamefont {Ramos-Buades}, \citenamefont
  {Mateu-Lucena},\ and\ \citenamefont {Jaume}}]{Garcia-Quiros:2020qpx}%
  \BibitemOpen
  \bibfield  {author} {\bibinfo {author} {\bibfnamefont {C.}~\bibnamefont
  {Garc{\'i}a-Quir{\'o}s}}, \bibinfo {author} {\bibfnamefont {M.}~\bibnamefont
  {Colleoni}}, \bibinfo {author} {\bibfnamefont {S.}~\bibnamefont {Husa}},
  \bibinfo {author} {\bibfnamefont {H.}~\bibnamefont {Estellés}}, \bibinfo
  {author} {\bibfnamefont {G.}~\bibnamefont {Pratten}}, \bibinfo {author}
  {\bibfnamefont {A.}~\bibnamefont {Ramos-Buades}}, \bibinfo {author}
  {\bibfnamefont {M.}~\bibnamefont {Mateu-Lucena}}, \ and\ \bibinfo {author}
  {\bibfnamefont {R.}~\bibnamefont {Jaume}},\ }\href@noop {} {\bibfield
  {journal} {\bibinfo  {journal} {ArXiv e-prints}\ } (\bibinfo {year}
  {2020}{\natexlab{b}})},\ \Eprint {http://arxiv.org/abs/2001.10914}
  {arXiv:2001.10914 [gr-qc]} \BibitemShut {NoStop}%
%%CITATION = ARXIV:2001.10914;%%
\bibitem [{\citenamefont {London}\ \emph {et~al.}(2018)\citenamefont {London},
  \citenamefont {Khan}, \citenamefont {Fauchon-Jones}, \citenamefont
  {Garc{\'i}a}, \citenamefont {Hannam}, \citenamefont {Husa}, \citenamefont
  {Jiménez-Forteza}, \citenamefont {Kalaghatgi}, \citenamefont {Ohme},\ and\
  \citenamefont {Pannarale}}]{London:2017bcn}%
  \BibitemOpen
  \bibfield  {author} {\bibinfo {author} {\bibfnamefont {L.}~\bibnamefont
  {London}}, \bibinfo {author} {\bibfnamefont {S.}~\bibnamefont {Khan}},
  \bibinfo {author} {\bibfnamefont {E.}~\bibnamefont {Fauchon-Jones}}, \bibinfo
  {author} {\bibfnamefont {C.}~\bibnamefont {Garc{\'i}a}}, \bibinfo {author}
  {\bibfnamefont {M.}~\bibnamefont {Hannam}}, \bibinfo {author} {\bibfnamefont
  {S.}~\bibnamefont {Husa}}, \bibinfo {author} {\bibfnamefont {X.}~\bibnamefont
  {Jiménez-Forteza}}, \bibinfo {author} {\bibfnamefont {C.}~\bibnamefont
  {Kalaghatgi}}, \bibinfo {author} {\bibfnamefont {F.}~\bibnamefont {Ohme}}, \
  and\ \bibinfo {author} {\bibfnamefont {F.}~\bibnamefont {Pannarale}},\ }\href
  {\doibase 10.1103/PhysRevLett.120.161102} {\bibfield  {journal} {\bibinfo
  {journal} {Phys. Rev. Lett.}\ }\textbf {\bibinfo {volume} {120}},\ \bibinfo
  {pages} {161102} (\bibinfo {year} {2018})},\ \Eprint
  {http://arxiv.org/abs/1708.00404} {arXiv:1708.00404 [gr-qc]} \BibitemShut
  {NoStop}%
%%CITATION = ARXIV:1708.00404;%%
\bibitem [{\citenamefont {Veitch}\ \emph {et~al.}(2015)\citenamefont {Veitch}
  \emph {et~al.}}]{Veitch:2014wba}%
  \BibitemOpen
  \bibfield  {author} {\bibinfo {author} {\bibfnamefont {J.}~\bibnamefont
  {Veitch}} \emph {et~al.},\ }\href {\doibase 10.1103/PhysRevD.91.042003}
  {\bibfield  {journal} {\bibinfo  {journal} {Phys. Rev.}\ }\textbf {\bibinfo
  {volume} {D91}},\ \bibinfo {pages} {042003} (\bibinfo {year} {2015})},\
  \Eprint {http://arxiv.org/abs/1409.7215} {arXiv:1409.7215 [gr-qc]}
  \BibitemShut {NoStop}%
%%CITATION = ARXIV:1409.7215;%%
\bibitem [{\citenamefont {Ashton}\ \emph {et~al.}(2019)\citenamefont {Ashton}
  \emph {et~al.}}]{Ashton:2018jfp}%
  \BibitemOpen
  \bibfield  {author} {\bibinfo {author} {\bibfnamefont {G.}~\bibnamefont
  {Ashton}} \emph {et~al.},\ }\href {\doibase 10.3847/1538-4365/ab06fc}
  {\bibfield  {journal} {\bibinfo  {journal} {Astrophys. J. Suppl.}\ }\textbf
  {\bibinfo {volume} {241}},\ \bibinfo {pages} {27} (\bibinfo {year} {2019})},\
  \Eprint {http://arxiv.org/abs/1811.02042} {arXiv:1811.02042 [astro-ph.IM]}
  \BibitemShut {NoStop}%
%%CITATION = ARXIV:1811.02042;%%
\bibitem [{\citenamefont {Vinciguerra}\ \emph {et~al.}(2017)\citenamefont
  {Vinciguerra}, \citenamefont {Veitch},\ and\ \citenamefont
  {Mandel}}]{Vinciguerra:2017ngf}%
  \BibitemOpen
  \bibfield  {author} {\bibinfo {author} {\bibfnamefont {S.}~\bibnamefont
  {Vinciguerra}}, \bibinfo {author} {\bibfnamefont {J.}~\bibnamefont {Veitch}},
  \ and\ \bibinfo {author} {\bibfnamefont {I.}~\bibnamefont {Mandel}},\ }\href
  {\doibase 10.1088/1361-6382/aa6d44} {\bibfield  {journal} {\bibinfo
  {journal} {Class. Quant. Grav.}\ }\textbf {\bibinfo {volume} {34}},\ \bibinfo
  {pages} {115006} (\bibinfo {year} {2017})},\ \Eprint
  {http://arxiv.org/abs/1703.02062} {arXiv:1703.02062 [gr-qc]} \BibitemShut
  {NoStop}%
%%CITATION = ARXIV:1703.02062;%%
\bibitem [{\citenamefont {Khan}\ \emph {et~al.}(2019)\citenamefont {Khan},
  \citenamefont {Chatziioannou}, \citenamefont {Hannam},\ and\ \citenamefont
  {Ohme}}]{Khan:2018fmp}%
  \BibitemOpen
  \bibfield  {author} {\bibinfo {author} {\bibfnamefont {S.}~\bibnamefont
  {Khan}}, \bibinfo {author} {\bibfnamefont {K.}~\bibnamefont {Chatziioannou}},
  \bibinfo {author} {\bibfnamefont {M.}~\bibnamefont {Hannam}}, \ and\ \bibinfo
  {author} {\bibfnamefont {F.}~\bibnamefont {Ohme}},\ }\href {\doibase
  10.1103/PhysRevD.100.024059} {\bibfield  {journal} {\bibinfo  {journal}
  {Phys. Rev.}\ }\textbf {\bibinfo {volume} {D100}},\ \bibinfo {pages} {024059}
  (\bibinfo {year} {2019})},\ \Eprint {http://arxiv.org/abs/1809.10113}
  {arXiv:1809.10113 [gr-qc]} \BibitemShut {NoStop}%
%%CITATION = ARXIV:1809.10113;%%
\bibitem [{\citenamefont {Khan}\ \emph {et~al.}(2020)\citenamefont {Khan},
  \citenamefont {Ohme}, \citenamefont {Chatziioannou},\ and\ \citenamefont
  {Hannam}}]{Khan:2019kot}%
  \BibitemOpen
  \bibfield  {author} {\bibinfo {author} {\bibfnamefont {S.}~\bibnamefont
  {Khan}}, \bibinfo {author} {\bibfnamefont {F.}~\bibnamefont {Ohme}}, \bibinfo
  {author} {\bibfnamefont {K.}~\bibnamefont {Chatziioannou}}, \ and\ \bibinfo
  {author} {\bibfnamefont {M.}~\bibnamefont {Hannam}},\ }\href {\doibase
  10.1103/PhysRevD.101.024056} {\bibfield  {journal} {\bibinfo  {journal}
  {Phys. Rev.}\ }\textbf {\bibinfo {volume} {D101}},\ \bibinfo {pages} {024056}
  (\bibinfo {year} {2020})},\ \Eprint {http://arxiv.org/abs/1911.06050}
  {arXiv:1911.06050 [gr-qc]} \BibitemShut {NoStop}%
%%CITATION = ARXIV:1911.06050;%%
\bibitem [{\citenamefont {Schmidt}\ \emph {et~al.}(2011)\citenamefont
  {Schmidt}, \citenamefont {Hannam}, \citenamefont {Husa},\ and\ \citenamefont
  {Ajith}}]{Schmidt:2010it}%
  \BibitemOpen
  \bibfield  {author} {\bibinfo {author} {\bibfnamefont {P.}~\bibnamefont
  {Schmidt}}, \bibinfo {author} {\bibfnamefont {M.}~\bibnamefont {Hannam}},
  \bibinfo {author} {\bibfnamefont {S.}~\bibnamefont {Husa}}, \ and\ \bibinfo
  {author} {\bibfnamefont {P.}~\bibnamefont {Ajith}},\ }\href {\doibase
  10.1103/PhysRevD.84.024046} {\bibfield  {journal} {\bibinfo  {journal} {Phys.
  Rev.}\ }\textbf {\bibinfo {volume} {D84}},\ \bibinfo {pages} {024046}
  (\bibinfo {year} {2011})},\ \Eprint {http://arxiv.org/abs/1012.2879}
  {arXiv:1012.2879 [gr-qc]} \BibitemShut {NoStop}%
%%CITATION = ARXIV:1012.2879;%%
\bibitem [{\citenamefont {Schmidt}\ \emph {et~al.}(2012)\citenamefont
  {Schmidt}, \citenamefont {Hannam},\ and\ \citenamefont
  {Husa}}]{Schmidt:2012rh}%
  \BibitemOpen
  \bibfield  {author} {\bibinfo {author} {\bibfnamefont {P.}~\bibnamefont
  {Schmidt}}, \bibinfo {author} {\bibfnamefont {M.}~\bibnamefont {Hannam}}, \
  and\ \bibinfo {author} {\bibfnamefont {S.}~\bibnamefont {Husa}},\ }\href
  {\doibase 10.1103/PhysRevD.86.104063} {\bibfield  {journal} {\bibinfo
  {journal} {Phys. Rev.}\ }\textbf {\bibinfo {volume} {D86}},\ \bibinfo {pages}
  {104063} (\bibinfo {year} {2012})},\ \Eprint {http://arxiv.org/abs/1207.3088}
  {arXiv:1207.3088 [gr-qc]} \BibitemShut {NoStop}%
%%CITATION = ARXIV:1207.3088;%%
\bibitem [{\citenamefont {{LIGO Scientific
  Collaboration}}(2020{\natexlab{a}})}]{lalsuite}%
  \BibitemOpen
  \bibfield  {author} {\bibinfo {author} {\bibnamefont {{LIGO Scientific
  Collaboration}}},\ }\href {\doibase 10.7935/GT1W-FZ16} {\enquote {\bibinfo
  {title} {{LIGO} {A}lgorithm {L}ibrary - {LALS}uite},}\ }\bibinfo
  {howpublished} {free software (GPL), \url{https://doi.org/10.7935/GT1W-FZ16}}
  (\bibinfo {year} {2020}{\natexlab{a}})\BibitemShut {NoStop}%
\bibitem [{\citenamefont {Gerosa}\ \emph {et~al.}(2015)\citenamefont {Gerosa},
  \citenamefont {Kesden}, \citenamefont {O'Shaughnessy}, \citenamefont {Klein},
  \citenamefont {Berti}, \citenamefont {Sperhake},\ and\ \citenamefont
  {Trifirò}}]{Gerosa:2015hba}%
  \BibitemOpen
  \bibfield  {author} {\bibinfo {author} {\bibfnamefont {D.}~\bibnamefont
  {Gerosa}}, \bibinfo {author} {\bibfnamefont {M.}~\bibnamefont {Kesden}},
  \bibinfo {author} {\bibfnamefont {R.}~\bibnamefont {O'Shaughnessy}}, \bibinfo
  {author} {\bibfnamefont {A.}~\bibnamefont {Klein}}, \bibinfo {author}
  {\bibfnamefont {E.}~\bibnamefont {Berti}}, \bibinfo {author} {\bibfnamefont
  {U.}~\bibnamefont {Sperhake}}, \ and\ \bibinfo {author} {\bibfnamefont
  {D.}~\bibnamefont {Trifirò}},\ }\href {\doibase
  10.1103/PhysRevLett.115.141102} {\bibfield  {journal} {\bibinfo  {journal}
  {Phys. Rev. Lett.}\ }\textbf {\bibinfo {volume} {115}},\ \bibinfo {pages}
  {141102} (\bibinfo {year} {2015})},\ \Eprint
  {http://arxiv.org/abs/1506.09116} {arXiv:1506.09116 [gr-qc]} \BibitemShut
  {NoStop}%
%%CITATION = ARXIV:1506.09116;%%
\bibitem [{\citenamefont {Bruegmann}\ \emph {et~al.}(2008)\citenamefont
  {Bruegmann}, \citenamefont {Gonzalez}, \citenamefont {Hannam}, \citenamefont
  {Husa},\ and\ \citenamefont {Sperhake}}]{Brugmann:2007zj}%
  \BibitemOpen
  \bibfield  {author} {\bibinfo {author} {\bibfnamefont {B.}~\bibnamefont
  {Bruegmann}}, \bibinfo {author} {\bibfnamefont {J.~A.}\ \bibnamefont
  {Gonzalez}}, \bibinfo {author} {\bibfnamefont {M.}~\bibnamefont {Hannam}},
  \bibinfo {author} {\bibfnamefont {S.}~\bibnamefont {Husa}}, \ and\ \bibinfo
  {author} {\bibfnamefont {U.}~\bibnamefont {Sperhake}},\ }\href {\doibase
  10.1103/PhysRevD.77.124047} {\bibfield  {journal} {\bibinfo  {journal} {Phys.
  Rev.}\ }\textbf {\bibinfo {volume} {D77}},\ \bibinfo {pages} {124047}
  (\bibinfo {year} {2008})},\ \Eprint {http://arxiv.org/abs/0707.0135}
  {arXiv:0707.0135 [gr-qc]} \BibitemShut {NoStop}%
%%CITATION = ARXIV:0707.0135;%%
\bibitem [{\citenamefont {Ramos-Buades}\ \emph {et~al.}(2020)\citenamefont
  {Ramos-Buades}, \citenamefont {Schmidt}, \citenamefont {Pratten},\ and\
  \citenamefont {Husa}}]{Ramos-Buades:2020noq}%
  \BibitemOpen
  \bibfield  {author} {\bibinfo {author} {\bibfnamefont {A.}~\bibnamefont
  {Ramos-Buades}}, \bibinfo {author} {\bibfnamefont {P.}~\bibnamefont
  {Schmidt}}, \bibinfo {author} {\bibfnamefont {G.}~\bibnamefont {Pratten}}, \
  and\ \bibinfo {author} {\bibfnamefont {S.}~\bibnamefont {Husa}},\ }\href@noop
  {} {\bibfield  {journal} {\bibinfo  {journal} {ArXiv e-prints}\ } (\bibinfo
  {year} {2020})},\ \Eprint {http://arxiv.org/abs/2001.10936} {arXiv:2001.10936
  [gr-qc]} \BibitemShut {NoStop}%
%%CITATION = ARXIV:2001.10936;%%
\bibitem [{\citenamefont {Thomas}\ \emph {et~al.}(2020)\citenamefont {Thomas},
  \citenamefont {Schmidt},\ and\ \citenamefont {Pratten}}]{Thomas:2020uqj}%
  \BibitemOpen
  \bibfield  {author} {\bibinfo {author} {\bibfnamefont {L.~M.}\ \bibnamefont
  {Thomas}}, \bibinfo {author} {\bibfnamefont {P.}~\bibnamefont {Schmidt}}, \
  and\ \bibinfo {author} {\bibfnamefont {G.}~\bibnamefont {Pratten}},\
  }\href@noop {} {\  (\bibinfo {year} {2020})},\ \Eprint
  {http://arxiv.org/abs/2012.02209} {arXiv:2012.02209 [gr-qc]} \BibitemShut
  {NoStop}%
\bibitem [{\citenamefont {Boh{\'e}}\ \emph {et~al.}(2016)\citenamefont
  {Boh{\'e}}, \citenamefont {Hannam}, \citenamefont {Husa}, \citenamefont
  {Ohme}, \citenamefont {Puerrer},\ and\ \citenamefont {Schmidt}}]{Bohe:PPv2}%
  \BibitemOpen
  \bibfield  {author} {\bibinfo {author} {\bibfnamefont {A.}~\bibnamefont
  {Boh{\'e}}}, \bibinfo {author} {\bibfnamefont {M.}~\bibnamefont {Hannam}},
  \bibinfo {author} {\bibfnamefont {S.}~\bibnamefont {Husa}}, \bibinfo {author}
  {\bibfnamefont {F.}~\bibnamefont {Ohme}}, \bibinfo {author} {\bibfnamefont
  {M.}~\bibnamefont {Puerrer}}, \ and\ \bibinfo {author} {\bibfnamefont
  {P.}~\bibnamefont {Schmidt}},\ }\href {https://dcc.ligo.org/LIGO-T1500602}
  {\emph {\bibinfo {title} {PhenomPv2 - Technical Notes for LAL
  Implementation}}},\ \bibinfo {type} {Tech. Rep.}\ \bibinfo {number}
  {{LIGO}-T1500602}\ (\bibinfo  {institution} {{LIGO} Project},\ \bibinfo
  {year} {2016})\BibitemShut {NoStop}%
\bibitem [{\citenamefont {Abbott}\ \emph {et~al.}(2017)\citenamefont {Abbott}
  \emph {et~al.}}]{Abbott:2016wiq}%
  \BibitemOpen
  \bibfield  {author} {\bibinfo {author} {\bibfnamefont {B.~P.}\ \bibnamefont
  {Abbott}} \emph {et~al.} (\bibinfo {collaboration} {LIGO Scientific,
  Virgo}),\ }\href {\doibase 10.1088/1361-6382/aa6854} {\bibfield  {journal}
  {\bibinfo  {journal} {Class. Quant. Grav.}\ }\textbf {\bibinfo {volume}
  {34}},\ \bibinfo {pages} {104002} (\bibinfo {year} {2017})},\ \Eprint
  {http://arxiv.org/abs/1611.07531} {arXiv:1611.07531 [gr-qc]} \BibitemShut
  {NoStop}%
%%CITATION = ARXIV:1611.07531;%%
\bibitem [{\citenamefont {Abbott}\ \emph {et~al.}(2016)\citenamefont {Abbott}
  \emph {et~al.}}]{Abbott:2016blz}%
  \BibitemOpen
  \bibfield  {author} {\bibinfo {author} {\bibfnamefont {B.~P.}\ \bibnamefont
  {Abbott}} \emph {et~al.} (\bibinfo {collaboration} {LIGO Scientific,
  Virgo}),\ }\href {\doibase 10.1103/PhysRevLett.116.061102} {\bibfield
  {journal} {\bibinfo  {journal} {Phys. Rev. Lett.}\ }\textbf {\bibinfo
  {volume} {116}},\ \bibinfo {pages} {061102} (\bibinfo {year} {2016})},\
  \Eprint {http://arxiv.org/abs/1602.03837} {arXiv:1602.03837 [gr-qc]}
  \BibitemShut {NoStop}%
%%CITATION = ARXIV:1602.03837;%%
\bibitem [{\citenamefont {Apostolatos}\ \emph {et~al.}(1994)\citenamefont
  {Apostolatos}, \citenamefont {Cutler}, \citenamefont {Sussman},\ and\
  \citenamefont {Thorne}}]{Apostolatos:1994mx}%
  \BibitemOpen
  \bibfield  {author} {\bibinfo {author} {\bibfnamefont {T.~A.}\ \bibnamefont
  {Apostolatos}}, \bibinfo {author} {\bibfnamefont {C.}~\bibnamefont {Cutler}},
  \bibinfo {author} {\bibfnamefont {G.~J.}\ \bibnamefont {Sussman}}, \ and\
  \bibinfo {author} {\bibfnamefont {K.~S.}\ \bibnamefont {Thorne}},\ }\href
  {\doibase 10.1103/PhysRevD.49.6274} {\bibfield  {journal} {\bibinfo
  {journal} {Phys. Rev.}\ }\textbf {\bibinfo {volume} {D49}},\ \bibinfo {pages}
  {6274} (\bibinfo {year} {1994})}\BibitemShut {NoStop}%
%%CITATION = PHRVA,D49,6274;%%
\bibitem [{\citenamefont {Kidder}(1995)}]{Kidder:1995zr}%
  \BibitemOpen
  \bibfield  {author} {\bibinfo {author} {\bibfnamefont {L.~E.}\ \bibnamefont
  {Kidder}},\ }\href {\doibase 10.1103/PhysRevD.52.821} {\bibfield  {journal}
  {\bibinfo  {journal} {Phys. Rev.}\ }\textbf {\bibinfo {volume} {D52}},\
  \bibinfo {pages} {821} (\bibinfo {year} {1995})},\ \Eprint
  {http://arxiv.org/abs/gr-qc/9506022} {arXiv:gr-qc/9506022 [gr-qc]}
  \BibitemShut {NoStop}%
%%CITATION = GR-QC/9506022;%%
\bibitem [{\citenamefont {Schmidt}\ \emph {et~al.}(2017)\citenamefont
  {Schmidt}, \citenamefont {Harry},\ and\ \citenamefont
  {Pfeiffer}}]{Schmidt:2017btt}%
  \BibitemOpen
  \bibfield  {author} {\bibinfo {author} {\bibfnamefont {P.}~\bibnamefont
  {Schmidt}}, \bibinfo {author} {\bibfnamefont {I.~W.}\ \bibnamefont {Harry}},
  \ and\ \bibinfo {author} {\bibfnamefont {H.~P.}\ \bibnamefont {Pfeiffer}},\
  }\href@noop {} {\bibfield  {journal} {\bibinfo  {journal} {ArXiv e-prints}\ }
  (\bibinfo {year} {2017})},\ \Eprint {http://arxiv.org/abs/1703.01076}
  {arXiv:1703.01076 [gr-qc]} \BibitemShut {NoStop}%
%%CITATION = ARXIV:1703.01076;%%
\bibitem [{\citenamefont {Thorne}(1980)}]{Thorne:1980ru}%
  \BibitemOpen
  \bibfield  {author} {\bibinfo {author} {\bibfnamefont {K.~S.}\ \bibnamefont
  {Thorne}},\ }\href {\doibase 10.1103/RevModPhys.52.299} {\bibfield  {journal}
  {\bibinfo  {journal} {Rev. Mod. Phys.}\ }\textbf {\bibinfo {volume} {52}},\
  \bibinfo {pages} {299} (\bibinfo {year} {1980})}\BibitemShut {NoStop}%
%%CITATION = RMPHA,52,299;%%
\bibitem [{\citenamefont {Goldberg}\ \emph {et~al.}(1967)\citenamefont
  {Goldberg}, \citenamefont {MacFarlane}, \citenamefont {Newman}, \citenamefont
  {Rohrlich},\ and\ \citenamefont {Sudarshan}}]{Goldberg:1966uu}%
  \BibitemOpen
  \bibfield  {author} {\bibinfo {author} {\bibfnamefont {J.~N.}\ \bibnamefont
  {Goldberg}}, \bibinfo {author} {\bibfnamefont {A.~J.}\ \bibnamefont
  {MacFarlane}}, \bibinfo {author} {\bibfnamefont {E.~T.}\ \bibnamefont
  {Newman}}, \bibinfo {author} {\bibfnamefont {F.}~\bibnamefont {Rohrlich}}, \
  and\ \bibinfo {author} {\bibfnamefont {E.~C.~G.}\ \bibnamefont {Sudarshan}},\
  }\href {\doibase 10.1063/1.1705135} {\bibfield  {journal} {\bibinfo
  {journal} {J. Math. Phys.}\ }\textbf {\bibinfo {volume} {8}},\ \bibinfo
  {pages} {2155} (\bibinfo {year} {1967})}\BibitemShut {NoStop}%
%%CITATION = JMAPA,8,2155;%%
\bibitem [{\citenamefont {Wiaux}\ \emph {et~al.}(2007)\citenamefont {Wiaux},
  \citenamefont {Jacques},\ and\ \citenamefont {Vandergheynst}}]{Wiaux:2005fm}%
  \BibitemOpen
  \bibfield  {author} {\bibinfo {author} {\bibfnamefont {Y.}~\bibnamefont
  {Wiaux}}, \bibinfo {author} {\bibfnamefont {L.}~\bibnamefont {Jacques}}, \
  and\ \bibinfo {author} {\bibfnamefont {P.}~\bibnamefont {Vandergheynst}},\
  }\href {\doibase 10.1016/j.jcp.2007.07.005} {\bibfield  {journal} {\bibinfo
  {journal} {J. Comput. Phys.}\ }\textbf {\bibinfo {volume} {226}},\ \bibinfo
  {pages} {2359} (\bibinfo {year} {2007})},\ \Eprint
  {http://arxiv.org/abs/astro-ph/0508514} {arXiv:astro-ph/0508514 [astro-ph]}
  \BibitemShut {NoStop}%
%%CITATION = ASTRO-PH/0508514;%%
\bibitem [{\citenamefont {Boyle}\ \emph {et~al.}(2011)\citenamefont {Boyle},
  \citenamefont {Owen},\ and\ \citenamefont {Pfeiffer}}]{Boyle:2011gg}%
  \BibitemOpen
  \bibfield  {author} {\bibinfo {author} {\bibfnamefont {M.}~\bibnamefont
  {Boyle}}, \bibinfo {author} {\bibfnamefont {R.}~\bibnamefont {Owen}}, \ and\
  \bibinfo {author} {\bibfnamefont {H.~P.}\ \bibnamefont {Pfeiffer}},\ }\href
  {\doibase 10.1103/PhysRevD.84.124011} {\bibfield  {journal} {\bibinfo
  {journal} {Phys.\ Rev.\ D}\ }\textbf {\bibinfo {volume} {84}},\ \bibinfo
  {pages} {124011} (\bibinfo {year} {2011})},\ \Eprint
  {http://arxiv.org/abs/1110.2965} {arXiv:1110.2965 [gr-qc]} \BibitemShut
  {NoStop}%
\bibitem [{\citenamefont {Marsat}\ and\ \citenamefont
  {Baker}(2018{\natexlab{a}})}]{Marsat:2018oam}%
  \BibitemOpen
  \bibfield  {author} {\bibinfo {author} {\bibfnamefont {S.}~\bibnamefont
  {Marsat}}\ and\ \bibinfo {author} {\bibfnamefont {J.~G.}\ \bibnamefont
  {Baker}},\ }\href@noop {} {\bibfield  {journal} {\bibinfo  {journal} {ArXiv
  e-prints}\ } (\bibinfo {year} {2018}{\natexlab{a}})},\ \Eprint
  {http://arxiv.org/abs/1806.10734} {arXiv:1806.10734 [gr-qc]} \BibitemShut
  {NoStop}%
%%CITATION = ARXIV:1806.10734;%%
\bibitem [{\citenamefont {Arun}\ \emph {et~al.}(2009)\citenamefont {Arun},
  \citenamefont {Buonanno}, \citenamefont {Faye},\ and\ \citenamefont
  {Ochsner}}]{Arun:2008kb}%
  \BibitemOpen
  \bibfield  {author} {\bibinfo {author} {\bibfnamefont {K.~G.}\ \bibnamefont
  {Arun}}, \bibinfo {author} {\bibfnamefont {A.}~\bibnamefont {Buonanno}},
  \bibinfo {author} {\bibfnamefont {G.}~\bibnamefont {Faye}}, \ and\ \bibinfo
  {author} {\bibfnamefont {E.}~\bibnamefont {Ochsner}},\ }\href {\doibase
  10.1103/PhysRevD.79.104023, 10.1103/PhysRevD.84.049901} {\bibfield  {journal}
  {\bibinfo  {journal} {Phys. Rev.}\ }\textbf {\bibinfo {volume} {D79}},\
  \bibinfo {pages} {104023} (\bibinfo {year} {2009})},\ \bibinfo {note}
  {[Erratum: Phys. Rev.D84,049901(2011)]},\ \Eprint
  {http://arxiv.org/abs/0810.5336} {arXiv:0810.5336 [gr-qc]} \BibitemShut
  {NoStop}%
%%CITATION = ARXIV:0810.5336;%%
\bibitem [{\citenamefont {Inc.}(2019)}]{Mathematica}%
  \BibitemOpen
  \bibfield  {author} {\bibinfo {author} {\bibfnamefont {W.}~\bibnamefont
  {Inc.}},\ }\href {https://www.wolfram.com/mathematica} {\enquote {\bibinfo
  {title} {Mathematica, {V}ersion 12.0},}\ } (\bibinfo {year} {2019}),\
  \bibinfo {note} {{C}hampaign, IL}\BibitemShut {NoStop}%
\bibitem [{\citenamefont {Finn}\ and\ \citenamefont
  {Chernoff}(1993)}]{Finn:1992xs}%
  \BibitemOpen
  \bibfield  {author} {\bibinfo {author} {\bibfnamefont {L.~S.}\ \bibnamefont
  {Finn}}\ and\ \bibinfo {author} {\bibfnamefont {D.~F.}\ \bibnamefont
  {Chernoff}},\ }\href {\doibase 10.1103/PhysRevD.47.2198} {\bibfield
  {journal} {\bibinfo  {journal} {Phys. Rev.}\ }\textbf {\bibinfo {volume}
  {D47}},\ \bibinfo {pages} {2198} (\bibinfo {year} {1993})},\ \Eprint
  {http://arxiv.org/abs/gr-qc/9301003} {arXiv:gr-qc/9301003 [gr-qc]}
  \BibitemShut {NoStop}%
%%CITATION = GR-QC/9301003;%%
\bibitem [{\citenamefont {Cutler}\ and\ \citenamefont
  {Flanagan}(1994)}]{Cutler:1994ys}%
  \BibitemOpen
  \bibfield  {author} {\bibinfo {author} {\bibfnamefont {C.}~\bibnamefont
  {Cutler}}\ and\ \bibinfo {author} {\bibfnamefont {E.~E.}\ \bibnamefont
  {Flanagan}},\ }\href {\doibase 10.1103/PhysRevD.49.2658} {\bibfield
  {journal} {\bibinfo  {journal} {Phys. Rev.}\ }\textbf {\bibinfo {volume}
  {D49}},\ \bibinfo {pages} {2658} (\bibinfo {year} {1994})},\ \Eprint
  {http://arxiv.org/abs/gr-qc/9402014} {arXiv:gr-qc/9402014 [gr-qc]}
  \BibitemShut {NoStop}%
%%CITATION = GR-QC/9402014;%%
\bibitem [{\citenamefont {Droz}\ \emph {et~al.}(1999)\citenamefont {Droz},
  \citenamefont {Knapp}, \citenamefont {Poisson},\ and\ \citenamefont
  {Owen}}]{Droz:1999qx}%
  \BibitemOpen
  \bibfield  {author} {\bibinfo {author} {\bibfnamefont {S.}~\bibnamefont
  {Droz}}, \bibinfo {author} {\bibfnamefont {D.~J.}\ \bibnamefont {Knapp}},
  \bibinfo {author} {\bibfnamefont {E.}~\bibnamefont {Poisson}}, \ and\
  \bibinfo {author} {\bibfnamefont {B.~J.}\ \bibnamefont {Owen}},\ }\href
  {\doibase 10.1103/PhysRevD.59.124016} {\bibfield  {journal} {\bibinfo
  {journal} {Phys. Rev.}\ }\textbf {\bibinfo {volume} {D59}},\ \bibinfo {pages}
  {124016} (\bibinfo {year} {1999})},\ \Eprint
  {http://arxiv.org/abs/gr-qc/9901076} {arXiv:gr-qc/9901076 [gr-qc]}
  \BibitemShut {NoStop}%
%%CITATION = GR-QC/9901076;%%
\bibitem [{\citenamefont {Abbott}\ \emph
  {et~al.}(2019{\natexlab{a}})\citenamefont {Abbott} \emph
  {et~al.}}]{LIGOScientific:2018mvr}%
  \BibitemOpen
  \bibfield  {author} {\bibinfo {author} {\bibfnamefont {B.~P.}\ \bibnamefont
  {Abbott}} \emph {et~al.} (\bibinfo {collaboration} {LIGO Scientific,
  Virgo}),\ }\href {\doibase 10.1103/PhysRevX.9.031040} {\bibfield  {journal}
  {\bibinfo  {journal} {Phys. Rev.}\ }\textbf {\bibinfo {volume} {X9}},\
  \bibinfo {pages} {031040} (\bibinfo {year} {2019}{\natexlab{a}})},\ \Eprint
  {http://arxiv.org/abs/1811.12907} {arXiv:1811.12907 [astro-ph.HE]}
  \BibitemShut {NoStop}%
%%CITATION = ARXIV:1811.12907;%%
\bibitem [{\citenamefont {Blanchet}\ \emph {et~al.}(2011)\citenamefont
  {Blanchet}, \citenamefont {Buonanno},\ and\ \citenamefont
  {Faye}}]{Blanchet:2011zv}%
  \BibitemOpen
  \bibfield  {author} {\bibinfo {author} {\bibfnamefont {L.}~\bibnamefont
  {Blanchet}}, \bibinfo {author} {\bibfnamefont {A.}~\bibnamefont {Buonanno}},
  \ and\ \bibinfo {author} {\bibfnamefont {G.}~\bibnamefont {Faye}},\ }\href
  {\doibase 10.1103/PhysRevD.84.064041} {\bibfield  {journal} {\bibinfo
  {journal} {Phys. Rev.}\ }\textbf {\bibinfo {volume} {D84}},\ \bibinfo {pages}
  {064041} (\bibinfo {year} {2011})},\ \Eprint {http://arxiv.org/abs/1104.5659}
  {arXiv:1104.5659 [gr-qc]} \BibitemShut {NoStop}%
%%CITATION = ARXIV:1104.5659;%%
\bibitem [{\citenamefont {Marsat}\ \emph {et~al.}(2013)\citenamefont {Marsat},
  \citenamefont {Blanchet}, \citenamefont {Bohe},\ and\ \citenamefont
  {Faye}}]{Marsat:2013wwa}%
  \BibitemOpen
  \bibfield  {author} {\bibinfo {author} {\bibfnamefont {S.}~\bibnamefont
  {Marsat}}, \bibinfo {author} {\bibfnamefont {L.}~\bibnamefont {Blanchet}},
  \bibinfo {author} {\bibfnamefont {A.}~\bibnamefont {Bohe}}, \ and\ \bibinfo
  {author} {\bibfnamefont {G.}~\bibnamefont {Faye}}\ }(\bibinfo {year} {2013})\
  \Eprint {http://arxiv.org/abs/1312.5375} {arXiv:1312.5375 [gr-qc]}
  \BibitemShut {NoStop}%
%%CITATION = ARXIV:1312.5375;%%
\bibitem [{\citenamefont {Estell\'es}\ \emph
  {et~al.}(2020{\natexlab{a}})\citenamefont {Estell\'es}, \citenamefont
  {Ramos-Buades}, \citenamefont {Husa}, \citenamefont {Garc\'ia-Quir\'os},
  \citenamefont {Colleoni}, \citenamefont {Haegel},\ and\ \citenamefont
  {Jaume}}]{phenomt}%
  \BibitemOpen
  \bibfield  {author} {\bibinfo {author} {\bibfnamefont {H.}~\bibnamefont
  {Estell\'es}}, \bibinfo {author} {\bibfnamefont {A.}~\bibnamefont
  {Ramos-Buades}}, \bibinfo {author} {\bibfnamefont {S.}~\bibnamefont {Husa}},
  \bibinfo {author} {\bibfnamefont {C.}~\bibnamefont {Garc\'ia-Quir\'os}},
  \bibinfo {author} {\bibfnamefont {M.}~\bibnamefont {Colleoni}}, \bibinfo
  {author} {\bibfnamefont {L.}~\bibnamefont {Haegel}}, \ and\ \bibinfo {author}
  {\bibfnamefont {R.}~\bibnamefont {Jaume}},\ }\href@noop {} {\enquote
  {\bibinfo {title} {{IMRPhenomTP: A phenomenological time domain model for
  dominant quadrupole gravitational wave signal of coalescing binary black
  holes}},}\ } (\bibinfo {year} {2020}{\natexlab{a}}),\ \Eprint
  {http://arxiv.org/abs/2004.08302} {arXiv:2004.08302 [gr-qc]} \BibitemShut
  {NoStop}%
\bibitem [{\citenamefont {Schmidt}\ \emph {et~al.}(2015)\citenamefont
  {Schmidt}, \citenamefont {Ohme},\ and\ \citenamefont
  {Hannam}}]{Schmidt:2014iyl}%
  \BibitemOpen
  \bibfield  {author} {\bibinfo {author} {\bibfnamefont {P.}~\bibnamefont
  {Schmidt}}, \bibinfo {author} {\bibfnamefont {F.}~\bibnamefont {Ohme}}, \
  and\ \bibinfo {author} {\bibfnamefont {M.}~\bibnamefont {Hannam}},\ }\href
  {\doibase 10.1103/PhysRevD.91.024043} {\bibfield  {journal} {\bibinfo
  {journal} {Phys. Rev.}\ }\textbf {\bibinfo {volume} {D91}},\ \bibinfo {pages}
  {024043} (\bibinfo {year} {2015})},\ \Eprint {http://arxiv.org/abs/1408.1810}
  {arXiv:1408.1810 [gr-qc]} \BibitemShut {NoStop}%
%%CITATION = ARXIV:1408.1810;%%
\bibitem [{\citenamefont {Bender}\ and\ \citenamefont
  {A.}(1999)}]{Bender:1999aa}%
  \BibitemOpen
  \bibfield  {author} {\bibinfo {author} {\bibfnamefont {C.~M.}\ \bibnamefont
  {Bender}}\ and\ \bibinfo {author} {\bibfnamefont {O.~S.}\ \bibnamefont
  {A.}},\ }\href@noop {} {\emph {\bibinfo {title} {{Advanced Mathematical
  Methods of Scientists and Engineers I, Asymptotic Methods and Perturbation
  Theory}}}}\ (\bibinfo  {publisher} {Springer},\ \bibinfo {address} {New
  York},\ \bibinfo {year} {1999})\BibitemShut {NoStop}%
\bibitem [{\citenamefont {Racine}(2008)}]{Racine:2008qv}%
  \BibitemOpen
  \bibfield  {author} {\bibinfo {author} {\bibfnamefont {E.}~\bibnamefont
  {Racine}},\ }\href {\doibase 10.1103/PhysRevD.78.044021} {\bibfield
  {journal} {\bibinfo  {journal} {Phys. Rev.}\ }\textbf {\bibinfo {volume}
  {D78}},\ \bibinfo {pages} {044021} (\bibinfo {year} {2008})},\ \Eprint
  {http://arxiv.org/abs/0803.1820} {arXiv:0803.1820 [gr-qc]} \BibitemShut
  {NoStop}%
%%CITATION = ARXIV:0803.1820;%%
\bibitem [{\citenamefont {Klein}\ \emph {et~al.}(2013)\citenamefont {Klein},
  \citenamefont {Cornish},\ and\ \citenamefont {Yunes}}]{Klein:2013qda}%
  \BibitemOpen
  \bibfield  {author} {\bibinfo {author} {\bibfnamefont {A.}~\bibnamefont
  {Klein}}, \bibinfo {author} {\bibfnamefont {N.}~\bibnamefont {Cornish}}, \
  and\ \bibinfo {author} {\bibfnamefont {N.}~\bibnamefont {Yunes}},\ }\href
  {\doibase 10.1103/PhysRevD.88.124015} {\bibfield  {journal} {\bibinfo
  {journal} {Phys. Rev.}\ }\textbf {\bibinfo {volume} {D88}},\ \bibinfo {pages}
  {124015} (\bibinfo {year} {2013})},\ \Eprint {http://arxiv.org/abs/1305.1932}
  {arXiv:1305.1932 [gr-qc]} \BibitemShut {NoStop}%
%%CITATION = ARXIV:1305.1932;%%
\bibitem [{\citenamefont {Kesden}\ \emph {et~al.}(2015)\citenamefont {Kesden},
  \citenamefont {Gerosa}, \citenamefont {O'Shaughnessy}, \citenamefont
  {Berti},\ and\ \citenamefont {Sperhake}}]{Kesden:2014sla}%
  \BibitemOpen
  \bibfield  {author} {\bibinfo {author} {\bibfnamefont {M.}~\bibnamefont
  {Kesden}}, \bibinfo {author} {\bibfnamefont {D.}~\bibnamefont {Gerosa}},
  \bibinfo {author} {\bibfnamefont {R.}~\bibnamefont {O'Shaughnessy}}, \bibinfo
  {author} {\bibfnamefont {E.}~\bibnamefont {Berti}}, \ and\ \bibinfo {author}
  {\bibfnamefont {U.}~\bibnamefont {Sperhake}},\ }\href {\doibase
  10.1103/PhysRevLett.114.081103} {\bibfield  {journal} {\bibinfo  {journal}
  {Phys. Rev. Lett.}\ }\textbf {\bibinfo {volume} {114}},\ \bibinfo {pages}
  {081103} (\bibinfo {year} {2015})},\ \Eprint {http://arxiv.org/abs/1411.0674}
  {arXiv:1411.0674 [gr-qc]} \BibitemShut {NoStop}%
%%CITATION = ARXIV:1411.0674;%%
\bibitem [{\citenamefont {Chatziioannou}\ \emph {et~al.}(2013)\citenamefont
  {Chatziioannou}, \citenamefont {Klein}, \citenamefont {Yunes},\ and\
  \citenamefont {Cornish}}]{Chatziioannou:2013dza}%
  \BibitemOpen
  \bibfield  {author} {\bibinfo {author} {\bibfnamefont {K.}~\bibnamefont
  {Chatziioannou}}, \bibinfo {author} {\bibfnamefont {A.}~\bibnamefont
  {Klein}}, \bibinfo {author} {\bibfnamefont {N.}~\bibnamefont {Yunes}}, \ and\
  \bibinfo {author} {\bibfnamefont {N.}~\bibnamefont {Cornish}},\ }\href
  {\doibase 10.1103/PhysRevD.88.063011} {\bibfield  {journal} {\bibinfo
  {journal} {Phys. Rev.}\ }\textbf {\bibinfo {volume} {D88}},\ \bibinfo {pages}
  {063011} (\bibinfo {year} {2013})},\ \Eprint {http://arxiv.org/abs/1307.4418}
  {arXiv:1307.4418 [gr-qc]} \BibitemShut {NoStop}%
%%CITATION = ARXIV:1307.4418;%%
\bibitem [{\citenamefont {Klein}\ \emph {et~al.}(2014)\citenamefont {Klein},
  \citenamefont {Cornish},\ and\ \citenamefont {Yunes}}]{Klein:2014bua}%
  \BibitemOpen
  \bibfield  {author} {\bibinfo {author} {\bibfnamefont {A.}~\bibnamefont
  {Klein}}, \bibinfo {author} {\bibfnamefont {N.}~\bibnamefont {Cornish}}, \
  and\ \bibinfo {author} {\bibfnamefont {N.}~\bibnamefont {Yunes}},\ }\href
  {\doibase 10.1103/PhysRevD.90.124029} {\bibfield  {journal} {\bibinfo
  {journal} {Phys. Rev.}\ }\textbf {\bibinfo {volume} {D90}},\ \bibinfo {pages}
  {124029} (\bibinfo {year} {2014})},\ \Eprint {http://arxiv.org/abs/1408.5158}
  {arXiv:1408.5158 [gr-qc]} \BibitemShut {NoStop}%
%%CITATION = ARXIV:1408.5158;%%
\bibitem [{\citenamefont {Cabero}\ \emph {et~al.}(2017)\citenamefont {Cabero},
  \citenamefont {Nielsen}, \citenamefont {Lundgren},\ and\ \citenamefont
  {Capano}}]{Cabero:2016ayq}%
  \BibitemOpen
  \bibfield  {author} {\bibinfo {author} {\bibfnamefont {M.}~\bibnamefont
  {Cabero}}, \bibinfo {author} {\bibfnamefont {A.~B.}\ \bibnamefont {Nielsen}},
  \bibinfo {author} {\bibfnamefont {A.~P.}\ \bibnamefont {Lundgren}}, \ and\
  \bibinfo {author} {\bibfnamefont {C.~D.}\ \bibnamefont {Capano}},\ }\href
  {\doibase 10.1103/PhysRevD.95.064016} {\bibfield  {journal} {\bibinfo
  {journal} {Phys. Rev.}\ }\textbf {\bibinfo {volume} {D95}},\ \bibinfo {pages}
  {064016} (\bibinfo {year} {2017})},\ \Eprint
  {http://arxiv.org/abs/1602.03134} {arXiv:1602.03134 [gr-qc]} \BibitemShut
  {NoStop}%
%%CITATION = ARXIV:1602.03134;%%
\bibitem [{\citenamefont {Blanchet}(2006)}]{Blanchet:2006zz}%
  \BibitemOpen
  \bibfield  {author} {\bibinfo {author} {\bibfnamefont {L.}~\bibnamefont
  {Blanchet}},\ }\href@noop {} {\bibfield  {journal} {\bibinfo  {journal}
  {Living Rev.\ Rel.}\ }\textbf {\bibinfo {volume} {9}},\ \bibinfo {pages} {4}
  (\bibinfo {year} {2006})}\BibitemShut {NoStop}%
\bibitem [{\citenamefont {Le~Tiec}\ \emph {et~al.}(2012)\citenamefont
  {Le~Tiec}, \citenamefont {Blanchet},\ and\ \citenamefont
  {Whiting}}]{LeTiec:2011ab}%
  \BibitemOpen
  \bibfield  {author} {\bibinfo {author} {\bibfnamefont {A.}~\bibnamefont
  {Le~Tiec}}, \bibinfo {author} {\bibfnamefont {L.}~\bibnamefont {Blanchet}}, \
  and\ \bibinfo {author} {\bibfnamefont {B.~F.}\ \bibnamefont {Whiting}},\
  }\href {\doibase 10.1103/PhysRevD.85.064039} {\bibfield  {journal} {\bibinfo
  {journal} {Phys. Rev.}\ }\textbf {\bibinfo {volume} {D85}},\ \bibinfo {pages}
  {064039} (\bibinfo {year} {2012})},\ \Eprint {http://arxiv.org/abs/1111.5378}
  {arXiv:1111.5378 [gr-qc]} \BibitemShut {NoStop}%
%%CITATION = ARXIV:1111.5378;%%
\bibitem [{\citenamefont {Bohe}\ \emph {et~al.}(2013)\citenamefont {Bohe},
  \citenamefont {Marsat}, \citenamefont {Faye},\ and\ \citenamefont
  {Blanchet}}]{Bohe:2012mr}%
  \BibitemOpen
  \bibfield  {author} {\bibinfo {author} {\bibfnamefont {A.}~\bibnamefont
  {Bohe}}, \bibinfo {author} {\bibfnamefont {S.}~\bibnamefont {Marsat}},
  \bibinfo {author} {\bibfnamefont {G.}~\bibnamefont {Faye}}, \ and\ \bibinfo
  {author} {\bibfnamefont {L.}~\bibnamefont {Blanchet}},\ }\href {\doibase
  10.1088/0264-9381/30/7/075017} {\bibfield  {journal} {\bibinfo  {journal}
  {Class. Quant. Grav.}\ }\textbf {\bibinfo {volume} {30}},\ \bibinfo {pages}
  {075017} (\bibinfo {year} {2013})},\ \Eprint {http://arxiv.org/abs/1212.5520}
  {arXiv:1212.5520 [gr-qc]} \BibitemShut {NoStop}%
%%CITATION = ARXIV:1212.5520;%%
\bibitem [{\citenamefont {Damour}\ \emph {et~al.}(2014)\citenamefont {Damour},
  \citenamefont {Jaranowski},\ and\ \citenamefont {Schäfer}}]{Damour:2014jta}%
  \BibitemOpen
  \bibfield  {author} {\bibinfo {author} {\bibfnamefont {T.}~\bibnamefont
  {Damour}}, \bibinfo {author} {\bibfnamefont {P.}~\bibnamefont {Jaranowski}},
  \ and\ \bibinfo {author} {\bibfnamefont {G.}~\bibnamefont {Schäfer}},\
  }\href {\doibase 10.1103/PhysRevD.89.064058} {\bibfield  {journal} {\bibinfo
  {journal} {Phys. Rev.}\ }\textbf {\bibinfo {volume} {D89}},\ \bibinfo {pages}
  {064058} (\bibinfo {year} {2014})},\ \Eprint {http://arxiv.org/abs/1401.4548}
  {arXiv:1401.4548 [gr-qc]} \BibitemShut {NoStop}%
%%CITATION = ARXIV:1401.4548;%%
\bibitem [{\citenamefont {Bernard}\ \emph {et~al.}(2018)\citenamefont
  {Bernard}, \citenamefont {Blanchet}, \citenamefont {Faye},\ and\
  \citenamefont {Marchand}}]{Bernard:2017ktp}%
  \BibitemOpen
  \bibfield  {author} {\bibinfo {author} {\bibfnamefont {L.}~\bibnamefont
  {Bernard}}, \bibinfo {author} {\bibfnamefont {L.}~\bibnamefont {Blanchet}},
  \bibinfo {author} {\bibfnamefont {G.}~\bibnamefont {Faye}}, \ and\ \bibinfo
  {author} {\bibfnamefont {T.}~\bibnamefont {Marchand}},\ }\href {\doibase
  10.1103/PhysRevD.97.044037} {\bibfield  {journal} {\bibinfo  {journal} {Phys.
  Rev.}\ }\textbf {\bibinfo {volume} {D97}},\ \bibinfo {pages} {044037}
  (\bibinfo {year} {2018})},\ \Eprint {http://arxiv.org/abs/1711.00283}
  {arXiv:1711.00283 [gr-qc]} \BibitemShut {NoStop}%
%%CITATION = ARXIV:1711.00283;%%
\bibitem [{\citenamefont {Blanchet}\ and\ \citenamefont
  {Le~Tiec}(2017)}]{Blanchet:2017rcn}%
  \BibitemOpen
  \bibfield  {author} {\bibinfo {author} {\bibfnamefont {L.}~\bibnamefont
  {Blanchet}}\ and\ \bibinfo {author} {\bibfnamefont {A.}~\bibnamefont
  {Le~Tiec}},\ }\href {\doibase 10.1088/1361-6382/aa79d7} {\bibfield  {journal}
  {\bibinfo  {journal} {Class. Quant. Grav.}\ }\textbf {\bibinfo {volume}
  {34}},\ \bibinfo {pages} {164001} (\bibinfo {year} {2017})},\ \Eprint
  {http://arxiv.org/abs/1702.06839} {arXiv:1702.06839 [gr-qc]} \BibitemShut
  {NoStop}%
%%CITATION = ARXIV:1702.06839;%%
\bibitem [{\citenamefont {Marsat}(2015)}]{Marsat:2014xea}%
  \BibitemOpen
  \bibfield  {author} {\bibinfo {author} {\bibfnamefont {S.}~\bibnamefont
  {Marsat}},\ }\href {\doibase 10.1088/0264-9381/32/8/085008} {\bibfield
  {journal} {\bibinfo  {journal} {Class. Quant. Grav.}\ }\textbf {\bibinfo
  {volume} {32}},\ \bibinfo {pages} {085008} (\bibinfo {year} {2015})},\
  \Eprint {http://arxiv.org/abs/1411.4118} {arXiv:1411.4118 [gr-qc]}
  \BibitemShut {NoStop}%
%%CITATION = ARXIV:1411.4118;%%
\bibitem [{\citenamefont {Vines}\ and\ \citenamefont
  {Steinhoff}(2018)}]{Vines:2016qwa}%
  \BibitemOpen
  \bibfield  {author} {\bibinfo {author} {\bibfnamefont {J.}~\bibnamefont
  {Vines}}\ and\ \bibinfo {author} {\bibfnamefont {J.}~\bibnamefont
  {Steinhoff}},\ }\href {\doibase 10.1103/PhysRevD.97.064010} {\bibfield
  {journal} {\bibinfo  {journal} {Phys. Rev.}\ }\textbf {\bibinfo {volume}
  {D97}},\ \bibinfo {pages} {064010} (\bibinfo {year} {2018})},\ \Eprint
  {http://arxiv.org/abs/1606.08832} {arXiv:1606.08832 [gr-qc]} \BibitemShut
  {NoStop}%
%%CITATION = ARXIV:1606.08832;%%
\bibitem [{\citenamefont {Siemonsen}\ \emph {et~al.}(2018)\citenamefont
  {Siemonsen}, \citenamefont {Steinhoff},\ and\ \citenamefont
  {Vines}}]{Siemonsen:2017yux}%
  \BibitemOpen
  \bibfield  {author} {\bibinfo {author} {\bibfnamefont {N.}~\bibnamefont
  {Siemonsen}}, \bibinfo {author} {\bibfnamefont {J.}~\bibnamefont
  {Steinhoff}}, \ and\ \bibinfo {author} {\bibfnamefont {J.}~\bibnamefont
  {Vines}},\ }\href {\doibase 10.1103/PhysRevD.97.124046} {\bibfield  {journal}
  {\bibinfo  {journal} {Phys. Rev.}\ }\textbf {\bibinfo {volume} {D97}},\
  \bibinfo {pages} {124046} (\bibinfo {year} {2018})},\ \Eprint
  {http://arxiv.org/abs/1712.08603} {arXiv:1712.08603 [gr-qc]} \BibitemShut
  {NoStop}%
%%CITATION = ARXIV:1712.08603;%%
\bibitem [{\citenamefont {{Johnson-McDaniel}}\ \emph
  {et~al.}(2016)\citenamefont {{Johnson-McDaniel}}, \citenamefont {Gupta},
  \citenamefont {Ajith}, \citenamefont {Keitel}, \citenamefont {Birnholtz},
  \citenamefont {Ohme},\ and\ \citenamefont {Husa}}]{T1600168}%
  \BibitemOpen
  \bibfield  {author} {\bibinfo {author} {\bibfnamefont {N.~K.}\ \bibnamefont
  {{Johnson-McDaniel}}}, \bibinfo {author} {\bibfnamefont {A.}~\bibnamefont
  {Gupta}}, \bibinfo {author} {\bibfnamefont {P.}~\bibnamefont {Ajith}},
  \bibinfo {author} {\bibfnamefont {D.}~\bibnamefont {Keitel}}, \bibinfo
  {author} {\bibfnamefont {O.}~\bibnamefont {Birnholtz}}, \bibinfo {author}
  {\bibfnamefont {F.}~\bibnamefont {Ohme}}, \ and\ \bibinfo {author}
  {\bibfnamefont {S.}~\bibnamefont {Husa}},\ }\href
  {https://dcc.ligo.org/T1600168/public} {\emph {\bibinfo {title} {Determining
  the final spin of a binary black hole system including in-plane spins: Method
  and checks of accuracy}}},\ \bibinfo {type} {Tech. Rep.}\ \bibinfo {number}
  {{LIGO}-T1600168}\ (\bibinfo  {institution} {LIGO},\ \bibinfo {year} {2016})\
  \bibinfo {note}
  {\url{https://dcc.ligo.org/LIGO-T1600168/public/main}}\BibitemShut {NoStop}%
\bibitem [{\citenamefont {Varma}\ \emph
  {et~al.}(2019{\natexlab{a}})\citenamefont {Varma}, \citenamefont {Gerosa},
  \citenamefont {Stein}, \citenamefont {Hébert},\ and\ \citenamefont
  {Zhang}}]{Varma:2018aht}%
  \BibitemOpen
  \bibfield  {author} {\bibinfo {author} {\bibfnamefont {V.}~\bibnamefont
  {Varma}}, \bibinfo {author} {\bibfnamefont {D.}~\bibnamefont {Gerosa}},
  \bibinfo {author} {\bibfnamefont {L.~C.}\ \bibnamefont {Stein}}, \bibinfo
  {author} {\bibfnamefont {F.}~\bibnamefont {Hébert}}, \ and\ \bibinfo
  {author} {\bibfnamefont {H.}~\bibnamefont {Zhang}},\ }\href {\doibase
  10.1103/PhysRevLett.122.011101} {\bibfield  {journal} {\bibinfo  {journal}
  {Phys. Rev. Lett.}\ }\textbf {\bibinfo {volume} {122}},\ \bibinfo {pages}
  {011101} (\bibinfo {year} {2019}{\natexlab{a}})},\ \Eprint
  {http://arxiv.org/abs/1809.09125} {arXiv:1809.09125 [gr-qc]} \BibitemShut
  {NoStop}%
%%CITATION = ARXIV:1809.09125;%%
\bibitem [{\citenamefont {Rezzolla}\ \emph {et~al.}(2008)\citenamefont
  {Rezzolla}, \citenamefont {Barausse}, \citenamefont {Dorband}, \citenamefont
  {Pollney}, \citenamefont {Reisswig}, \citenamefont {Seiler},\ and\
  \citenamefont {Husa}}]{Rezzolla:2007rz}%
  \BibitemOpen
  \bibfield  {author} {\bibinfo {author} {\bibfnamefont {L.}~\bibnamefont
  {Rezzolla}}, \bibinfo {author} {\bibfnamefont {E.}~\bibnamefont {Barausse}},
  \bibinfo {author} {\bibfnamefont {E.~N.}\ \bibnamefont {Dorband}}, \bibinfo
  {author} {\bibfnamefont {D.}~\bibnamefont {Pollney}}, \bibinfo {author}
  {\bibfnamefont {C.}~\bibnamefont {Reisswig}}, \bibinfo {author}
  {\bibfnamefont {J.}~\bibnamefont {Seiler}}, \ and\ \bibinfo {author}
  {\bibfnamefont {S.}~\bibnamefont {Husa}},\ }\href {\doibase
  10.1103/PhysRevD.78.044002} {\bibfield  {journal} {\bibinfo  {journal} {Phys.
  Rev.}\ }\textbf {\bibinfo {volume} {D78}},\ \bibinfo {pages} {044002}
  (\bibinfo {year} {2008})},\ \Eprint {http://arxiv.org/abs/0712.3541}
  {arXiv:0712.3541 [gr-qc]} \BibitemShut {NoStop}%
%%CITATION = ARXIV:0712.3541;%%
\bibitem [{\citenamefont {Boyle}\ \emph {et~al.}(2019)\citenamefont {Boyle}
  \emph {et~al.}}]{Boyle:2019kee}%
  \BibitemOpen
  \bibfield  {author} {\bibinfo {author} {\bibfnamefont {M.}~\bibnamefont
  {Boyle}} \emph {et~al.},\ }\href {\doibase 10.1088/1361-6382/ab34e2}
  {\bibfield  {journal} {\bibinfo  {journal} {Class. Quant. Grav.}\ }\textbf
  {\bibinfo {volume} {36}},\ \bibinfo {pages} {195006} (\bibinfo {year}
  {2019})},\ \Eprint {http://arxiv.org/abs/1904.04831} {arXiv:1904.04831
  [gr-qc]} \BibitemShut {NoStop}%
%%CITATION = ARXIV:1904.04831;%%
\bibitem [{\citenamefont {{SXS Collaboration}}(2019)}]{SXS:catalog}%
  \BibitemOpen
  \bibfield  {author} {\bibinfo {author} {\bibnamefont {{SXS Collaboration}}},\
  }\href@noop {} {\enquote {\bibinfo {title} {{SXS Gravitational Waveform
  Database}},}\ }\bibinfo {howpublished}
  {\url{https://www.black-holes.org/waveforms}} (\bibinfo {year}
  {2019})\BibitemShut {NoStop}%
\bibitem [{\citenamefont {Varma}\ \emph
  {et~al.}(2019{\natexlab{b}})\citenamefont {Varma}, \citenamefont {Field},
  \citenamefont {Scheel}, \citenamefont {Blackman}, \citenamefont {Gerosa},
  \citenamefont {Stein}, \citenamefont {Kidder},\ and\ \citenamefont
  {Pfeiffer}}]{Varma:2019csw}%
  \BibitemOpen
  \bibfield  {author} {\bibinfo {author} {\bibfnamefont {V.}~\bibnamefont
  {Varma}}, \bibinfo {author} {\bibfnamefont {S.~E.}\ \bibnamefont {Field}},
  \bibinfo {author} {\bibfnamefont {M.~A.}\ \bibnamefont {Scheel}}, \bibinfo
  {author} {\bibfnamefont {J.}~\bibnamefont {Blackman}}, \bibinfo {author}
  {\bibfnamefont {D.}~\bibnamefont {Gerosa}}, \bibinfo {author} {\bibfnamefont
  {L.~C.}\ \bibnamefont {Stein}}, \bibinfo {author} {\bibfnamefont {L.~E.}\
  \bibnamefont {Kidder}}, \ and\ \bibinfo {author} {\bibfnamefont {H.~P.}\
  \bibnamefont {Pfeiffer}},\ }\href {\doibase 10.1103/PhysRevResearch.1.033015}
  {\bibfield  {journal} {\bibinfo  {journal} {Phys. Rev. Research.}\ }\textbf
  {\bibinfo {volume} {1}},\ \bibinfo {pages} {033015} (\bibinfo {year}
  {2019}{\natexlab{b}})},\ \Eprint {http://arxiv.org/abs/1905.09300}
  {arXiv:1905.09300 [gr-qc]} \BibitemShut {NoStop}%
%%CITATION = ARXIV:1905.09300;%%
\bibitem [{\citenamefont {Aasi}\ \emph {et~al.}(2015)\citenamefont {Aasi} \emph
  {et~al.}}]{TheLIGOScientific:2014jea}%
  \BibitemOpen
  \bibfield  {author} {\bibinfo {author} {\bibfnamefont {J.}~\bibnamefont
  {Aasi}} \emph {et~al.} (\bibinfo {collaboration} {LIGO Scientific}),\ }\href
  {\doibase 10.1088/0264-9381/32/7/074001} {\bibfield  {journal} {\bibinfo
  {journal} {Class. Quant. Grav.}\ }\textbf {\bibinfo {volume} {32}},\ \bibinfo
  {pages} {074001} (\bibinfo {year} {2015})},\ \Eprint
  {http://arxiv.org/abs/1411.4547} {arXiv:1411.4547 [gr-qc]} \BibitemShut
  {NoStop}%
%%CITATION = ARXIV:1411.4547;%%
\bibitem [{\citenamefont {Barsotti}\ \emph {et~al.}(2018)\citenamefont
  {Barsotti}, \citenamefont {Fritschel}, \citenamefont {Evans},\ and\
  \citenamefont {Gras}}]{adligopsd}%
  \BibitemOpen
  \bibfield  {author} {\bibinfo {author} {\bibfnamefont {L.}~\bibnamefont
  {Barsotti}}, \bibinfo {author} {\bibfnamefont {P.}~\bibnamefont {Fritschel}},
  \bibinfo {author} {\bibfnamefont {M.}~\bibnamefont {Evans}}, \ and\ \bibinfo
  {author} {\bibfnamefont {S.}~\bibnamefont {Gras}},\ }\href
  {https://dcc.ligo.org/T1800044/public} {\emph {\bibinfo {title} {{The updated
  Advanced LIGO design curve}}}},\ \bibinfo {type} {Tech. Rep.}\ \bibinfo
  {number} {LIGO-T1800044}\ (\bibinfo  {institution} {{LIGO Project}},\
  \bibinfo {year} {2018})\BibitemShut {NoStop}%
\bibitem [{\citenamefont {Harry}\ \emph
  {et~al.}(2016{\natexlab{a}})\citenamefont {Harry}, \citenamefont {Privitera},
  \citenamefont {Boh\'{e}},\ and\ \citenamefont
  {Buonanno}}]{PhysRevD.94.024012}%
  \BibitemOpen
  \bibfield  {author} {\bibinfo {author} {\bibfnamefont {I.}~\bibnamefont
  {Harry}}, \bibinfo {author} {\bibfnamefont {S.}~\bibnamefont {Privitera}},
  \bibinfo {author} {\bibfnamefont {A.}~\bibnamefont {Boh\'{e}}}, \ and\
  \bibinfo {author} {\bibfnamefont {A.}~\bibnamefont {Buonanno}},\ }\href
  {\doibase 10.1103/PhysRevD.94.024012} {\bibfield  {journal} {\bibinfo
  {journal} {Phys. Rev. D}\ }\textbf {\bibinfo {volume} {94}},\ \bibinfo
  {pages} {024012} (\bibinfo {year} {2016}{\natexlab{a}})}\BibitemShut
  {NoStop}%
\bibitem [{\citenamefont {{Virtanen}}\ \emph {et~al.}(2020)\citenamefont
  {{Virtanen}}, \citenamefont {{Gommers}}, \citenamefont {{Oliphant}},
  \citenamefont {{Haberland}}, \citenamefont {{Reddy}}, \citenamefont
  {{Cournapeau}}, \citenamefont {{Burovski}}, \citenamefont {{Peterson}},
  \citenamefont {{Weckesser}}, \citenamefont {{Bright}}, \citenamefont {{van
  der Walt}}, \citenamefont {{Brett}}, \citenamefont {{Wilson}}, \citenamefont
  {{Jarrod Millman}}, \citenamefont {{Mayorov}}, \citenamefont {{Nelson}},
  \citenamefont {{Jones}}, \citenamefont {{Kern}}, \citenamefont {{Larson}},
  \citenamefont {{Carey}}, \citenamefont {{Polat}}, \citenamefont {{Feng}},
  \citenamefont {{Moore}}, \citenamefont {{VanderPlas}}, \citenamefont
  {{Laxalde}}, \citenamefont {{Perktold}}, \citenamefont {{Cimrman}},
  \citenamefont {{Henriksen}}, \citenamefont {{Quintero}}, \citenamefont
  {{Harris}}, \citenamefont {{Archibald}}, \citenamefont {{Ribeiro}},
  \citenamefont {{Pedregosa}}, \citenamefont {{van Mulbregt}},\ and\
  \citenamefont {{Contributors}}}]{2020SciPy-NMeth}%
  \BibitemOpen
  \bibfield  {author} {\bibinfo {author} {\bibfnamefont {P.}~\bibnamefont
  {{Virtanen}}}, \bibinfo {author} {\bibfnamefont {R.}~\bibnamefont
  {{Gommers}}}, \bibinfo {author} {\bibfnamefont {T.~E.}\ \bibnamefont
  {{Oliphant}}}, \bibinfo {author} {\bibfnamefont {M.}~\bibnamefont
  {{Haberland}}}, \bibinfo {author} {\bibfnamefont {T.}~\bibnamefont
  {{Reddy}}}, \bibinfo {author} {\bibfnamefont {D.}~\bibnamefont
  {{Cournapeau}}}, \bibinfo {author} {\bibfnamefont {E.}~\bibnamefont
  {{Burovski}}}, \bibinfo {author} {\bibfnamefont {P.}~\bibnamefont
  {{Peterson}}}, \bibinfo {author} {\bibfnamefont {W.}~\bibnamefont
  {{Weckesser}}}, \bibinfo {author} {\bibfnamefont {J.}~\bibnamefont
  {{Bright}}}, \bibinfo {author} {\bibfnamefont {S.~J.}\ \bibnamefont {{van der
  Walt}}}, \bibinfo {author} {\bibfnamefont {M.}~\bibnamefont {{Brett}}},
  \bibinfo {author} {\bibfnamefont {J.}~\bibnamefont {{Wilson}}}, \bibinfo
  {author} {\bibfnamefont {K.}~\bibnamefont {{Jarrod Millman}}}, \bibinfo
  {author} {\bibfnamefont {N.}~\bibnamefont {{Mayorov}}}, \bibinfo {author}
  {\bibfnamefont {A.~R.~J.}\ \bibnamefont {{Nelson}}}, \bibinfo {author}
  {\bibfnamefont {E.}~\bibnamefont {{Jones}}}, \bibinfo {author} {\bibfnamefont
  {R.}~\bibnamefont {{Kern}}}, \bibinfo {author} {\bibfnamefont
  {E.}~\bibnamefont {{Larson}}}, \bibinfo {author} {\bibfnamefont
  {C.}~\bibnamefont {{Carey}}}, \bibinfo {author} {\bibfnamefont
  {{\.I}.}~\bibnamefont {{Polat}}}, \bibinfo {author} {\bibfnamefont
  {Y.}~\bibnamefont {{Feng}}}, \bibinfo {author} {\bibfnamefont {E.~W.}\
  \bibnamefont {{Moore}}}, \bibinfo {author} {\bibfnamefont {J.}~\bibnamefont
  {{VanderPlas}}}, \bibinfo {author} {\bibfnamefont {D.}~\bibnamefont
  {{Laxalde}}}, \bibinfo {author} {\bibfnamefont {J.}~\bibnamefont
  {{Perktold}}}, \bibinfo {author} {\bibfnamefont {R.}~\bibnamefont
  {{Cimrman}}}, \bibinfo {author} {\bibfnamefont {I.}~\bibnamefont
  {{Henriksen}}}, \bibinfo {author} {\bibfnamefont {E.~A.}\ \bibnamefont
  {{Quintero}}}, \bibinfo {author} {\bibfnamefont {C.~R.}\ \bibnamefont
  {{Harris}}}, \bibinfo {author} {\bibfnamefont {A.~M.}\ \bibnamefont
  {{Archibald}}}, \bibinfo {author} {\bibfnamefont {A.~H.}\ \bibnamefont
  {{Ribeiro}}}, \bibinfo {author} {\bibfnamefont {F.}~\bibnamefont
  {{Pedregosa}}}, \bibinfo {author} {\bibfnamefont {P.}~\bibnamefont {{van
  Mulbregt}}}, \ and\ \bibinfo {author} {\bibnamefont {{Contributors}}},\
  }\href {\doibase https://doi.org/10.1038/s41592-019-0686-2} {\bibfield
  {journal} {\bibinfo  {journal} {Nature Methods}\ }\textbf {\bibinfo {volume}
  {17}},\ \bibinfo {pages} {261} (\bibinfo {year} {2020})}\BibitemShut
  {NoStop}%
\bibitem [{\citenamefont {Harry}\ \emph
  {et~al.}(2016{\natexlab{b}})\citenamefont {Harry}, \citenamefont {Privitera},
  \citenamefont {Boh\'e},\ and\ \citenamefont {Buonanno}}]{Harry:2016aa}%
  \BibitemOpen
  \bibfield  {author} {\bibinfo {author} {\bibfnamefont {I.}~\bibnamefont
  {Harry}}, \bibinfo {author} {\bibfnamefont {S.}~\bibnamefont {Privitera}},
  \bibinfo {author} {\bibfnamefont {A.}~\bibnamefont {Boh\'e}}, \ and\ \bibinfo
  {author} {\bibfnamefont {A.}~\bibnamefont {Buonanno}},\ }\href {\doibase
  10.1103/PhysRevD.94.024012} {\bibfield  {journal} {\bibinfo  {journal} {Phys.
  Rev. D}\ }\textbf {\bibinfo {volume} {94}},\ \bibinfo {pages} {024012}
  (\bibinfo {year} {2016}{\natexlab{b}})}\BibitemShut {NoStop}%
\bibitem [{\citenamefont {Calder\'on~Bustillo}\ \emph
  {et~al.}(2017)\citenamefont {Calder\'on~Bustillo}, \citenamefont {Laguna},\
  and\ \citenamefont {Shoemaker}}]{PhysRevD.95.104038}%
  \BibitemOpen
  \bibfield  {author} {\bibinfo {author} {\bibfnamefont {J.}~\bibnamefont
  {Calder\'on~Bustillo}}, \bibinfo {author} {\bibfnamefont {P.}~\bibnamefont
  {Laguna}}, \ and\ \bibinfo {author} {\bibfnamefont {D.}~\bibnamefont
  {Shoemaker}},\ }\href {\doibase 10.1103/PhysRevD.95.104038} {\bibfield
  {journal} {\bibinfo  {journal} {Phys. Rev. D}\ }\textbf {\bibinfo {volume}
  {95}},\ \bibinfo {pages} {104038} (\bibinfo {year} {2017})}\BibitemShut
  {NoStop}%
\bibitem [{\citenamefont {Varma}\ \emph
  {et~al.}(2019{\natexlab{c}})\citenamefont {Varma}, \citenamefont {Field},
  \citenamefont {Scheel}, \citenamefont {Blackman}, \citenamefont {Kidder},\
  and\ \citenamefont {Pfeiffer}}]{Varma:2018mmi}%
  \BibitemOpen
  \bibfield  {author} {\bibinfo {author} {\bibfnamefont {V.}~\bibnamefont
  {Varma}}, \bibinfo {author} {\bibfnamefont {S.~E.}\ \bibnamefont {Field}},
  \bibinfo {author} {\bibfnamefont {M.~A.}\ \bibnamefont {Scheel}}, \bibinfo
  {author} {\bibfnamefont {J.}~\bibnamefont {Blackman}}, \bibinfo {author}
  {\bibfnamefont {L.~E.}\ \bibnamefont {Kidder}}, \ and\ \bibinfo {author}
  {\bibfnamefont {H.~P.}\ \bibnamefont {Pfeiffer}},\ }\href {\doibase
  10.1103/PhysRevD.99.064045} {\bibfield  {journal} {\bibinfo  {journal} {Phys.
  Rev.}\ }\textbf {\bibinfo {volume} {D99}},\ \bibinfo {pages} {064045}
  (\bibinfo {year} {2019}{\natexlab{c}})},\ \Eprint
  {http://arxiv.org/abs/1812.07865} {arXiv:1812.07865 [gr-qc]} \BibitemShut
  {NoStop}%
%%CITATION = ARXIV:1812.07865;%%
\bibitem [{\citenamefont {Cotesta}\ \emph {et~al.}(2020)\citenamefont
  {Cotesta}, \citenamefont {Marsat},\ and\ \citenamefont
  {Pürrer}}]{Cotesta:2020qhw}%
  \BibitemOpen
  \bibfield  {author} {\bibinfo {author} {\bibfnamefont {R.}~\bibnamefont
  {Cotesta}}, \bibinfo {author} {\bibfnamefont {S.}~\bibnamefont {Marsat}}, \
  and\ \bibinfo {author} {\bibfnamefont {M.}~\bibnamefont {Pürrer}},\
  }\href@noop {} {\bibfield  {journal} {\bibinfo  {journal} {ArXiv e-prints}\ }
  (\bibinfo {year} {2020})},\ \Eprint {http://arxiv.org/abs/2003.12079}
  {arXiv:2003.12079 [gr-qc]} \BibitemShut {NoStop}%
%%CITATION = ARXIV:2003.12079;%%
\bibitem [{\citenamefont {Cotesta}\ \emph {et~al.}(2018)\citenamefont
  {Cotesta}, \citenamefont {Buonanno}, \citenamefont {Bohé}, \citenamefont
  {Taracchini}, \citenamefont {Hinder},\ and\ \citenamefont
  {Ossokine}}]{Cotesta:2018fcv}%
  \BibitemOpen
  \bibfield  {author} {\bibinfo {author} {\bibfnamefont {R.}~\bibnamefont
  {Cotesta}}, \bibinfo {author} {\bibfnamefont {A.}~\bibnamefont {Buonanno}},
  \bibinfo {author} {\bibfnamefont {A.}~\bibnamefont {Bohé}}, \bibinfo
  {author} {\bibfnamefont {A.}~\bibnamefont {Taracchini}}, \bibinfo {author}
  {\bibfnamefont {I.}~\bibnamefont {Hinder}}, \ and\ \bibinfo {author}
  {\bibfnamefont {S.}~\bibnamefont {Ossokine}},\ }\href {\doibase
  10.1103/PhysRevD.98.084028} {\bibfield  {journal} {\bibinfo  {journal} {Phys.
  Rev.}\ }\textbf {\bibinfo {volume} {D98}},\ \bibinfo {pages} {084028}
  (\bibinfo {year} {2018})},\ \Eprint {http://arxiv.org/abs/1803.10701}
  {arXiv:1803.10701 [gr-qc]} \BibitemShut {NoStop}%
%%CITATION = ARXIV:1803.10701;%%
\bibitem [{\citenamefont {Buonanno}\ \emph {et~al.}(2009)\citenamefont
  {Buonanno}, \citenamefont {Iyer}, \citenamefont {Ochsner}, \citenamefont
  {Pan},\ and\ \citenamefont {Sathyaprakash}}]{Buonanno:2009zt}%
  \BibitemOpen
  \bibfield  {author} {\bibinfo {author} {\bibfnamefont {A.}~\bibnamefont
  {Buonanno}}, \bibinfo {author} {\bibfnamefont {B.}~\bibnamefont {Iyer}},
  \bibinfo {author} {\bibfnamefont {E.}~\bibnamefont {Ochsner}}, \bibinfo
  {author} {\bibfnamefont {Y.}~\bibnamefont {Pan}}, \ and\ \bibinfo {author}
  {\bibfnamefont {B.~S.}\ \bibnamefont {Sathyaprakash}},\ }\href {\doibase
  10.1103/PhysRevD.80.084043} {\bibfield  {journal} {\bibinfo  {journal} {Phys.
  Rev.}\ }\textbf {\bibinfo {volume} {D80}},\ \bibinfo {pages} {084043}
  (\bibinfo {year} {2009})},\ \Eprint {http://arxiv.org/abs/0907.0700}
  {arXiv:0907.0700 [gr-qc]} \BibitemShut {NoStop}%
%%CITATION = ARXIV:0907.0700;%%
\bibitem [{\citenamefont {Ossokine}\ \emph {et~al.}(2020)\citenamefont
  {Ossokine} \emph {et~al.}}]{SEOBNRv4PHM:inprep}%
  \BibitemOpen
  \bibfield  {author} {\bibinfo {author} {\bibfnamefont {S.}~\bibnamefont
  {Ossokine}} \emph {et~al.},\ }\href {\doibase 10.1103/PhysRevD.102.044055}
  {\bibfield  {journal} {\bibinfo  {journal} {Phys. Rev. D}\ }\textbf {\bibinfo
  {volume} {102}},\ \bibinfo {pages} {044055} (\bibinfo {year} {2020})},\
  \Eprint {http://arxiv.org/abs/2004.09442} {arXiv:2004.09442 [gr-qc]}
  \BibitemShut {NoStop}%
\bibitem [{\citenamefont {Nagar}\ and\ \citenamefont
  {Rettegno}(2019)}]{Nagar:2018gnk}%
  \BibitemOpen
  \bibfield  {author} {\bibinfo {author} {\bibfnamefont {A.}~\bibnamefont
  {Nagar}}\ and\ \bibinfo {author} {\bibfnamefont {P.}~\bibnamefont
  {Rettegno}},\ }\href {\doibase 10.1103/PhysRevD.99.021501} {\bibfield
  {journal} {\bibinfo  {journal} {Phys. Rev.}\ }\textbf {\bibinfo {volume}
  {D99}},\ \bibinfo {pages} {021501} (\bibinfo {year} {2019})},\ \Eprint
  {http://arxiv.org/abs/1805.03891} {arXiv:1805.03891 [gr-qc]} \BibitemShut
  {NoStop}%
%%CITATION = ARXIV:1805.03891;%%
\bibitem [{\citenamefont {Pürrer}(2016)}]{Purrer:2015tud}%
  \BibitemOpen
  \bibfield  {author} {\bibinfo {author} {\bibfnamefont {M.}~\bibnamefont
  {Pürrer}},\ }\href {\doibase 10.1103/PhysRevD.93.064041} {\bibfield
  {journal} {\bibinfo  {journal} {Phys. Rev.}\ }\textbf {\bibinfo {volume}
  {D93}},\ \bibinfo {pages} {064041} (\bibinfo {year} {2016})},\ \Eprint
  {http://arxiv.org/abs/1512.02248} {arXiv:1512.02248 [gr-qc]} \BibitemShut
  {NoStop}%
%%CITATION = ARXIV:1512.02248;%%
\bibitem [{\citenamefont {Antil}\ \emph {et~al.}(2013)\citenamefont {Antil},
  \citenamefont {Field}, \citenamefont {Herrmann}, \citenamefont {Nochetto},\
  and\ \citenamefont {Tiglio}}]{Antil:2012wf}%
  \BibitemOpen
  \bibfield  {author} {\bibinfo {author} {\bibfnamefont {H.}~\bibnamefont
  {Antil}}, \bibinfo {author} {\bibfnamefont {S.~E.}\ \bibnamefont {Field}},
  \bibinfo {author} {\bibfnamefont {F.}~\bibnamefont {Herrmann}}, \bibinfo
  {author} {\bibfnamefont {R.~H.}\ \bibnamefont {Nochetto}}, \ and\ \bibinfo
  {author} {\bibfnamefont {M.}~\bibnamefont {Tiglio}},\ }\href {\doibase
  10.1007/s10915-013-9722-z} {\bibfield  {journal} {\bibinfo  {journal} {J.
  Sci. Comput.}\ }\textbf {\bibinfo {volume} {57}},\ \bibinfo {pages} {604}
  (\bibinfo {year} {2013})},\ \Eprint {http://arxiv.org/abs/1210.0577}
  {arXiv:1210.0577 [cs.NA]} \BibitemShut {NoStop}%
%%CITATION = ARXIV:1210.0577;%%
\bibitem [{\citenamefont {Canizares}\ \emph {et~al.}(2013)\citenamefont
  {Canizares}, \citenamefont {Field}, \citenamefont {Gair},\ and\ \citenamefont
  {Tiglio}}]{Canizares:2013ywa}%
  \BibitemOpen
  \bibfield  {author} {\bibinfo {author} {\bibfnamefont {P.}~\bibnamefont
  {Canizares}}, \bibinfo {author} {\bibfnamefont {S.~E.}\ \bibnamefont
  {Field}}, \bibinfo {author} {\bibfnamefont {J.~R.}\ \bibnamefont {Gair}}, \
  and\ \bibinfo {author} {\bibfnamefont {M.}~\bibnamefont {Tiglio}},\ }\href
  {\doibase 10.1103/PhysRevD.87.124005} {\bibfield  {journal} {\bibinfo
  {journal} {Phys. Rev.}\ }\textbf {\bibinfo {volume} {D87}},\ \bibinfo {pages}
  {124005} (\bibinfo {year} {2013})},\ \Eprint {http://arxiv.org/abs/1304.0462}
  {arXiv:1304.0462 [gr-qc]} \BibitemShut {NoStop}%
%%CITATION = ARXIV:1304.0462;%%
\bibitem [{\citenamefont {Canizares}\ \emph {et~al.}(2015)\citenamefont
  {Canizares}, \citenamefont {Field}, \citenamefont {Gair}, \citenamefont
  {Raymond}, \citenamefont {Smith},\ and\ \citenamefont
  {Tiglio}}]{Canizares:2014fya}%
  \BibitemOpen
  \bibfield  {author} {\bibinfo {author} {\bibfnamefont {P.}~\bibnamefont
  {Canizares}}, \bibinfo {author} {\bibfnamefont {S.~E.}\ \bibnamefont
  {Field}}, \bibinfo {author} {\bibfnamefont {J.}~\bibnamefont {Gair}},
  \bibinfo {author} {\bibfnamefont {V.}~\bibnamefont {Raymond}}, \bibinfo
  {author} {\bibfnamefont {R.}~\bibnamefont {Smith}}, \ and\ \bibinfo {author}
  {\bibfnamefont {M.}~\bibnamefont {Tiglio}},\ }\href {\doibase
  10.1103/PhysRevLett.114.071104} {\bibfield  {journal} {\bibinfo  {journal}
  {Phys. Rev. Lett.}\ }\textbf {\bibinfo {volume} {114}},\ \bibinfo {pages}
  {071104} (\bibinfo {year} {2015})},\ \Eprint {http://arxiv.org/abs/1404.6284}
  {arXiv:1404.6284 [gr-qc]} \BibitemShut {NoStop}%
%%CITATION = ARXIV:1404.6284;%%
\bibitem [{\citenamefont {Smith}\ \emph {et~al.}(2016)\citenamefont {Smith},
  \citenamefont {Field}, \citenamefont {Blackburn}, \citenamefont {Haster},
  \citenamefont {P\"urrer}, \citenamefont {Raymond},\ and\ \citenamefont
  {Schmidt}}]{Smith:2016qas}%
  \BibitemOpen
  \bibfield  {author} {\bibinfo {author} {\bibfnamefont {R.}~\bibnamefont
  {Smith}}, \bibinfo {author} {\bibfnamefont {S.~E.}\ \bibnamefont {Field}},
  \bibinfo {author} {\bibfnamefont {K.}~\bibnamefont {Blackburn}}, \bibinfo
  {author} {\bibfnamefont {C.-J.}\ \bibnamefont {Haster}}, \bibinfo {author}
  {\bibfnamefont {M.}~\bibnamefont {P\"urrer}}, \bibinfo {author}
  {\bibfnamefont {V.}~\bibnamefont {Raymond}}, \ and\ \bibinfo {author}
  {\bibfnamefont {P.}~\bibnamefont {Schmidt}},\ }\href {\doibase
  10.1103/PhysRevD.94.044031} {\bibfield  {journal} {\bibinfo  {journal} {Phys.
  Rev. D}\ }\textbf {\bibinfo {volume} {94}},\ \bibinfo {pages} {044031}
  (\bibinfo {year} {2016})}\BibitemShut {NoStop}%
\bibitem [{\citenamefont {Veitch}\ and\ \citenamefont
  {Vecchio}(2008)}]{Veitch:2008wd}%
  \BibitemOpen
  \bibfield  {author} {\bibinfo {author} {\bibfnamefont {J.}~\bibnamefont
  {Veitch}}\ and\ \bibinfo {author} {\bibfnamefont {A.}~\bibnamefont
  {Vecchio}},\ }\bibfield  {booktitle} {\emph {\bibinfo {booktitle}
  {{Proceedings, 12th Workshop on Gravitational wave data analysis (GWDAW-12):
  Cambridge, USA, December 13-16, 2007}}},\ }\href {\doibase
  10.1088/0264-9381/25/18/184010} {\bibfield  {journal} {\bibinfo  {journal}
  {Class. Quant. Grav.}\ }\textbf {\bibinfo {volume} {25}},\ \bibinfo {pages}
  {184010} (\bibinfo {year} {2008})},\ \Eprint {http://arxiv.org/abs/0807.4483}
  {arXiv:0807.4483 [gr-qc]} \BibitemShut {NoStop}%
%%CITATION = ARXIV:0807.4483;%%
\bibitem [{\citenamefont {Veitch}\ and\ \citenamefont
  {Vecchio}(2010)}]{Veitch:2009hd}%
  \BibitemOpen
  \bibfield  {author} {\bibinfo {author} {\bibfnamefont {J.}~\bibnamefont
  {Veitch}}\ and\ \bibinfo {author} {\bibfnamefont {A.}~\bibnamefont
  {Vecchio}},\ }\href {\doibase 10.1103/PhysRevD.81.062003} {\bibfield
  {journal} {\bibinfo  {journal} {Phys. Rev.}\ }\textbf {\bibinfo {volume}
  {D81}},\ \bibinfo {pages} {062003} (\bibinfo {year} {2010})},\ \Eprint
  {http://arxiv.org/abs/0911.3820} {arXiv:0911.3820 [astro-ph.CO]} \BibitemShut
  {NoStop}%
%%CITATION = ARXIV:0911.3820;%%
\bibitem [{\citenamefont {{Skilling}}(2004)}]{Skilling:2004ns}%
  \BibitemOpen
  \bibfield  {author} {\bibinfo {author} {\bibfnamefont {J.}~\bibnamefont
  {{Skilling}}},\ }in\ \href {\doibase 10.1063/1.1835238} {\emph {\bibinfo
  {booktitle} {24th International Workshop on Bayesian Inference and Maximum
  Entropy Methods in Science and Engineering}}},\ \bibinfo {series} {AIP Conf.
  Proc.}, Vol.\ \bibinfo {volume} {735},\ \bibinfo {editor} {edited by\
  \bibinfo {editor} {\bibfnamefont {R.}~\bibnamefont {{Fischer}}}, \bibinfo
  {editor} {\bibfnamefont {R.}~\bibnamefont {{Preuss}}}, \ and\ \bibinfo
  {editor} {\bibfnamefont {U.~V.}\ \bibnamefont {{Toussaint}}}}\ (\bibinfo
  {year} {2004})\ pp.\ \bibinfo {pages} {395--405}\BibitemShut {NoStop}%
\bibitem [{\citenamefont {Speagle}(2020)}]{Speagle:2020spe}%
  \BibitemOpen
  \bibfield  {author} {\bibinfo {author} {\bibfnamefont {J.~S.}\ \bibnamefont
  {Speagle}},\ }\href {\doibase 10.1093/mnras/staa278} {\bibfield  {journal}
  {\bibinfo  {journal} {Monthly Notices of the Royal Astronomical Society}\
  }\textbf {\bibinfo {volume} {493}},\ \bibinfo {pages} {3132–3158} (\bibinfo
  {year} {2020})}\BibitemShut {NoStop}%
\bibitem [{\citenamefont {Smith}\ and\ \citenamefont
  {Ashton}(2019)}]{Smith:2019ucc}%
  \BibitemOpen
  \bibfield  {author} {\bibinfo {author} {\bibfnamefont {R.}~\bibnamefont
  {Smith}}\ and\ \bibinfo {author} {\bibfnamefont {G.}~\bibnamefont {Ashton}},\
  }\href@noop {} {\bibfield  {journal} {\bibinfo  {journal} {ArXiv e-prints}\ }
  (\bibinfo {year} {2019})},\ \Eprint {http://arxiv.org/abs/1909.11873}
  {arXiv:1909.11873 [gr-qc]} \BibitemShut {NoStop}%
%%CITATION = ARXIV:1909.11873;%%
\bibitem [{\citenamefont {{LIGO Scientific Collaboration, Virgo
  Collaboration}}(2019{\natexlab{a}})}]{GWOSC}%
  \BibitemOpen
  \bibfield  {author} {\bibinfo {author} {\bibnamefont {{LIGO Scientific
  Collaboration, Virgo Collaboration}}},\ }\href@noop {} {\enquote {\bibinfo
  {title} {{Gravitational Wave Open Science Center}},}\ }\bibinfo
  {howpublished}
  {\href{https://www.gw-openscience.org}{https://www.gw-openscience.org}}
  (\bibinfo {year} {2019}{\natexlab{a}})\BibitemShut {NoStop}%
\bibitem [{\citenamefont {Abbott}\ \emph
  {et~al.}(2019{\natexlab{b}})\citenamefont {Abbott} \emph
  {et~al.}}]{Abbott:2019ebz}%
  \BibitemOpen
  \bibfield  {author} {\bibinfo {author} {\bibfnamefont {R.}~\bibnamefont
  {Abbott}} \emph {et~al.} (\bibinfo {collaboration} {LIGO Scientific,
  Virgo}),\ }\href@noop {} {\bibfield  {journal} {\bibinfo  {journal} {ArXiv
  e-prints}\ } (\bibinfo {year} {2019}{\natexlab{b}})},\ \Eprint
  {http://arxiv.org/abs/1912.11716} {arXiv:1912.11716 [gr-qc]} \BibitemShut
  {NoStop}%
%%CITATION = ARXIV:1912.11716;%%
\bibitem [{\citenamefont {{LIGO Scientific Collaboration, Virgo
  Collaboration}}(2019{\natexlab{b}})}]{gwtc1calib}%
  \BibitemOpen
  \bibfield  {author} {\bibinfo {author} {\bibnamefont {{LIGO Scientific
  Collaboration, Virgo Collaboration}}},\ }\href
  {https://dcc.ligo.org/LIGO-P1900040/public} {\enquote {\bibinfo {title}
  {{Calibration uncertainty envelope release for GWTC-1}},}\ } (\bibinfo {year}
  {2019}{\natexlab{b}}),\ \Eprint
  {http://arxiv.org/abs/https://dcc.ligo.org/LIGO-P1900040/public}
  {https://dcc.ligo.org/LIGO-P1900040/public} \BibitemShut {NoStop}%
\bibitem [{\citenamefont {{LIGO Scientific Collaboration, Virgo
  Collaboration}}(2019{\natexlab{c}})}]{gwtc1psd}%
  \BibitemOpen
  \bibfield  {author} {\bibinfo {author} {\bibnamefont {{LIGO Scientific
  Collaboration, Virgo Collaboration}}},\ }\href
  {https://dcc.ligo.org/LIGO-P1900011/public} {\enquote {\bibinfo {title}
  {{Power Spectral Densities (PSD) release for GWTC-1}},}\ } (\bibinfo {year}
  {2019}{\natexlab{c}}),\ \Eprint
  {http://arxiv.org/abs/https://dcc.ligo.org/LIGO-P1900011/public}
  {https://dcc.ligo.org/LIGO-P1900011/public} \BibitemShut {NoStop}%
\bibitem [{\citenamefont {{Cahillane}}\ \emph {et~al.}(2017)\citenamefont
  {{Cahillane}}, \citenamefont {{Betzwieser}}, \citenamefont {{Brown}},
  \citenamefont {{Goetz}}, \citenamefont {{Hall}}, \citenamefont {{Izumi}},
  \citenamefont {{Kandhasamy}}, \citenamefont {{Karki}}, \citenamefont
  {{Kissel}}, \citenamefont {{Mendell}}, \citenamefont {{Savage}},
  \citenamefont {{Tuyenbayev}}, \citenamefont {{Urban}}, \citenamefont
  {{Viets}}, \citenamefont {{Wade}},\ and\ \citenamefont
  {{Weinstein}}}]{Cahillane:2017vkb}%
  \BibitemOpen
  \bibfield  {author} {\bibinfo {author} {\bibfnamefont {C.}~\bibnamefont
  {{Cahillane}}}, \bibinfo {author} {\bibfnamefont {J.}~\bibnamefont
  {{Betzwieser}}}, \bibinfo {author} {\bibfnamefont {D.~A.}\ \bibnamefont
  {{Brown}}}, \bibinfo {author} {\bibfnamefont {E.}~\bibnamefont {{Goetz}}},
  \bibinfo {author} {\bibfnamefont {E.~D.}\ \bibnamefont {{Hall}}}, \bibinfo
  {author} {\bibfnamefont {K.}~\bibnamefont {{Izumi}}}, \bibinfo {author}
  {\bibfnamefont {S.}~\bibnamefont {{Kandhasamy}}}, \bibinfo {author}
  {\bibfnamefont {S.}~\bibnamefont {{Karki}}}, \bibinfo {author} {\bibfnamefont
  {J.~S.}\ \bibnamefont {{Kissel}}}, \bibinfo {author} {\bibfnamefont
  {G.}~\bibnamefont {{Mendell}}}, \bibinfo {author} {\bibfnamefont {R.~L.}\
  \bibnamefont {{Savage}}}, \bibinfo {author} {\bibfnamefont {D.}~\bibnamefont
  {{Tuyenbayev}}}, \bibinfo {author} {\bibfnamefont {A.}~\bibnamefont
  {{Urban}}}, \bibinfo {author} {\bibfnamefont {A.}~\bibnamefont {{Viets}}},
  \bibinfo {author} {\bibfnamefont {M.}~\bibnamefont {{Wade}}}, \ and\ \bibinfo
  {author} {\bibfnamefont {A.~J.}\ \bibnamefont {{Weinstein}}},\ }\href
  {\doibase 10.1103/PhysRevD.96.102001} {\bibfield  {journal} {\bibinfo
  {journal} {Physical Review D}\ }\textbf {\bibinfo {volume} {96}},\ \bibinfo
  {pages} {102001} (\bibinfo {year} {2017})},\ \Eprint
  {http://arxiv.org/abs/1708.03023} {arXiv:1708.03023 [astro-ph.IM]}
  \BibitemShut {NoStop}%
%%CITATION = ARXIV:1708.03023;%%
\bibitem [{\citenamefont {Viets}\ \emph {et~al.}(2018)\citenamefont {Viets}
  \emph {et~al.}}]{Viets:2017yvy}%
  \BibitemOpen
  \bibfield  {author} {\bibinfo {author} {\bibfnamefont {A.}~\bibnamefont
  {Viets}} \emph {et~al.},\ }\href {\doibase 10.1088/1361-6382/aab658}
  {\bibfield  {journal} {\bibinfo  {journal} {Class. Quant. Grav.}\ }\textbf
  {\bibinfo {volume} {35}},\ \bibinfo {pages} {095015} (\bibinfo {year}
  {2018})},\ \Eprint {http://arxiv.org/abs/1710.09973} {arXiv:1710.09973
  [astro-ph.IM]} \BibitemShut {NoStop}%
%%CITATION = ARXIV:1710.09973;%%
\bibitem [{\citenamefont {{Cahillane}}\ \emph {et~al.}(2018)\citenamefont
  {{Cahillane}}, \citenamefont {{Hulko}}, \citenamefont {{Kissel}} \emph
  {et~al.}}]{Kissel:2018cal}%
  \BibitemOpen
  \bibfield  {author} {\bibinfo {author} {\bibfnamefont {C.}~\bibnamefont
  {{Cahillane}}}, \bibinfo {author} {\bibfnamefont {M.}~\bibnamefont
  {{Hulko}}}, \bibinfo {author} {\bibfnamefont {J.~S.}\ \bibnamefont
  {{Kissel}}},  \emph {et~al.},\ }\href
  {https://dcc.ligo.org/LIGO-G1800319/public} {\emph {\bibinfo {title} {{O2 C02
  Calibration Uncertainty}}}},\ \bibinfo {type} {Tech. Rep.}\ \bibinfo {number}
  {{LIGO-G1800319}}\ (\bibinfo  {institution} {{LIGO Scientific
  Collaboration}},\ \bibinfo {year} {2018})\BibitemShut {NoStop}%
\bibitem [{\citenamefont {Chatziioannou}\ \emph {et~al.}(2019)\citenamefont
  {Chatziioannou} \emph {et~al.}}]{Chatziioannou:2019dsz}%
  \BibitemOpen
  \bibfield  {author} {\bibinfo {author} {\bibfnamefont {K.}~\bibnamefont
  {Chatziioannou}} \emph {et~al.},\ }\href {\doibase
  10.1103/PhysRevD.100.104015} {\bibfield  {journal} {\bibinfo  {journal}
  {Phys. Rev.}\ }\textbf {\bibinfo {volume} {D100}},\ \bibinfo {pages} {104015}
  (\bibinfo {year} {2019})},\ \Eprint {http://arxiv.org/abs/1903.06742}
  {arXiv:1903.06742 [gr-qc]} \BibitemShut {NoStop}%
%%CITATION = ARXIV:1903.06742;%%
\bibitem [{\citenamefont {Payne}\ \emph {et~al.}(2019)\citenamefont {Payne},
  \citenamefont {Talbot},\ and\ \citenamefont {Thrane}}]{Payne:2019wmy}%
  \BibitemOpen
  \bibfield  {author} {\bibinfo {author} {\bibfnamefont {E.}~\bibnamefont
  {Payne}}, \bibinfo {author} {\bibfnamefont {C.}~\bibnamefont {Talbot}}, \
  and\ \bibinfo {author} {\bibfnamefont {E.}~\bibnamefont {Thrane}},\ }\href
  {\doibase 10.1103/PhysRevD.100.123017} {\bibfield  {journal} {\bibinfo
  {journal} {Phys. Rev.}\ }\textbf {\bibinfo {volume} {D100}},\ \bibinfo
  {pages} {123017} (\bibinfo {year} {2019})},\ \Eprint
  {http://arxiv.org/abs/1905.05477} {arXiv:1905.05477 [astro-ph.IM]}
  \BibitemShut {NoStop}%
%%CITATION = ARXIV:1905.05477;%%
\bibitem [{\citenamefont {Abbott}\ \emph
  {et~al.}(2020{\natexlab{a}})\citenamefont {Abbott} \emph
  {et~al.}}]{Abbott:2020niy}%
  \BibitemOpen
  \bibfield  {author} {\bibinfo {author} {\bibfnamefont {R.}~\bibnamefont
  {Abbott}} \emph {et~al.} (\bibinfo {collaboration} {LIGO Scientific,
  Virgo}),\ }\href@noop {} {\  (\bibinfo {year} {2020}{\natexlab{a}})},\
  \Eprint {http://arxiv.org/abs/2010.14527} {arXiv:2010.14527 [gr-qc]}
  \BibitemShut {NoStop}%
\bibitem [{\citenamefont {Abbott}\ \emph
  {et~al.}(2020{\natexlab{b}})\citenamefont {Abbott} \emph
  {et~al.}}]{Abbott:2020gyp}%
  \BibitemOpen
  \bibfield  {author} {\bibinfo {author} {\bibfnamefont {R.}~\bibnamefont
  {Abbott}} \emph {et~al.} (\bibinfo {collaboration} {LIGO Scientific,
  Virgo}),\ }\href@noop {} {\  (\bibinfo {year} {2020}{\natexlab{b}})},\
  \Eprint {http://arxiv.org/abs/2010.14533} {arXiv:2010.14533 [astro-ph.HE]}
  \BibitemShut {NoStop}%
\bibitem [{\citenamefont {{LIGO-Virgo Collaboration}}(2016)}]{PSDs_NRinj}%
  \BibitemOpen
  \bibfield  {author} {\bibinfo {author} {\bibnamefont {{LIGO-Virgo
  Collaboration}}},\ }\href {https://dcc.ligo.org/LIGO-T2000012/public} {\emph
  {\bibinfo {title} {Noise curves used for Simulations in the update of the
  Observing Scenarios Paper}}},\ \bibinfo {type} {Tech. Rep.}\ \bibinfo
  {number} {{LIGO}-T2000012-v1}\ (\bibinfo  {institution} {{LIGO} Project},\
  \bibinfo {year} {2016})\BibitemShut {NoStop}%
\bibitem [{\citenamefont {Marsat}\ and\ \citenamefont
  {Baker}(2018{\natexlab{b}})}]{marsat2018fourierdomain}%
  \BibitemOpen
  \bibfield  {author} {\bibinfo {author} {\bibfnamefont {S.}~\bibnamefont
  {Marsat}}\ and\ \bibinfo {author} {\bibfnamefont {J.~G.}\ \bibnamefont
  {Baker}},\ }\href@noop {} {\enquote {\bibinfo {title} {Fourier-domain
  modulations and delays of gravitational-wave signals},}\ } (\bibinfo {year}
  {2018}{\natexlab{b}}),\ \Eprint {http://arxiv.org/abs/1806.10734}
  {arXiv:1806.10734 [gr-qc]} \BibitemShut {NoStop}%
\bibitem [{\citenamefont {Hamilton}(2020)}]{eleanor}%
  \BibitemOpen
  \bibfield  {author} {\bibinfo {author} {\bibfnamefont {E.}~\bibnamefont
  {Hamilton}},\ }\emph {\bibinfo {title} {{Towards a precision description of
  precessing black-hole-binaries}}},\ \href@noop {} {\bibinfo {type} {{PhD}
  dissertation}},\ \bibinfo  {school} {Cardiff University} (\bibinfo {year}
  {2020})\BibitemShut {NoStop}%
\bibitem [{\citenamefont {Estell\'es}\ \emph
  {et~al.}(2020{\natexlab{b}})\citenamefont {Estell\'es}, \citenamefont {Husa},
  \citenamefont {Colleoni}, \citenamefont {Keitel}, \citenamefont
  {Mateu-Lucena}, \citenamefont {Garc\'ia-Quir\'os}, \citenamefont
  {Ramos-Buades},\ and\ \citenamefont {Borchers}}]{phenomthm}%
  \BibitemOpen
  \bibfield  {author} {\bibinfo {author} {\bibfnamefont {H.}~\bibnamefont
  {Estell\'es}}, \bibinfo {author} {\bibfnamefont {S.}~\bibnamefont {Husa}},
  \bibinfo {author} {\bibfnamefont {M.}~\bibnamefont {Colleoni}}, \bibinfo
  {author} {\bibfnamefont {D.}~\bibnamefont {Keitel}}, \bibinfo {author}
  {\bibfnamefont {M.}~\bibnamefont {Mateu-Lucena}}, \bibinfo {author}
  {\bibfnamefont {C.}~\bibnamefont {Garc\'ia-Quir\'os}}, \bibinfo {author}
  {\bibfnamefont {A.}~\bibnamefont {Ramos-Buades}}, \ and\ \bibinfo {author}
  {\bibfnamefont {A.}~\bibnamefont {Borchers}},\ }\href@noop {} {\enquote
  {\bibinfo {title} {Time domain phenomenological model of gravitational wave
  subdominant harmonics for quasi-circular non-precessing binary black hole
  coalescences},}\ } (\bibinfo {year} {2020}{\natexlab{b}}),\ \Eprint
  {http://arxiv.org/abs/2012.11923} {arXiv:2012.11923 [gr-qc]} \BibitemShut
  {NoStop}%
\bibitem [{\citenamefont {Colleoni}\ \emph {et~al.}(2021)\citenamefont
  {Colleoni}, \citenamefont {Mateu-Lucena}, \citenamefont {Estell\'es},
  \citenamefont {Garc\'ia-Quir\'os}, \citenamefont {Keitel}, \citenamefont
  {Pratten}, \citenamefont {Ramos-Buades},\ and\ \citenamefont
  {Husa}}]{GW190412_uib}%
  \BibitemOpen
  \bibfield  {author} {\bibinfo {author} {\bibfnamefont {M.}~\bibnamefont
  {Colleoni}}, \bibinfo {author} {\bibfnamefont {M.}~\bibnamefont
  {Mateu-Lucena}}, \bibinfo {author} {\bibfnamefont {H.}~\bibnamefont
  {Estell\'es}}, \bibinfo {author} {\bibfnamefont {C.}~\bibnamefont
  {Garc\'ia-Quir\'os}}, \bibinfo {author} {\bibfnamefont {D.}~\bibnamefont
  {Keitel}}, \bibinfo {author} {\bibfnamefont {G.}~\bibnamefont {Pratten}},
  \bibinfo {author} {\bibfnamefont {A.}~\bibnamefont {Ramos-Buades}}, \ and\
  \bibinfo {author} {\bibfnamefont {S.}~\bibnamefont {Husa}},\ }\href {\doibase
  10.1103/PhysRevD.103.024029} {\bibfield  {journal} {\bibinfo  {journal}
  {Phys. Rev. D}\ }\textbf {\bibinfo {volume} {103}},\ \bibinfo {pages}
  {024029} (\bibinfo {year} {2021})}\BibitemShut {NoStop}%
\bibitem [{\citenamefont {Dietrich}\ \emph {et~al.}(2017)\citenamefont
  {Dietrich}, \citenamefont {Bernuzzi},\ and\ \citenamefont
  {Tichy}}]{Dietrich:2017aum}%
  \BibitemOpen
  \bibfield  {author} {\bibinfo {author} {\bibfnamefont {T.}~\bibnamefont
  {Dietrich}}, \bibinfo {author} {\bibfnamefont {S.}~\bibnamefont {Bernuzzi}},
  \ and\ \bibinfo {author} {\bibfnamefont {W.}~\bibnamefont {Tichy}},\ }\href
  {\doibase 10.1103/PhysRevD.96.121501} {\bibfield  {journal} {\bibinfo
  {journal} {Phys. Rev. D}\ }\textbf {\bibinfo {volume} {96}},\ \bibinfo
  {pages} {121501} (\bibinfo {year} {2017})}\BibitemShut {NoStop}%
\bibitem [{\citenamefont {Dietrich}\ \emph {et~al.}(2019)\citenamefont
  {Dietrich}, \citenamefont {Samajdar}, \citenamefont {Khan}, \citenamefont
  {Johnson-McDaniel}, \citenamefont {Dudi},\ and\ \citenamefont
  {Tichy}}]{Dietrich:2019kaq}%
  \BibitemOpen
  \bibfield  {author} {\bibinfo {author} {\bibfnamefont {T.}~\bibnamefont
  {Dietrich}}, \bibinfo {author} {\bibfnamefont {A.}~\bibnamefont {Samajdar}},
  \bibinfo {author} {\bibfnamefont {S.}~\bibnamefont {Khan}}, \bibinfo {author}
  {\bibfnamefont {N.~K.}\ \bibnamefont {Johnson-McDaniel}}, \bibinfo {author}
  {\bibfnamefont {R.}~\bibnamefont {Dudi}}, \ and\ \bibinfo {author}
  {\bibfnamefont {W.}~\bibnamefont {Tichy}},\ }\href {\doibase
  10.1103/PhysRevD.100.044003} {\bibfield  {journal} {\bibinfo  {journal}
  {Phys. Rev.}\ }\textbf {\bibinfo {volume} {D100}},\ \bibinfo {pages} {044003}
  (\bibinfo {year} {2019})},\ \Eprint {http://arxiv.org/abs/1905.06011}
  {arXiv:1905.06011 [gr-qc]} \BibitemShut {NoStop}%
%%CITATION = ARXIV:1905.06011;%%
\bibitem [{\citenamefont {Thompson}\ \emph {et~al.}(2020)\citenamefont
  {Thompson}, \citenamefont {Fauchon-Jones}, \citenamefont {Khan},
  \citenamefont {Nitoglia}, \citenamefont {Pannarale}, \citenamefont
  {Dietrich},\ and\ \citenamefont {Hannam}}]{Thompson:2020nei}%
  \BibitemOpen
  \bibfield  {author} {\bibinfo {author} {\bibfnamefont {J.~E.}\ \bibnamefont
  {Thompson}}, \bibinfo {author} {\bibfnamefont {E.}~\bibnamefont
  {Fauchon-Jones}}, \bibinfo {author} {\bibfnamefont {S.}~\bibnamefont {Khan}},
  \bibinfo {author} {\bibfnamefont {E.}~\bibnamefont {Nitoglia}}, \bibinfo
  {author} {\bibfnamefont {F.}~\bibnamefont {Pannarale}}, \bibinfo {author}
  {\bibfnamefont {T.}~\bibnamefont {Dietrich}}, \ and\ \bibinfo {author}
  {\bibfnamefont {M.}~\bibnamefont {Hannam}},\ }\href@noop {} {\bibfield
  {journal} {\bibinfo  {journal} {ArXiv e-prints}\ } (\bibinfo {year}
  {2020})},\ \Eprint {http://arxiv.org/abs/2002.08383} {arXiv:2002.08383
  [gr-qc]} \BibitemShut {NoStop}%
%%CITATION = ARXIV:2002.08383;%%
\bibitem [{\citenamefont {Brown}\ \emph {et~al.}(2007)\citenamefont {Brown},
  \citenamefont {Fairhurst}, \citenamefont {Krishnan}, \citenamefont {Mercer},
  \citenamefont {Kopparapu}, \citenamefont {Santamaria},\ and\ \citenamefont
  {Whelan}}]{Brown:2007jx}%
  \BibitemOpen
  \bibfield  {author} {\bibinfo {author} {\bibfnamefont {D.}~\bibnamefont
  {Brown}}, \bibinfo {author} {\bibfnamefont {S.}~\bibnamefont {Fairhurst}},
  \bibinfo {author} {\bibfnamefont {B.}~\bibnamefont {Krishnan}}, \bibinfo
  {author} {\bibfnamefont {R.}~\bibnamefont {Mercer}}, \bibinfo {author}
  {\bibfnamefont {R.}~\bibnamefont {Kopparapu}}, \bibinfo {author}
  {\bibfnamefont {L.}~\bibnamefont {Santamaria}}, \ and\ \bibinfo {author}
  {\bibfnamefont {J.}~\bibnamefont {Whelan}},\ }\href@noop {} {\bibfield
  {journal} {\bibinfo  {journal} {ArXiv e-prints}\ } (\bibinfo {year}
  {2007})},\ \Eprint {http://arxiv.org/abs/0709.0093} {arXiv:0709.0093 [gr-qc]}
  \BibitemShut {NoStop}%
\bibitem [{\citenamefont {Blanchet}\ \emph {et~al.}(2008)\citenamefont
  {Blanchet}, \citenamefont {Faye}, \citenamefont {Iyer},\ and\ \citenamefont
  {Sinha}}]{Blanchet:2008je}%
  \BibitemOpen
  \bibfield  {author} {\bibinfo {author} {\bibfnamefont {L.}~\bibnamefont
  {Blanchet}}, \bibinfo {author} {\bibfnamefont {G.}~\bibnamefont {Faye}},
  \bibinfo {author} {\bibfnamefont {B.~R.}\ \bibnamefont {Iyer}}, \ and\
  \bibinfo {author} {\bibfnamefont {S.}~\bibnamefont {Sinha}},\ }\href
  {\doibase 10.1088/0264-9381/25/16/165003, 10.1088/0264-9381/29/23/239501}
  {\bibfield  {journal} {\bibinfo  {journal} {Class. Quant. Grav.}\ }\textbf
  {\bibinfo {volume} {25}},\ \bibinfo {pages} {165003} (\bibinfo {year}
  {2008})},\ \bibinfo {note} {[Erratum: Class. Quant. Grav.29,239501(2012)]},\
  \Eprint {http://arxiv.org/abs/0802.1249} {arXiv:0802.1249 [gr-qc]}
  \BibitemShut {NoStop}%
%%CITATION = ARXIV:0802.1249;%%
\bibitem [{\citenamefont {{LIGO Scientific
  Collaboration}}(2020{\natexlab{b}})}]{laldoxygen}%
  \BibitemOpen
  \bibfield  {author} {\bibinfo {author} {\bibnamefont {{LIGO Scientific
  Collaboration}}},\ }\href@noop {} {\enquote {\bibinfo {title}
  {{LALSimulation: Inspiral Simulation Packages}},}\ }\bibinfo {howpublished}
  {\url{https://lscsoft.docs.ligo.org/lalsuite/lalsimulation/group__lalsimulation__inspiral.html}}
  (\bibinfo {year} {2020}{\natexlab{b}})\BibitemShut {NoStop}%
\bibitem [{\citenamefont {Beazley}(1996)}]{Beazley:1996swig}%
  \BibitemOpen
  \bibfield  {author} {\bibinfo {author} {\bibfnamefont {D.~M.}\ \bibnamefont
  {Beazley}},\ }in\ \href {http://dl.acm.org/citation.cfm?id=1267498.1267513}
  {\emph {\bibinfo {booktitle} {Proc. 4th Conf. USENIX Tcl/Tk Workshop}}}\
  (\bibinfo  {publisher} {USENIX Association},\ \bibinfo {address} {Berkeley,
  CA, USA},\ \bibinfo {year} {1996})\ pp.\ \bibinfo {pages}
  {15--15}\BibitemShut {NoStop}%
\bibitem [{\citenamefont {Beazley}\ \emph {et~al.}(2019)\citenamefont {Beazley}
  \emph {et~al.}}]{SWIG:code}%
  \BibitemOpen
  \bibfield  {author} {\bibinfo {author} {\bibfnamefont {D.~M.}\ \bibnamefont
  {Beazley}} \emph {et~al.},\ }\href@noop {} {\enquote {\bibinfo {title} {{SWIG
  - Simplified Wrapper and Interface Generator}},}\ }\bibinfo {howpublished}
  {\url{https://www.swig.org}} (\bibinfo {year} {2019})\BibitemShut {NoStop}%
\bibitem [{\citenamefont {Wette}(2020)}]{Wette:2020swig}%
  \BibitemOpen
  \bibfield  {author} {\bibinfo {author} {\bibfnamefont {K.}~\bibnamefont
  {Wette}},\ }\href {\doibase https://doi.org/10.1016/j.softx.2020.100634}
  {\bibfield  {journal} {\bibinfo  {journal} {SoftwareX}\ }\textbf {\bibinfo
  {volume} {12}},\ \bibinfo {pages} {100634} (\bibinfo {year} {2020})},\
  \bibinfo {note} {https://dcc.ligo.org/P2000094/}\BibitemShut {NoStop}%
\end{thebibliography}%

% ~~~~~~~~~~ Archived Code ~~~~~~~~~~ %

%\graphicspath{%
%  {Plots/}%
%  % More directories are added in braces, without commas between
%}

%\begin{figure}
%  \includegraphics[scale=0.5]{./Plots/PlaceholderPlot}
%  \caption{ \label{fig:Placeholder} %
%    \CapName{Caption title.} 
%    Example of figure taken from first PhenomD paper.
%  }
%\end{figure}

% ~~~~~~~~~~ END DOCUMENT ~~~~~~~~~~ %

\end{document}